\documentclass[11pt,a4paper,notitlepage]{article}

\usepackage[all]{xy}

\usepackage{cmap} 
\usepackage{refcount}
\usepackage[pdfpagemode=None,
pagebackref=true,
bookmarksopen=false,
hyperindex=true,
colorlinks=true,
linkcolor=blue,
citecolor=blue,
pdftitle={Local quantum geometric dynamics},
pdfauthor={Ryszard Pawel Kostecki}]{hyperref}

\usepackage[LGR,OT2,T1]{fontenc}

\usepackage[polutonikogreek,russian,english]{babel}
\usepackage[utf8]{inputenc}
\usepackage[russian,english]{babel}

				\usepackage{rpk}

\usepackage{makeidx}
\usepackage{idxlayout}
\usepackage{sectsty}
\usepackage{bibspacing}
\usepackage{setspace} 

\usepackage[toc]{multitoc}            
\usepackage{tocloft}
\usepackage[pdftex,final]{graphicx}
\usepackage{mathabx} 

\renewcommand*{\backref}[1]{}
\renewcommand*{\backrefalt}[4]{%
  \ifcase #1 %
    \relax
  \or
    $\uparrow$~#2.
  \else
    $\uparrow$~#2.
  \fi%
}

\newcommand{\cyrrm}[1]{\mbox{\fontencoding{OT2}\fontfamily{wncyr}\selectfont#1}} 
\makeatletter
\renewcommand\tableofcontents{%
    \@starttoc{toc}%
}
\makeatother

\addto\captionsenglish{
  \renewcommand{\contentsname}%
    {\ }%
}

\makeindex
\begin{document}
\thispagestyle{empty}
\setlength\cftparskip{0pt}
\setlength\cftbeforesecskip{1pt}
\setlength{\cftbeforetoctitleskip}{-20pt}
\setlength{\cftaftertoctitleskip}{-50pt}
\thispagestyle{empty}
\selectlanguage{english}
\setcounter{tocdepth}{2}
\setcounter{secnumdepth}{4}
\thispagestyle{empty}
\begin{center}
{\vspace{-30pt}}
{\Large\textbf{Local quantum information dynamics}}{\tiny\\\ \\\ \\}
{\large Ryszard Pawe{\l} Kostecki{\tiny\\\ \\}{\small Perimeter Institute for Theoretical Physics\\\textit{31 Caroline Street North, Waterloo, Ontario N2L 2Y5, Canada}{\vskip 0.15cm}{\small Basic Research Community for Physics\\\textit{Mariannenstra\ss{}e 89, 04315 Leipzig, Germany}}%
}
}
{\vskip 0.15cm}
{\small\texttt{ryszard.kostecki@fuw.edu.pl}}
{\small\\\ \\} 
{June 10, 2016}
\end{center}
\newif\ifvargaugecompile
\vargaugecompiletrue
\begin{spacing}{0.85}
\begin{center}{\small\textbf{Abstract}}
\end{center}{\small
\noindent  This paper is intended to: 1) show how the local smooth geometry of spaces of normal quantum states over W$^*$-algebras (generalised spaces of density matrices) may be used to substantially enrich the description of quantum dynamics in the algebraic and path integral approaches; 2) provide a framework for construction of quantum information theories beyond quantum mechanics, such that quantum mechanical linearity holds only locally, while the nonlocal multi-user dynamics exhibits some similarity with general relativity. In the algebraic setting, we propose a method of incorporating nonlinear Poisson and relative entropic local dynamics, as well as local gauge and local source structures, into an effective description of local temporal evolution of quantum states by using fibrewise perturbations of liouvilleans in the fibre bundle of Hilbert spaces over the quantum state manifold. We apply this method to construct an algebraic generalisation of Savvidou's action operator. In the path integral setting, motivated by the Savvidou--Anastopoulous analysis of the role of K\"{a}hler space geometry in the Isham--Linden quantum histories, we propose to incorporate local geometry by means of a generalisation of the Daubechies--Klauder coherent state phase space propagator formula. Finally, we discuss the role of Br\`{e}gman relative entropy in the Jaynes--Mitchell--Favretti renormalisation scheme. Using these tools we show that: 1) the propagation of quantum particles (in Wigner's sense) can be naturally explained as a free fall along the trajectories locally minimising the quantum relative entropy; 2) the contribution of particular trajectories to the global path integral is weighted by the local quantum entropic prior, measuring user's lack of information; 3) the presence of nonlinear quantum control variables results in the change of the curvature of the global quantum state space; 4) the behaviour of zero-point energy under renormalisation of local entropic dynamics is maintained by local redefinition of information mass (prior), which encodes the curvature change. We conclude this work with a proposal of a new framework for nonequilibrium quantum statistical mechanics based on quantum Orlicz spaces, quantum Br\`{e}gman distances and Banach Lie algebras.}

\begin{center}
\textit{dedicated to Professor Stanis{\l}aw Woronowicz on the occasion of  his 75$^{\,th}$ birthday}
\end{center}
\end{spacing}
{\vskip 0.2cm}
{\small 
\addtocontents{toc}{\protect\begin{multicols}{2}}
\tableofcontents
\addtocontents{toc}{\protect\end{multicols}}
\vspace{20pt}
\addtocontents{toc}{\vspace{-30pt}}
}
\section{Introduction}
\begin{flushright}
{\scriptsize \textit{%
Like a brush in hand\\
To paint a picture\\
Of what we would like to see}\\
C.M. Schuldiner}
\end{flushright}
In this Section we will motivate the approach of this paper by discussing how it is related to the structures and problems of K\"{a}hler geometric and C$^*$-algebraic approaches to quantum kinematics, as well as hamiltonian and path integral approaches to quantum dynamics. Our goal is to overview and explain the merits of the constructions that we pursue on the subsequent pages. The in-depth analysis of an approach to foundations of quantum theory\footnote{We distinguish between the quantum mechanics, understood as a framework defined in \cite{vonNeumann:1932:grundlagen}, and the quantum theory, understood as a (currently unknown) framework that should be capable of providing mathematically exact nonperturbative foundations for relativistic quantum field theory and nonequilibrium quantum statistical mechanics.} that we briefly postulate and use here is provided in another paper \cite{Kostecki:2016:microhauptwerk}.
\subsection{Quantum information geometric foundations: global postulates\label{qig.foundations.intro.section}}
The basic kinematic postulates of our framework are:
\begin{itemize}
\item \textbf{Postulate 1:} \textit{The underlying spaces of inquiry are W$^*$-algebras $\N$, instead of sample spaces and  Hilbert spaces.}
\item \textbf{Postulate 2:} \textit{The state spaces of quantified knowledge are sets $\M(\N)$ of positive normal states on W$^*$-algebras, instead of probabilistic models and spaces of density matrices.}
\item \textbf{Postulate 3:} \textit{The observables are arbitrary real valued functions $f:\M(\N)\ra\RR$ of normal states, instead of arbitrary real valued functions on sample spaces and self-adjoint operators.}
\item \textbf{Postulate 4:} \textit{Given an experimental configuration space $\Theta$, a method of model construction defining a mapping $\Theta\ni\theta\mapsto\phi(\theta)\in\M(\N)$, and a choice of the set of functions $\tilde{f}:\Theta\ra\RR$ that one is interested in, the set of observables that are relevant for a given problem is given by $\{f:\M(\N)\ra\RR\mid f\circ\theta=\tilde{f}\}$.}
\end{itemize}

While C$^*$-algebras\footnote{A C$^*$-algebra is defined as an algebra $\C$ over $\CC$, equipped with: an operation $^*:\C\ra\C$ that algebraically abstracts the properties of a complex conjugation of complex numbers, and a norm $\n{\cdot}:\C\ra\RR^+$ that turns $\C$ into a Banach space satisfying $\n{x^*x}=\n{x}^2$ $\forall x\in\C$. W$^*$-algebras are characterised as such C$^*$-algebras for which there exists a Banach space, denoted $\C_\star$ and called a (Banach) predual, satisfying $(\C_\star)^\star\iso\C$. Given a Banach space $X$ over $\CC$, the operation $^\star:X\ra X^\star$ forms a Banach space of all continuous linear $\CC$-valued functionals, equipped with a supremum norm.} generalise algebras of complex continuous functions on compact topological spaces, W$^*$-algebras generalise $L_\infty$ spaces over localisable boolean algebras (or, equivalently, localisable measure spaces), so the problem of choice between them depends not only on the mathematical properties of a specific application but also on the general interpretation assigned to the quantum theoretic formalism. From the mathematical perspective of general integration theory (including integration on noncommutative W$^*$-algebras, on nonassociative Jordan algebras, and on spectral convex sets), it is completely natural to extend considerations from the sets of density matrices to the subsets of positive parts of Banach preduals of arbitrary W$^*$-algebras. The set of all (not necessarily normalised) density matrices is characterised as a positive part $\ell_1(\BH)^+$ of a noncommutative $\ell_1$ space associated to the W$^*$-algebra $\BH$ of bounded linear operators on a Hilbert space $\H$ (more precisely, $\ell_1(\BH)\iso\schatten_1(\H)$, where $\schatten_1(\H)$ is a Banach space of all trace class operators, equipped with a trace norm), while the Banach preduals $\N_\star$ of arbitrary W$^*$-algebras $\N$ are characterised as a noncommutative $L_1$ spaces associated to these algebras, and this association is functorial with respect to W$^*$-isomorphisms \cite{Falcone:Takesaki:2001}. Hence, if one considers quantum mechanics and probability theory as two instances of a more general class of information theories, then the use of W$^*$-algebras $\N$ and elements of $L_1(\N)^+\iso\N_\star^+$ for the mathematical foundations of quantum mechanics is a \textit{natural} and \textit{exact} generalisation of mathematical formulation of probability theory in terms of a normalised measure theory (as proposed by Steinhaus \cite{Steinhaus:1923} and developed by Kolmogorov \cite{Kolmogorov:1933}). This leads us: 1) to chose the framework of W$^*$-algebras $\N$ and normal positive states $\omega\in\N^+_\star$ instead the framework of C$^*$-algebras $\C$ and positive states $\omega\in\C^{\star+}$; 2) to consider subsets $\M(\N)\subseteq\N^+_\star$ as spaces of quantum states that are setting the arena for quantum kinematics. We view them as natural generalisation of probabilistic models $\M(\X,\mho(\X),\tmu)\subseteq L_1(\X,\mho(\X),\tmu)^+$. The observables in our framework are defined as \textit{arbitrary} real valued functions $f:\M(\N)\ra\RR$. The observables in the standard sense of quantum mechanics are precisely determined as an affine subset of the observables in our sense:
\begin{equation}
		f_x(\phi):=\phi(x)\;\;\forall x\in\N^\sa\;\forall\phi\in\N^+_\star.
\end{equation}
Postulates 1-4 do not yet determine how we are going to define specific kinematic and dynamic models. Following Chencov's geometric approach to foundations of statistical inference theory \cite{Chencov:1964,Chencov:1965,Chencov:1966,Chencov:1967,Chencov:1968,Chencov:1969,Chencov:1972,Chencov:1978,Chencov:1980,Chencov:1987,Morozova:Chencov:1991} (developed later by Amari and others \cite{Amari:1968,Amari:1980,Amari:1982,Amari:1985,Amari:1987,Amari:Nagaoka:1993}), Jaynes information theoretic approach to foundations of statistical mechanics \cite{Jaynes:1957,Jaynes:1957:2,Jaynes:1963,Jaynes:Scalapino:1963,Jaynes:1965,Jaynes:1967,Jaynes:1979:where:do:we:stand,Jaynes:1985:macropred,Jaynes:1986,Jaynes:1993} (developed later in the geometric direction by Ingarden and others \cite{Ingarden:1963,Ingarden:Kossakowski:1975,Ingarden:1974:1976,Ingarden:1978,ISSK:1979,Ingarden:1981,Ingarden:Janyszek:1982,IJKK:1982,IKO:1997}), and the program of smooth geometrisation of quantum mechanics \cite{Strocchi:1966,Marsden:1968:generalized,Kibble:1979,Cirelli:Lanzavecchia:Mania:1983,Simon:1983,Cirelli:Lanzavecchia:1984,Heslot:1985,Page:1987,Aharonov:Anandan:1987,Anandan:Aharonov:1990,Cirelli:Mania:Pizzocchero:1990,Gibbons:1992,Cirelli:Mania:Pizzocchero:1994,Hughston:1995,Hughston:1996,Schilling:1996,Field:1996,Field:1997,Brody:Hughston:1998,Ashtekar:Schilling:1999,Cirelli:Gatti:Mania:1999,Brody:Hughston:2001,Chruscinski:Jamiolkowski:2004,Bengtsson:Zyczkowski:2006}, we propose the following

\begin{itemize}
\item \textbf{Postulate 5:} \textit{The construction of specific models of kinematics and dynamics is based upon the geometric structures over state spaces, provided by quantum relative entropies and Banach Lie algebras, instead of scalar product of Hilbert space.}
\end{itemize}

In order to investigate the possible generalisations of quantum mechanical prescriptions of dynamics we want first to understand the geometric structures on state spaces. For a given Hilbert space $\H$ equipped with a scalar product $\s{\cdot,\cdot}_\H:\H\times\H\ra\CC$, the projection $\PP:\H\setminus\{0\}\ni\xi\mapsto\frac{\xi}{\n{\xi}_\H}\in\PP\H$ induces a manifold structure on $\PP\H$, with tangent spaces given by $\H$, riemannian metric $\gbold^\H$ and symplectic form $\wbold^\H$ determined uniquely by a decomposition \cite{Abraham:Marsden:1978}
\begin{equation}
	\s{\cdot,\cdot}_\H=\frac{1}{2}\gbold^\H(\cdot,\cdot)+\ii\frac{1}{2}\wbold^\H(\cdot,\cdot),
\end{equation}
and complex structure defined by $\s{\cdot,\jbold^\H(\cdot)}_\H=\ii\s{\cdot,\cdot}_\H$. The tangent bundle of $\H$ over $\PP\H$ can be viewed as a principal $U(1)$-bundle equipped with a $U(1)$ connection $1$-form $\nabla^\PP:\H\times\H\ni(\xi,\zeta)\mapsto\ii\s{\xi,\ddd\zeta}_\H\in\CC$. In our case, the lack of a unique global Hilbert space implies the lack of unique specification of riemannian metric and symplectic form derived from a scalar product. In order to facilitate a well-defined generalisation of riemannian and symplectic structure, the sets $\M(\N)$ can be equipped with two different smooth real Banach manifold structures. On one hand, an information manifold structure on the set $\N^+_{\star0}$ of all faithful (strictly positive) elements of $\N^+_\star$ is constructed by a choice of a quantum relative entropy functional on $\N^+_\star$, and has tangent spaces defined as noncommutative Orlicz spaces that provide suitable convergence behaviour of neighbourhoods of states as measured by constrained relative entropy minimisation (see Section \ref{relat.entr.MCP.bundle}). On the other hand, a Banach Lie--Poisson manifold structure is constructed by the choice of a Banach Lie algebra (such as the real Banach Lie algebra $\N^\sa$ of all self-adjoint elements of a W$^*$-algebra $\N$), and has tangent spaces defined as copies of a predual of this algebra (see Section \ref{W.star.BLP.section}). Given the information distance manifold structure on $\M(\N)$, the first two nonzero orders of Taylor expansion for a wide class of relative entropy functionals $D(\cdot,\cdot)$ on $\M(\N)$ give rise to the torsion-free smooth Norden--Sen geometries $(\M(\N),\gbold^D,\nabla^D,(\nabla^D)^\nsdual)$. The Norden--Sen geometry is defined by the condition \eqref{geometric.duality}, which directly generalises the condition characterising the Levi--Civita affine connection $\nabla^\gbold$ in riemannian geometries $(\M,\gbold)$. The Fubini--Study riemannian metric $\gbold^\H$ becomes recovered as an extension to the boundary of pure states for a wide class of geometries $(\M(\N),\gbold^D)$ \cite{Petz:Sudar:1999}. On the other hand, given the choice of a Banach--Lie algebra $\B$ such that $\M(\N)$ is a real Banach submanifold\footnote{For a discussion why an injective immersion of $\M(\N)$ into $\B_\star$ is not sufficient, see \cite{Bona:2004}.} of $\B_\star$ with $\T_\phi\B_\star\iso\B_\star$ $\forall \phi\in\B_\star$ (or, more generally, $\ad^\star_x(\B_\star)\subseteq\B_\star$ $\forall x\in\B$), the coadjoint action of $\B$ on $\B_\star$ induces a Poisson structure on Fr\'{e}chet smooth real valued functions on $\M(\N)$. If $\B$ is a Lie algebra of a group $G$, then the Banach Lie--Poisson manifolds $\M(\N)$ are symplectic if they are coadjoint orbits of $G$. In particular, if $\B\iso\N^\sa$, then $G\iso\N^\uni$ (a group of all unitary elements of $\N$), and $\B_\star\iso\N^\sa_\star:=\{\phi\in\N_\star\mid\phi(x^*)=\phi(x)^*\}\iso(\N^\sa)_\star$. The example of such case is given by the orbit of a group of unitary operators of density matrices with finite fixed number $n\in\NN$ of nonzero eigenvalues \cite{Bona:2000}. For $n=1$, one recovers precisely the symplectic structure $\wbold^\H$ on a projective Hilbert space $\PP\H$ that is induced from $\s{\cdot,\cdot}_\H$ on $\H$ \cite{Bona:1991}.

The resulting description of quantum geometry can be summarised as follows:
\begin{enumerate}
\item[(a)] In the general setting of W$^*$-algebras $\N$ and quantum states defined as elements of $\N_\star^+$ the smooth manifold structure required to implement the infinite-dimensional quantum generalisation of Poisson geometry does not match with the smooth manifold structure required to implement the infinite-dimensional quantum generalisation of riemannian geometry (this issue is discussed in more details in \cite{Kostecki:2016:microhauptwerk}). As a result,  the geometry of a Hilbert space $\H$ (consisting of pure states), formulated in terms of  riemannian metric $\gbold^\H$ and symplectic structure $\wbold^\H$ defined over the same real Hilbert smooth manifold $\PP\H$, becomes generalised to the geometry of spaces $\M(\N)$ equipped with two different real Banach smooth manifold structures: of a Banach Lie--Poisson manifold and of a quantum information geometric (relative entropic) manifold. The former is determined by the choice of a Banach Lie algebra $\B$, and in the special cases reduces to a symplectic space. The latter is determined by the choice of a relative entropy functional $D:\M(\N)\times\M(\N)\ra[0,\infty]$, and in the special cases reduces to a torsion free Norden--Sen manifold, or just a riemannian space.
\item[(b)] Apart from the above two alternative systems of tangent, cotangent, and higher jet bundles, one can also introduce a bundle of Hilbert spaces over $\M(\N)\subseteq\N^+_\star$, that can serve as as an ambient framework to represent different geometrical objects. A natural candidate is a Gel'fand--Na\u{\i}mark--Segal bundle $\H\M(\N)$ of Hilbert spaces \cite{Odzijewicz:Slizewska:2011}. Because the bundle $\H\M(\N)$ is defined by states of an underlying manifold (as opposed to a projection $\PP$), there is no $\nabla^{\PP}$ connection $U(1)$ action in fibers. However, for $\M(\N)\subseteq\N^+_{\star0}$, each fibre of $\H\M(\N)$ bundle is equipped with a strongly continuous $U(1)$ action of a modular (Tomita--Takesaki) automorphism $\sigma^\omega:\RR\ni t\mapsto\Ad(\Delta_\omega^{\ii t})\in\Aut(\pi_\omega(\N))$. We will study the role of this automorphism in Section \ref{algebraic.action.operator.section}, showing that it resembles some interesting similarities with $\nabla^{\PP}$ when considered over a trajectory $\RR\ra\M(\N)\subseteq\N^+_{\star0}$. Yet, in Section \ref{quant.info.geom.section} we will show that, for any $\M(\N)\subseteq\N^+_{\star0}$, the bundle $\H\M(\N)$ carries a natural connection structure, with parallel transport given by the standard unitary transition operators $V_{\phi,\omega}:=J_{\phi,\phi}J_{\phi,\omega}$, where $J_{\phi,\omega}$ is a relative modular conjugation between two faithful normal GNS representations, while $J_{\phi,\phi}$ is a Tomita modular conjugation. We observe that this connection is Levi-Civita with respect to the Wigner--Yanase riemannian metric, and its local geodesic free fall corresponds to constrained minimisation of the Hilbert space norm (projective measurement).
\item[(c)] Because two manifold structures mentioned in (a) do not coincide, there seems to be no obvious candidate for a `complex structure' on a general quantum model $\M(\N)$. However, we notice that the use of complex Hilbert spaces of dimension $n$ instead of real Hilbert spaces of dimension $2n$ is crucially associated with the requirement that the generators of unitary transformations of these spaces should be represented by self-adjoint operators (identified with observables). Observing further that the standard quantum mechanical method of defining relevant observables proceeds by representations of Lie algebra $\glie$ of Lie group $G$ on the complex Hilbert space, we introduce the structure of a principal $G$-bundle $E$ over $\M(\N)$, equipped with a family of representations of an associated Lie algebra in the fibers of the GNS Hilbert space bundle $\H\M(\N)$. This leads to a rise of a $\glie$-valued connection form on $E$, and a fiberwise family of its representations on the fibers of $\H\M(\N)$. This structure exhibits some interesting features of the  relationship between Berry connection, complex structure of a Hilbert space, and construction of observables by means of representations of Lie algebras. The bundle structures of $E$ and $\H\M(\N)$ do not require $\M(\N)$ to be a Banach smooth manifold, but they require it to be a (Hausdorff and paracompact) topological space. We do not assume any \textit{a priori} relationship between the geometries of local causal dynamics, as provided by $\B$, and the transformations used to identify locally the relevant observables, as provided by $\glie$, because we consider it to be a model-dependent feature.
\item[(d)] Using the $\glie$-valued connection $\nabla^\glie$, we can define a kinematic propagator (in Prugove\v{c}ki sense) of the particles (in Wigner sense) as the holonomy of $\nabla^\glie$ along the geodesics of $\nabla^D$ (or $(\nabla^D)^\nsdual$ or $\nabla^{\gbold^D}$) affine connection. 
\item[(e)] In Section \ref{MCP.bundle.section} we propose a construction of what we call the Morozova--Chencov--Petz Hilbert space bundle $\H^{\hhh}\M(\N)$. Its purpose is to use the riemannian metrics $\gbold^D$, associated with a class $D_\fff$ of quantum distances, in order to determine Hilbert spaces and corresponding representations that are different from the GNS construction, and include the information about the local riemannian geometry of a model (this is inspired by the ideas of \cite{Sukhanov:Rudoy:2005,Rudoy:Sukhanov:2005,Rudoy:2009}). In principle, it can be used as an alternative to $\H\M(\N)$ (especially for the purpose of the tasks (c)-(d)), however it is essentially harder to deal with mathematically. We consider this as an indication that the natural framework for a simultaneous implementation of geometric and algebraic tools used in this paper (entropic Norden--Sen geometries, Banach Lie--Poisson structure, perturbations of liouvilleans) are Banach dual pairs of noncommutative Orlicz spaces, used as tangent and cotangent spaces. However, the technical implementation of this idea requires one to develop a standard construction of Orlicz spaces for any (countably finite) W$^*$-algebra, and for a large class of quantum relative entropies, as well as to develop the theory of perturbations of ``Orlicz liouvilleans'' (by an analogy to $L_p$-liouvilleans of Jak\v{s}i\'{c} and Pillet \cite{JOPP:2012}). While these tasks are beyond the scope of this paper, it can be understood as a testing ground for them, so we will discuss more precisely the above ideas in Sections \ref{local.infogeometry.histories.intro.section} and \ref{effective.local.quantum.dynamics.section}.
\end{enumerate} 

With all these tools on the stage, we can approach the problem of construction of quantum dynamics.

\begin{itemize}
\item \textbf{Postulate 6:} \textit{The elementary form of causal dynamics is given by a Poisson flow generated by a smooth observable on a state space, as opposed to a unitary evolution on a Hilbert space. The elementary form of inferential dynamics is given by a (nonlinear) constrained quantum relative entropy minimisation, as opposed to a (linear) projection in a Hilbert space or L\"{u}ders' rule.}
\end{itemize}

Unitary evolution is a special case of a hamiltonian evolution on the self-adjoint part of a predual of a W$^*$-algebra, understood as a BLP space \cite{Bona:1991,Bona:2000,Odzijewicz:Ratiu:2003}. Derivation of L\"{u}ders' rules (selective and nonselective) as a special case of constrained quantum relative entropy minimisation\footnote{Technically, $\PPP^D_\Q(\phi)$ is a nonlinear projection in a positive cone of a noncommutative $L_1(\N)$ space, constrained by the data represented as a convex closed set $\Q_\Upsilon$ in a noncommutative Orlicz space $L_\Upsilon(\N)$, associated by means of a nonlinear bijective mapping $\ell:L_1(\N)\ra L_\Upsilon(\N)$, so that $\Q=\ell^{-1}(\Q_\Upsilon)$ forms a codomain of a projection. While in general we are interested in entropic projections for a wide class of quantum distance functionals, the results cited in this paragraph were proven for the most recognised example: the Umegaki--Araki noncommutative generalisation of the Kullback--Leibler distance.} (in short: entropic projection) for $D$ given by the Umegaki--Araki distance \eqref{Araki.std.rep.distance} was provided in \cite{Hellmann:Kaminski:Kostecki:2016} and \cite{Kostecki:2014}.\footnote{Derivation of the nonselective L\"{u}ders' rule from minimisation of the Hilbert--Schmidt norm distance was provided much earlier in \cite{Herbut:1969}, while derivations of some special cases of the selective L\"{u}ders' rule using some symmetric quantum information distances were obtained in \cite{Hadjisavvas:1978,Hadjisavvas:1981,Dieks:Veltkamp:1983,Raggio:1984}. See \cite{Kostecki:2014} for more discussion.} Hence, the basic dynamical setting of quantum mechanical evolution of quantum states, which is a unitary evolution followed by a projective measurement can be completely recovered as a special case of a \textit{causal inference instrument} given by the map\footnote{The composition in the reverse order can also be studied.}
\begin{equation}
	\M_1(\N)\ni\phi\mapsto\PPP^D_\Q\circ w^{\B,h}_t(\phi)\in\M_2(\N),
\label{global.hamiltonian.entropic.map}
\end{equation}
where 
\begin{equation}
	\PPP^D_\Q(\psi):=\arginff{\psi\in\Q}{D(\omega,\psi)}
\end{equation}
is an entropic projection onto constrained set $\Q$ for a quantum distance functional $D$, while $w^{\B,h}_t$ is a Poisson flow generated by a Banach Lie algebra $\B$ and a hamiltonian function $h$ for a time range $[0,t]$, corresponding uniquely to an integral line of a vector field
\begin{equation}
	\XXX_h(\phi):=-\ad^\star_{\DF_\phi h}(\phi)\;\;\forall\phi\in\B_\star,
\end{equation}
where $\DF_\phi h$ is a Fr\'{e}chet derivative of $h$ at $\phi$, implementing the differential form $\ddd h(\phi)$.

In \cite{MunkNielsen:2015} it was shown that partial trace is also a special case of entropic projection. Hence, all linear completely positive maps can be considered as a special case of the maps formed by composition of tensor products, Poisson flows, and entropic projections \cite{Kostecki:MunkNielsen:2016,Kostecki:2016:microhauptwerk}. This way the kinematic and dynamic setting of quantum mechanics and nonrelativistic quantum information theory becomes fully recovered as a special case of the framework specified by Postulates 1-6 above.

This leads us to:
\begin{itemize}
\item \textbf{Open problem:} \textit{Reconstruct (some aspects of) dynamics of quantum field theory and nonequilibrium quantum statistical mechanics using the framework specified by Postulates 1-6.}
\end{itemize}
Unlike in quantum mechanics, the dynamics of both these theories is sensitive to local geometric features of the kinematic structure of a quantum model. In consequence, we are lead to investigate how, and to what extent, the above geometric structures, and the corresponding nonlinear dynamical maps, can give account of the local structures in QFT and NQSM.

\subsection{Local quantum information dynamics in algebraic and path integral approaches\label{local.quantum.dynamics.intro.section}}

The main questions underlying the constructions carried on in this paper are: what if the correct setting for bridging the gap between algebraic and path integral setting is to use quantum state spaces and their geometry:
\begin{enumerate}
\item[1)] to define local evolution in the algebraic approach by means of \textit{locally} defined and perturbed liouvilleans?
\item[2)] \textit{instead} of using phase space geometry in the continuous time coherent state path integral ``quantisation''? 
\item[3)] to describe renormalisation as purely information theoretic procedure?
\end{enumerate}
The discussion below is intended to show that the proposal of the geometric framework for locally quantum information theories that we provide in Section \ref{locally.quantum.information.relativity.section} is remarkably grounded in the insights coming from \textit{three} very distinct theoretical frameworks: 1) a geometric extension of algebraic hamiltonian dynamics with local gauge and local sources by means of local perturbation of liouvilleans; 2) a generalisation of the Daubechies--Klauder path integration to an algebraic setting; 3) a geometric Jaynes--Mitchell--Favretti renormalisation, applicable in nonequilibrium quantum statistical mechanics.

The standard Haag--Kastler \cite{Haag:Kastler:1964,Haag:1992} setting of an algebraic approach to quantum field theory is widely considered as being unable to incorporate the local gauge principle\footnote{E.g. \cytat{The Lagrangean and the Feynman path integral are at present indispensable tools in the characterization and study of a specific theory. Together with the local gauge principle they pose questions which in the algebraic approach are not understood and should be tackled.} \cite{Haag:1992}.} (the global gauge principle has been partially incorporated to an algebraic approach by means of the Doplicher--Haag--Roberts theory \cite{Doplicher:Haag:Roberts:1969:1,Doplicher:Haag:Roberts:1969:2}). Apart from renormalisation techniques, this principle is a fundamental tool in the construction of the predictively sound models in quantum field theory. Its maintenance by the lagrangean/path-integral approach leads to an abandonment of the algebraic approach by most of the practitioners of QFT, but this is provided at the price of replacing mathematically well-defined objects by symbolic (and usually perturbative) techniques of calculations. This makes QFT very different from quantum mechanics, because the latter facilitates construction of predictive models without the expense of mathematical precision. In this paper we intend to show that the consideration of geometric structures on the spaces $\N^+_\star$ of normal states over W$^*$-algebras $\N$, as well as construction of effective local dynamics by means of local perturbations of liouvilleans, may provide an extension of an algebraic approach capable of dealing with the local gauge principle and the use of `external sources', typical in the path integral formalism. Our point of departure from the Haag--Kastler perspective is to consider locality in the Prugove\v{c}ki sense \cite{Prugovecki:1992}, being associated with the fiber at a given point of an underlying space (so the Lorentz or Poincar\'{e} covariance condition is to be applied fiberwisely), as opposed to a neighbourhood of this point (e.g. as given by the special relativistic diamond). This allows to think of local GNS Hilbert space associated to the manifold of quantum states as (a model of, or as a container of) local tangent space, corresponding to a local quantum mechanical description provided by a single (quantum bayesian) user. Further extension from the bundle of (self-dual) Hilbert spaces to the  bundle of dual pairs of noncommutative Orlicz spaces is necessary to allow the geometry of local quantum inference to be governed by a wide class of conventions of estimation, beyond the Wigner--Yanase riemannian metric (so that the quantum nonequilibrium thermodynamic Kubo--Mori--Bogolyubov metric, as well as the quantum estimation theoretic Helstrom--Uhlmann--Bures metric can be included on the equivalent mathematical footing). The construction of an underlying manifold structure commits to the principle of equivalence between local inference by means of constrained maximisation of relative entropy, and the free fall along the geodesics of the dually flat local geometry, derived from this entropy.

On the path integral side, our approach is directly inspired by the Daubechies--Klauder \cite{Daubechies:Klauder:1985,Klauder:1988,Klauder:1995,Klauder:Maraner:1997,Bodmann:2003,Klauder:2011} continuous-time regularised coherent state phase space approach to path integration, and the closely related Ana\-sto\-pou\-los--Savvidou \cite{Anastopoulos:2001,Anastopoulos:Savvidou:2002,Anastopoulos:Savvidou:2003} analysis of decoherence functional in the Isham--Linden quantum histories approach. Both have shown that one can think of the underlying dynamical objects of respective theories (path integrals and decoherence functionals) as consisting of the hamiltonian evolution perturbed by the geometric structures on the space of quantum states. Our goal here is to follow Klauder's remark \cytat{If there is ever any hope to define path integrals rigorously as path integrals over a set of paths (functions of time), then it is \textit{essential} to give up the notion that the paths involved are sharp value paths and replace that with another interpretation of which the expectation value paths is a completely satisfactory example} \cite{Klauder:2003} by extending these approaches to the state spaces over W$^*$-algebras, and relating them with the local liouvillean approach to algebraic dynamics. The virtue of the Daubechies--Klauder approach is that it provides a mathematically rigourous continuous time regularisation of the functional integral in a way that gives the same results under arbitrary canonical transformations of the underlying phase space. This is not true for most of other approaches to quantisation, not only path integral based. The restrictions on the class of hamiltonians that are allowed in order to maintain this procedure to be well defined are quite mild. The key ingredient of this approach is introducing a regulariser that represents a riemannian metric on the phase space (corresponding to a Fubini--Study metric on coherent quantum states), and determines a pinned Wiener measure of the Brownian process on the phase space.


The heuristic ideas underlying our treatment of quantum dynamics are: 1) Quantum kinematics and dynamics should be defined without recourse to classical models and their quantisation; 2) Classical (phase space, but also space-time) geometry should be considered as a locally emergent feature describing particular properties of the multi-agent information processing systems and not as a fundamental structure (background); 3) Local spatial (phase space or space-time) variables should arise as epistemic (e.g. operational) parameters of information states (see e.g. \cite{Rodriguez:1999,Duch:Kostecki:2011}). This heuristics is in a disagreement with the perspectives of the orthodox algebraic and path integral approaches (yet, there are some exceptions\footnote{In particular, \cytat{the interpretation of the formal path integral (...) in terms of paths $\mathrm{p}(t)$ and $\mathrm{q}(t)$ for which the meaning of the variables is that of expectation values is far more acceptable than the one in which the meaning is that of both sharp position and sharp momentum (eigen)values. (...) One is almost tempted to assert that the usual interpretation in terms of sharp eigenvalues is ``wrong'', because it cannot be consistently maintained, while the interpretation in terms of [expectation] values is ``right'', because it can be consistently maintained} \cite{Klauder:1997}.}), however we consider this disagreement as a virtue, because it allows us to learn something new.\footnote{See \cite{Kostecki:2007:qht,Kostecki:2008:aqh,Kostecki:2010:aqh,Kostecki:2010:AIP} for some wi(l)der heuristic ideas that have lead to the current work.}

The main mathematical tools used in what follows are: the Hilbert space bundle over $\N^+_\star$ arising from the Gel'fand--Na\u{\i}mark--Segal representation, introduced recently by Odzijewicz and Sli\.zewska \cite{Odzijewicz:Slizewska:2011}, the Derezi\'{n}ski--Jak\v{s}i\'{c}--Pillet theory \cite{DJP:2003} of unbounded perturbation of standard liouvilleans, the B\'{o}na--Odzijewicz--Ratiu construction \cite{Bona:1991,Bona:1993,Bona:2000,Odzijewicz:Ratiu:2003} of the Banach--Lie--Poisson manifold structure on the self-adjoint part $\N^\sa_\star$ of the Banach predual of $\N$, Jen\v{c}ov\'{a}'s construction \cite{Jencova:2006,Jencova:2010} of real Banach smooth manifold over the spaces $\N^+_{\star01}$ of normalised faithful (strictly positive) elements of $\N^+_\star$, based on quantum relative entropic perturbations of states and noncommutative Orlicz spaces, the Daubechies--Klauder continuous time regularised coherent state path integrals \cite{Daubechies:Klauder:1985,Klauder:1988,Klauder:1997,Watson:Klauder:2002,Klauder:2011}, and Favretti's geometrisation \cite{Favretti:2007} of the Jaynes--Mitchell source theory \cite{Mitchell:1967,Jaynes:1985:scattering,Jaynes:1993,Grandy:1987,Grandy:2008}. Section \ref{geometric.structures.section} provides an introduction to most of these mathematical tools. More specifically, Section \ref{W.star.BLP.section} is intended as an introduction to the Banach--Lie--Poisson structure of preduals of W$^*$-algebras. This material is mostly based on papers \cite{Bona:2000,Odzijewicz:Ratiu:2003}. In Section \ref{standard.liouvilleans.section} we review the construction of a standard liouvillean (including standard representation and Haagerup's theorem), some relative modular theory, as well as the Odzijewicz--Sli\.zewska construction of the GNS bundle of Hilbert spaces. In Section \ref{relat.entr.MCP.bundle} we discuss some results from quantum information geometry which we will use in the subsequent sections, including Jen\v{c}ov\'{a}'s construction of a smooth quantum manifold structure. The Daubechies--Klauder approach is discussed in Section \ref{Hilbert.space.geometry.path.integrals}, while Favretti's approach is discussed in Section \ref{Favretti.section}.
\subsubsection{Locally perturbed liouvilleans\label{locally.perturbed.liouvilleans.intro.section}}
Section \ref{algebraic.hamiltonian} provides an elementary analysis of the relationship between W$^*$-dynamical systems, hamiltonian flows on BLP spaces, and standard liouvilleans of the W$^*$-dynamical systems $(\N,\RR,\alpha)$. In order to study the relationship of the BLP structure with the usual usage of standard liouvilleans\footnote{By the Haagerup theorem \cite{Haagerup:1975:standard:form} for standard representations of W$^*$-algebras, for every pair of a W$^*$-dynamical system $(\N,\RR,\alpha)$ and a standard representation $(\H,\pi,J,\stdcone)$ of a W$^*$-algebra $\N$, there exists a unique unitary evolution on $\H$ that represents $\alpha$ leaving $\stdcone$ unchanged. Its generator is an unbounded operator, called the \textit{standard liouvillean}.}, we begin with characterisation of the class of weakly-$\star$ continuous representations $\alpha:\RR\ra\Aut(\N)$ whose predualised actions on $\N_\star^{\sa}$ can be described as Poisson flows of some hamiltonian vector field. 

Our main conclusion from this analysis is that the relationship between Poisson flows and standard liouvilleans should be \textit{localised}: instead of requiring a Poisson flow to globally agree with a family of norm continuous isometries arising from a predetermined W$^*$-dynamical system, we can start from a \textit{quantum Poisson system} (defined as a set $\M(\N)\subseteq\N^+_\star$ equipped with \textit{some} Banach Lie--Poisson manifold structure, not necessarily determined by the coadjoint action of the Lie algebra of self-adjoint elements of $\N$), and determine a fiberwise family of  local W$^*$-dynamical systems generated by a $1$-form corresponding to a hamiltonian vector field on the state space. This way we consider the fiberwise family of local standard liouvilleans as a Hilbert space/algebraic counterpart of the smooth manifold/geometric hamiltonian vector field of the Poisson flow. 

The main technique used in Section \ref{from.local.gauge.to.lagrangeans.section} is: 1) to represent a (possibly, nonlinear) local Poisson flow \textit{on the state space manifold} in each fibre of the GNS Hilbert space bundle by constructing a \textit{local} standard liouvillean, generating unitary evolution uniquely corresponding to the hamiltonian vector field of this flow, and then: 2) to perturb it using objects that represent additional geometric structures on the state space. The resulting structure is shown to determine a nonlinear instrument on $\N^+_\star$ (in the sense of \cite{Davies:Lewis:1970}), which we call a \textit{local liouvillean instrument}. It describes the temporal evolution of quantum states determined by the postulated `internal' dynamics (a W$^*$-dynamical system, a Poisson flow, or a globally defined vector field) perturbed by the \textit{geometric} structures on the quantum model. In other words, the local liouvillean instrument encodes the effective dynamics, that takes into account a nontrivial geometry of the space of quantum states. In addition, we discuss the possible expressions for time dependent $n$-point correlation functions that can be constructed using the above structures. Both local liouvillean instruments and correlation functions are understood as tools as quantification of the effective dynamics.

%
%

Noticing that both the GNS bundle and the tangent bundle of the manifold of quantum states can allow in principle for introduction of a nontrivial action of some Lie group $G$ on fibres, we propose to consider a specific relationship between local gauge (principal $G$-bundle connection) structure and the GNS bundle. We begin with incorporation of the (fiberwise representation of the) action of the nontrivial gauge connection $\mathbf{A}$ (one-form valued in the Lie algebra $\glie$ of $G$) into the perturbation of the local liouvillean, discussing also the possible relationship between affine connection on the tangent bundle of the quantum manifold and propagation of quantum particles (in Wigner sense) in the GNS fibre bundle. 

Apart from the gauge connection, we study also the class of objects that, from the perspective of the BLP structure, could be considered as nonlinear quantum fields. These are introduced as the additional source/sink terms, representing the Lie algebra valued differential one-forms on the base quantum manifold. (The idea of using source-based approach is inspired by Schwinger's \cite{Schwinger:1969,Schwinger:1970} and the Mitchell--Jaynes \cite{Mitchell:1967,Jaynes:1993} approaches.) 



As a result, we construct a setting that allows to define various nonlinear quantum models equipped with smooth geometric structures that can be represented directly in terms of the families of operators acting locally on the fibres of the GNS bundle of Hilbert spaces over the model. It seems that this framework covers quite well some of the components of the lagrangean framework (under nonorthodox assumption that space-time/phase space geometry is emergent from the geometry of quantum state spaces). The investigated correspondences between geometric and algebraic structures can be briefly summarised as:
\begin{center}
\begin{tabular}{c|c}
\textit{$C^\infty$-geometric} & \textit{GNS-bundle-algebraic}\\\hline
principal $G$-bundle sections & gauge propagators\\
one-forms & local quantum field source operators\\
Lie algebra valued one-forms & local gauge quantum fields\\
global charges & global source strengths\\\hline
\end{tabular}
\end{center}
Under some additional conditions we can establish also some relationships between structures of tangent and the GNS bundles:
\begin{center}
\begin{tabular}{c|c|c}
\textit{$C^\infty$-geometric} & \textit{GNS-bundle-algebraic} & \textit{extra condition}\\\hline
hamiltonian vector fields & standard liouvilleans & (PC$_2$)\\
geodesic trajectories & gauge geodesic propagations & (QP$_1$)\\\hline
\end{tabular}
\end{center}
There is also a correspondence between the local liouvillean instruments acting on $\M(\N)$ and local liouvillean operators acting on the fibres of the GNS bundle. These instruments might be nonsmooth. Two main ideas regarding the local quantum dynamics contained in Section \ref{from.local.gauge.to.lagrangeans.section} can be summarised as:
\begin{align}
	\mbox{local gauge dynamics}&=\mbox{local Poisson dynamics}+\mathbf{A}\mbox{-propagation},\\
	\mbox{local liouvillean dynamics}&=\mbox{local Poisson dynamics}+\mathbf{A}\mbox{-propagation}+\mbox{action of sources}.
\end{align}

An especially interesting possibility for introducing an affine connection $\nabla$ on a tangent bundle $\T\M(\N)$ is a third order Taylor expansion of a quantum relative entropy functional $D$. In such case the gauge geodesic propagation of quantum particles can be carried precisely along the lines of local information flow, defined by a constrained maximisation of a relative entropy, and equivalent to the local $\nabla^D$-geodesic free fall. This particular application shows a virtue of using the Hilbert space bundle combined with the technique of local perturbation of standard liouvilleans: it allows to accomodate different smooth manifold structures on the space of quantum states into a single fiber-wise operator formulation. See \cite{Kostecki:2016:microhauptwerk} for further discussion.

In Section \ref{algebraic.action.operator.section} we provide another example of application of this technique, constructing an algebraic generalisation of Savvidou's action operator. It is specified by perturbation of a standard liouvillean $L_\alpha$ of a weak-$\star$ continuous $*$-automorphism $\alpha:\RR\ra\Aut(\N)$ of a W$^*$-algebra $\N$ by the generator $K_\omega:=-\log\Delta_\omega$ of the Tomita--Takesaki modular automorphism $\sigma^\omega:\RR\ra\Aut(\N)$. This can be tentatively interpreted as incorporation of an action of a $U(1)$-connection on a fibre bundle of Hilbert spaces over a real line of a trajectory of $\alpha_\star$ on $\N^+_\star$. For any W$^*$-algebra $\N$, the Falcone--Takesaki theory \cite{Falcone:Takesaki:2001} functorially associates a `core' von Neumann algebra $\core$. If $\N$ is equipped with a faithful normal algebraic state $\omega$, then there exists a canonical unitary isomorphism $\core\iso\N\rtimes_{\sigma^\omega}\RR$ with the crossed product corresponding to a W$^*$-dynamical system $(\N,\RR,\sigma^\omega)$ formed by a modular $*$-automorphism $\sigma^\omega$ of $\N$. This crossed product is a von Neumann algebra generated by the operators $\pi_{\sigma^\omega}(x)$ and $u_\RR(t)=\ee^{-\ii t\tilde{V}}$ acting on the space $\H_\omega\otimes L_2(\RR,\dd t)$ by means of the equations \eqref{pi.sigma} and \eqref{lambda.tau}. The covariance equation  \eqref{modular.histories.covariance}, where a self-adjoint linear operator $K_\omega$ is equal to the Tomita--Takesaki modular hamiltonian, turns the `Liouville' (in Savvidou's sense) action of $\ee^{-\ii t\tilde{V}}$ on $L_2(\RR,\dd t)$ into the action of $\ee^{-\ii tK_\omega}$ on the space $\H_\omega$. Hence, one can say that it `internalises the description of external unitary kinematics'. The perturbed operator $L_\alpha+K_\omega=L_\alpha-\log\Delta_\omega$ provides an algebraic replacement of the quantum histories description of action operator given by equations \eqref{s.h.auto} and \eqref{s.h.auto2}. 
\subsubsection{Local information geometry in quantum histories\label{local.infogeometry.histories.intro.section}}
By analysis of the virtues and drawbacks of the above formulation of an algebraic action operator, we come to a conclusion that the proper candidate for a description of the geometric perturbation of a dynamics due to the local change of state in a projective measurement is not $K_\omega$, but a standard unitary transition operator $V_{\phi,\omega}$ (which is not easily incorporable into the local liouvillean framework). It is a parallel transport of the Levi-Civita connection $\nabla^{1/2}$ of the Wigner--Yanase metric $\gbold^{1/2}$, and projections along its ``free fall'' geodesics are equal to the linear projections in a (standard representation) Hilbert space. This observation leads us to revisit the use of a Fubini--Study metric $\gbold^{\mathrm{FS}}$ on the space of coherent states (which coincides, up to a multiplicative scalar factor $4$, with $\gbold^{1/2}$, when the latter is extended to the boundary of the pure states) for the purposes of regulation of the propagator of a quantum dynamics defined by means of functional integration. As a result, we propose a suitable generalisation of the Daubechies--Klauder expression \eqref{DK.Wiener.expression} for regularised continous time path integration, replacing the coherent states over phase space by all states in the given quantum model $\M(\N)\subseteq\N^+_{\star0}$. In what follows, we will discuss the conceptual aspects of the mathematical formulation that we provide.

Motivated by the Anastopoulos--Savvidou analysis of the term $\ee^{\ii\int p\dot{q}}$ in the Daubechies--Klauder formula \eqref{DK.Wiener.expression} as a holonomy of the Berry connection, we replace it by
\begin{equation}
	\exp\left(\ii\int\dd t\s{\Omega_{\phi(t)},\ddd_{\nabla^{1/2}}(\phi(t))\Omega_{\phi(t)}}_{\H_{\phi(t)}}\right),
\label{one.form.nabla.onehalf.holonomy}
\end{equation}
where $\H_{\phi(t)}$ is the GNS Hilbert space associated with $\phi(t)\in\M(\N)$, $\Omega_{\phi(t)}$ is its representing vector, while $\ddd_{\nabla^{1/2}}$ is a $\nabla^{1/2}$ connection $1$-form. This is equivalent to a local integral of an infinitesimal entropic projection generated by a quantum Br\`{e}gman distance $D_{1/2}$ on $\M(\N)$. While the necessary mathematical background describing the relationship between entropic projections and geodesic free falls is discussed in Section \ref{distances.NS.geom.section}, let us briefly explain the conceptual perspective behind using it to define the dynamics in quantum theory. 

In general, the entropic projections $\PPP^D_{\Q}$ can be used to generate the global temporal evolution of quantum models following the ideas of Jaynes \cite{Jaynes:Scalapino:1963,Mitchell:1967,Jaynes:1985:scattering,Jaynes:1993}, promoted from an absolute to relative entropy by Schl\"{o}gl \cite{Schloegl:1966,Schloegl:1967:produzierte,Schloegl:1967:foundation,Schloegl:1971:stability,Schloegl:1971:fluctuations,Schloegl:1971:near,Schloegl:1971:produced} and Hobson \cite{Hobson:1971,Hobson:Cheng:1973} (see \cite{Grandy:1987,Grandy:2008,Streater:2009:book,Caticha:2012} for the recent account on further developments of these approaches).\footnote{While the definition of $\int\mu p\log\frac{p}{q}$ as well as its conceptualisation as a measure of relative information gain is due to Kullback \cite{Kullback:Leibler:1951,Kullback:1959}, the use of this object for defining information dynamics can be credited to the above authors.} Given a time dependent set of constraints $\Q(s)$, the map
\begin{equation}
	\phi_0\mapsto\PPP^D_{\Q(s)}(\phi_0)
\label{entropic.time.evolution}
\end{equation}
selects a unique trajectory of quantum states, if for each $s$ the set $\Q(s)$ is such that it gives a unique solution to the corresponding minimisation problem (in order to recover the typical formulation of dynamical problems, one may additionally require the map $s\mapsto\PPP^D_{\Q(s)}(\phi_0)$ to be continuous, and $\PPP^D_{\Q(0)}(\phi_0)=\phi_0$). 

However, this construction is not the same as local re-updating of the state in time $s$ to the state in time $s+\delta s$ by the new data. While in principle it is nothing wrong with it (after all, the classical action principle $\delta S=0$ is an inherently nonlocal construction), it is interesting to see whether a \textit{local} entropic dynamics can be proposed. The equivalence of entropic projections with geodesic projections for the class of \textit{Br\`{e}gman distances} $D_\Psi$ provides such a possibility. In such case, instead of consideration of subsequent stages of a relative entropy driven evolution that is nonlocally determined by an initial state $\phi_0$, one can just follow the $\nabla^{D_\Psi}$-geodesics of the $\nabla^{D_\Psi}$-connection (derived from $D_\Psi$, as a third order  Taylor expansion, by means of the Eguchi equations, see Section \ref{distances.NS.geom.section}), maintaining the condition of the $(\nabla^{D_\Psi})^\nsdual$-convexity and $(\nabla^{D_\Psi})^\nsdual$-affinity of the local constraints, as well as their $(\gbold^{D_\Psi},\nabla^{D_\Psi},(\nabla^{D_\Psi})^\nsdual)$-orthogonality with respect to the local $\nabla^{D_\Psi}$-geodesic trajectory. This way the local user's inference, based on smooth reminimisation of an information distance (locally $D_\Psi$-optimal learning process) becomes equivalent with a free fall along $\nabla^{D_\Psi}$-geodesics. One can call it a \textit{local equivalence principle} of an ``information gravity''. The restriction of an arbitrary $D_\Psi$ and $\nabla^{D_\Psi}$ to $D_{1/2}$ and $\nabla^{1/2}$, as expressed by \eqref{one.form.nabla.onehalf.holonomy}, is caused by two reasons: requirement of showing explicit backwards compatibility with the Daubechies--Klauder path integrals, and also the structural restrictions of the GNS bundle.\footnote{Using Hasegawa \cite{Hasegawa:1993} representation of a tangent space in terms of functions of density operator, we could generalise the use of the GNS bundle at least to the case of $\nabla^\gamma$ connections derived from $D_\gamma$. However, this would be restricted to the finite dimensional case. Moreover, we are interested here more in the search of an appropriate analytic setting for the general theory than in the explicit calculations of special finite dimensional cases.} In order to use other $\nabla^{D_\Psi}$ connections, we would have to systematically apply noncommutative Orlicz spaces, which is beyond the scope of this paper.

In addition, we replace an affine function $h(z(t))$ in the Daubechies--Klauder formula \eqref{Klauder.path.integral}, which is corresponding to a Killing hamiltonian vector field and is generated by a coherent state expectation value of a self-adjoint hamiltonian operator, by any smooth function on $\M(\N)$, understood as a hamiltonian function on a BLP manifold. This leads us to ask how one can relate local entropic and local hamiltonian dynamics in the histories context.

Combined with the causal inference Ansatz \eqref{global.hamiltonian.entropic.map}, the principle \eqref{entropic.time.evolution} leads to a \textit{global} evolution
\begin{equation}
	\phi(t,s)=\PPP^D_{\Q(s)}\circ w^{\B,h}_t(\phi_0).
\label{two.times.global.causal.inference.instrument}
\end{equation}
While one can chose $t\in[r_0,r_1]$ and $s\in[r_1,r_2]$, $r_0,r_1,r_2\in\RR$, there is no obligation to do so. In general, $w^{\B,h}_t$ represents a causal evolution governed by the principle of a \textit{local conservation of absolute energy} (undestood as an element of the local space of quantitative effects, such as the self-adjoint observables in quantum mechanics), while $\PPP^D_{\Q(s)}$ represents an inferential evolution governed by the principle of a \textit{global growth of relative entropy} (understood as a function on the global space of states). It was first observed by K\k{e}pi\'{n}ski \cite{Kepinski:1972:Rytm,Kepinski:1972:Schizofrenia} that each of those dynamical processes carries its own notion of time. Our work grew out from consideration of this duality, and \textit{a priori} independence of two associated notions of time, as the fundamental principle of physical dynamics. In the context of the present paper, as we discuss it below, we postulate that the `energetic' time of causality and the `entropic' time of inference are equal, but only \textit{infinitesimally}.\footnote{Due to incoherence between the standard use of the word `local' in physics and in mathematics, it is hard to propose any universally optimal terminology for distinguishing between different regimes. In this paper we use the terms: \textit{global} to refer to objects acting on all space $\M$; \textit{local} and (equivalently) \textit{infinitesimal} to refer to objects acting at $\phi\in\M$; \textit{nonlocal} to refer to objects acting in some neighbourhood of $\phi\in\M$ (maybe \textit{quasi-local} would be a better term). Within our setting, the local regime corresponds to a single user system, defined by the states-and-effects kinematics equipped with the causal-inferential dynamics, global regime corresponds to a multi-user framework, while nonlocality corresponds to the issues of construction of effective multi-user kinematics and dynamics, based on the choice of specific criteria of synchronisation between individual user's systems, at the expense of some individual properties being no longer maintained at the effective level. See Sections \ref{curvature.intro.section} and \ref{locally.quantum.information.relativity.section} for more discussion.} This forms the first principle of local information dynamics. More precisely, we consider the local causal dynamics, governed by $\ddd h_\B$,\footnote{The notation $\ddd h_{\{\cdot,\cdot\}_\B}$ would be completely precise, and symmetric with the notation $\ddd_{\nabla^{D_\Psi}}$, but also quite expensive visually.} and the local inferential dynamics, governed by $\ddd_{\nabla^{D_{\Psi}}}$, as \textit{two} independent fundamental dynamical processes that should be treated on the equal footing as components generating jointly an infinitesimal temporal evolution in a \textit{single} time. In other words, we postulate that the complete description of the local dynamics should be governed by a differential $1$-form
\begin{equation}
	{\cal F}_{D_\Psi,h_\B}:=\ddd h_\B+\ddd_{\nabla^{D_\Psi}}.
\label{local.causal.inferential.form}
\end{equation}

However, the existing constructions of a quantum information manifold use different tangent-cotangent space structure than the quantum Poisson spaces, so, as a result, the addition operation in \eqref{local.causal.inferential.form} is precisely as meaningful, as is adding the element of the noncommutative Orlicz space to the element of a Banach Lie subalgebra $\B$ of $\N^\sa$ (or of $\N$, if one wants to use some nonstandard possibilities). The best solution to this problem would be to represent the action of $\B$ on the cotangent space, given by the noncommutative Orlicz space $L_\Upsilon(\N)$. Yet, even in the simple case, when $\B$ is a Lie algebra of a group of unitaries $\N^\uni$, a corresponding functional analytic study requires to prove the analogue of the Haagerup theorem \cite{Haagerup:1975:standard:form} for the noncommutative Orlicz spaces, resulting in the $L_\Upsilon$-liouvillean (a generalisation of the Jak\v{s}i\'{c}--Pillet $L_{1/\gamma}$-liouvilleans \cite{JOPP:2012}). This is beyond the scope of this paper. As a result, the natural functional analytic representation of the above differential geometric framework is not yet available, and we need to carry our investigations by means of a bit less canonical tools, as reviewed in Section \ref{qig.foundations.intro.section}. The analysis of local perturbations of standard (i.e., $L_{1/2}$-) liouvilleans associated with the GNS Hilbert bundle, carried in Section \ref{from.local.gauge.to.lagrangeans.section} and briefly reviewed in Section \ref{locally.perturbed.liouvilleans.intro.section}, is intended to be a warm-up investigation preceding the (currently unavailable) theory of local perturbations of $L_\Upsilon$-liouvilleans, associated with the tangent and cotangent spaces of quantum information manifolds based on quantum Br\`{e}gman distances. In this sense, the current paper can be considered as an investigation of the mathematical framework for the local Ansatz \eqref{local.causal.inferential.form}, as well as a family of related dynamical problems.\footnote{In future work, we will also consider a replacement of $\ddd h_\B$ by the connection $1$-form of the quantum relative free energy, as generated by the Fenchel--Legendre conjugate of a relative entropy.}

\subsection{Effective local quantum dynamics\label{effective.local.quantum.dynamics.section}}
Comparing the path integral propagator \eqref{RPK.propagator} with the local liouvillean propagator \eqref{liouv.prop}, we can see that both consist essentially of the subtraction of the local free fall along $\nabla^{1/2}$ geodesics from the local hamiltonian flow. This free fall corresponds to the local $D_{1/2}$-projection, which can be in turn interpreted as a continuous projective measurement. This leads us to revisit the problem of analytical implementation of the postulate that the \textit{local} geometric dynamics should be generated by the $1$-form \eqref{two.times.global.causal.inference.instrument}. In particular, applying the infinitesimal approximation $\phi=\omega+\delta\omega$ in the local liouvillean propagator \eqref{liouv.prop} results in the generator $\pi_\omega(\DF_\omega h)-\log(J_{\omega+\delta\omega}J_{\omega+\delta\omega,\omega})$ , which can be considered as an implementation of \eqref{local.causal.inferential.form}.

As a warm-up, consider the case $\N\iso\BH$, $\B\iso\BH^\sa$, $\M(\N)=\schatten_1(\H)^+_{1}$. Combining the insights of B\'{o}na \cite{Bona:1993,Bona:2000} and Grandy \cite{Grandy:2004:1,Grandy:2008}, one can propose the following generalisation of the von Neumann equation,
\begin{equation}
		\ii\frac{\dd}{\dd t}\rho(t)=
		[\ddd h(\rho(t)),\rho(t)]-\frac{\partial}{\partial t}\rho(t),
\label{dissipative.generalisation.of.vonNeumann.eq}
\end{equation}
where
\begin{equation}
	\frac{\partial}{\partial t}\rho(t)=\sum_{i\in I}P_i\frac{\partial}{\partial t}p_i(t)
\end{equation}
is determined by the nonhamiltonian change of probabilities $\{p_1(t),\ldots,p_n(t),\ldots\}$ that determine $\rho(t)$ by means of $\rho(t)=\sum_{i\in I}P_ip_i(t)$, given $\sum_{i\in I}P_i=\II$, $P_iP_j=\dirac_{ij}P_i=\dirac_{ij}P_j$, and $P_i\in\Proj(\BH)$ $\forall i\in I$. Grandy, following Jaynes \cite{Jaynes:Scalapino:1963,Jaynes:1979:where:do:we:stand,Jaynes:1986}, proposes to use maximum absolute relative entropy to construct the evolution $p(t)$. As compared to Jaynes' approach, we propose to replace the use of absolute entropy on probability densities by the use of relative quantum entropies on quantum states. The resulting geometrisation of a contribution $\frac{\partial}{\partial t}\rho(t)$ by means of a connection $1$-form $\ddd_{\nabla^{1/2}}(\phi(t))$ can be provided by a choice of a frame (ordered list of vector fields) $\xi(t):=(\xi_1(t),\ldots,\xi_n(t),\ldots)\in\H_{\rho(t)}$, such that $\ab{\xi_i(t)}^2=p_i(t)$ $\forall i\in I$, and evaluation 
\begin{equation}
	(\ddd_{\nabla^{1/2}})^j_{\;i}(\rho(t))=\sum_k({\Gamma^{\nabla^{1/2}}})^j_{\;\;ki}(\xi(t))P^k.
\label{d.gamma.two.sum.christo}
\end{equation}
Contracting the missing indices with $\xi^i(t)$, we derive the explicit representation of \eqref{one.form.nabla.onehalf.holonomy} as
\begin{equation}
	\exp\left(\ii\int\dd t\,\xi_j(t)\sum_k({\Gamma^{\nabla^{1/2}}})^j_{\;\;ki}(\xi(t))P^k\xi^i(t)\right).
\end{equation}
Hence, \textit{if} one implements the principle \eqref{two.times.global.causal.inference.instrument} as a formal equation 
\begin{equation}
	\dot{\rho}(t)=-\ii\left[{\cal F}_{h,\nabla^{1/2}}(\rho(t)),\rho(t)\right],
\label{formal.effective.form.perturbation}
\end{equation}
\textit{then} the latter can be represented in the above situation as
\begin{equation}
	\ii\frac{\dd}{\dd t}\rho=[\ddd h(\rho(t)),\rho(t)]-\sum_{k,i}({\Gamma^{\nabla^{1/2}}})^j_{\;\;ki}(\xi(t))P^k(P^i-J_{\rho(t)}P^iJ_{\rho(t)}),
\label{RPK.local.equation}
\end{equation}
where the modular conjugations $J_{\rho(t)}$ arise from a commutator of $\ddd_{\nabla^{1/2}}$ with $\rho(t)$. 

Alternatively, taking into account our earlier observation that the standard transition unitary $V_{\omega,\phi}=J_{\phi}J_{\phi,\omega}$ is \textit{exactly} a parallel transport of $\nabla^{1/2}$, we can begin from the $\nabla^{1/2}$-parallel transport equation of a vector $v^a$ along the trajectory $\rho(t)$,
\begin{equation}
\frac{\dd}{\dd t}v^a(t)=-\sum_{b,c}({\Gamma^{\nabla^{1/2}}})^a_{\;\;bc}(\rho(t))v^b(t)\left(\frac{\dd}{\dd t}\rho(t)\right)^c.
\end{equation}
Substituting $v=\dot{\rho}(t)$, and integrating out, we get
\begin{equation}
\ii\frac{\dd}{\dd t}\rho(t)=
-\int_{-\infty}^t\dd t \sum_{b,c}({\Gamma^{\nabla^{1/2}}})^a_{\;\;bc}(\rho(t))\left(\frac{\dd}{\dd t}\rho(t)\right)^b\left(\frac{\dd}{\dd t}\rho(t)\right)^c.
\label{geodesic.free.fall}
\end{equation}
This equation, describes the equation of motion of the free fall along the $\nabla^{1/2}$-geodesic trajectory $\rho(t)$, when represented in the GNS Hilbert space bundle by means of $\H_{\rho(t)}\iso\schatten_2(\H)$. 

An infinitesimal transformation $\rho\mapsto\rho+\der\rho$ can be decomposed as \cite{Hasegawa:1993}
\begin{equation}
\der\rho:=\widetilde{\der}\rho+[\rho,W]=\sum_{i=1}^n\left(\frac{\partial\rho(\theta)}{\partial\theta^i}+[\rho,W_i]\right)\ddd\theta^i,
\label{orthogonal.decomp.of.rho}
\end{equation}
where $\widetilde{\der}\rho=\sum_{i=1}^n\frac{\partial\rho(\theta)}{\partial\theta^i}\ddd\theta^i$ is defined by $[\widetilde{\der}\rho,\rho]=0$, and $W=\sum_{i=1}^nW_i\ddd\theta^i$ is an antiself-adjoint operator (hence, $k_i^*=k_i:=\ii W_i$). The mappings $\der$, $\widetilde{\der}$ and $[\,\cdot\,,W]$ are derivations on $\BH$. This determines a decomposition of tangent space at $\rho$ into the direct product of the corresponding subspaces. An explicit representation of the tangent space in terms of $\schatten_2(\H)$ space by means of finite dimensional coordinate parametrisation $\RR^n\supsetneq\Theta\ni\theta\mapsto\rho(\theta)\in\schatten_1(\H)^+$ reads \cite{Hasegawa:1993}
\begin{equation}
\T_\rho\ell_{1/2}(u)=\sum_{i=1}^nu^i\left(\sqrt{\rho}\frac{\partial\rho}{\partial\theta^i}+2[\sqrt{\rho},W_i]\right).
\end{equation}
As a result of these considerations, if we interpret the principle \eqref{two.times.global.causal.inference.instrument} as a statement that the effective local dynamics is generated by the sum of vectors $\dot{\rho}(t)$ arising independently from the hamiltonian flow and the geodesic free fall \eqref{geodesic.free.fall}, then we should use the equation 
\begin{align}
\ii\frac{\dd}{\dd t}\rho(t)&=[\ddd h(\rho(t)),\rho(t)]\label{RPK.effective}\\
&-\int_{-\infty}^t\dd t \sum_{b,c}({\Gamma^{\nabla^{1/2}}})^a_{\;\;bc}(\rho(t))\left(\sum_{i}u^i\left(\sqrt{\rho}\frac{\partial\rho}{\partial\theta^i}+2\left[\sqrt{\rho},k_i\right]\right)\right)^b\left(\sum_{i}u^i\left(\sqrt{\rho}\frac{\partial\rho}{\partial\theta^i}+2\left[\sqrt{\rho},k_i\right]\right)\right)^c.\nonumber
\end{align}
This equation describes the \textit{effective local dynamics}, including causality and inference effects on the equal footing (thus, \textit{paralelly processing} them). Comparing the nonhamiltonian parts of the equations  \eqref{RPK.local.equation} and \eqref{RPK.effective}, we see that the equation \eqref{RPK.local.equation} can be at best some sort of approximation of \eqref{RPK.effective}. We interpret it as an indication of the weakness of the implementation \eqref{formal.effective.form.perturbation}, as compared with \eqref{RPK.effective}.

The interpretation of \eqref{local.causal.inferential.form} an infinitesimal analogue (but not a generator) of the ``entropic inference following causal Poisson evolution'' global W$^*$-geometric dynamics \eqref{global.hamiltonian.entropic.map}, and the fact that the latter allows to reconstruct CPTP maps as a special case \cite{Kostecki:MunkNielsen:2016,Kostecki:2016:microhauptwerk}, suggests to intepret \eqref{RPK.effective} as a nonlinear geometric analogue of the Lindblad--Gorini--Kossakowski--Sudarshan equation \cite{Lindblad:1976,Gorini:Kossakowski:Sudarshan:1976}.
\subsection{Curvature measures desynchronisation in the multi-user inference\label{curvature.intro.section}}
In this Section we will consider the problem of the effective \textit{nonlocal} quantum information dynamics. As opposed to effective local dynamics, which provides the infinitesimal description of causality and inference from the perspective of a single user (thus, allowing for an immediate subjective bayesian interpretation), nonlocality is intended to describe the multi-user (intersubjective) conventions of relating causal and inferential dynamics of individual users. In our opinion the geometric space is an emergent property of the specific intersubjective conventions of information (causal and inferential) dynamics, which is, in turn, relative to a specific choice of (class of) users. In this text we are focused on an analysis of a specific example, rooted in quantum information geometry. More foundational discussion will be postponed to another paper \cite{Kostecki:2016:microhauptwerk}.

In principle, given two or more different users, each with his/her own method of providing inferences and evaluating causal evolution, nothing can be said about how their information dynamics is related. If only the inferential aspect of information dynamics is taken into account, then such situation can be understood as incommensurability of inferences provided by different subjective bayesians with their arbitrary choices of respective priors and of methods of updating. In order to relate these different instances of local information dynamics, one needs to introduce some method of translation between the respective evolutions, as well as their initial assumptions. Each such method represents a specific intersubjective convention, which allows to translate between individual instances of information dynamics at the expense of constraining its possible forms to such that are subjectible to a given convention. In particular, for inferential part of the information dynamics, there should be a way of identification of a given state of information as `the same' state for all users under consideration. Note that there is no need for a such identification being made globally for all \textit{possible} users: it is sufficient if one can do it for different sets of users that are under the scope of interest.

The general setting for these considerations can be defined as follows. Given an abstract set $\M$ of users, with a single user represented as a point $\phi\in\M$ equipped with a vector space of \textit{local states} $V(\phi)$ and a Banach dual vector space $V^\dual(\phi)$ of \textit{local effects} (one can think of them in terms somewhat similar to \cite{Ludwig:1985:1987}, but the duality does not have to be carried by Banach space structure, but e.g. by convenient vector space structure).\footnote{By the reasons discussed above and below, we postulate that \textit{if} $\M$ is given by $\M(\N)$, \textit{then} $V(\phi)$ should be specified as $L_{\Upsilon(\phi)}(\N)$. However, in order to distinguish the conceptual and the representational aspects of our approach, we state it in more general terms.} In order to model causality and inference, each user can chose locally his/her own `causal' Banach Lie algebra $\B$ acting on $V^\dual(\phi)$, as well as its own `inferential' \textit{Br\`{e}gman functional} $\tilde{D}_\Psi$ on $V(\phi)$, the latter determined by the duality between $V(\phi)$ and $V^\dual(\phi)$ and the choice of a function $\Psi:V(\phi)\ra\RR$. Given a set $U(\phi)\subseteq\M$ of users, such that $\phi\in U(\phi)$, a choice of a function $\ell_\phi:U(\phi)\ra V(\phi)$ allows to use the Br\`{e}gman \textit{functional} $\tilde{D}_\Psi:V(\phi)\times V(\phi)\ra[0,\infty]$ in order to construct the Br\`{e}gman \textit{distance} $D_\Psi:U(\phi)\times U(\phi)\ra[0,\infty]$ by means of
\begin{equation}
D_\Psi(\omega,\psi):=\tilde{D}_\Psi(\ell_\phi(\omega),\ell_\phi(\psi))
\end{equation}
Thus, the choice of the function $\ell_\phi$ defines how the local user interprets `inferentially' the subset $U(\phi)$ of the set $\M$, while the choice of a hamiltonian function $h(\phi)$ defines how he/she interprets it `causally'. The forms $\ddd h_\B(\phi)$ and $\ddd_{\nabla^{D_\Psi}}(\phi)$, constructed as elements of $V^\dual(\phi)$, are encoding the corresponding infinitesimal dynamics.\footnote{On the level of implementation, it is sufficient to model $V(\phi)$ and $V^\dual(\phi)$ as convenient vector spaces in order to guarantee that the infinitesimal calculus is well defined.} The translation between different users in the set $\Q\subseteq\M$ requires, within this model, to specify relationship between $\ell_\phi$ and $\ell_\psi$, as well as $h(\phi)$ and $h(\psi)$, for all elements $\phi,\psi\in\Q$.

The \textit{local} state-effect kinematics is completely described using the pair $(V(\phi),V^\dual(\phi))$ at a given point $\phi\in\M$, while the resulting \textit{local} causal-inferential dynamics is generated by \eqref{local.causal.inferential.form}. This leads to a question to what extent, and at what expense, this dynamics can be extended to a larger \textit{nonlocal} area of $\M$, for example allowing to interconnect the dual pairs of vector spaces of different users by means of the (not necessarily global) sheaf of tangent and cotangent spaces. On the conceptual level, this corresponds to the question how the local (individual) state-and-effect dual pairs of different users $\phi\in\M$ can be mutually related, how their respective causal and inferential local dynamics can be synchronised, and what is the price to pay for it? For example, even if all users $(\phi,V(\phi),V^\dual(\phi))$ agree to use the same generating objects for their respective local system of causality (e.g., a Banach Lie algebra $\B$) and local system of inference (e.g., a discrimination function $\Psi$), their individual choices of functions $h_\B(\phi)$ and $\ell_\phi$ may be not extendible to a set $\M$, resulting in sheaves of different effective local dynamics \eqref{local.causal.inferential.form}. The problem of conditions for integrability of \eqref{local.causal.inferential.form} is thus directly related to the issue of  nonlocal (multi-user) synchronisation of local systems of causality and inference about local states and effects.


In general, the convention defining emergent multi-user inference can be arbitrarily different from the convention defining the emergent multi-user causality, and they both can differ from any of the local instances of inference and causality that are amalgamated into the emergent structure. As a result, the emergence of a nonlocal spatio-temporal causal-inferential dynamics is be provided at the expense of its departure from the local causal-inferential dynamics. In the context of the structures analysed in this paper, it is represented by the appearance of the system of local entropic priors associated with a specific integral line of a vector field on a model $\M$. (From the closely related point of view, one can notice that the postulate of Section \ref{local.infogeometry.histories.intro.section} identifies the local inferential time with local causal time for each individual user \textit{separately}, but it does not say anything about mutual relationships of those time parameters for different users. In principle, one could also study theories for which the local causal and inferential time of each user would \textit{not} be identified, yet the multi-user synchronisation of those two temporal structures would be considered. We find this perspective very attractive, but it is beyond the scope of the current paper.)

On the technical level, we observe that the Fubini--Study riemannian metric, used in the regularising term in the Daubechies--Klauder and the Anastopoulos--Savvidou approaches, can be replaced by the second order Taylor expansion $\gbold^{1/2}$ of a quantum relative entropy $D_{1/2}$. This leads us to postulate to use a quantum entropic prior \textit{localised} to a neighbourhood of a given state as a general geometric form of the regulariser (see Section \ref{entropic.priors.section} for a discussion of the notion of an entropic prior in the commutative case). Integration of a local entropic prior along a given trajectory on a state manifold constructs a regularised weight for this trajectory. 

On the conceptual level, we interpret the appearance of local entropic prior as a measure of (impossibility of) synchronisation of the causal-inferential dynamics of the subsequent local users at a given spatio-temporal trajectory. More specifically, the choice of a \textit{single} nonlocal (multi-agent) time trajectory sets up the nonlocal vector field, corresponding to a specific system of synchronisation (=~nonlocal/noninertial observation frame) for the local forms ${\cal F}_{D_\Psi,h_\B}$. This choice is arbitrary, but with each such choice, the passage from point to point on the corresponding nonlocal trajectory of a \textit{single spatialised time} adds an additional term to an effective dynamics \textit{along this trajectory}. Only some specific conventions of the multi-user inference avoid the path-dependence of the synchronisation of the inference: in the example that we study here they are given by such models $(\M(\N),D_\Psi)$ for which the scalar curvature $\kappa(\nabla^{\gbold^\Psi})$ is constant over $\M(\N)$.

One can wonder why this phenomenon of ``breaking of symmetry'' between causality and inference results in an entropic (hence inferential), as opposed to a hamiltonian (hence causal) contribution. If all users along the trajectory (as well as in the neighbourhood of this trajectory, in order to have a situation that is more canonical than that of a $1$-dimensional manifold) agree to have the same choice of $\B$ and $\Psi$, then in principle they should share the same nontrivial inferential and causal geometry, so there is yet no reason for ``breaking of the symmetry''. However, the Carath\'{e}odory--Jacobi--Lie theorem (which generalises the Darboux theorem) implies that for any symplectic manifold with a hamiltonian function $h$, the function $h$ is a conserved quantity in an open neighbourhood for any $\phi$ for which $\ddd h(\phi)\neq0$. An analogous statement is not true in riemannian geometry, as measured by the curvature. The generalisation from symplectic to Poisson geometry and from riemannian to Norden--Sen geometry does not change qualitatively this difference. Thus, the passage from infinitesimal/local (jets) to neighbourhood/nonlocal (germs) is trivial for causality, but not for inference. So, while it is possible to nonlocally synchronise the local causal structure along a trajectory of users (whenever it is modelled by the Poisson geometry), it is impossible to do this with the local inferential structure (whenever it is modelled by the Norden--Sen geometry). It is a very interesting phenomenon, which may be interpreted by saying that while the global synchronisation of local systems of causality is a trivial consequence of choice of the same system of locally conserved quantities for each user, the global synchronisation of local systems of inference is impossible, with the system of local entropic priors measuring the scale of this impossibility for each of the local users along the trajectory. The local entropic priors could be in principle replaced by any other local priors. Each of such choices represents a different \textit{initial system of assumptions} that are used by local users for the purposes to measure the desynchronisation between their local/jet inferences, as provided by means of $\tilde{D}_\Psi$ on $V(\phi)$, and their nonlocal/germ inferences, as provided by means of $D_\Psi$ on $\M$. The local priors are just measures on the local neighbourhoods in the model $\M$, and they have to be \textit{postulated independently} from the choice of ${\cal F}_{D_\Psi,h_\B}$.\footnote{While a detailed discussion of this phenomenon is beyond the scope of the current paper, we want to note that a somewhat similar situation (roughly speaking, a necessity of using additional geometric structures on the pre-state spaces in order to select a specific sheaf of states) was discovered by Bostelmann \cite{Bostelmann:2000} in the Haag--Ojima approach \cite{Haag:Ojima:1996} to construction of germs of algebraic states corresponding to a choice of a specific predictive dynamical theory in the algebraic approach.}

Thus, while the equation ${\cal F}_{D_\Psi,h_\B}=0$ (whenever valid) can be understood as a causal-inferential analogue of Einstein's version of Newton's first law of motion (\textit{for a single user}, from his/her own perspective), the local priors are somewhat similar to the second law of thermodynamics or to the universality of the gravitation: the globalisation of inferences is provided at the expense of inevitability of making those inferences dependent on additional arbitrary assumptions, that are in principle different for each user, and \textit{are nonobservable in the infinitesimal causal-inferential reference frame}. The shadow of dependence of the multi-userr inferences, as well as of the effective \textit{nonlocal} causal-inferential dynamics, on arbitrary additional assumptions (\textit{attributed} to other users by the user residing at the end point of the integrated trajectory) is a price paid for a requirement of global spatialisation of the inferential dynamics. The lack of the similar effect in the case of causal dynamics is in essence a result of a local linearity of the Banach Lie--Poisson spaces. If other mathematical structure would be chosen to model causality, the similar dialectics may occur. For example, one could model causal dynamics by means of extremum of quantum relative free energies, defined as the Legendre--Fenchel conjugates of quantum information distances. The nonlocal trivialisation of the causal structure by means of the CJL theorem would not be applicable in such case. As a result, the dual, local relative free energy priors, should be also included, as an additional regularising term, into integration giving rise to an effective dynamics. In general, we propose to consider the priors on the model $\M$ as an information theoretic analogue of the notion of mass, so the local prior at the neighbourhood $U(\phi)$ of $\phi\in\M$ can be interpreted as an information theoretic analogue of the mass of the user $\phi$ distributed over the set $U(\phi)$ of users. The special case in which the integration against the local relative free energy priors would cancel the contribution arising from the local entropic priors can be then considered to be an analogue of Einstein's principle of equivalence of gravitational (inferential) and inertial (causal) mass. 

As observed in \cite{Maraner:1992,Alicki:Klauder:Lewandowski:1993,Watson:Klauder:2002}, if the riemannian geometries of the phase space used for the Wiener measure regularisation have nonconstant scalar curvature, then the weighting of the phase space paths is nonuniform, corresponding to the phase space point dependency of the zero-point energy. On the other hand, in a completely different context, Jaynes has argued \cite{Jaynes:1957,Jaynes:1978:electrotoday,Jaynes:1990} that the zero-point energy should be interpreted not as an ontic feature of the system, but as observer's measure of uncertainty regarding his/her own prediction of the value of energy, as based on his/her own \textit{prior information}. Quite independently from these considerations, Rodr\'{\i}guez has developed the theory of entropic priors \cite{Rodriguez:1991} interpreting them as the \cytat{statistical representation of the vacuum of information in a given hypothesis space} \cite{Rodriguez:1999}. Our formulation recombines the above insights, integrating them into a single statement: zero-point energy's point-dependence is a manifestation of a dependence of nonlocal integrability of local inferential structure on the local geometry of user's prior knowledge (or, equivalently, prior ignorance).\footnote{This gives also some justice to an otherwise quite cryptic remark of Jaynes on the book \cite{Popov:1987} on functional integration: \cytat{A useful start on understanding of these phenomena, but still lacking any coherent theoretical basis -- which we think is supplied only by the principle of maximum entropy as a method of reasoning.} \cite{Jaynes:2003}.} 

However, while our proposal is well founded on the geometric side, and has also an interesting information theoretic interpretation, it has to be regarded as heuristic from the perspective of stochastic process based foundations for path integration. More specifically, the Daubechies--Klauder formulation relies on the interpretation of $\int\dd t p\dot{q}=\int p \dd q$ as the Stratonovich integral (so $\dd q$ is considered as a Stratonovich differential $\dd_{\mathrm{S}}q$), and on the interpretation of (roughly) $\int\mathcal{D}p\mathcal{D}q\ee^{-\frac{1}{2\upsilon}\int\dd t(\dot{p}^2+\dot{q}^2)}$ as the pinned Wiener measure (see \cite{Bodmann:2003} for a systematic mathematical treatment of these objects in terms of the Berezin--Toeplitz operators). The generalisation to a wide class of connections $\nabla^{D_\Psi}$ and local entropic priors (even if kept at the second order riemannian level of $\gbold^D$) asks for a systematic development of a technique of stochastic integration of random walks $\mathbf{X}$ on $\RR^n$ associated with the Brownian motions on smooth manifolds $\M$, $\dim\M=n$, that could systematically address the  functional integration of the above geometric structures beyond the level of heuristic treatment that is standard for physicists. In particular, following \cite{Ikeda:Watanabe:1981,Kendall:1987}, let us consider the system of stochastic differential equations
\begin{align}
	\dd_\mathrm{S}\mathbf{X}&=\mathbf{e}\,\dd_\mathrm{S}\mathbf{B},\\
	\dd_\mathrm{S}\mathbf{e}&=H_\mathbf{e}\dd_\mathrm{S}\mathbf{X},
\end{align}
where $\dd_\mathrm{S}$ are Stratonovich differentials, $\mathbf{B}$ is an euclidean brownian motion on $\RR^n$, frame $\mathbf{e}$ is a map from $\RR^n$ to the tangent space of $\M$, and $H$ is a horizontal lift of a tangent space at $\mathbf{X}$ to the tangent space at $\mathbf{e}$, dependent on the choice of an affine connection $\nabla$ on $\M$. If this connection is not Levi-Civita, then the process $\mathbf{X}$ will be not markovian. Furthermore, a suitable riemannian metric reproducing the Laplace--Beltrami operator can be uniquely constructed by an appropriate choice of an elliptic diffusion. Hence, the exact mathematical foundation for our generalisation of the Daubechies--Klauder formula is possible at least for the second order Taylor expansion of the entropic prior and for different connections $\nabla^{D_\Psi}$. While we consider the task of a systematic treatment of this topic to be of high importance, it will be left beyond the scope of this paper.

The issue of renormalisation cannot be omitted in any foundational discussion of local quantum dynamics. Section \ref{information.theoretic.renormalisation} is dedicated to the study how the tools of quantum information geometry can be used in order to deal with the tasks of renormalisation. In Sections \ref{section.JM.source.theory} and \ref{Favretti.section} we briefly review the Jaynes--Mitchell source theory \cite{Mitchell:1967,Jaynes:1985:scattering,Jaynes:1993,Grandy:1987,Grandy:2008} and its geometric generalisation by Favretti \cite{Favretti:2007}, respectively. This approach provides a geometric implementation of the idea of renormalisation of dynamics by reduction of dimensionality of the model by fixing the control parameter, which is specified as a constraint on the space of information states (as opposed to the space of functions or operators). We observe that the Jaynes--Mitchell--Favretti approach is canonically related to the use of Br\`{e}gman distances, and that it can be used to provide a \textit{local} description of entropic information dynamics and multiparameter nonlinear quantum control problems on an arbitrary quantum manifold. The key insight of this approach can be summarised as: renormalisation of the action of control parameters (sources) leads to departure of the geometry of a model $\M$ from a dually flat one. Hence, the appearance of nonzero curvature is an indicator of a nontrivial constraints for information dynamics. Moreover, the nonconstancy of this curvature indicated local dependence of these constraints. From the perspective of our approach to the Daubechies--Klauder path integrals, we postulate that the renormalised description of dynamics should use renormalised riemannian metric $\tilde{\gbold}$ instead of $\gbold^{D}$ in the regulariser. This corresponds to replacement of a ``vacuum of information'' by the ``vacuum of information storing the shadow of the knowledge about the sources that were renormalised out''.

We also introduce another type of information geometric renormalisation of inferential dynamics of quantum states, which describes situations where none of  specific control (covariate) parameter is fixed, but the quantum model is subjected to the action of completely positive maps. This procedure is based on the use of $D_\fff$ distances as well as associated \textit{contraction coefficients}, introduced by Ruskai et al \cite{CIRRSZ:1993,Choi:Ruskai:Seneta:1993,Ruskai:1994,Lesniewski:Ruskai:1999}. For an alternative (and essentially more developed) approach to renormalisation based on $D_\fff$ on $\N_\star^+$, see \cite{Beny:Osborne:2012,Beny:Osborne:2013,Beny:Osborne:2014} (c.f. \cite{DeBrota:2015} for a pedagogical introduction).
\subsection{Locally quantum multi-user information relativity\label{locally.quantum.information.relativity.section}}
The discussion in this paper is aimed at the construction of the framework of the multi-user (intersubjective) information relativity with emergent spaces. While most of discussion is kept in the framework build upon the W$^*$-algebras, it is more a useful testing ground (and a verifying constraint for backwards compatibility), then a desired property of the framework. In other words, our intention is to get rid of the \textit{Postulate 1} of Section \ref{qig.foundations.intro.section} by a suitable reformulation of the \textit{Postulate 5} and \textit{6}. We propose to consider the following tentative structure:
\begin{enumerate}
\item[1)] Consider arbitrary set $\M$ of users. 
\item[2)] each user $\phi\in\M$ is equipped with his/her own vector space $V(\phi)$ containing his/her descripton of `states' (`configurations', `preparations', `inputs') and a dual vector space $V^\dual(\phi)$, containing his/her description of `effects' (`registrations', `results', `outputs').
\item[3)] User's notion of inference (respectively, causality) is modelled by a choice of set of of endomorphisms of $V(\phi)$ (respectively, $V^\dual(\phi)$).
\item[4)] In particular, inferences on $V(\phi)$ can be provided by means of the Br\`{e}gman functional $\tilde{D}_{\Psi,\phi}:V(\phi)\times V(\phi)\ra[0,\infty]$, while the causality on $V^\dual(\phi)$ can be provided either by the Legendre--Fenchel conjugate of $\tilde{D}_{\Psi,\phi}$, or by a representation of some Lie algebra.
\item[5)] More specifically, if one wants to do statistical inference, it is necessary to make clear how the behaviour of finite data sets is associated with the specific idealisations used in the theoretic framework, as represented by the `ideal' theoretical states and effects. In order to assert such relation, it is necessary to admit some statistical tools that are representing the control over a `convergence to an ideal form of a data set', for any given nonideal form of a real data set. Large number estimation and asymptotic estimation are two typical tools (on the geometric level, they correspond, respectively to relative entropy maximisation, and the local linearity). If a user introduces a discrimination function (information potential, absolute entropy) $\Psi_\phi:V(\phi)\ra\,]-\infty,+\infty]$, then he/she is able to quantify the rates of convergence of sequences of data. The key property of the Br\`{e}gman functional is that it allows for a generalised pythagorean theorem \eqref{generalised.pythagore.Bregman}, which is a nonlinear generalisation of the fundamental property of euclidean and Hilbert spaces. Yet, it is doing it without necessity of assuming that $V(\phi)$ is normed, or even metrisable. As a result, user's local inferences on $V(\phi)$ based on entropic projections $\PPP^{\tilde{D}_{\Psi,\phi}}$ allow to decompose the information distance to an `ideal inference' ($\tilde{D}_{\Psi,\phi}(x,y)$) into a sum of a distance to an effective solution (satisfying given constraints and minimising the distance) and an uncertainty within the constrained space. (By the Legendre--Fenchel duality, completely parallel considerations are applicable to causality on $V^\dual(\phi)$ determined by the relative free energy $\tilde{D}_{\Psi^\lfdual,\phi}:V^\dual(\phi)\times V^\dual(\phi)\ra[0,\infty]$.)
\item[6)] The geometric structures on the set $\M$ are introduced as quantitative means of relating (synchronising) inferences and causality of different users.
\item[7)] Each user $\phi$ provides his/her own mappings from $\M$ into $V(\phi)$, given by bijective embeddings $\ell_\phi:U(\phi)\ra V(\phi)$, where $U(\phi)\subseteq\M$ is the subset of users in $\M$ that a user $\phi$ considers as representable in terms of his/her `configuration' space $V(\phi)$.
\item[8)] Using the embeddings $\ell_\phi$, each user can relate his/her individual inferences on $V(\phi)$ with other users using (a part of) the same set $U$. In particular, the relative Br\`{e}gman entropy $D_\Psi$ on $\M$ is induced by  
\begin{equation}
	\tilde{D}_{\Psi,\phi}(\ell_\phi(\omega_1),\ell_\phi(\omega_2))=D_{\Psi,U(\phi)\cap U(\psi)}(\omega_1,\omega_2)=\tilde{D}_{\Psi,\psi}(\ell_\psi(\omega_1),\ell_\psi(\omega_2))\;\;\forall \omega_1,\omega_2\in U(\phi)\cap U(\psi).
\end{equation}
This provides the means to relate the \textit{large number} inferential dynamics of different users (such as in Sanov's theorem).
\item[8)] The local smooth manifold structure induced by $D_\Psi$ on $\M$ provides the means to relate \textit{asymptotic} inferential dynamics of different users. In particular, the equivalence of local geodesic projection and local $D_\Psi$ projections can be understood as a method of locally linear synchronisation of inferences of different users. In such case the curvature of the manifold $\M$ measures impossibility of ideal synchronisation of inferences between different users. The relationships with the notions of causality, effective dynamics, and renormalisation were discussed in the previous Section. They are dependent only on the notion of Br\`{e}gman distance on an arbitrary set $\M$, and Eguchi equations on an arbirary smooth manifold $\M$, hence they hold in general, without assuming that $\M$ is a subset of $\N^+_\star$ for some W$^*$-algebra $\N$.
\item[9)] The reconstruction of the special case, when $\M$ is equal to the set $\M(\N)$ of states over a globally defined W$^*$-algebra $\N$ is an open problem. Our conjecture is that quantum mechanics can be characterised as a set $\M$ equipped with an induced structure of the riemannian manifold $(\M,\gbold^{D_\Psi})$ and a Poisson manifold such that the set of extremal points of the convex hull of $\M$ admits an induced metric and induced symplectic structure that determine a K\"{a}hler manifold satisfying the standard properties of the quantum mechanical K\"{a}hler manifolds.
\end{enumerate}
In principle, the above scheme is applicable to a wide class of postquantum information theoretical settings, such as general probabilistic theories. In what follows, having in mind the possible applications in nonequilibrium quantum statistical mechanics, we will focus on its implementation in the context of W$^*$-algebras and the associated functional analytic spaces.

 While \eqref{SS} considers only the case of local $\nabla^{1/2}$/$D_{1/2}$-projections, \cite{Hellmann:Kaminski:Kostecki:2016,Kostecki:2014,Kostecki:MunkNielsen:2016} consider only the case of global $D_0$-projections. Yet, we think that the natural geometric objects for construction of local  and nonlocal quantum kinematics and dynamics are arbitrary Banach Lie algebras (to describe causality) and arbitrary quantum Br\`{e}gman distances (to describe inference), connected together via dual pairs of noncommutative Orlicz spaces. More specifically, following a discussion in Section \ref{local.infogeometry.histories.intro.section} we think that the problems considered in this paper should be readdressed in a more general  foundational framework, based on the following principles:
\begin{enumerate}
\item[1)] The use of GNS Hilbert bundle should be replaced by a suitable bundle of noncommutative Orlicz spaces $L_{\Upsilon(\phi)}(\N)$ playing the role of tangent spaces $\T_\phi\M(\N)$, understood as the spaces of local `configurations' (e.g. $\phi(\theta)\mapsto\theta\mapsto\left(\frac{\partial}{\partial\theta_i}\right)$), with their Banach duals $L_{\Upsilon(\phi)}(\N)^\star$ playing the role of cotangent spaces $\T^\ct_\phi\M(\N)$, understood as the spaces of local `effects' ($f(\phi)\mapsto\ddd f\mapsto(\ddd f_i)$). Somewhat similar ideas were considered earlier in \cite{Streater:2009:book,Majewski:Labuschagne:2014}, but only in a global context, restricted to a single `tangent' and `cotangent' space.
\item[2)] There should be provided a canonical construction of a quantum Br\`{e}gman distance $D_\Psi$ associated with a Banach dual pair $(L_{\Upsilon(\phi)}(\N),L_{\Upsilon(\phi)}(\N)^\star)$, a function $\Psi:L_{\Upsilon(\phi)}(\N)\ra\RR$, and a family of embeddings $\ell_\phi:\N^+_\star\ra L_{\Upsilon(\phi)}(\N)$, such that the second order G\^{a}teaux derivative of $D_\Psi$ determines a map 
\begin{equation}
\gbold_\phi^{D_\Psi}:L_{\Upsilon(\phi)}(\N)\times L_{\Upsilon(\phi)}(\N)^\star\ra[0,\infty],
\end{equation}
 while the third order G\^{a}teaux derivatives determine the connections $\nabla^{D_\Psi}$ and $(\nabla^{D_\Psi})^\nsdual$, with the respective parallel transports equal to the isometric transition operators 
\begin{align}
\transport^{\nabla^{D_\Psi}}_{\phi,\omega}&:\T_\phi\M(\N)\ra\T_\omega\M(\N),\\
\transport^{(\nabla^{D_\Psi})^\nsdual}_{\phi,\omega}&:\T^\ct_\phi\M(\N)\ra\T^\ct_\omega\M(\N),
\end{align}
and satisfying the generalised Norden--Sen duality
\begin{equation}
	\gbold^{D_\Psi}_\phi(\transport^{\nabla^{D_\Psi}}_{\phi,\omega}(x),\transport^{(\nabla^{D_\Psi})^\nsdual}_{\phi,\omega}(y))=\gbold^{D_\Psi}_\omega(x,y).
\end{equation}
This is intended to implement the principle discussed in Section \ref{Orlicz.Br\`{e}gman.section}: the local structure of an information manifold should implement the local equivalence of an entropic projection (user's inference) and a  geodesic free fall (an absence of geometric `gravity').
\item[3)] Lie algebras $\glie$ and Banach Lie algebras $\B$ should be represented on $L_{\Upsilon(\phi)}(\N)^\star$, the latter giving rise to a Banach Lie--Poisson manifold structure on $\M(\N)$ determined by the action of $\B_\star$ on $L_{\Upsilon(\phi)}(\N)$. 
\item[4)] As a result, both $\ddd_{\nabla^{D_\Psi}}(\phi)$ and $\ddd h_\B(\phi)$ become the elements of the same operator space $L_{\Upsilon(\phi)}(\N)^\star$, so the local effective dynamics can be formulated in terms of a $1$-form ${\cal F}=\ddd h_\B+\ddd_{\nabla^{D_\Psi}}$, treating local causality and local inference on an equal footing. Note that, in face of the discussion in Sections \ref{locally.perturbed.liouvilleans.intro.section}, \ref{local.infogeometry.histories.intro.section}, and \ref{Orlicz.Br\`{e}gman.section}, the condition ${\cal F}=0$ can be understood as an information-theoretic analogue of Einstein's version of Newton's first law of motion: a rest in a local causal-inference frame (provided by user's choice of generators $\B$ for causal dynamics in a cotangent space of effects, and a discrimination function $\Psi$ for inferential dynamics in a tangent space of states).
\item[5)] The local tangent BLP hamiltonian vector field, generating a local W$^*$-dynamical system associated with a local fiber of a GNS Hilbert bundle, should be represented in terms of $L_\Upsilon$-liouvillean, acting on the $L_{\Upsilon(\phi)}(\N)$ tangent space. The incorporation of the contribution of the local free fall along $\nabla^{D_\Psi}$-connection geodesics should be provided by the perturbation of this $L_\Upsilon$-liouvillean, defined in an analogy with the Jak\v{s}i\'{c}--Pillet $L_{1/\gamma}$-liouvillean \cite{JOPP:2012}.
\item[6)] The local entropic prior should be constructed using $D$ and $\gbold^D$ representing the \textit{effective geometry} of a model $\M(\N)$, after the renormalisation of all contributions from control sources. One can interpret the local entropic prior at a point $\phi\in\M$ as an information theoretic analogue of the (inferential) mass of the user at $\phi$. This interpretation gives a particularly neat meaning to the observation \cite{Maraner:1992,Alicki:Klauder:Lewandowski:1993}, discussed in Section \ref{local.quantum.dynamics.intro.section}, that the regularising term in the Daubechies--Klauder formula leads to a point-dependence of a zero-point energy if{}f the curvature of the Fubini--Study metric is nonconstant. From our point of view, it means: the curvature $\kappa^{\nabla^{D}}$ of a quantum model is a measure of desynchronisation of the ideal multi-user inference (as given locally by $\nabla^{D_\Psi}$-geodesics, or, equivalently, $\ddd_{\nabla^{D_\Psi}}$), reflected in the influence of a local mass on the local zero-point energy. The case of information model with a constant curvature is an inferential analogue of the Carath\'{e}odory--Jacobi--Lie theorem for symplectic manifold, allowing for a global trivial synchronisation of local causality systems.
\item[7)] Given these constructions, the dynamical Ansatz \eqref{SS} should be generalised by means of the replacement of the GNS Hilbert bundle by a corresponding dual bundle of noncommutative Orlicz spaces. 
\item[8)] In principle, every quantum Br\`{e}gman functional $\tilde{D}_\Psi$ on $L_{\Upsilon(\phi)}(\N)$ determines the Legendre--Fenchel conjugate functional $\tilde{D}^\lfdual_{\Psi}$ on $L_{\Upsilon(\phi)}(\N)$, which can be naturally interpreted as a relative free energy. If one would replace the use of the (nonlocally trivially synchronisable) $1$-form $\ddd h_\B$ by the form $\ddd_{\nabla^{D^\lfdual_\Psi}}$, then it would be necessary to introduce a corresponding local entropic prior (causal mass), measuring the influence of desynchronisation of local systems of causality, along the spatial trajectory on $\M$, on the effective quantitative nonlocal evolution. The postulate of the local cancellation of effects of those two (inferential and causal) priors, when integrated together, would be an information theoretic analogue of the postulate of equality of gravitational and inertial mass.
\item[9)] In the JMF source theoretic approach the local departure of riemannian metric from the hessian geometry is described by the equation \eqref{JMF.renormalisation.of.metric}. Applying the change of geometry $\gbold^{D_\Psi}\mapsto\tilde{\gbold}$ to a definition of a prior $\exp\left(-k\frac{\epsilon^2}{2}\int_\pathgamma\dd t\,\gbold_{ab}(\phi)\dot{\phi}^a\dot{\phi}^b\right)\sqrt{\det(\gbold(\phi))}$, we can see that the change of the scalar curvature $\kappa(\phi)$ of the model is reflected in the redefinition of the local prior (information mass) $P$. This leads to a more general problem. Given any neighbourhood $U\subseteq\M$ of a user $\phi\in\M$, one can ask how to determine the departure of a prior $P$ on $U$ (from the originally postulated one) that is caused by the changes of curvature $\kappa$ of a model $\M$ associated with a given system of inference, when the transformation of geometry of a model $\M$ (e.g., due to the Jaynes--Mitchell renormalisation) is considered. The analogy with general relativity that we were pursuing in this paper, as well as an observation \cite{Watson:Klauder:2002} that the uniform weighting of the trajectories requires constant scalar curvature, leads us to conjecture that
\begin{equation}
	\delta\int_UP=\delta\int_U\kappa.
\end{equation}
\end{enumerate}
We hope that this mathematical framework will allow for a unified treatment of the foundations of nonequilibrium quantum statistical mechanics, as discussed from different perspectives in \cite{Grandy:2008}, \cite{Streater:2009:book}, and \cite{JOPP:2012} (see also \cite{Caticha:2012}). More generally, we consider it to be a testing ground for a construction of a predictive dynamical theory that would unify several concepts of general relativity with quantum information theory into a single framework of multi-agent (post)quantum information relativity. Maybe this sounds a bit like a fountain of conjectures and high hopes, but we believe that the science fiction of today is just an advanced propagator of a scientific research of tomorrow.
\ifvargaugecompile 
\section{Quantum geometry \& global dynamics\label{geometric.structures.section}}
\else 
\subsection{Quantum geometry\label{geometric.structures.section}}
\fi 

\ifvargaugecompile 
Sections \ref{W.star.BLP.section}, \ref{standard.liouvilleans.section}, \ref{distances.NS.geom.section}, \ref{quant.info.geom.section} do not contain new results or constructions, except of the definition of the quantum Poisson system.
\subsection{Quantum Banach Lie--Poisson spaces\label{W.star.BLP.section}}
\else 

\fi 

\subsubsection{Banach--Lie--Poisson spaces}

Let $\KK\in\{\RR,\CC\}$. A vector space $X$ over $\KK$ is called a \df{Lie algebra} if{}f it is equipped with a function $[\cdot,\cdot]:X\times X\ra X$ such that for all $f_1,f_2,f_3\in X$ and for all $\lambda\in\KK$
\begin{enumerate}
\item[1)] $[f_1,f_2]=-[f_2,f_1]$ (antisymmetry),
\item[2)] $[f_1,f_2+\lambda f_3]=[f_1,f_2]+\lambda[f_1,f_3]$ (linearity),
\item[3)] $[f_1,[f_2,f_3]]+[f_3,[f_1,f_2]]+[f_2,[f_3,f_1]]=0$ (Jacobi identity).
\end{enumerate}
The function $[\cdot,\cdot]$ is called the \df{Lie bracket}. If $(X,[\cdot,\cdot])$ satisfies also the Leibniz's rule
\begin{enumerate}
\item[4)] $\{f_1,f_2f_3\}=\{f_1,f_2\}f_3+f_2\{f_1,f_3\}\;\;\forall f_1,f_2,f_3\in X$,
\end{enumerate}
then $[\cdot,\cdot]$ is called the \df{Lie--Poisson bracket} \cite{Lie:1890}, while $(X,[\cdot,\cdot])$ is called a \df{Lie--Poisson algebra}. If $G$ is a Lie group, then a Lie algebra of its generators will be denoted $\Lie(G)$. A vector space $X$ over $\KK$ is called a \df{Banach Lie algebra} if{}f it is a Banach space with a norm $\n{\cdot}$, a Lie algebra, and its Lie bracket $[\cdot,\cdot]:X\times X\ra X$ is bilinear and continuous in the topology of $\n{\cdot}$. If $\glie$ is a Banach Lie algebra, then the \df{adjoint map} $\rpktarget{ADMAP}\ad_x:\glie\ni y\mapsto[x,y]\in\glie$ and \df{coadjoint map} $\ad_x^\banach:\glie^\banach\ra\glie^\banach$,
\begin{equation}
        \duality{y,\ad^\banach_x(z)}_{\glie\times\glie^\banach}:=\duality{\ad_x(y),z}_{\glie\times\glie^\banach}\;\;\forall z\in\glie^\banach,
\end{equation}
are norm continuous for each $x\in\glie$. 

Let $M$ be a real Banach smooth manifold, and $\rpktarget{CIF}\CIF(M;\RR)$ denotes the space of all infinitely Fr\'{e}chet differentiable $\RR$-valued functions on $M$. Then a real \df{Poisson structure} on $M$ is defined as a function $\rpktarget{POISSON}\{\cdot,\cdot\}:\CIF(M;\RR)\times\CIF(M;\RR)\ra\CIF(M;\RR)$ such that $(\CIF(M;\RR),\{\cdot,\cdot\})$ is a Lie algebra \cite{Lie:1890}. %
\ifvargaugecompile 
%
\else 
\footnote{Note that, on an algebraic level, the conditions i.2) and 4) on $\{f,\cdot\}$ are the same as the conditions 1. and 2. on $\der(\cdot)$ in Section \ref{derivations.section}.}
\fi 
If $M$ above is \textit{finite dimensional}, then $(M,\{\cdot,\cdot\})$ is called a real \df{Poisson manifold} \cite{Lichnerowicz:1977,Weinstein:1983,Vaisman:1994,Weinstein:1998}. If $M$ is a real Banach smooth manifold, then the cotangent space at $x$ can be defined by $\T_x^\ct M:=(\T_xM)^\banach$, and each element of $\T_x^\ct M$ has a form
\begin{equation}\rpktarget{DDDf}
        \ddd f(x)\equiv\ddd_x f:\T_xM\ni v\mapsto\ddd_x f(v):=v(f)\in\RR
\end{equation}
for some $f\in\CIF(M;\RR)$. Let $\T^\ct{}^\ct M:=\bigcup_{x\in M}\T_x^\ct{}^\ct M$, where $\T_x^\ct{}^\ct M:=(\T^\ct_x M)^\banach=(\T_x M)^\banach{}^\banach$. If $(M,\{\cdot,\cdot\})$ is a real Poisson manifold, then every $k\in\CIF(M;\RR)$ determines a unique vector field $\XXX_k\in\T M$ by\rpktarget{HAMILVF}
\begin{equation}
        \XXX_k(f):=\bigcup_{x\in M}\left\{x\mapsto\duality{\XXX_k(x),\ddd f(x)}_{\T_xM\times\T_x^\ct M}\right\}=\{f,k\}\;\forall f\in\CIF(M;\RR).
\label{hamiltonian.vector.field}
\end{equation}
Such $\XXX_k$ is called a \df{hamiltonian vector field}, while the corresponding $k$ is called the \df{Hamilton function}. If $M$ is infinite dimensional, then the value of $\{\cdot,\cdot\}$ at $x\in M$ depends only on the differentials $\ddd f(x),\ddd k(x)\in\T_x^\ct M$. Hence, there exists a section $\rpktarget{VARPI}\varpi$ of the vector bundle $\bigwedge^2\T^\ct{}^\ct M$ such that
\begin{equation}
        \{f,k\}=\varpi(\ddd f,\ddd k).
\end{equation}
Let $\varpi$ be a \textit{smooth section}, meaning that for each $x\in M$ there exists a continuous bilinear antisymmetric function $\varpi_x:\T_x^\ct M\times\T_x^\ct M\ra\RR$ such that $x\mapsto\varpi_x$ is smooth. Let the function $\natural:\T^\ct M\ra\T^\ct{}^\ct M$ be the bundle map covering identity, isometric on fibres, and satisfying
\begin{equation}
        \natural_x(\ddd f(x)):=\varpi_x(\cdot,\ddd f),
\end{equation}
which means that 
\begin{equation}
        (\natural_x(\ddd f(x)))(\ddd k(x))=\{k,f\}(x)\;\;\forall f,k\in\CIF(M;\RR).
\end{equation}
Then 
\begin{equation}
        \XXX_k:=\varpi(\cdot,\ddd k)=\natural(\ddd k)=\{\cdot,k\}
\label{smooth.section.ttwom}
\end{equation}
is a smooth section of $\T^\ct{}^\ct M$, but may not be a vector field on $M$, because $\T M\subseteq\T^\ct{}^\ct M$ and $\T M\not\iso\T^\ct{}^\ct M$ in general. In order to solve this problem, Odzijewicz and Ratiu \cite{Odzijewicz:Ratiu:2003} proposed (see also further discussion and results in \cite{Odzijewicz:Ratiu:2004,Beltita:Ratiu:2005,Beltita:Ratiu:Tumpach:2008,Odzijewicz:Ratiu:2008,Tumpach:2009,Ratiu:2011,Odzijewicz:2011}) to define a real \df{Banach Poisson manifold} as a pair $(M,\{\cdot,\cdot\})$ of a real Banach smooth manifold $M$ and a Poisson structure $\rpktarget{POISSON.TWO}\{\cdot,\cdot\}$ on it such that the function $\natural:\T^\ct M\ra\T^\ct{}^\ct M$ defined above satisfies $\natural(\T^\ct M)\subseteq\T M$. 

If $(M,\{\cdot,\cdot\})$ is a real Banach Poisson manifold, then every $k\in\CIF(M;\RR)$ determines a unique vector field $\XXX_k\in\T M$ defined by \eqref{smooth.section.ttwom}, and called a \df{hamiltonian vector field}. Such $k$ is then called the \df{Hamilton function}. Every real Poisson manifold is a real Banach Poisson manifold, so this terminology is consistent. Odzijewicz and Ratiu \cite{Odzijewicz:Ratiu:2003} define also a \textit{holomorphic} Banach Poisson manifold which provides an analogous setting for $\KK=\CC$, but we will not use its specific properties here, so we omit its definition. If a Banach space $X$ over $\KK$ is equipped with a Poisson structure $\{\cdot,\cdot\}$ that turns it into a (real or holomorphic) Banach Poisson manifold, then $X^\banach\subseteq\CIF(X;\KK)$. Moreover, $\{\cdot,\cdot\}$ is linear on $\CIF(X;\KK)$ if $\{X^\star,X^\star\}\subseteq X^\star$.

A \df{Banach--Lie--Poisson space} is defined \cite{Odzijewicz:Ratiu:2003} as a pair $(X,\{\cdot,\cdot\})$ such that
\begin{enumerate}
\item[1)] $X$ is a Banach space over $\KK$,
\item[2)] $(X,\{\cdot,\cdot\})$ is a real (if $\KK=\RR$) or holomorphic (if $\KK=\CC$) Banach Poisson manifold,
\item[3)] $X^\banach\subseteq\CIF(X;\KK)$ is a Banach Lie algebra with respect to $\{\cdot,\cdot\}$.
\end{enumerate}
In such case, the restriction of $\{\cdot,\cdot\}$ on $\CIF(X;\KK)$ to $X^\banach$ will be denoted by $[\cdot,\cdot]$. As proved in \cite{Odzijewicz:Ratiu:2003}, the Banach space $X$ is a BLP space $(X,\{\cdot,\cdot\})$ if{}f $X^\banach$ is a Banach Lie algebra $(X^\banach,[\cdot,\cdot])$ satisfying 
\begin{equation}
	\ad_x^\banach(X)\subseteq X\subseteq X^\banach{}^\banach\;\;\forall x\in X^\banach,
\end{equation}
and in such case $\{\cdot,\cdot\}$ is given by
\begin{equation}
        \{f,k\}(z)=\duality{z,[\DF_zf,\DF_zk]}_{X\times X^\banach}\;\;\forall f,k\in\CIF(X;\KK)\;\forall z\in X.
\label{BLP.three}
\end{equation}
Moreover, under those conditions the hamiltonian vector field associated to \textit{any} $k\in\CIF(X;\KK)$ reads\rpktarget{HAMILT.ZWEI}
\begin{equation}
        \XXX_k(z)=-\ad^\banach_{\DF_zk}(z)\;\;\forall z\in X.
\label{BLP.four}
\end{equation}

If $(X,\{\cdot,\cdot\})$ is a BLP space, and if $X\iso\T_xX$ $\forall x\in X$, then $\T_x^\ct X\iso(\T_x X)^\banach\iso X^\banach$, so if $f\in X^\banach$, then one can identify $\ddd_x f\in\T_x^\ct X$ with $\DF_xf\in X^\banach$. As a result, for any $z\in X$ and $y\in X^\banach$ the linearity of $y$ gives $\DF_zy=y$, and for every $x\in X$ one has \cite{Bona:2000,Odzijewicz:Ratiu:2003}
\begin{align}
        \duality{y,\ad^\banach_x(z)}_{X^\banach\times X^\banach{}^\banach}&=\duality{z,[x,y]}_{X\times X^\banach}=\{x,y\}(z)=-\{y,x\}(z)\\
        &=(\XXX_x(y))(z)=\duality{\DF_zy,\XXX_x(z)}_{X^\banach\times X^\banach{}^\banach}=-\duality{y,\XXX_x(z)}_{X^\banach\times X^\banach{}^\banach}.
\end{align}
Hence,
\begin{equation}
        \XXX_x(z)=-\ad^\banach_x(z)\;\;\;\forall z\in X\;\forall x\in X^\banach.
\end{equation}

The notion of the BLP space can be viewed as a suitable generalisation of the important properties of \textit{strong} symplectic manifold to infinite dimensional situation which need not admit decomposition into symplectic leaves. If $(M_1,\{\cdot,\cdot\}_1)$ and $(M_2,\{\cdot,\cdot\}_2)$ are BLP spaces, then a smooth function $w:M_1\ra M_2$ is called a \df{Poisson map} if{}f
\begin{equation}
        \{f\circ w,k\circ w\}_1=\{f,k\}_2\circ w\;\;\forall f,k\in\CIF(M_2;\KK).
\label{BLP.five}
\end{equation}
As shown in \cite{Marsden:Ratiu:1994}, the condition iii) above makes \eqref{BLP.five} equivalent to
\begin{equation}
        \XXX_k\circ w=\T w\circ\XXX_{k\circ w}\;\;\forall k\in\CIF(M_2;\KK).
\end{equation}
If $(M,\{\cdot,\cdot\})$ is a BLP space and $h\in\CIF(M;\KK)$, then the \df{Hamilton equation}
\begin{equation}
        \frac{\dd}{\dd t}f(w^h_t(x))=\{h,f(w^h_t)\}(x)\;\;\forall f\in\CIF(M;\KK)\;\;\forall t\in\RR\;\;\forall x\in M
\label{hamilton.eqn}
\end{equation}
determines a unique \textit{local} map $\rpktarget{WHTM}w^h_t:M\ra M$, called a \df{hamiltonian flow} of $h$, which is a Poisson map. The solutions of the equation $x(t)=w^h_t(x)$ with $x(0)=x$ need not exist \textit{globally}, that is, for all $t\in\RR$ and all $x\in M$. If they exist globally, then the hamiltonian vector field $\{\cdot,h\}$ is called \df{complete}.

\ifvargaugecompile 
\subsubsection{W$^*$-algebra predual as a BLP space}
\else 
\subsubsection{W$^*$-algebra predual as a BLP space}
\fi 
If $M$ is a Banach space, and a Banach smooth manifold modelled on itself by means of an identity mapping then for each $x\in M$ there is a Banach space isomorphism $\T_xM\iso M$. If $\N$ is a W$^*$-algebra, then the Banach Lie algebra structure of $(\N_\star)^\banach\iso\N$ is given by its commutator $[x,y]:=xy-yx$, while $\rpktarget{AD.SECOND}\ad_x:=[x,\,\cdot\,]=\LLL_x-\RRR_x$ and $\ad_x^\banach=\LLL^\banach_x-\RRR^\banach_x$ are defined by weakly-$\star$ continuous maps $\LLL_x:\N\ni y\mapsto xy\in\N$, $\RRR_x:\N\ni y\mapsto yx\in\N$. The condition $\ad_x^\banach(\N_\star)\subseteq\N_\star$ holds for all $x\in\N$, so $\N_\star$ is a BLP space that is a holomorphic Banach Poisson manifold (modelled on itself by the atlas consisting of one chart, an identity mapping $\id_{\N_\star}$) with the Poisson structure given by \eqref{BLP.three},\rpktarget{POISSON.DREI}
\begin{equation}
        \{f,k\}(\phi)=\phi([\DF_\phi f,\DF_\phi k])\;\;\forall f,k\in\CIF(\N_\star;\CC)\;\forall\phi\in\N_\star.
\label{BLP.six}
\end{equation}
As a result, the hamiltonian vector field associated to every $k\in\CIF(\N_\star;\CC)$ by means of \eqref{BLP.four} takes a form\rpktarget{HAMILT.DREI}
\begin{equation}
        \XXX_f(\phi)=-\ad^\banach_{\DF_\phi f}(\phi)=\LLL^\banach_{\DF_\phi f}(\phi)-\RRR^\banach_{\DF_\phi f}(\phi)\;\;\forall\phi\in\N_\star.
\label{BLP.seven}
\end{equation}
These results, including the BLP space structure of $\N_\star$, were discovered by B\'{o}na \cite{Bona:1991,Bona:1993,Bona:2000} in the $\N_\star=\schatten_1(\H)=\BH_\star$ case, and were generalised to arbitrary W$^*$-algebras by Odzijewicz and Ratiu \cite{Odzijewicz:Ratiu:2003}. We will call \eqref{BLP.seven} the \df{B\'{o}na--Odzijewicz--Ratiu equation}. 
If $\N=\BH$ then $\N_\star\iso\schatten_1(\H)$ and for every $\rho\in\schatten_1(\H)$ and every $x,y\in\BH$
\begin{equation}
        \duality{y,-\ad^\banach_x(\rho)}_{\BH\times\BH^\banach}=-\duality{[x,y],\rho}_{\BH\times\BH^\banach}=-\tr_\H([x,y]\rho)=\duality{[x,\rho],y}_{\schatten_1(\H)\times\BH},
\label{BLP.eight}
\end{equation}
which follows from the fact that $\schatten_1(\H)$ is an ideal in $\BH$. As a result,
\begin{equation}
        -\ad_x^\banach(\rho)=[x,\rho],
\end{equation}  
and the BOR equation \eqref{BLP.seven} turns to the Lax equation \cite{Lax:1968}
\begin{equation}
        \XXX_f(\rho)=[\DF_\rho f,\rho]\;\;\forall\rho\in\schatten_1(\H).
\end{equation}
In particular, a choice of the Hamilton function $h(\rho):=\tr_\H(H\rho)$, where $H\in\BH$ but is not necessarily self-adjoint, turns \eqref{hamilton.eqn} to
\begin{equation}
        \frac{\dd}{\dd t}\rho(t)=-\ad^\banach_{\DF_\rho h}(\rho)=[H,\rho].
\end{equation}
\subsubsection{Quantum Poisson systems}
The above equation is derived for a Poisson structure on $\N_\star=\schatten_1(\H)$ viewed as a holomorphic Banach Poisson manifold. However, the standard construction of unitary dynamics in quantum mechanics makes us to be more interested in the real Banach Poisson manifold $\N_\star^\sa$ (and submanifolds of it that are also subsets of $\N^+_\star$), equipped with the Poisson structure coinduced by the action of the Lie algebra $\N^\asa$ of anti-selfadjoint elements of $\N$. More precisely, the set $\N^\uni$ of all unitary elements of a W$^*$-algebra $\N$ is a real Banach Lie group and has a real Lie Banach algebra $\Lie(\N^\uni)=\N^\asa:=\{x\in\N\mid x=-x^*\}$ with $[x,y]:=xy-yx$ \cite{Bourbaki:1972}. The elements of the Banach Lie algebra $\N^\asa$ can be represented by $x\in\N^\sa=\ii\N^\asa$, using the Lie bracket $\N^\sa\times\N^\sa\ni(x,y)\mapsto\ii[x,y]\in\N^\sa$, which corresponds to the commutator $[\ii x,\ii y]=\ii z$ in $\N^\asa$. This algebra has a unique Banach predual, given by $\N_\star^\sa\iso L_1(\N)^\sa:=\{\phi\in L_1(\N)\mid\phi=\phi^*\}$, with an isomorphism $(\N_\star^\sa)^\banach\iso\N^\sa$ defined by duality
\begin{equation}
        \N_\star^\sa\times\N^\sa\ni(\phi,x)\mapsto\duality{\phi,x}_{\N_\star^\sa\times\N^\sa}:=\phi(x)\in\RR.
\label{duality.sa.sa}
\end{equation}
The adjoint representation $\Ad(\N^\uni)$ of a Banach Lie group $\N^\uni$ on $\Lie(\N^\uni)=\ii\N^\sa=\N^\asa$,
\begin{equation}\rpktarget{AD.BIG}
        \Ad(u)x:=uxu^*\;\;\forall u\in\N^\uni\;\forall x\in\N^\sa
\end{equation}
determine the coadjoint representation $\Ad^\banach(\N^\uni)$ on $(\N^\banach)^\sa$,
\begin{equation}
        \duality{x,\Ad^\banach(u)\phi}_{\N^\sa\times(\N^\banach)^\sa}:=\duality{\Ad(u^{-1})x,\phi}_{\N^\sa\times(\N^\banach)^\sa}\;\;\forall x\in\N^\sa\;\forall\phi\in(\N^\banach)^\sa.
\end{equation}
Using these properties, one can show that \cite{Bona:2000,Odzijewicz:Ratiu:2003}
\begin{equation}
        \ad^\banach_x(\N_\star^\sa)\subseteq\N_\star^\sa\subseteq(\N^\asa)^\banach\;\;\forall x\in\N^\asa.
\end{equation}
The space $\N_\star^\sa$ can be equipped with a real Banach smooth manifold structure modelled on itself by the atlas consisting of one chart, which is determined by the identity mapping on $\N_\star^\sa$. As a result, $\T_\phi(\N_\star^\sa)\iso\N_\star^\sa$ $\forall \phi\in\N_\star^\sa$. So, it is possible to use \eqref{BLP.three} and \eqref{duality.sa.sa} to \textit{define} the BLP structure on $\N_\star^\sa$ by
\begin{equation}
        \{f,k\}(\phi):=\ii\phi([\DF_\phi f,\DF_\phi k])\;\;\forall f,k\in\CIF(\N_\star^\sa;\RR)\;\forall\phi\in\N_\star^\sa.
\label{BLP.on.N.sa}
\end{equation}
As a result, the Hamilton equation \eqref{hamilton.eqn} for $h\in\CIF(\N_\star^\sa;\RR)$ reads 
\begin{equation}
        \frac{\dd}{\dd t}f(\phi(t))=\{h,f\}(\phi(t))=\ii(\phi(t))\left([\DF_{\phi(t)}h,\DF_{\phi(t)}f]\right).
\label{BLP.nine}
\end{equation}
The spaces $\N^+_{\star1}$, $\N^+_\star$ and $(\N_\star^\sa,\{\cdot,\cdot\}_{\N_\star^\sa})$ are subsets of $(\N^\banach)^\sa$ that are invariant with respect to $\Ad^\banach(\N^\uni)$. As shown in \cite{Beltita:Ratiu:2005}, they decompose into union of orbits of $\Ad^\banach(\N^\uni)$, which in turn are weak symplectic manifolds, which provides the symplectic foliation of the BLP space $(\N_\star^\sa,\{\cdot,\cdot\}_{\N_\star^\sa})$. Similarly, $(\N_\star,\{\cdot,\cdot\}_{\N_\star})$ is invariant with respect to the action of the Banach Lie group $\N^\inv$ of all invertible elements of $\N$. If $\N=\BH$ and $\rho\in\schatten_1(\H)^\sa$, then the calculation analogous to \eqref{BLP.eight} gives
\begin{equation}
        \ad_x^\banach(\rho)=[\rho,x]\;\;\forall x\in\BH^\asa\;\forall \rho\in\schatten_1(\H)^\sa.
\end{equation}
As a result, the BOR equation \eqref{BLP.seven} on $\N_\star^\sa=\schatten_1(\H)^\sa$ takes the form
\begin{equation}
        \XXX_f(\rho)=[\rho,\DF_\rho f]\;\;\forall\rho\in\schatten_1(\H),
\end{equation}
while the Hamilton equation \eqref{BLP.nine} becomes \cite{Bona:2000}
\begin{equation}
        \frac{\dd}{\dd t}f(\rho(t))=\ii\,\tr_\H\left([\rho(t),\DF_{\rho(t)}h]\DF_{\rho(t)}f\right).
\label{BLP.ten}
\end{equation}
Because of the identity
\begin{equation}
        \frac{\dd}{\dd t}f(\rho(t))=\tr_\H\left((\DF_{\rho(t)}f)\frac{\dd}{\dd t}\rho(t)\right),
\end{equation}
the equation \eqref{BLP.ten} is equivalent to the \df{B\'{o}na equation} \cite{Bona:1991,Jordan:1993,Bona:2000},
\begin{equation}
        \ii\frac{\dd}{\dd t}\rho(t)=[\DF_{\rho(t)}h,\rho(t)].
\label{gen.vN}
\end{equation}
The solutions of \eqref{gen.vN} are state-dependent unitary operators $U(\rho,t)$. They do not form a group, but satisfy a cocycle relationship:
\begin{equation}
	U(\rho,t+s)=U((\Ad(U(\rho,t)))(\rho),s)U(\rho,t)\;\;\forall t,s\in\RR.
\end{equation}
In the special case, when $h(\rho)=\tr_\H(\rho H)$ for $H\in\BH^\sa=\ii\BH^\asa$, \eqref{gen.vN} turns to the \df{von Neumann equation}
\begin{equation}
        \ii\frac{\dd}{\dd t}\rho(t)=[H,\rho(t)].
\label{vN.equation.from.BLP}
\end{equation}

So far we have followed the B\'{o}na--Odzijewicz--Ratiu approach, hence our main object of interest was the real Banach manifold $\N^\sa_\star$, equipped with the BLP space structure coinduced by the Banach--Lie algebra $\N^\asa$ via $\N^\sa$, corresponding to the group $\N^\uni$ of unitary elements of $\N$. However, in principle, a geometric setting for nonlinear dynamics of quantum models can be generated by an arbitrary Banach Lie algebra $\B$ over $\RR$ such that:
\begin{enumerate}
\item[(i)] its Banach predual space $\B_\star$ exists, is unique, and is a real Banach Poisson manifold,
\item[(ii)] $\ad^\star_x(\B_\star)\subseteq\B_\star\subseteq\B^\star$ $\forall x\in\B$,
\item[(iii)] there exists a nonempty set $\M(\N,\B)\subseteq\N^+_\star$ that is a real BLP submanifold of $\B_\star$.
\end{enumerate}
In what follows, we will call such Banach Lie algebras $\B$ to be \df{well-adapted}. For any choice of $h\in\CIF(\M(\N,\B);\RR)$, the pair $(\M(\N,\B),h)$ will be called a \df{quantum Poisson system} whenever 
\begin{equation}
	w^h_t(\phi)\in\M(\N,\B)\;\;\forall \phi\in\M(\N,\B)\;\;\forall t\in\RR.
\label{quantum.Poisson.system.condition}
\end{equation}
Hence, each quantum Poisson system is determined by a choice of: a space of quantum states $\M(\N)\subseteq\N^+_\star$, a Banach Lie algebra $\B$, a tangent bundle (real Banach manifold) structure on $\M(\N)$, and a real Fr\'{e}chet smooth function on $\M(\N)$, satisfying the conditions (i), (ii), (iii), and \eqref{quantum.Poisson.system.condition}. The assumptions $\B\iso\N^\sa$ and $\T_\phi\N^\sa_\star:=\N_\star^\sa$ $\forall\phi\in\N^\sa_\star$ recover the BOR setting completely. Note that a general quantum Poisson system $(\M(\N,\B),h)$ does not have to be related to any group, so in particular to a group of unitary operators $\N^\uni$. The only shared property (securing the backwards compatibility with quantum mechanical setting) is implementation of the Poisson flow on the predual by means of a coinduced action of a Banach--Lie algebra.

\ifvargaugecompile 
\else 
\subsubsection{Symplectic submanifolds}
[[write about symplectic orbits (and maybe also lagrangeans!)]]
[[write also about the solutions to B\'{o}na equation etc!]]
\fi 
%

\ifvargaugecompile 
\subsection{Relative modular operators, standard liouvilleans and the GNS bundle\label{standard.liouvilleans.section}}
A \df{weight} on a W$^*$-algebra $\N$ is defined as a function $\omega:\N^+\ra[0,+\infty]$ such that $\omega(0)=0$, $\omega(x+y)=\omega(x)+\omega(y)$, and $\lambda\geq0\limp\omega(\lambda x)=\lambda\omega(x)$, with the convention $0\cdot(+\infty)=0$. A weight is called: \df{faithful} if{}f $\omega(x)=0\limp x=0$; \df{finite} if{}f $\omega(\II)<\infty$; \df{semi-finite} if{}f a left ideal in $\N$ given by $\nnn_\phi:=\{x\in\N\mid\phi(x^*x)<\infty\}$ is weakly-$\star$ dense in $\N$; \df{normal} if{}f $\omega(\sup\{x_\iota\})=\sup\{\omega(x_\iota)\}$ for any uniformly bounded increasing net $\{x_\iota\}\subseteq\N^+$. A space of all normal semi-finite weights on a W$^*$-algebra $\N$ is denoted $\W(\N)$, while the subset of all faithful elements of $\W(\N)$ is denoted $\W_0(\N)$. Hence, $\N_\star^+\subset\W(\N)$ and $\N_{\star0}^+\subset\W_0(\N)$.  For $\psi\in\W(\N)$, $\supp(\psi)=\II-\sup\{P\in\Proj(\N)\mid\psi(P)=0\}$. An element $\omega\in\N^+_\star$ is faithful if{}f $\supp(\omega)=\II$. 

A \df{representation} of a W$^*$-algebra $\N$ is defined as a pair $(\H,\pi)$ of a Hilbert space $\H$ and a $*$-homomorphism $\rpktarget{pi}\pi:\N\ra\BH$. A representation $\pi:\N\ra\BH$ is called: \df{nondegenerate} if{}f $\{\pi(x)\xi\mid (x,\xi)\in\N\times\H\}$ is dense in $\H$; \df{normal} if{}f it is continuous with respect to the weak-$\star$ topologies of $\N$ and $\BH$; \df{faithful} if{}f $\ker(\pi)=\{0\}$. An element $\xi\in\H$ is called \df{cyclic} for a W$^*$-algebra $\N\subseteq\BH$ if{}f $\N\xi:=\bigcup_{x\in\N}\{x\xi\}$ is norm dense in $\BH$. A representation $\pi:\N\ra\BH$ is called \df{cyclic} if{}f there exists $\Omega\in\H$ that is cyclic for $\pi(\N)$. According to the Gel'fand--Na\u{\i}mark--Segal theorem \cite{Gelfand:Naimark:1943,Segal:1947:irreducible} for every pair $(\N,\omega)$ of a W$^*$-algebra $\N$ and $\omega\in\N^{\banach+}$ there exists a triple $(\H_\omega,\pi_\omega,\Omega_\omega)$ of a Hilbert space $\H_\omega$ and a cyclic representation $\pi_\omega:\N\ra\BH$ with a cyclic vector $\Omega_\omega\in\H_\omega$\rpktarget{h.omega}, and this triple is unique up to unitary equivalence. An analogue of this theorem for weights follows the similar construction, but lacks cyclicity. If $\omega$ is a weight on a W$^*$-algebra $\N$, then there exists the Hilbert space $\H_\omega$, defined as the completion of $\nnn_\omega/\ker(\omega)$ in the topology of a norm generated by the scalar product $\s{\cdot,\cdot}_{\omega}:\nnn_\omega\times\nnn_\omega\ni(x,y)\mapsto\omega(x^*y)\in\CC$,
\begin{equation}\rpktarget{h.omega.zwei}
        \H_\omega:=\overline{\nnn_\omega/\ker(\omega)}=\overline{\{x\in\N\mid\omega(x^*x)<\infty\}/\{x\in\N\mid\omega(x^*x)=0\}}=\overline{\nnn_\omega/\I_\omega},
\end{equation}
and there exist the maps\rpktarget{rep.omega.zwei}\rpktarget{pi.omega.zwei} 
\begin{align}
        [\cdot]_\omega:\nnn_\omega\ni x&\mapsto [x]_\omega\in\H_\omega,
        \label{GNS.class.weight}\\
        \pi_\omega:\N\ni x&\mapsto([y]_\omega\mapsto[xy]_\omega)\in\BBB(\H_\omega),
        \label{GNS.rep.weight}
\end{align}
such that $[\cdot]_\omega$ is linear, $\ran([\cdot]_\omega)$ is dense in $\H_\omega$, and $(\H_\omega,\pi_\omega)$ is a representation of $\N$. If $\omega\in\W(\N)$ then $(\H_\omega,\pi_\omega)$ is nondegenerate and normal. It is also faithful if $\omega\in\W_0(\N)$.

A standard representation \cite{Haagerup:1975:standard:form} of a W$^*$-algebra $\N$ is defined as a quadruple $(\H,\pi,J,\stdcone)$ of a Hilbert space $\H$, a nondegenerate faithful weakly-$\star$ continuous representation $\pi:\N\ra\BH$, a conjugation $J:\H\ra\H$, and a self-polar cone\footnote{A subspace $\D$ of a Hilbert space $\H$ is called a self-polar cone if{}f $\lambda\xi\in\D$ $\forall\xi\in\D$ $\forall\lambda\geq0$ and $\D=\{\zeta\in\H\mid\s{\xi,\zeta}_\H\geq0\;\forall\xi\in\D\}$.}  $\stdcone\subseteq\H$, satisfying the conditions
\begin{equation}
	J\pi(\N) J=\pi(\N)^\comm,\;\;
	\xi\in\stdcone\;\limp\;J\xi=\xi,\;\;
	\pi(x)J\pi(x)\stdcone\subseteq\stdcone,\;\;
	y\in\zentr(\pi(\N))\;\limp\;JyJ=y^*.
\end{equation}
For any standard representation 
\begin{equation}
	\forall\phi\in\N^+_\star\;\;\exists!\stdembed_\pi(\phi)\in\stdcone\;\;\forall x\in\N\;\;\phi(x)=\s{\stdembed_\pi(\phi),\pi(x)\stdembed_\pi(\phi)}_\H.
\label{std.vector.representative}
\end{equation}
The map $\stdembed_\pi:\N^+_\star\ra\stdcone$ is order preserving. Moreover, $\stdembed^\natural_\pi:\stdcone\ra\N_\star^+$, defined by $(\stdembed^\natural_\pi(\xi))(x)=\s{\xi,\pi(x)\xi}_\H$ $\forall x\in\N$, is a bijective norm continuous homomorphism with $(\stdembed^\natural_\pi)^{-1}=\stdembed_\pi$. For any two standard representations $(\H_1,\pi_1,J_1,\H^\natural_1)$ and $(\H_2,\pi_2,J_2,\H^\natural_2)$ of a W$^*$-algebra $\N$ and a given $^*$-isomorphism $\varsigma:\pi_1(\N)\ra\pi_2(\N)$, there exists a unique unitary $u_\varsigma:\H_1\ra\H_2$ such that $\varsigma(x)=u_\varsigma xu_\varsigma^*$ $\forall x\in\pi_1(\N)$, $J_2=u_\varsigma J_1u_\varsigma^*$, $\H^\natural_2=u_\varsigma\H^\natural_1$. Such $u_\varsigma$ will be called a standard unitary equivalence. If $\phi\in\N^+_{\star0}$ then, by means of the Tomita--Takesaki theory \cite{Tomita:1967:b,Takesaki:1970}, the GNS representation associated with $\phi$ determines a unique conjugation $J_\phi$, and a weakly-$\star$ continuous group homomorphism $\sigma^\omega:\RR\ra\Aut(\N)$. An associated $\stdcone$ is given by \cite{Connes:1974,Araki:1974:modular:conjugation} 
\begin{equation}
	\H^\natural_\phi:=\overline{\bigcup_{x\in\nnn_\phi\cap\nnn_\phi^*}\{\pi_\phi(x)J_\phi\pi_\phi(x)J_\phi\Omega_\phi\}}^{\H_\phi}.
\label{Connes.Araki.natural.cone}
\end{equation}
If $\N\iso\BBB(\K)$ for some Hilbert space $\K$ and $\phi\in\W_0(\N)$ is given by $\tr_\K$, then the corresponding GNS Hilbert space $\H_\phi$ is given by the space of all Hilbert--Schmidt operators, 
\begin{equation}
	\schatten_2(\K):=\{x\in\BBB(\K)\mid(\tr_\K(x^*x))^{1/2}<\infty\}=\nnn_{\tr_\K},
\end{equation}
equipped with a scalar product $\s{x,y}_{\schatten_2(\K)}:=\tr_\K(x^*y)$, so that $\schatten_2(\K)\iso\K\otimes\K^\banach$ as Hilbert spaces. Moreover, $\H^\natural_\phi=\schatten_2(\K)^+$, $\pi_\phi(x)=\LLL_x$ (which denotes left multiplication by $x$), while $\stdembed_\pi:\schatten_1(\K)^+\ni\rho\mapsto\rho^{1/2}\in\schatten_2(\K)^+$. 
\
For a given W$^*$-algebra $\N$, $\phi\in\W(\N)$, and $\omega\in\W_0(\N)$ the map
\begin{equation}
        R_{\phi,\omega}:[x]_\omega\mapsto[x^*]_\phi\;\;\forall x\in\nnn_\omega\cap\nnn_\phi^*
        \label{relative.modular.weights}
\end{equation}
is a densely defined, closable antilinear operator. Its closure admits a unique polar decomposition
\begin{equation}
        \overline{R}_{\phi,\omega}=J_{\phi,\omega}\Delta^{1/2}_{\phi,\omega},
\end{equation}
where $\rpktarget{JREL}J_{\phi,\omega}$ is a conjugation operator, called \df{relative modular conjugation}, while $\rpktarget{DELTAREL}\Delta_{\phi,\omega}$ is a positive self-adjoint operator on $\dom(\Delta_{\phi,\omega})\subseteq\H_\omega$ with $\supp(\Delta_{\phi,\omega})=\supp(\phi)\H_\omega$, called a \df{relative modular operator} \cite{Araki:1973:relative:hamiltonian,Connes:1974,Digernes:1975}. Given $\omega\in\W_0(\N)$, $\Delta_{\omega,\omega}=:\Delta_\omega$ implements the action of the Tomita--Takesaki (modular) automorphism $\sigma^\omega$ by
\begin{equation}
	\pi_\omega(\sigma^\omega_t(x))=\Delta_\omega^{\ii t}\pi_\omega(x)\Delta_\omega^{-\ii t}.
\end{equation}
If $\N\iso\BH$, $\phi=\tr_\H(\rho_\phi\,\cdot)$, $\omega=\tr_\H(\rho_\omega\,\cdot)$, and $\RRR_x$ denotes right multiplication by $x\in\BH$, then $\Delta_{\phi,\omega}=\LLL_{\rho_\phi}\RRR_{\rho_\omega^{-1}}$.

For every $\phi,\omega\in\W_0(\N)$ the relative modular conjugation $J_{\phi,\omega}$ determines a unique unitary operator $\rpktarget{STDUNITRANS}J_{\phi,\phi}J_{\phi,\omega}=:V_{\phi,\omega}:\H_\omega\ra\H_\phi$, such that
\begin{align}
        \pi_\phi(x)&=V_{\phi,\omega}\pi_\omega(x)V_{\phi,\omega}^*,\label{std.uni.trans.eq.one}\\
        V_{\phi,\omega}(\stdcone_\omega)&=\stdcone_\phi,\label{std.uni.trans.eq.two}\\
        V_{\phi,\omega}J_{\omega,\omega}&=J_{\phi,\phi}V_{\phi,\omega}.\label{std.uni.trans.eq.three}
\end{align}
We will call $V_{\phi,\omega}$ \df{standard unitary transition} between $\H_\omega$ and $\H_\phi$. It is a standard unitary equivalence of a $^*$-isomorphism $\varsigma_{\phi,\omega}:\pi_\omega(\N)\ra\pi_\phi(\N)$ determined by the condition $\varsigma_{\phi,\omega}\circ\pi_\omega=\pi_\phi$. Thus, if $\phi,\omega\in\N^+_{\star0}$, then $V_{\phi,\omega}$ provides a default unitary mapping between the corresponding GNS Hilbert spaces and representations.

Given any group 
 $G$, a \df{representation} of $G$ in the group $\Aut(\N)$ of $*$-automorphisms of a W$^*$-algebra $\N$ is a map $\alpha:G\ni g\mapsto\alpha(g)=:\alpha_g\in\Aut(\N)$ which is a \df{group homomorphism}, that is,
\begin{enumerate}
        \item[1)] $\alpha(e)=\id_\N$,
        \item[2)] $\alpha(g_1)\circ\alpha(g_2)=\alpha(g_1\circ g_2)$ $\forall g_1,g_2\in G$,
\end{enumerate}
where $e$ denotes the neutral element of $G$. A group $G$ is called: \df{topological} if{}f it is also a topological space and a map $G\times G\ni(g_1,g_2)\mapsto g_1\circ g_2^{-1}\in G$ is continuous for all $g_1,g_2\in G$; \df{locally compact} if{}f it is topological and $e\in G$ has a compact 
 topological neighbourhood. For any W$^*$-algebra $\N$, $\Aut(\N)$ is a topological group with respect to \df{weak-$\star$ topology} on $\Aut(\N)$, defined by the collection of neighbourhoods \cite{Takesaki:1983}
\begin{equation}
        N_{\{\omega_i\}}(\alpha):=
        \{\varsigma\in\Aut(\N)\mid
        \n{\omega_i\circ\alpha-\omega_i\circ\varsigma}_{\N_\star}<1,\;
        \n{\omega_i\circ\alpha^{-1}-\omega_i\circ\varsigma^{-1}}_{\N_\star}<1\},
\end{equation}
where $\{\omega_i\}\subseteq\N_\star$, $i\in\{1,\ldots,n\}$, $n\in\NN$. A triple $(\N,G,\alpha)$ of a W$^*$-algebra, locally compact group $G$, and a representation $\alpha:G\ra\Aut(\N)$ is called a \df{W$^*$-dynamical system} (or a \df{W$^*$-covariant system}) if{}f $\alpha$ is continuous in the weak-$\star$ topology of $\Aut(\N)$. This condition is equivalent to the continuity of the map $G\ni g\mapsto\alpha_g(x)\in\N$ in the weak-$\star$ topology of $\N$ for any $x\in\N$, that is, to 
\begin{equation}
        G\ni g\mapsto\phi(\alpha_g(x))\in\CC\mbox{ is a continuous function}\;\;\forall x\in\N,
\label{weak.star.continuous.group}
\end{equation}
and such $\alpha$ is called a \df{weakly-$\star$ continuous} representation. Uniqueness of a predual of a W$^*$-algebra $\N$ allows to define isometries $\alpha_\star$ of $\N_\star$ that uniquely correspond to the elements $\alpha\in\Aut(\N)$, and to define the isometries of $\N_\star$ uniquely corresponding to representations $\alpha:G\ra\Aut(\N)$:
\begin{equation}
        \duality{(\alpha_g)_\star(\phi),x}_{\N_\star\times\N}=\duality{\phi,\alpha_g(x)}_{\N_\star\times\N}=\phi(\alpha_g(x))\;\;\forall x\in\N\;\forall\phi\in\N_\star.
\end{equation}
The above equivalence can be shown (see e.g. \cite{Sakai:1991}) by proving that \eqref{weak.star.continuous.group} implies continuity of $\alpha_\star$ in the norm of $\N_\star$,
\begin{equation}
        \lim_{g\ra e}\n{(\alpha_g)_\star(\phi)-\phi}_{\N_\star}=0\;\;\forall\phi\in\N_\star\;\forall g\in G.
\end{equation}

A \df{unitary implementation} of a representation $\alpha:G\ra\Aut(\N)$ in a given representation $\pi:\N\ra\BH$ is defined as a map $u:G\ni g\mapsto u(g)\in\BH^\uni$ that determines a family $\{u(g)\mid g\in G\}$ of unitary operators satisfying the \df{covariance equation}
\begin{equation}
        \pi(\alpha_g(x))=u(g)\pi(x)u(g)^*\;\;\forall x\in\N\;\forall g\in G.
\label{covariance.equation.group.G}
\end{equation}
The condition \eqref{covariance.equation.group.G} alone does not determine $\{u(g)\mid g\in G\}$ uniquely. The setting of W$^*$-algebras admits a remarkable solution to this problem: every pair of a W$^*$-dynamical system $(\N,\RR,\alpha)$ and a standard representation $(\H,\pi,J,\stdcone)$ determines uniquely a corresponding unitary implementation \textit{together with} a unique self-adjoint generator of this family of unitaries. This generator is called a \textit{standard liouvillean}\footnote{It would be however more precise to call it \textit{quantum koopmanian}, because in the commutative setting (of statistical mechanics and probability measures) the `liouvillean operator' (defined by the Poisson bracket) acts on elements of $L_1(\X,\mho(\X),\tmu)$, while it is the `koopmanian operator' \cite{Koopman:1931,vonNeumann:1932:zur,vonNeumann:1932:zusaetze} that acts on the positive cone of $L_1(\X,\mho(\X),\tmu)$.}. It is not called `hamiltonian', because in general its spectrum may be not bounded from any side, while the notion of  `hamiltonian' is usually understood as referring to a self-adjoint operator that generates a strongly continuous group of unitary operators \textit{and} has a nonnegative (or at least bounded from below) spectrum\footnote{E.g., \cytat{one of the most important principles of quantum field theory, ensuring the stability, demands that the energy should have a lower bound} \cite{Haag:1992}.}. Moreover, as opposed to hamiltonian, the construction of standard liouvillean for a given W$^*$-dynamical system does not require any additional analytic conditions that constrain derivation to an `integrable' infinitesimal generator. This way the W$^*$-algebraic approach makes the notion of a hamiltonian less relevant than the notion of a liouvillean. 

For any W$^*$-algebra $\N$, the unique predualisation of action of $\alpha\in\Aut(\N)$ can be connected with the uniqueness property of representation of elements of $\N_\star^+$ in terms of a standard cone of a standard representation $(\H,\pi,J,\stdcone)$ of $\N$: any $\alpha\in\Aut(\N)$ defines a unique map $u:\stdcone\ra\stdcone$ by 
\begin{equation}
        u\stdembed_\pi(\phi):=\stdembed_\pi(\alpha_\star(\phi))\;\;\;\forall\phi\in\N_\star^+.
\end{equation} 
This map is linear, can be extended to a unitary operator on all $\H$, and satisfies 
\begin{equation}
        u\pi(x)u^*=\pi(\alpha(x))\;\;\forall x\in\N.
\end{equation}
This leads to a question, whether it is possible to generate this way a \textit{standard} unitary implementation of a given representation $\alpha:G\ra\Aut(\N)$. The answer is in the affirmative, and was established by Haagerup \cite{Haagerup:1975:standard:form} (the special cases of this result were obtained earlier in \cite{Kadison:1965,Kallman:1971,Henle:1970,Halpern:1972,Pedersen:Takesaki:1973}). If $(\H,\pi,J,\stdcone)$ is a standard representation of a W$^*$-algebra $\N$, then there exists a unique  strongly continuous unitary implementation $\rpktarget{V.ALPHA}V_\alpha(g)$ of $\alpha$ satisfying
\begin{align}
        V_\alpha(g)\stdcone&=\stdcone,\\
        JV_\alpha(g)&=V_\alpha(g)J.
\end{align}
Such family $\{V_\alpha(g)\mid g\in G\}$ is called a \df{standard} unitary implementation of $\alpha$.

Thus, if $(\N,\RR,\alpha)$ is a W$^*$-dynamical system with $\N$ in standard form $(\H,\pi(\N),J,\stdcone)$, then from the theorems of Haagerup and Stone \cite{Stone:1930,Stone:1932,vonNeumann:1932:Stone} it follows that there exists a unique strongly continuous group of unitaries $\{V_\alpha(t)\mid t\in\RR\}\subseteq\BH^\uni$, and a unique self-adjoint operator $\rpktarget{K.ALPHA}K^\alpha$ on $\H$, called \df{standard liouvillean}, such that $V_\alpha(t)$ is a strongly continuous unitary implementation of $\alpha$ and for every $t\in\RR$
\begin{enumerate}
\item[i)] $V_\alpha(t)=\ee^{-\ii tK^\alpha}$,
\item[ii)] $\ee^{-\ii tK^\alpha}\stdcone=\stdcone$,
\item[iii)] $JK^\alpha+K^\alpha J=0$.
\end{enumerate}
The definition of a standard liouvillean $K^\alpha$ does not depend on any choice of $\omega\in\N^+_\star$ or $\omega\in\W(\N)$: it depends only on a W$^*$-dynamical system and a standard representation of W$^*$-algebra. If $\N$ is semi-finite, $(\H,\pi,J,\stdcone)$ is its standard representation, $(\N,\RR,\alpha)$ is a W$^*$-dynamical system, $H\in\pi(\N)^\sa$, and $\{U(t):=\ee^{-\ii t H}\in\pi(\N)\mid t\in\RR\}$ is a strongly continuous group of unitary operators such that 
\begin{equation}
	\ee^{\ii tH}\pi(x)\ee^{-\ii tH}=\pi(\alpha_t(x))\;\;\forall x\in\N\;\forall t\in\RR,
\end{equation}
then the standard liouvillean reads \cite{Bratteli:Robinson:1979,JOPP:2012}
\begin{equation}
	K^\alpha=H-JHJ=[H,\,\cdot\,].
\label{std.liouvillean.for.BH}
\end{equation}

\else 

\fi 
Following Odzijewicz and Sli\.zewska \cite{Odzijewicz:Slizewska:2011}, consider a bundle $\eee:V\ra\N_\star^+$, where
\begin{equation}
        V:=\{(x,\omega)\in\N\times\N_\star^+\mid x\,\supp(\omega)=x\},
\end{equation}
and the bundle projection $\eee$ is given by a restriction of the cartesian product projection $\N\times\N_\star^+\ra\N_\star^+$ to $V$. Because $V_\omega:=\eee^{-1}(\omega)=\N\supp(\omega)$ $\forall\omega\in\N_\star^+$, the scalar product
\begin{equation}
        V_\omega\times V_\omega\ni(x,y)\mapsto\s{x,y}_\omega:=\duality{\omega,x^*y}_{\N_\star\times\N}\in\CC
\end{equation}
is nondegenerate. Moreover, $\s{x,x}_\omega=0$ $\iff$ $x\in\N(\II-\supp(\omega))$. The completion $\bar{V}_\omega$ of $V_\omega$ under the norm generated by $\s{\cdot,\cdot}_\omega$ determines a bundle $\rpktarget{H.MN}\H\N_\star^+:=\bar{V}\ra\N_\star^+$ of Hilbert spaces, which the authors of \cite{Odzijewicz:Slizewska:2011} call the \df{Gel'fand--Na\u{\i}mark--Segal bundle} (the notion of the GNS bundle was earlier alluded in \cite{Jadczyk:1988,Chruscinski:Marmo:2009}).

While \eqref{Connes.Araki.natural.cone} secures that the GNS representation is a standard representation whenever $\phi\in\N^+_{\star0}$, it doesn't have to be in more general case. Thus, in order to be sure that our use of GNS bundle coincides with the necessary conditions for Haagerup's theorem, we will restrict our discussion in multiple places of this paper to subsets and submanifolds of $\N^+_{\star0}$. We consider this restriction to be nonoptimal, but in order to work it out in larger generality, we would have to work with a different bundle of Hilbert spaces. Restriction to $\N^+_{\star0}$ allows us to use standard unitary transitions $V_{\phi,\omega}$ to map between Hilbert spaces, at the expense of consideration of unitarily equivalent representations only. Whenever the assumption of restriction to $\N^+_{\star0}$ is made, it implies restriction of considerations to countably finite W$^*$-algebras, because only for them $\N^+_{\star0}\neq\varnothing$.
\ifvargaugecompile 
\subsection{Case study: Algebraic hamiltonian vector fields\label{algebraic.hamiltonian}}
\else 
\subsubsection{Algebraic hamiltonian vector fields\label{algebraic.hamiltonian}}
\fi 
The BLP structure of $\N_\star$ and $\N_\star^\sa$ allows to introduce and analyse the temporal evolution on $\N_\star^+$ by means of the hamiltonian vector field and the Hamilton equation. On the other hand, %
\ifvargaugecompile 
%
\else 
as discussed in Sections \ref{automorphisms.section} and \ref{standard.liouvilleans.section},
\fi 
 for any W$^*$-dynamical system $(\N,\RR,\alpha)$ one can predualise the representation $\alpha:
\RR\ni t\mapsto\alpha_t\in\Aut(\N)$ obtaining the family $\{\alpha_\star^t\mid t\in\RR\}$ of norm continuous isometries $\alpha_\star^t:=(\alpha_t)_\star:\N_\star\ra\N_\star$, which in turn can be analysed by means of a unique self-adjoint standard liouvillean operator $K^\alpha$ that generates a unitary evolution in $L_2(\N)$ leaving $L_2(\N)^+$ invariant. The virtue of a geometric description in terms of hamiltonian vector field is that it allows for an analysis of the local differential structure of temporal evolution in terms of a local Poisson flow and tangent space. However, it does not guarantee the existence of global flow. On the other hand, an algebraic description in terms of a predualised representation $\alpha_\star$ and an associated standard liouvillean $K^\alpha$ guarantees the existence of a global flow on $\N_\star^+$, but it is not necessarily a Poisson flow and it is not related to a tangent space, thus it does not allow (in general) for a refined smooth geometric description. This leads us to single out the class of evolutions on $\N_\star^\sa$ (and $\N_\star^+$) that satisfy both conditions. 

The isometries $\alpha_\star^t$ of $\N_\star$ that are also the Poisson flows leaving $\N_\star^\sa$ invariant are characterised as solutions of the equation \eqref{BLP.five}
\begin{equation}
        \{f\circ\alpha_\star^t,k\circ\alpha_\star^t\}_{\N_\star^\sa}=\{f,k\}_{\N_\star^\sa}\circ\alpha_\star^t\;\;\forall f,k\in\CIF(\N_\star^\sa;\RR)\;\;\forall t\in\RR,
\end{equation}
which gives, by \eqref{BLP.on.N.sa},
\begin{align}
        \phi([\DF_\phi(f\circ\alpha_\star^t),\DF_\phi(k\circ\alpha_\star^t)])=(\alpha_\star^t(\phi))([\DF_{\alpha_\star^t(\phi)},\DF_{\alpha^t_\star(\phi)}k]),\\
        0=\phi([\DF_\phi(f\circ\alpha_\star^t),\DF_\phi(k\circ\alpha_\star^t)]-\alpha_t([\DF_{\alpha_\star^t(\phi)}f,\DF_{\alpha_\star^t(\phi)}k])).
\end{align}
Hence, the predualisation $\alpha_\star$ of a weakly-$\star$ continuous representation $\alpha:\RR\ra\Aut(\N)$ is a Poisson flow on $(\N_\star^\sa,\{\cdot,\cdot\})$ if{}f $\alpha$ satisfies
\begin{equation}
        \phi((\id_\N-\alpha_t)([\DF_{\alpha_\star^t(\phi)}f,\DF_{\alpha_\star^t(\phi)}k]))=0\;\;\;\forall\phi\in\N_\star^\sa\;\;\forall f,k\in\CIF(\N_\star^\sa;\RR)\;\;\forall t\in\RR.
\label{poisson.flow.compatibility}
\end{equation}
We will call \eqref{poisson.flow.compatibility} the \df{Poisson compatibility condition} (PC$_1$). Let $(\N,\RR,\alpha)$ be a W$^*$-dynamical system satisfying the Poisson compatibility condition. Then the Poisson flow $\alpha_\star|_{\N_\star^\sa}$ is generated by the Hamilton function $\rpktarget{HA.ALPH}h^\alpha\in\CIF(\N_\star^\sa;\RR)$ according to \eqref{hamilton.eqn},
\begin{equation}
        \frac{\dd}{\dd t}f_t=\{h^\alpha,f_t\},\;\;f_t(x):=f(\alpha^t_\star(x))\;\;\forall t\in\RR\;\forall x\in\N_\star\;\forall f\in\CIF(\N_\star^\sa;\RR),
\end{equation}
which determines the corresponding unique hamiltonian vector field $\XXX_{h^\alpha}\in\T\N_\star^\sa$ by means of the BOR equation \eqref{BLP.seven},
\begin{equation}
        \XXX_{h^\alpha}(\phi)=-\ad^\banach_{\DF_\phi h^\alpha}(\phi)=\LLL^\banach_{\DF_\phi h^\alpha}(\phi)-\RRR^\banach_{\DF_\phi h^\alpha}(\phi)\;\;\forall\phi\in\N_\star^\sa.
\end{equation}
We will call $\XXX_{h^\alpha}$ an \df{algebraic hamiltonian vector field}.

Now, let us recall from Section \ref{standard.liouvilleans.section} that, by the Haagerup theorem, each pair of a W$^*$-dynamical system $(\N,\RR,\alpha)$ and a standard representation $(\H,\pi,J,\stdcone)$ of $\N$ determines a unique self-adjoint generator $K^\alpha$ of a unitary implementation of $\alpha$ in $\BH$ satisfying the condition $\ee^{-\ii tK^\alpha}\stdcone\subseteq\stdcone$. By means of \eqref{std.vector.representative}, this condition expresses the requirement that $\alpha_\star(\N_\star^+)\subseteq\N_\star^+$ (if formulated in terms of Kosaki's canonical representation, for which $\stdcone= L_2(\N)^+$, this condition is just an $L_2(\N)$ version of $L_1(\N)^+=\N_\star^+$ invariance under $\alpha$). If $\omega\in\N^+_{\star0}$, then the GNS representation, $(\H_\omega,\pi_\omega,\Omega_\omega)$ is also a standard representation $(\H_\omega,\pi_\omega,J_\omega,\H^+_\omega)$, so we can apply Haagerup's theorem to the fibres of the GNS bundle $\H\N_\star^+$ restricted to the submanifold $\N^+_{\star0}$, $\eee:\H\N^+_{\star0}\ra\N^+_{\star0}$. Because $\supp(\omega)=\II$ for each $\omega\in\N^+_{\star0}$, the bundle projection $\eee$ reduces in this case to a cartesian product projection. Orbits of any Poisson flow leave $\N^+_{\star0}\subseteq\N_\star^\sa$ invariant \cite{Bona:2000,Beltita:Ratiu:2005,Ratiu:2011}, while $\alpha_\star^t$ is norm preserving, so the restrictions of Poisson compatible isometries $\alpha^t_\star$ to $\N^+_{\star0}$ are automorphisms of this space. As a result, we obtain a remarkable geometric correspondence: every weakly-$\star$ continuous representation $\alpha:\RR\ni t\mapsto\alpha_t\in\Aut(\N)$ satisfying the Poisson compatibility condition \eqref{poisson.flow.compatibility} determines a unique globally integrable hamiltonian vector field $\XXX_{h^\alpha}\in\T\N_\star^\sa$ \textit{and} a family of standard liouvillean operators $\rpktarget{K.OMEGA.ALPHA}\N^+_{\star0}\ni\omega\mapsto K_\omega^\alpha\in(\Lin(\H_\omega))^\sa$ acting pointwise on the GNS bundle of Hilbert spaces. In other words, the family $\{\alpha_t\in\Aut(\N)\mid t\in\RR\}$ of Poisson compatible, weakly-$\star$ continuous automorphisms of a W$^*$-algebra $\N$ is uniquely represented \textit{in} the tangent vector bundle $\T\N_\star^\sa\ra\N_\star^\sa$ (and, by linearity, also in $\T\N_\star^+\ra\N_\star^+$), as well as \textit{on} the GNS Hilbert bundle $\H\N^+_{\star0}\ra\N^+_{\star0}$.

Due to uniqueness property \eqref{std.vector.representative} of the embedding $\stdembed_\pi:\N_\star^+\ni\omega\mapsto\stdembed_\pi(\omega)\in\stdcone$ of any standard representation $(\H,\pi,J,\stdcone)$, this means that the embedding of a trajectory generated by $\alpha_\star^t$ on $\N^+_{\star0}$ to $\H_\phi$ for any $\phi\in\N^+_{\star0}$ coincides with the evolution in $\H_\phi^+$ generated by $\ee^{-\ii tK_\phi^\alpha}$. Hence, the hamiltonian flow $w_t^{h^\alpha}$ of $\XXX_{h^\alpha}$ on $\N^+_{\star0}$ from $\phi(0)$ to $\phi(t)$ can be always represented as liouvillean evolution
\begin{equation}
        \ee^{-\ii tK_{\phi(t)}^\alpha}\stdembed_{\phi(t)}(\phi(0))=\stdembed_{\phi(t)}(\phi(t))=:\Omega_{\phi(t)},
\label{liovillean.movement}
\end{equation}
where by $\stdembed_\psi(\omega)$ we denote the standard representative of $\omega\in\N^+_\star$ in the positive cone $\H^+_\psi$ of the GNS representation Hilbert space of $\psi\in\N^+_{\star0}$.

If $\N=\BH$, $\rho\in\schatten_1(\H)^+_0$, and the weakly-$\star$ continuous representation $\alpha:\RR\ra\Aut(\BH)$ is unitary (that is, $\alpha_t=\Ad(u(t))$ with $u(t)\in\BH^\uni$ $\forall t\in\RR$), then the algebraic hamiltonian vector field can be expressed by the von Neumann equation \eqref{vN.equation.from.BLP}, while the corresponding liouvillean evolution $\rho(t)=\ee^{-\ii tK^\alpha}\rho(0)$ in $\schatten_2(\H)\iso\H\otimes\H^\banach$ is a solution of the equation
\begin{equation}
\frac{\dd}{\dd t}\rho(t)=-\ii K^\alpha\rho(t),
\end{equation}
which gives \eqref{std.liouvillean.for.BH}%
\ifvargaugecompile 
.
\else 
, in agreement with \eqref{hamiltonian.liouvillean.J.eqn}.
\fi 
 In this sense, the Poisson compatibility condition \eqref{poisson.flow.compatibility} extends the equivalence between the algebraic (liouvillean operator) and geometric (hamiltonian vector) descriptions of temporal evolution of quantum states to the general W$^*$-dynamical systems. In the next Section we will investigate how standard liouvilleans can be used to encode the perturbation of Poisson flow by additional geometric structures over state space, beyond the realms of W$^*$-dynamical systems.

\subsection{Relative entropy, Norden--Sen geometry, and noncommutative Orlicz spaces\label{relat.entr.MCP.bundle}}
\subsubsection{Distances, Norden--Sen geometries, and geodesic free falls as entropic projections\label{distances.NS.geom.section}}
A pair $(\nabla,\nabla^\nsdual)$ of two affine connections over a smooth manifold $\M$ will be called \df{Norden--Sen dual} with respect to a riemannian metric $\gbold$ on $\M$, if{}f \cite{Sen:1944,Sen:1945,Sen:1946,Norden:1945,Norden:1948,Norden:1949}\rpktarget{NORDEN.SEN}
\begin{equation}
        \gbold(\nabla_uv,w)+\gbold(v,\nabla^\nsdual_uw)=u(\gbold(v,w))\;\forall u,v,w\in\T\M,
\label{geometric.duality}
\end{equation}
which is equivalent to
\begin{equation}
        \gbold(\transport_c^\nabla u,\transport_c^{\nabla^\nsdual}v)=\gbold(u,v)
\label{geometric.duality.two}
\end{equation}
for all $u,v\in\T\M$ and for all curves $c:\RR\supset[r_1,r_2]\ra\M$ (the symbol $\transport_c^\nabla$ denotes a parallel transport along $c$ that is determined by an affine connection $\nabla$). The quadruple $(\M,\gbold,\nabla,\nabla^\nsdual)$ is called a \df{Norden--Sen geometry}. A riemannian geometry is characterised as a Norden--Sen geometry with $\nabla=\nabla^\nsdual$.

Given a set $M$, we define a \df{distance} on $M$ as a function $D:M\times M\ra[0,\infty]$ such that $D(\phi,\omega)=0$ $\iff$ $\omega=\phi$. Eguchi \cite{Eguchi:1983,Eguchi:1985,Eguchi:1992} showed that for any smooth manifold $\M$ and any smooth distance $D$ on $\M$ that satisfies
\begin{equation}
        \DG_v|_p\DG_v|_pD(p,q)|_{q=p}\in\,]0,\infty[\;\;\;\forall p\in\M\;\forall v\in\T_p\M\setminus\{0\},
\label{Eguchi.condition}
\end{equation}
where $\DG_v|_p$ denotes here the G\^{a}teaux derivative at $p\in\M$ in the direction $v\in\T_p\M$, the distance $D$ determines a riemannian metric $\gbold$ and a pair of affine connections $(\nabla,\nabla^\nsdual)$ on $\M$, given by the \df{Eguchi equations}
\begin{align}
                \gbold_\phi(u,v)&:=-\DG_u|_\phi\DG_v|_\omega D(\phi,\omega)|_{\omega=\phi},\label{Eguchi.metric}\\
                \gbold_\phi((\nabla_u)_\phi v,w)&:=-\DG_u|_\phi\DG_v|_\phi\DG_w|_\omega D(\phi,\omega)|_{\omega=\phi},\label{Eguchi.connection}\\
                \gbold_\phi(v,(\nabla_u^\nsdual)_\phi w)&:=-\DG_u|_\omega\DG_w|_\omega\DG_v|_\phi D(\phi,\omega)|_{\omega=\phi}\label{Eguchi.dualconnection}.
\end{align}
Every quadruple $(\M,\gbold,\nabla,\nabla^\nsdual)$ determined in this way is a Norden--Sen geometry such that both $\nabla$ and $\nabla^\nsdual$ are torsion-free. A torsion-free Norden--Sen geometry will be called an \df{Eguchi geometry}. While in riemannian geometry the affine connection is determined by the riemannian metric, in the Eguchi geometry the triple of riemannian metric and two Norden--Sen dual affine connections are determined by the distance. The Levi-Civita connection $\bar{\nabla}$ of an associated riemannian geometry $(\M,\gbold)$ satisfies $\bar{\nabla}=(\nabla+\nabla^\nsdual)/2$. In this sense, the Eguchi geometry (based on the nonsymmetric distance) provides a generalisation of a riemannian geometry and cartesian geometry, including all of their main notions: distance, length, parallelity and orthogonality. Generalisation of the cartesian distance is provided by the distance $D$, the induced riemannian metric $\gbold$ provides the generalisation of orthogonality and length, while the induced torsion-free Norden--Sen dual connections $(\nabla,\nabla^\nsdual)$ provide a generalisation of parallelity.\footnote{The idea that $D$ should be considered as generalisation of the cartesian distance, while the connection $\nabla$ associated to a projection by means of $D$ should be considered as a proper generalisation of parallelity (at least in the setting of statistical manifolds) is due to Chencov \cite{Chencov:1964,Chencov:1968}.} The invariance of length under parallel transport that characterises riemannian geometry is weakened to covariance in the sense of \eqref{geometric.duality.two}.

If both affine connections of a Norden--Sen geometry are flat and torsion-free, then it is called a \df{dually flat geometry} \cite{Nagaoka:Amari:1982,Amari:1985,Amari:Nagaoka:1993}. If $\dim\M=:n<\infty$, then every dually flat geometry $(\M,\gbold,\nabla,\nabla^\nsdual)$ determines a unique pair of affine immersions $\Psi:\M\ra\RR$ and $\Psi^\lfdual:\M\ra\RR$ such that 
\begin{align}
	\gbold_{ij}(\rho(\theta))&=\frac{\partial^2\Psi(\rho(\theta))}{\partial\theta^i\partial\theta^j},\\
	\gbold_{ij}(\rho(\eta))&=\frac{\partial^2\Psi^\lfdual(\rho(\eta))}{\partial\eta^i\partial\eta^j},
\end{align}
where $\{\theta^i\}$ is a coordinate system such that $\Gamma^\nabla_{ijk}(\rho(\theta))=0$ $\forall\rho\in\M$ and $\Gamma^{\nabla^\nsdual}_{ijk}(\rho(\eta))=0$ $\forall\rho\in\M$ \cite{Dillen:Nomizu:Vracken:1990,Kurose:1990,Kurose:1993,Matsuzuoe:1998}. Conversely \cite{Amari:1985}, if there exists a convex function $\Psi$ such that its hessian (matrix of second derivatives) determines pointwise a riemannian metric, then there exists a pair of coordinate systems $\{\theta^i\}$ and $\{\eta^i\}$ and a convex function $\Psi^\lfdual:\M\ra\RR$ satisfying the above properties.  The dual flatness of a pair $(\theta,\eta)$ of coordinate systems is equivalent to the orthogonality of their tangent vectors at $q$ with respect to the riemannian metric $\gbold$ at $q$,
\begin{equation}
        \gbold_q\left((\T_q\theta)^{-1}\left(\frac{\partial}{\partial\theta^i}\right),(\T_q\eta)^{-1}\left(\frac{\partial}{\partial\eta_j}\right)\right)=\dirac_i^j\;\;\;\forall q\in\M.
\label{dual.metric.flatness}
\end{equation}
 The transition between these two formulations in the real finite dimensional case is provided by means of bijective \df{Legendre transformation} $\rpktarget{LFTRAFO}\LFtrafo_\Psi:\Theta\ra\Xi$, which acts between suitable open subsets $\Theta\subset\RR^n$ and $\Xi\subset\RR^n$, and is given by the gradient, 
\begin{equation}
        \LFtrafo_\Psi:\Theta\ni\theta\mapsto\eta:=\grad\Psi(\theta)\in\Xi.
\label{Legendre.gradient}
\end{equation}
In the coordinate-dependent form this reads
\begin{align}
        \eta_i&=(\LFtrafo_\Psi(\theta))_i:=\frac{\partial\Psi(\theta)}{\partial\theta^i},\label{LFgrad.one}\\
        \theta^i&=(\LFtrafo^{-1}_\Psi(\eta))^i:=\frac{\partial\Psi^\lfdual(\eta)}{\partial\eta_i},\label{LFgrad.two}
\end{align}
whenever the duality pairing is given by
\begin{equation}
        \duality{\cdot,\cdot}_{\RR^n\times\RR^n}:\RR^n\times\RR^n
        \ni(\theta,\eta)
        \mapsto
        \theta\cdot\eta^\top:=\sum_{i=1}^n\theta^i\eta_i
        \in\RR.
\label{selfduality.Rn}
\end{equation}

The Eguchi equations applied to the distance 
\begin{equation}
	D_\Psi(\rho,\sigma):=\Psi(\rho)+\Psi^\lfdual(\sigma)-\sum_{i=1}^n\theta^i(\rho)\eta^i(\sigma)
\end{equation}
yield $(\M,\gbold^{D_\Psi},\nabla^\Psi,{\nabla^{\Psi}}^\nsdual)=(\M,\gbold,\nabla,\nabla^\nsdual)$. We will call such $D_\Psi$ a canonical Br\`{e}gman distance of a dually flat geometry $(\M,\gbold,\nabla,\nabla^\nsdual)$. A riemannian metric $\gbold$ on an affine manifold $(\M,\nabla)$ with flat $\nabla$ is said to be \df{hessian}, and denoted $\gbold^\Psi$, if{}f there exists a smooth function $\Psi:\M\ra\RR$ such that \cite{Shima:1976,Shima:1977,Cheng:Yau:1982}
\begin{equation}
        \gbold(u,v)=(\nabla_u\ddd\Psi)(v)\;\;\forall u,v\in\T\M.
\label{hessian.metric}
\end{equation}
Such triple $(\M,\gbold,\nabla)$ will be called a \df{hessian geometry} \cite{Shima:2007} (see also \cite{Koszul:1961,Vinberg:1963,Duistermaat:2001}). The function $\Psi$ in \eqref{hessian.metric} is the same as in the representation of $\gbold_{ij}(\rho(\theta))$ above, and 
\begin{equation}
	\gbold(u,v)=(\nabla^\nsdual_u\ddd\Psi^\lfdual)(v)\;\;\forall u,v\in\T\M.
\end{equation}
Hence, given a riemannian manifold $(\M,\gbold)$ and an affine connection $\nabla$ on $\M$ the following conditions are equivalent \cite{Shima:1986,Shima:Yagi:1997,Shima:2007}: (1) $(\M,\gbold,\nabla)$ is a hessian geometry; (2) $(\M,\gbold,\nabla,2\bar{\nabla}-\nabla)$ is a dually flat geometry.

Let $(\M,\gbold,\nabla,\nabla^\nsdual)$ be a dually flat geometry, and let $\Q\subseteq\M$ be $\nabla^\nsdual$-affine (i.e., there exists a coordinate system $\{\eta^i\}$ on $\Q$ such that $\Gamma^{\nabla^\nsdual}_{ijk}(\rho(\eta))=0$ $\forall\rho\in\Q$) and $\nabla^\nsdual$-convex (i.e. $\forall\rho_1,\rho_2\in\Q$ $\exists!$ $\nabla^\nsdual$-geodesics in $\Q$ connecting them). Then there exists a unique entropic projection 
\begin{equation}
	\M\in\rho\mapsto\PPP^{D_\Psi}_\Q(\rho):=\arginff{\sigma\in\Q}{D_\Psi(\sigma,\rho)}\in\Q,
\end{equation}
and it is equal to a unique projection $\rho_\Q$ of $\rho$ onto $\Q$ along a $\nabla$-geodesic that is $(\gbold,\nabla,\nabla^\nsdual)$-orthogonal at $\Q$ \cite{Amari:1985,Amari:Nagaoka:1993}. More precisely, the projection $\rho_\Q$ is defined as such element of $\Q$ that 
\begin{equation}
	\gbold_{\rho_\Q}(\dot{c}^\nabla(t),\dot{c}^{\nabla^\nsdual}(s))=0\;\;\forall c^{\nabla^\nsdual},
\end{equation}
where $c^\nabla(t)$ is a $\nabla$-geodesic connecting $\rho$ and $\rho_\Q$, while $c^{\nabla^\nsdual}$ varies over all $\nabla^\nsdual$-geodesics intersecting $\rho_\Q$ and entirely included in $\Q$. 
\begin{center}
\includegraphics[width=8cm]{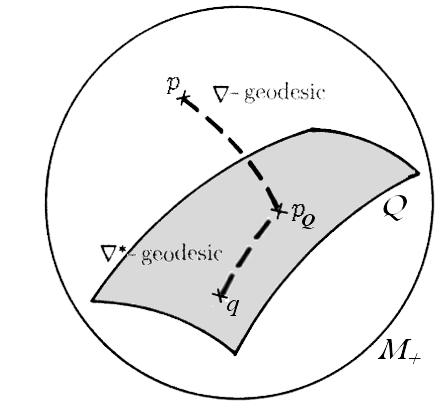}\\
{\small Picture 1. \textit{$\nabla$-geodesic projection onto $\nabla^\nsdual$-affine $\nabla^\nsdual$-convex set $\Q$.}}
\end{center}
Hence, for dually flat geometries, projections onto $\nabla^\nsdual$-affine $\nabla^\nsdual$-convex sets along $\nabla$-geodesics coincide with the entropic projections of the associated Br\`{e}gman relative entropies. In consequence, one can consider a local $\nabla$-geodesic flow to be an infinitesimal version of information dynamics defined by constrained relative entropy minimisation.
\subsubsection{Quantum information geometries\label{quant.info.geom.section}}
Given a standard representation $(\H,\pi,J,\stdcone)$ of a W$^*$-algebra $\N$, consider a family of distances $D_\fff:\N_\star^+\times\N_\star^+\ra[0,\infty]$ defined by \cite{Kosaki:1982:interpolation,Petz:1985:quasientropies}
\begin{equation}
	D_\fff(\omega,\phi):=\s{\stdembed_{\pi}(\phi),\fff(\Delta_{\omega,\phi})\stdembed_{\pi}(\phi)}_{\H}
\label{Df.definition}
\end{equation}
if $\supp(\omega)\leq\supp(\phi)$ and $D_\fff(\omega,\phi):=+\infty$ otherwise, where $\fff:\RR^+\ra\RR$ is any operator convex function (i.e. $\fff(tx+(1-t)y)\leq t\fff(x)+(1-t)\fff(y)$ $\forall x,y\in\N^+$ $\forall t\in[0,1]$ \cite{Kraus:1936}) satisfying $\fff(0)\leq0$ and $\fff(1)=0$. As proved in \cite{Petz:1985:quasientropies,Tomamichel:Colbeck:Renner:2009}, all $D_\fff$ satisfy the condition
\begin{equation}
	D(\omega,\phi)\geq D(T_\star(\omega),T_\star(\phi)) \forall\omega,\phi\in\M(\N)\;\;\forall T_\star:\N_\star^+\ra\N_\star^+,
\label{markovian.monotonicity.D}
\end{equation}
where $T_\star$ denotes a Banach predualisation of a weakly-$\star$ continuous unital completely positive map $T:\N\ra\N$, and $\M(\N)\subseteq\N^+_\star$ is arbitrary. Moreover, the equality is attained if{}f $T_\star$ is an isomorphism \cite{HMPB:2011}. 


For
\begin{equation}
        \fff_\gamma(t):=
        \left\{
                \begin{array}{ll}
                        \frac{1}{\gamma}+\frac{1}{1-\gamma}t-\frac{1}{\gamma(1-\gamma)}t^\gamma&:\gamma\in\RR\setminus\{0,1\}\\
                        t\log t-(t-1)&:\gamma=1\\
                        -\log t+(t-1)&:\gamma=0,
                \end{array}
        \right.
\label{f.gamma.function}
\end{equation}
the restriction to $\gamma=1$ and $\M(\N)\subseteq\N^+_{\star1}$ gives \cite{Araki:1976:relative:entropy:I,Araki:1977:relative:entropy:II}
\begin{equation}
	D_1|_{\N^+_{\star1}}(\omega,\phi)=\s{\stdembed_\pi(\phi),\log(\Delta_{\omega,\phi})\stdembed_\pi(\phi)}_\H,
\label{Araki.std.rep.distance}
\end{equation}
which, for $\N\iso\BH$, turns to \cite{Umegaki:1961,Umegaki:1962}
\begin{equation}
	D_1|_{\schatten_1(\H)^+_1}(\omega,\phi)=\tr_\H(\rho_\omega\log(\rho_\omega)-\rho_\omega\log(\rho_\phi)).
\end{equation}

Jen\v{c}ov\'{a} \cite{Jencova:2006,Jencova:2010} proposed to consider a (Young) function 
\begin{equation}
	\Orlicz_\phi:\N^\sa\ni h\mapsto\frac{1}{2}(\widetilde{\phi^h}(\II)+\widetilde{\phi^{-h}}(\II))-1\in\RR^+,
\end{equation}
where $\N$ is an arbitrary W$^*$-algebra, and 
\begin{equation}
	\widetilde{\phi^h}(\II):=\sup_{\omega\in\N^+_\star}\{-D_1|_{\N^+_{\star1}}(\omega,\phi)+\omega(h)+\omega(\II)\}
\end{equation}
With this function, she defined a noncommutative Orlicz\footnote{More precisely, it is a noncommutative analogue of a Morse--Transue--Krasnosel'ski\u{\i}--Ruticki\u{\i} space, see \cite{Kostecki:2013} for details.} space $L_{\Orlicz_\phi}(\N)$ as a completion of $\{x\in\N^\sa\mid\exists\lambda>0\;\;\Orlicz_\phi(\lambda x)<\infty\}$ in the norm $\n{x}_{\Orlicz_\phi}:=\inf\{\lambda>0\mid\Orlicz_\phi(\lambda^{-1}x)\leq1\}$. This space satisfies 
\begin{equation}
	\N^\sa\sqsubseteq L_{\Orlicz_\phi}(\N):=\overline{\N^\sa}^{\n{\cdot}_{\Orlicz_\phi}}.
\end{equation}

Given the choice of a Hilbert space $\H$ with $\dim\H=:n<\infty$ and a smooth bijective parametrisation $\Theta\ni\theta\mapsto\rho(\theta)\in\schatten_1(\H)^+_0$ by the elements of an open set $\Theta\subseteq\RR^m$, $m\in\NN$, the \textit{parametric quantum manifold} is defined as a quantum model
\begin{equation}
        \M(\H)=\left\{\rho(\theta)\in\schatten_1(\H)^+_0\mid\theta\in\Theta\subseteq\RR^m\right\}\subseteq\schatten_1(\CC^n)^+_0\iso\MNC^+_0.
\label{manif.dens.matrix.hilb.sp}
\end{equation}
Usually, the additional condition $\tr_\H(\rho(\theta))=1$ is imposed on the elements of $\M(\H)$. A tangent space $\T_\rho\MNC^+_0$ is the real vector space of all Fr\'{e}chet derivatives in the directions of smooth curves in $\MNC^+_0$ that pass through $\rho$, so it can be identified with a restriction of $\MNC^\sa$. A restriction of $\rho$ to $\MNC^+_{01}$ implies a restriction of the tangent vectors to the space $\{x\in\MNC^\sa\mid\tr_{\CC^n}(x)=0\}$. A Banach smooth manifold structure on $\N_{\star01}^+$ for an arbitrary countably finite\footnote{The W$^*$-algebras $\N$ which are not countably finite do not allow faithful quantum states: $\N^+_{\star0}=\varnothing$.} W$^*$-algebra $\N$ was introduced by Jen\v{c}ov\'{a} \cite{Jencova:2006,Jencova:2010}. She proved that the quantum model $\N^+_{\star01}$ can be equipped with the smooth Banach manifold structure modeled on a family of Banach spaces 
\begin{equation}
	L^0_{\Orlicz_\phi}(\N):=\{x\in L_{\Orlicz_\phi}(\N)\mid\phi(x)=0\}=\overline{\{x\in\N^\sa\mid\phi(x)=0\}}^{\n{\cdot}_{\Orlicz_\phi}}.
\end{equation}
This structure is introduced by means of the smooth atlas $\{(w_\phi^{-1}(U(\phi)),w_\phi)\mid\phi\in\N^+_{\star01}\}$, where $U(\phi):=\{x\in L^0_{\Orlicz_\phi}(\N)\mid\n{x}_{\Orlicz_\phi}<1\}$ and 
\begin{equation}
	w_\phi^{-1}:L^0_{\Orlicz_\phi}(\N)\supseteq U(\phi)\ni h\mapsto\phi^h\in\N^+_{\star01}
\end{equation}
is a diffeomorphism. If $\N\iso L_\infty(\X,\mho(\X),\tmu)$, then this construction reduces to a smooth Banach manifold structure on $L_1(\X,\mho(\X),\tmu)^+_{01}$ introduced in \cite{Pistone:2001,Grasselli:2001:PhD}.\footnote{A closely related approach to construction of smooth information manifold, utilising Orlicz spaces of unbounded operators/functions instead of the MTKR spaces of bounded elements, was developed in \cite{Pistone:Sempi:1995} for $L_1(\X,\mho(\X),\tmu)^+_{01}$ and in \cite{Streater:2000:bounded,Streater:2004:Orlicz,Streater:2008,Streater:2009:book,Streater:2010:Banach} for (a subspace of) $\schatten_1(\H)_{01}$.} We conjecture that (analogously to the extension of this smooth manifold structure from $L_1(\X,\mho(\X),\tmu)^+_{\star01}$ to $L_1(\X,\mho(\X),\tmu)^+_{\star0}$, provided in \cite{AJLS:2012}) Jen\v{c}ov\'{a}'s construction can be extended to $\N^+_{\star0}$. Under this conjecture, we define a \df{nonparametric quantum manifold} as a quantum model $\M(\N)\subseteq\N^+_{\star0}$ equipped with a Banach smooth manifold structure induced from $\N^+_{\star0}$, by replacing $L_{\Orlicz_\phi}^0(\N)$ with $L_{\Orlicz_\phi}(\N)$.

Given any countably finite W$^*$-algebra $\N$, a finite dimensional quantum model $\M(\N)\subseteq\N^+_\star$ that is a Banach smooth submanifold of $\N^+_{\star0}$, and a quantum distance $D$ on $\M(\N)$ that is smooth and satisfies \eqref{Eguchi.condition}, one can derive the corresponding quantum Norden--Sen geometry $(\M(\N),\gbold^D,\nabla^D,{\nabla^D}^\nsdual)$. In particular, given any distance $D_\fff$, for $\N=\BH$ and $\M(\H)=\schatten_1(\H)^+_{01}$, the corresponding riemannian metric $\gbold^{D_\fff}$ takes the form \cite{Morozova:Chencov:1989,Petz:1996:monotone,Lesniewski:Ruskai:1999}
\begin{equation}
	\gbold^{D_\fff}_\rho(u,v)=\s{u,\left(\hhh_\fff(\LLL_\rho\RRR_\rho^{-1})\RRR_\rho\right)^{-1}(v)}_{\schatten_2(\H)},
\label{g.D.f.eq}
\end{equation}
where $u,v\in\{x\in\BH\mid\tr_\H(x)=0\}$, while $\hhh_\fff:[0,\infty[\,\ra[0,\infty[$ is an operator monotone increasing function\footnote{A function $\hhh:\RR^+\ra\RR$ is called operator monotone increasing if{}f $0\leq x\leq y$ $\limp$ $\hhh(x)\leq\hhh(y)$ $\forall x,y\in\BH$ \cite{Loewner:1934}.} defined by
\begin{equation}
	\hhh_\fff(\lambda):=\frac{(\lambda-1)^2}{\fff(\lambda)-\lambda\fff(\frac{1}{\lambda})}.
\label{hhh.fff.eqn}
\end{equation}
This implies that several different $\fff$ lead to the same $\hhh_\fff$. Hence, for any riemannian metric given by \eqref{g.D.f.eq} there is a family of distances $D_\fff$ that have it as its second order Taylor term.

Using an integral representation 
\begin{equation}
        \fff(\lambda)=c_1(\lambda-1)+c_2(\lambda-1)^2+c_3\frac{(\lambda-1)^2}{\lambda}+\int_0^\infty\tmu(t)(\lambda-1)^2\frac{1+t}{\lambda+t},
\end{equation}
where $c_2,c_3\geq0$, $c_1\in\RR$, and $\tmu:\,]0,\infty[\,\ra\RR^+$ is a measure satisfying $\int_0^\infty\tmu(t)\in\RR$ \cite{Lesniewski:Ruskai:1999}, Jen\v{c}ov\'{a} \cite{Jencova:2004:entropies} showed that the \df{$\fff$-connections}, defined by the Eguchi equation \eqref{Eguchi.connection} applied to $D_\fff$ distance, have the form
\begin{equation}
        \gbold^{D_\fff}_\rho(\nabla^{D_\fff}_xy,z)=2\int_0^\infty\tmu(\lambda)\re(\widetilde{C}(\lambda,z,x,y))-2\int_0^\infty\tmu(\lambda^{-1})\left(\re(\widetilde{C}(\lambda,y,x,z)+\re(\widetilde{C}(\lambda,y,x,z))\right),
\end{equation}
where 
\begin{equation}
        \widetilde{C}(\lambda,x,y,z):=(1+\lambda)\tr\left(x\frac{1}{\lambda\RRR_\rho+\LLL_\rho}(y)\frac{1}{\RRR_\rho+\lambda\LLL_\rho}(z)\right).
\end{equation}
The connections $\nabla^{D_\fff}$ are torsion-free. Moreover,  the family of quantum Norden--Sen smooth geometries $(\MNC^+_0,\gbold^{D_\gamma},\nabla^{D_\gamma},(\nabla^{D_{\gamma}})^\nsdual)$ for $\gamma\in[-1,2]$ is characterised as the dually flat Eguchi geometry arising from the $D_\fff$ distances \cite{Jencova:2003:flat,Jencova:2004:entropies}. This result corresponds to the class $D_{\fff_\gamma}$ of quantum distances determined by \eqref{f.gamma.function} belonging to both families: $D_\Psi$ and $D_\fff$ \cite{Jencova:2005,Kostecki:2013}. The relationships between various information geometric objects on quantum state spaces $\M(\N)$ can be summarised in the following diagram:
\begin{center}
\includegraphics[width=15cm]{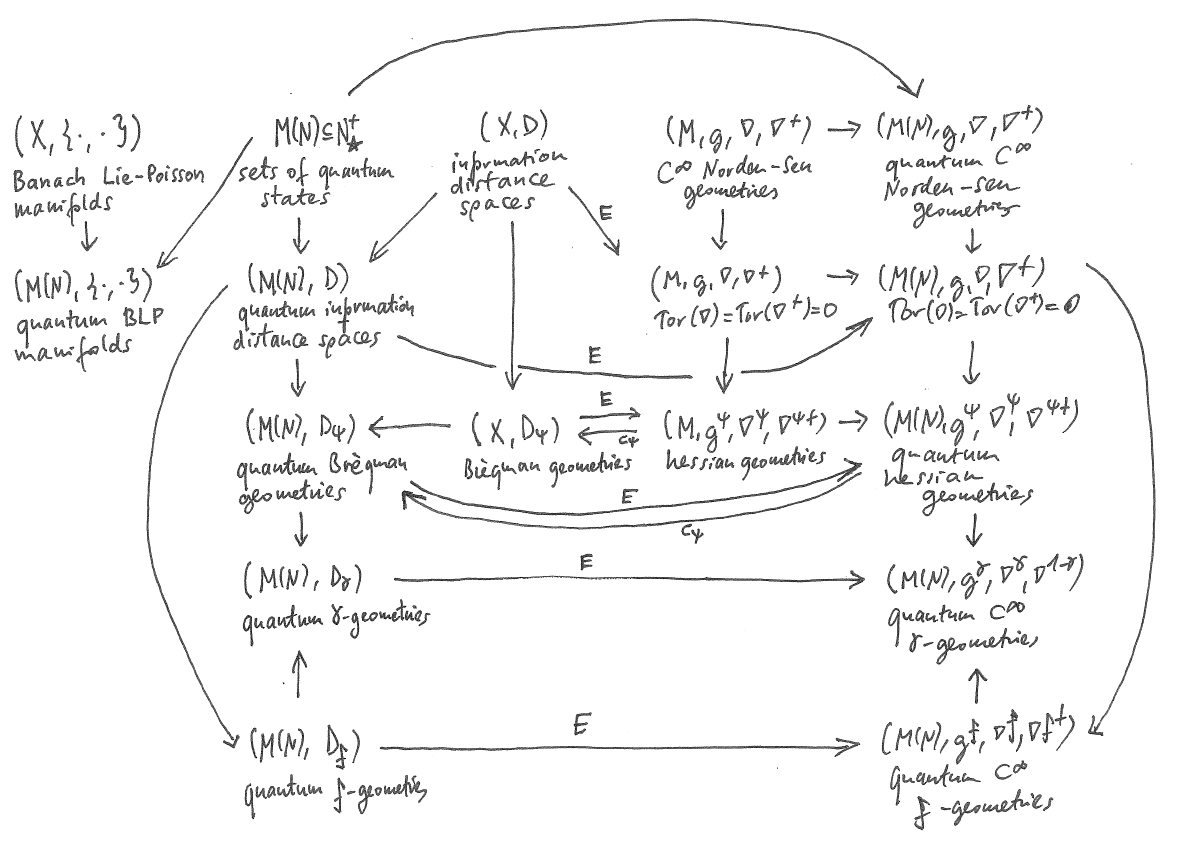}\\
{\small Picture 2. \textit{Relationships between different quantum geometries.} $E$ denotes an application of the Eguchi equations. $c_\Psi$ denotes the construction of an associated canonical Br\`{e}gman distance.}
\end{center}

\subsubsection{Orlicz spaces and Br\`{e}gman projections\label{Orlicz.Br\`{e}gman.section}}

As a consequence of the above results, if $\M(\N)$ is a dually flat manifold with respect to the triple $(\gbold^{D_\gamma},\nabla^{D_\gamma},(\nabla^{D_\gamma})^\nsdual)$, then $D_\gamma$-entropic projections onto ${(\nabla^{D_\gamma})^\nsdual}$-affine-and-convex subsets are \textit{locally equivalent} to $\nabla^{D_\gamma}$-geodesic ``free fall''. The construction of families of $\nabla^{D_\gamma}$-connections in infinite dimensional noncommutative case was provided in \cite{Gibilisco:Isola:1999,Grasselli:2001:PhD,Streater:2004:duality,Jencova:2005,Jencova:2006} using the linear structure of noncommutative $L_{1/\gamma}(\N)$ spaces, and in such case this statement also holds \cite{Jencova:2005}.

More generally (going a bit beyond the scope of the current paper), the quantum Br\`{e}gman distance $D_\Psi$ is defined via nonlinear embeddings $(\ell_{L_\Upsilon(\N)(\phi)},\ell_{(L_\Upsilon(\N))^\star(\phi)})$ into noncommutative Orlicz spaces $L_\Upsilon(\N)$ and $(L_\Upsilon(\N))^\star$, respectively. These spaces play the role of a tangent space `of states' and the cotangent space `of effects', respectively (and similarly to the commutative case of \cite{Gibilisco:Pistone:1998,Gibilisco:Isola:1999}). The ${\nabla^{D_\Psi}}^\nsdual$-affinity and ${\nabla^{D_\Psi}}^\nsdual$-convexity are defined as linear affinity and linear convexity in the Orlicz space $(L_\Upsilon(\N))^\star$ of effects. Thus, the global flatness of the connection on $L_\Upsilon(\N)$ understood as a tangent space $\T_\phi\M(\N)$ corresponds to its parallel transport being given by a family of isomorphisms $U_{\phi,\omega}:\T_\phi\M(\N)\ra\T_\omega\M(\N)$ satisfying $U_{\phi,\phi}=\id_{\T_\phi\M(\N)}$ and $U_{\phi,\omega}U_{\omega,\psi}=U_{\phi,\psi}$ \cite{Jencova:2005}. From this point of view, the standard unitary transitions $V_{\phi,\omega}:\H_\omega\ra\H_\phi$ can be understood as parallel transports of the connection defined naturally by the linear structure of the fibers in the GNS Hilbert bundle. This allows us to understand $V_{\phi,\omega}$ as a ``free fall'' along the geodesics of the Levi-Civita connection of $\gbold^\gamma$ for $\gamma=\frac{1}{2}$, known as the Wigner--Yanase metric \cite{Wigner:Yanase:1963}. The riemannian distance of $\gbold^{1/2}$ for $\N\iso\BH$ reads \cite{Gibilisco:Isola:2003:WY}
\begin{equation}
        d_{\gbold^{1/2}}(\rho_1,\rho_2)=2\arccos\left(\tr_\H(\sqrt{\rho_1}\sqrt{\rho_2})\right).
\label{Wigner.Yanase.distance}
\end{equation}
On the boundary of pure spaces $\gbold^{1/2}$ reduces to the Fubini--Study metric $\gbold^{\mathrm{FS}}$ \eqref{Fubini.Study.equation} multiplied by the scalar factor  $4$ \cite{Petz:Sudar:1999}, hence \eqref{Wigner.Yanase.distance} divided by $2$ reduces to \eqref{Fubini.Study.distance}. A generalisation of $\gbold^{1/2}$ to countably additive W$^*$-algebras was provided by Connes and St{\o}rmer \cite{Connes:Stormer:1978}. For a given standard representation $(\H,\pi,J,\stdcone)$, it reads 
\begin{equation}
	\gbold^{1/2}_\phi(x,y)=\frac{1}{2}\n{(J\pi(x^*)J-\pi(y))\stdembed_\pi(\phi)}_\H^2.
\label{CS.WY.eq}
\end{equation}
For any $\M(\N)\subseteq\N^+_{\star0}$, the GNS construction equipped with the Tomita--Takesaki theory defines a corresponding standard representation $(\H_\phi,\pi_\phi,J_\phi,\stdcone_\phi)$. In such case, a direct calculation based on the properties \eqref{std.uni.trans.eq.one}-\eqref{std.uni.trans.eq.three}, using $V_{\phi,\omega}$ in the role of $\transport^{\nabla^{1/2}}_{\phi,\omega}$ applied with respect to \eqref{CS.WY.eq}, shows that these objects satisfy the Levi-Civita version of the equation \eqref{geometric.duality.two}. The corresponding relative entropy is \cite{Jencova:2005} 
\begin{equation}
	D_{1/2}(\phi,\psi)=2\n{u_\phi\stdembed_\pi(\phi)-u_\psi\stdembed_\pi(\psi)}^2_\H,
\end{equation}
where $u_\phi$ and $u_\psi$ are unique unitary operators arising from the polar decomposition of relative modular operators $\Delta_{\phi,\omega}$ and $\Delta_{\psi,\omega}$, respectively, where $\omega\in\N^+_{\star0}$ is arbitrary. When expressed as a Br\`{e}gman distance on the standard representation Hilbert space $\H$, this relative entropy takes the form $D_{1/2}(x,y)=\frac{1}{2}\n{x-y}^2_\H$ \cite{Jencova:2005}. Hence, the local (infinitesimal) action of the operators $V_{\phi,\omega}$ can be understood as a geodesic free fall that is locally equivalent to the minimisation of the Hilbert space norm, which in turn corresponds to a continuous linear projection operator onto a convex closed subset. In what follows, we will use the GNS Hilbert bundle having in mind the above observations.

In face of presence of other approaches to construction of smooth manifold structure on the space $L_1(\X,\mho(\X),\tmu)^+_{01}$ \cite{Burdet:Combe:Nencka:2001,AJLS:2012,Newton:2012}, one may ask for the specific motivation of the Orlicz space based approach. The main reason is to guarantee that the local neighbourhoods of an information state (which are identified with the tangent space) are accessible by means of entropic projection. More precisely, each tangent vector is identified as an equivalence class of one dimensional exponential models (i.e., $p\exp(\lambda f-\log Z(p,\lambda f))$ in the neighbourhood of $p\in L_1(\X,\mho(\X),\tmu)^+_{01}$ and $\phi^{\lambda h}$ in the neighbourhood of $\phi\in\N^+_{\star01}$). This can be viewed as a localised version of Jaynes' maximum entropy principle \cite{Jaynes:1957,Jaynes:1979:where:do:we:stand} (cf. \cite{Streater:2011}) of model construction. We consider it as a step that is conceptually similar to localisation of Minkowski space in the passage from special to general relativity theory: instead of working with information models that are globally exponential, we assume only that they are locally (infinitesimally) exponential. This leads to a question whether one can postulate local approximation by means of some other models (corresponding to minimisation of some different information distance functional), and construct smooth information manifold structure out of this postulate. In the commutative case this question has been answered in the affirmative by the recent work \cite{Vigelis:Cavalcante:2013,Vigelis:Cavalcante:2013:delta,Vigelis:Cavalcante:2014}, who have generalised the construction of \cite{Pistone:Sempi:1995} to a large family of Orlicz (more precisely, Musielak--Orlicz) spaces. The Young functions that define these spaces define the corresponding Br\`{e}gman distances. We conjecture that the similar construction can be carried out in the noncommutative case. In the case when the information distance used for the construction of the smooth information manifold belongs to the Br\`{e}gman class, the resulting information manifold is no longer locally exponential, but it is locally dually flat (locally hessian). Because dually flat manifolds can be thought of as a generalisation of cartesian space, this construction strengthens analogy to relationship between Minkowski space-time and general lorentzian manifold. From the perspective of applications of entropic projections, one can say that the generalisation of Jen\v{c}ov\'{a}'s construction of a manifold structure from one based on $D_1$ to one based on Br\`{e}gman distances $D_\Psi$ would allow for local representations of entropic $D$-projections in terms of projections along $\nabla^D$-geodesics. If entropic projections are regarded as a form of information dynamics, then one can say that such construction of the smooth information manifold facilitates the possibility of introducing local (infinitesimal) representation of entropic information dynamics. Turning it to a slogan: \textit{introducing the structure of quantum information manifold based on a Young function $\Psi$ and a corresponding Br\`{e}gman distance $D_\Psi$ amounts to postulating that information flows locally along $\nabla^{D_\Psi}$-geodesics}.

In principle, one can construct smooth information manifold structure of $\M(\N)$ using some distance $D$, and then consider the geometric structures and information dynamics on $\M(\N)$ using some other distance $\tilde{D}$, or even using some class of distances, $\{\tilde{D}^i\mid i\in I\}$. However, using the same distance on both levels allows for stronger optimality results (concerning, for example, asymptotic estimation). More specifically, the same asymptotic results (up to third order) will be obtained for any $\widetilde{D}$ that locally generates a dually flat geometry that agrees with a dually flat geometry of a Br\`{e}gman distance $D_\Psi$. Hence, the above slogan can be equipped with a user's notice: \textit{a local Norden--Sen geometry and information dynamics of such manifold can be described by an arbitrary distance $D$ that has the same Taylor expansion, up to third order, as $D_\Psi$}. Given an arbitrary quantum model $\M(\N)$ and a distance $D$ on $\M(\N)$, the pair $(\M(\N),D)$ can be called \textit{dually flat localisable} quantum geometry if{}f $\M(\N)$ can be equipped with a smooth manifold structure based on some Br\`{e}gman distance $D_\Psi$ that agrees with $D$ up to third order. Such quantum geometry can be considered as a proper information geometric analogue of a lorentzian manifold: while global geometry (and dynamics) of $\M(\N)$ is described in terms of $D$, locally it is equivalent with the description in terms of $D_\Psi$, which is equivalent with the description in terms of a dually flat geometry. See Section \ref{section.dynamics.renormalisation} for an application of these considerations for the problem of geometric nonperturbative renormalisation in quantum nonequilibrium statistical mechanics.

Let W$^*$-algebra $\N$ admit a trace $\tau\in\W_0(\N)$. A closed densely defined linear operator $x:\dom(x)\ra\H$ is called \df{$\tau$-measurable} \cite{Segal:1953,Nelson:1974} if{}f 
\begin{equation}
	\exists\lambda>0\;\;\tau(P^{\ab{x}}(]\lambda,+\infty]))<\infty.
\end{equation}
Let $\MMM(\N,\tau)$ denote the space of all $\tau$-measurable operators affiliated with $\pi_\tau(\N)$. Let $\Upsilon:[0,\infty[\,\ra[0,\infty]$ be an \df{Orlicz function}, i.e. convex, continuous, nondecreasing, $\Upsilon(0)=0$, $\lambda>0$ $\limp$ $\Upsilon(\lambda)>0$, and $\lim_{\lambda\ra+\infty}\Upsilon(0)=+\infty$. A noncommutative Orlicz space over $\N$ associated with $\Upsilon$ is defined as \cite{Kunze:1990}
\begin{equation}
	L_\Upsilon(\N,\tau):=\Span_\CC\{x\in\MMM(\N,\tau)\mid\tau(\Upsilon(\ab{x}))\leq1\},
\end{equation}
equipped with a norm
\begin{equation}
	\n{\cdot}_\Upsilon:\MMM(\N,\tau)\ni x\mapsto\inf\{\lambda>0\mid\tau(\Upsilon(\lambda^{-1}\ab{x}))\leq1\}
\end{equation}
which turn it into a Banach space. It follows that 
\begin{equation}
	L_\Upsilon(\N,\tau)=\{x\in\MMM(\N,\tau)\mid\exists\lambda>0\;\;\tau(\Upsilon(\lambda\ab{x}))<\infty\}.
\end{equation}
An issue of canonical generalisation of the notion of a noncommutative Orlicz space to an arbitrary W$^*$-algebra is a matter of a current research, see \cite{Kostecki:2014:Orlicz} for a discussion.

The construction of a general notion of a quantum Br\`{e}gman distance for arbitrary spaces $\N^+_\star$ is an open problem. Let $X$ be a reflexive Banach space, let $\Psi:X\ra\,]-\infty,+\infty]$ be convex, lower semi-continuous, and Legendre (see \cite{Bauschke:Borwein:Combettes:2001} for a definition). Let $C\subseteq X$ be nonempty and convex, $C\cap\INT(\efd(\Psi))\neq\varnothing$, where $\efd(\Psi):=\{x\in X\mid\Psi(x)\neq+\infty\}$, and let $y\in\INT(\efd(\Psi))$. Then the \df{Br\`{e}gman functional} on $X$, defined by
\begin{equation}
	 \widetilde{D}_\Psi(x,y):=\Psi(x)-\Psi(y)-\duality{x-y,\DG_y\Psi(y)}_{X\times X^\star}
\end{equation}
for $y\in\INT(\efd(\Psi))$, and $\widetilde{D}_\Psi(x,y)=+\infty$ otherwise, satisfies \cite{Bauschke:Borwein:Combettes:2001}:
\begin{enumerate}
\item[1)] $\widetilde{D}_\Psi(\cdot,y)$ is convex and lower semi-continuous,
\item[2)] $\efd(\widetilde{D}_\Psi(\cdot,y))=\efd(\Psi)$,
\item[3)] $\widetilde{D}_\Psi(x,y)=0$ $\iff$ $x=y$,
\item[4)] $\PPP^{\widetilde{D}_\Psi}_C(y)=\{*\}\in\INT(\efd(\Psi))$,
\item[5)] $\PPP^{\widetilde{D}_\Psi}_C\circ\PPP^{\widetilde{D}_\Psi}_C(y)=\PPP^{\widetilde{D}_\Psi}_C(y)$,
\item[6)] if $K$ is a vector subspace of $X$, then Chencov's generalised pythagorean theorem \cite{Chencov:1968,Chencov:1972} holds:
\begin{equation}
	\widetilde{D}_\Psi(x,y)=\widetilde{D}_\Psi(x,\PPP^{\widetilde{D}_\Psi}_K(y))+\widetilde{D}_\Psi(\PPP^{\widetilde{D}_\Psi}_K(y),y)\;\;\forall(x,y)\in K\times X.
\label{generalised.pythagore.Bregman}
\end{equation}
\end{enumerate}
Let $X\iso L_\Upsilon(\N)$, and consider a map $\ell_{L_\Upsilon(\N)}:\N_\star\ra L_\Upsilon(\N)$ satisfying $\ell_{L_\Upsilon(\N)}(\N_\star^+)\subseteq(L_\Upsilon(\N))^+$, $\ell_{L_\Upsilon(\N)}(\N_\star)\subseteq\INT(\efd(\Psi))$, and bijective on its codomain. Then the \df{quantum Br\`{e}gman distance} is defined as \cite{Kostecki:2014:qig}
\begin{equation}
	D_\Psi(\phi,\psi):=\widetilde{D}_\Psi(\ell_{L_\Upsilon(\N)}(\phi),\ell_{L_\Upsilon(\N)}(\psi))\;\;\forall(\phi,\psi)\in\N^+_\star\times\N^+_\star.
\end{equation}
The nontrivial open problem consists of finding the minimal additional conditions on $\Psi$ and $\ell_{L_\Upsilon(\N)}$ that are necessary and sufficient to prove that $D_\Psi(\phi,\psi)$ is smooth (or at least triple G\^{a}teaux differentiable) and satisfies the infinite dimensional analogue of the property \eqref{geometric.duality.two} as well as an equivalence of $D_\Psi$-projections onto linear convex closed subspaces $\Q$ of $L_\Upsilon(\N)$ with $(\gbold^{D_\Psi},\nabla^{D_\Psi},(\nabla^{D_\Psi})^\nsdual)$-orthogonal projections along $\nabla^{D_\Psi}$-geodesics onto $\Q$. These conditions will necessarily intertwine the properties of $\Psi$, $\Upsilon$, and $\ell_{L_\Upsilon(\N)}$. See \cite{Kostecki:2014:qig} for an additional discussion.

\subsubsection{Conjecture: a Morozova--Chencov--Petz bundle\label{MCP.bundle.section}}
Let $(\H_\phi,\pi_\phi,\Omega_\phi)$ be a GNS representation of a W$^*$-algebra $\N$ for $\phi\in\N^+_{\star0}$. Consider a scalar product on $\H_\omega$, defined by
\begin{align}
	\s{[x]_\phi,[y]_\phi}_{\hhh,\phi}&:=\s{[x^*]_\phi,\JJJ^\hhh_{\phi,\phi}([y]_\phi)}_\phi,\label{MCP.scalar.product}\\
	\JJJ^\hhh_{\phi,\psi}&:=\frac{1}{\hhh(\Delta_{\phi,\psi})}\RRR(\psi)^{-1},
\end{align}
where $\RRR(\psi)$ is a right multiplication by $\psi\in\N^+_{\star0}$ \cite{Sherman:2001}, while $\hhh:\RR^+\ra\RR^+$ is an operator monotone increasing function satisfying \cite{Petz:1996:monotone} $\hhh(\lambda)=\lambda\hhh(\lambda^{-1})$ $\forall\lambda>0$ (all functions $\hhh_\fff$ given by \eqref{hhh.fff.eqn} satisfy this property). We conjecture that:
\begin{itemize}
\item[1)] the completion of the vector space $\pi_\phi(\N)/\ker(\JJJ^\hhh_\phi)$ in the scalar product \eqref{MCP.scalar.product} is a Hilbert space (denoted below as $\H_{\hhh,\phi}$, with elements denoted by $[x]_{\hhh,\phi}$ for any $x\in\nnn_\phi$),
\item[2)] $\pi_{\hhh,\phi}(x):[y]_{\hhh,\phi}\mapsto[xy]_{\hhh,\phi}$ defines a nondegenerate faithful normal representation $\N\ra\BBB(\H_{\hhh,\phi})$,
\item[3)] a unique extension of the antilinear isometry $J_{\hhh,\phi}:[x]_{\hhh,\phi}\mapsto[x^*]_{\hhh,\phi}$ $\forall x\in\nnn_\phi$ determines a conjugation on $\H_{\hhh,\phi}$, turning a quadruple
\begin{equation}
	\left(\H_{\hhh,\phi},\pi_{\hhh,\phi},J_{\hhh,\phi},\bigcup_{x\in\nnn_\phi\cap\nnn_\phi^*}\overline{\left\{\pi_{\hhh,\phi}(x)J_{\hhh,\phi}[x]_{\hhh,\phi}\right\}}^{\H_{\hhh,\phi}}\right)
\end{equation}
to a standard representation.
\end{itemize}
If this conjecture holds, then the above construction determines a bundle of Hilbert spaces over any topological space $\M(\N)\subseteq\N^+_{\star0}$. We will refer to it as a Morozova--Chencov--Petz bundle (it could be defined equivalently using the Morozova--Chencov functions $c$ instead of Petz's functions $\hhh$, see e.g. \cite{Kostecki:2014:qig}). Clearly, it provides an alternative to the GNS bundle. A virtue of the MCP bundle is that it encodes the local riemannian geometry of the state space in the variability of changes of the scalar product. Yet, it remains an open problem whether the above conjecture is true.



\ifvargaugecompile 
\subsection{Local gauge and geodesic propagation\label{local.geodesic.propagation}}
\else 
\subsubsection{Local gauge and geodesic propagation\label{local.gauge.propagation}}
\fi 
If $\glie$ is a Banach Lie algebra with a Lie bracket  $[\cdot,\cdot]$, then a \df{representation} of $\glie$ on a dense subset $\D\subseteq\H$ of a Hilbert space $\H$ is defined as a linear function $\aaa$ mapping each $x\in\glie$ to an anti-selfadjoint operator $\rpktarget{AAA}\aaa(x):\D\ra\H$ such that
\begin{equation}
        \aaa([x,y])=\aaa(x)\aaa(y)-\aaa(y)\aaa(x)\;\;\forall x,y\in\glie.
\end{equation}
Hence, for a given representation $\aaa$ of $\glie$ on $\D\subseteq\H$, every $x\in\glie$ determines a unique self-adjoint, and generally unbounded, operator $\ii\aaa(x)$. By definition, $\D=\dom(\ii\aaa(x))$.


Let $G$ be a Lie group, and let $\M(\N)\subseteq\N^+_\star$ be equipped with a principal $G$-bundle $E\ra\M(\N)$, and a $\glie$-valued connection one-form $\rpktarget{AAAA}\mathbf{A}$ on $E$, where $\glie$ is a Lie algebra of $G$. Moreover, assume that the GNS bundle $\H\M(\N)$ is equipped with the family $\aaa$ of the representations of the Lie algebra $\glie$,\footnote{This definition covers also the representations of $\glie$ in the well-adapted Banach--Lie subalgebras $\B\subseteq\eee^{-1}(\omega)$ (thus, within the fibers of $\H\M(\N)$) as the special case.}
\begin{equation}
	\aaa:=\{\aaa_\omega:\glie\ra(\Lin(\H_\omega))^\asa\mid\omega\in\M(\N)\}.
\end{equation}
The triple $(\M(\N),\mathbf{A},\aaa)$ satisfying the above conditions will be called a \df{local gauge model}, while the pair $(\mathbf{A},\aaa)$ will be called a \df{local gauge structure} on $\M(\N)$.\footnote{The term \df{gauge} means the section of a principal $G$-bundle \cite{Weyl:1918,Weyl:1929}. The \df{local gauge} means the local section, while the \df{global gauge} means the global section. A particularly interesting example of a local gauge structure is provided by the choice of a locally compact and connected Lie group $G=\mathrm{SO}^\uparrow(1,3)\ltimes\RR^4$, known as \df{ortochronous Poincar\'{e} group}.} In principle, a given manifold $\M(\N)$ can admit various different local gauge structures.

If the model $\M(\N)$ is equipped with the local gauge structure, then any curve $c:\RR\ni t\mapsto\phi(t)\in\M(\N)$ corresponds also to a specific choice of a section of the principal $G$-bundle $E$ along this trajectory, which can be expressed by means of integral of a $\glie$-valued connection 1-form $\mathbf{A}$.

If $\aaa$ is determined by setting $\ii\aaa_\omega(\glie)$ to be equal to the generators of the irreducible unitary representation of an action of $G$ on $\H_\omega$, then one can apply Wigner's theorem \cite{Wigner:1939} to each fibre of $\H\M(\N)$ \textit{separately}, classifying the elements of $\H_\omega$ into subsets by means of their transformation properties. According to Wigner's interpretation of this mathematical property (which became widely accepted afterwards), an element of $\H_\omega$ transforming under the above representation of $G$ shall be understood as a pure state of a `quantum particle', where pure state means a vector in a Hilbert space.

Our framework allows to enrich this interpretation by considering a propagation of a `quantum particle' state over the trajectory on the manifold $\M(\N)$, using the $\glie$-valued connection one-form $\mathbf{A}$. As discussed in Section
\ifvargaugecompile 
 \ref{standard.liouvilleans.section}%
\else 
 \ref{relative.modular.theory.section}%
\fi 
, if $\M(\N)\subseteq\N^+_{\star0}$, then every two standard representations determine a unique standard unitary transition between them that preserves the standard cone. Hence, one can map uniquely between the elements of the fibres $\H_{\phi_1}$ and $\H_{\phi_2}$, whenever $\phi_1,\phi_2\in\N^+_{\star0}$, by means of the standard unitary transition operator $V_{\phi_1,\phi_2}$. Let $\M(\N)\subseteq\N_\star^+$, let $c:[0,t]\ra\M(\N)$ be a curve with $c(0)=\omega$ and $c(t)=\phi$, let $\xi\in\H_\omega$ and $\zeta\in\H_\phi$. Then $\zeta$ will be called an \df{$\mathbf{A}$-propagation} of $\xi$ along $c$ if{}f $\ii\aaa_\phi(\int_c\mathbf{A})-\ii J_\phi\aaa_\phi(\int_c\mathbf{A})J_\phi$ is essentially self-adjoint on $\dom(\ii\aaa_\phi(\int_c\mathbf{A}))\cap\dom(\ii J_\phi\aaa_\phi(\int_c\mathbf{A})J_\phi)$ and
\begin{equation}
        \zeta=U^{\mathbf{A}}_{c,\phi}(t)V_{\phi,\omega}\xi
				:=\ee^{-\ii t(\ii\aaa_\phi(\int_{c(t)}\mathbf{A})+\ii J_\phi\aaa_\phi(\int_{c(t)}\mathbf{A})J_\phi)}V_{\phi,\omega}\xi
				=\ee^{t([(\aaa_\phi(\int_{c(t)}\mathbf{A})),\,\cdot\,])}V_{\phi,\omega}\xi.
\label{parallel.zeta}
\end{equation}
The operator $V_{\phi,\omega}$ is a parallel transport associated with the natural connection in the GNS Hilbert bundle determined by the linear structure of the Hilbert space. Hence, the equation \eqref{parallel.zeta} can be understood as an updating map $\xi\mapsto\zeta$ along the trajectory $c(t)$ that takes into account both $\mathbf{A}$ and $\nabla^{\gbold^{1/2}}$ connections.

The above construction suggests introducing more tight relationship between the connection structures of $\T\M(\N)$ and $\H\M(\N)$ for \textit{any} local gauge model $(\M(\N),\mathbf{A},\aaa)$ such that $\M(\N)$ can be equipped with a smooth manifold structure, and with $\M(\N)\subseteq\N^+_{\star0}$. Two quite intriguing possibilities are:
\begin{enumerate}
\item[(QP$_1$)] introduce an affine connection $\nabla$ on $\T\M(\N)$ and define \df{gauge geodesic propagation} as an $\mathbf{A}$-propagation along $\nabla$-geodesic in $\M(\N)$;
\item[(QP$_2$)] introduce: i) an action of $G$ on $\T\M(\N)$ turning it to a tangent $G$-bundle, ii) a $\glie$-valued connection 1-form $\mathbf{A}^\glie_\T$ on $\T\M(\N)$, and iii) a $\glie$-valued connection 1-form $\mathbf{A}_\H^\glie$ on $\H\M(\N)$, such that the $\mathbf{A}_\H^\glie$-parallel transports along $\mathbf{A}_\T^\glie$-geodesics in $\M(\N)$ are equal to $U^\mathbf{A}_{c,\phi}(t)V_{\phi,\omega}$ or $V_{\phi,\omega}U^\mathbf{A}_{c,\omega}(t)$ (where $t$ is an affine parameter of an $\mathbf{A}_\T^\glie$-geodesic $c(t)\in\M(\N)$), and define \df{gauge geodesic propagation} as a horizontal lift of an $\mathbf{A}_\T^\glie$-geodesic in $\M(\N)$ with respect to $\mathbf{A}_\H^\glie$. These gauge geodesic propagations are precisely the $\mathbf{A}$-propagations along $\mathbf{A}_\T^\glie$-geodesics in $\M(\N)$. So, this what we gain by such definition is an additional structure on $\H\M(\N)$ that allows for further study of a relationship between $(G,\glie,\mathbf{A}^\glie_\cdot)$-structures of $\T\M(\N)$ and $\H\M(\N)$. On the other hand, the price paid is the requirement that the $\mathbf{A}_\H^\glie$-parallel transport along $\mathbf{A}_\T^\glie$-geodesic depends only on its endpoint, which holds if $\M(\N)$ is simply connected and $\mathbf{A}_\H^\glie$ is flat.
\end{enumerate}
If the definition (QP$_1$) is used, and $G$ and $\aaa$ are chosen as for Wigner's `quantum particle' classification discussed above, then the gauge geodesic propagation has a direct interpretation as an $\mathbf{A}$-propagation of a `quantum particle' due to ``free fall'' along $\nabla$-geodesic. Note that (despite `dynamical' feeling associated with the word `fall') this propagation has no `dynamical' (causal) content: it is an extension of the description of gauge transformation properties of a `quantum particle' state from a single Hilbert space to a Hilbert space fibre bundle over a quantum model equipped with a local gauge structure. In such approach, a `quantum particle' becomes identified with a (not necessarily global) section of a fibre $G$-bundle represented in terms of the fibre bundle of the GNS Hilbert spaces, so one can discuss its quantum propagation between some `source model' and some `sink model', defined as suitable submanifolds of $\M(\N)$. Under the choice of a definition (QP$_1$), we will define a \df{gauge geodesic propagation model} as a local gauge model $(\M(\N),\mathbf{A},\aaa)$ equipped with an affine connection $\nabla$.

The motivation and interpretation of these definitions echoes Einstein's postulate \cite{Einstein:Grossmann:1913,Einstein:1916} of identification of geodesic lines with the \cytat{world-lines of freely moving point-particles}, but generalised from the pseudo-riemannian geometry, for which the geodesics of the Levi-Civita connection coincide with the curves of extreme distance of a pseudo-riemannian metric, to the setting of general affine connections \cite{Schouten:Haantjes:1934} (see e.g. \cite{Sa:1997,Manoff:2000} and references therein for a comparative discussion of these two approaches). The above construction of geodesic propagation of `quantum particles' is partially influenced by the works of Drechsler \cite{Drechsler:1984,Drechsler:1992,Drechsler:Tuckey:1996} and Prugove\v{c}ki \cite{Prugovecki:1985,Prugovecki:1987,Prugovecki:1991,Prugovecki:1992,Prugovecki:1994,Prugovecki:1995,Prugovecki:1996:I,Prugovecki:1996:II} (see also \cite{Graudenz:1994,Graudenz:1996}). As opposed to them, we do not require any pseudo-riemannian metric on the base manifold, so we do not introduce soldered Poincar\'{e} frame bundles, and we also consider the GNS Hilbert spaces (which may be unitarily inequivalent, if $\omega\notin\N^+_{\star0}$) varying over the base manifold instead of pasting fibre bundle from identical copies of a single Hilbert space. Moreover, our manifold is a space of quantum states over W$^*$-algebras, as opposed to \textit{a priori} postulated background space-time. On the other hand, similarly to Prugove\v{c}ki (see \cite{Prugovecki:1992,Prugovecki:1995,Prugovecki:1996:II}), and as opposed to the approaches of Wightman \cite{Wightman:1956,Wightman:1957,Wightman:1964,Wightman:Gaerding:1964} and Haag--Kastler \cite{Haag:1959,Haag:Kastler:1964}, we impose the requirement of Poincar\'{e} covariance not on the \textit{topological} subsets of the base manifold and on the (presheaves of) algebras of operators associated (functorially \cite{Brunetti:Fredenhagen:Verch:2003}) to those subsets, but on the fibres of \textit{geometric} fibre bundle and on the fibre bundle of Hilbert spaces over this manifold. (Note that, in addition, one can also introduce an independent group covariance requirement on the elements of an underlying W$^*$-algebra, determining this way their transformation properties at each fibre by means of the GNS representation.)

If $\M(\N)$ is equipped with the structure of quantum information manifold, and with a quantum information distance $D_\fff$, then (at least in the finite dimensional case) $\nabla$ can be chosen to be given by $\nabla^{D_\fff}$, $(\nabla^{D_\fff})^\nsdual$, or $(\nabla^{D_\fff}+(\nabla^{D_\fff})^\nsdual)/2$ (the latter is a Levi--Civita of the riemannian geometry $(\M(\N),\gbold^{D_\fff})$). Moreover, if $\M(\N)$ is flat with respect to $\nabla^{D_\fff}$ and $(\nabla^{D_\fff})^\nsdual$, then one can identify the entropic projection of the associated Br\`{e}gman distance $D_\Psi$ with the geodesic projection by $\nabla^{D_\fff}$. If $\fff=\fff_\gamma$, then $D_\fff$ is also a Br\`{e}gman distance. In such case, an assumption (QP$_1$) and identification of $\nabla$ with $\nabla^{D_\gamma}$, allows to state that the propagation of a `quantum particles' is due to the geodesic ``free fall'' along the (local) information dynamics, determined by a constrained minimisation of a relative entropy. 

\ifvargaugecompile 
\section{Locally perturbed liouvilleans\label{from.local.gauge.to.lagrangeans.section}}
\else 
\subsection{Locally perturbed liouvilleans\label{from.local.gauge.to.lagrangeans.section}}
\fi 

In this Section we will use the approach to unbounded `perturbations' of liouvilleans, developed by the Derezi\'{n}ski, Jak\v{s}i\'{c} and Pillet in \cite{DJP:2003}, and based on earlier results of Araki \cite{Araki:1973:relative:hamiltonian,Araki:1973:golden:thompson}. According to \cite{DJP:2003}, if
\begin{enumerate}
        \item $\C\subseteq\BH$ is a von Neumann algebra,
        \item $(\C,\RR,\varsigma)$ is a W$^*$-dynamical system with a standard liouvillean $K^\varsigma$ associated with a standard form $(\H,\C,J,\stdcone)$ of $\C$,
        \item $Q$ is a self-adjoint linear operator affiliated 
to $\C$,
        \item $K^\varsigma+Q$ is an essentially self-adjoint linear operator on $\dom(K^\varsigma)\cap\dom(Q)\subseteq\H$,
\end{enumerate}
then for 
\begin{equation}\rpktarget{DJP.PERT.ALPH}
        \varsigma_t^Q(x):=\ee^{\ii t(K^\varsigma+Q)}x\ee^{-\ii t(K^\varsigma+Q)}\;\;\forall x\in\C\;\forall t\in\RR
\end{equation}
the following statements are true:
\begin{enumerate}
        \item $(\C,\RR,\varsigma^Q)$ is a W$^*$-dynamical system.
        \item the operator
        \begin{equation}
                E_{\varsigma,Q}(t):=\ee^{\ii t(K^\varsigma+Q)}\ee^{-\ii tK^\varsigma}\in\C,
        \end{equation}
        called an \df{expansional}, is unitary and for all $t,t_1,t_2\in\RR$ and all $x\in\C$ it satisfies the following cocycle conditions:
\begin{align}                           \varsigma_t^Q(x)&=E_{\varsigma,Q}(t)\varsigma_t(x)E_{\varsigma,Q}(t)^{-1},\\
                                E_{\varsigma,Q}(t)^{-1}&=E_{\varsigma,Q}(t)^*=\varsigma_t(E_{\varsigma,Q}(-t)),\\
                                E_{\varsigma,Q}(t_1+t_2)&=E_{\varsigma,Q}(t_1)\varsigma_{t_1}(E_{\varsigma,Q}(t_2)).
        \label{EQcocycle}
        \end{align}
\item if\rpktarget{DJP.PERT}
\begin{equation}
        K^{\varsigma,Q}:=K^\varsigma+Q-JQJ
\label{pert.Q}
\end{equation}
is an essentially self-adjoint linear operator on $\dom(K^\varsigma)\cap\dom(Q)\cap\dom(JQJ)$, then a unique self-adjoint extension of $K^{\varsigma,Q}$, denoted (with an abuse of notation) by the same symbol, is a standard liouvillean of $\varsigma^Q$ in $(\H,\C,J,\stdcone)$. We will call $K^{\varsigma,Q}$ and $\varsigma^Q$ a \df{Derezi\'{n}ski--Jak\v{s}i\'{c}--Pillet perturbation} of $K^\varsigma$ and $\varsigma$, respectively.
\item if $Q$ is bounded, then, for any $x\in\C$ and $t\in\RR$, the Dyson--Feynman--Fujiwara--Araki perturbative expansions \cite{Dyson:1949,Feynman:1951,Fujiwara:1952,Araki:1973:expansionals},
\begin{align}
                \varsigma_t^Q(x)&=\sum\limits_{n=0}^\infty \ii^n\int\limits_{0\leq t_n\leq\cdots\leq t_1\leq t}\dd t_1\cdots\dd t_n[\varsigma_{t_n}(Q),[\ldots,               {[}\varsigma_{t_1}(Q),\varsigma_t(x)]\ldots]],\\
                E_{\varsigma,Q}(t)&=\sum\limits_{n=0}^\infty\ii^n\int\limits_{0\leq t_n\leq\cdots\leq t_1\leq t}\dd t_1\cdots \dd t_n\varsigma_{t_n}(Q)\cdots\varsigma_{t_1}(Q).
\end{align}
are convergent in weak-$\star$ topology and define a norm convergent series of bounded operators. Moreover, the associated generators $\der$ of $\varsigma$ and $\der_Q$ of $\varsigma^Q$ have in such case the same domain, and are related by 
\begin{equation}
        \der_{\varsigma^Q}(x)=\der_\varsigma(x)+\ii[Q,x]\;\;\forall x\in\dom(\der_\varsigma).
\label{bounded.perturbation.of.generator}
\end{equation}
\end{enumerate}

\ifvargaugecompile 
\subsection{Local gauge liouvilleans\label{local.gauge}}
\else 
\subsubsection{Local gauge liouvilleans\label{local.gauge}}
\fi 
In Section \ref{algebraic.hamiltonian} we have observed that the Poisson flow of an algebraic hamiltonian vector field on $\N^+_{\star0}$ can be always represented by a unitary evolution in a fibre of the GNS bundle $\H\N_{\star0}^+\ra\N^+_{\star0}$, generated by a standard liouvillean operator on this fibre. In what follows, we will abstract this relationship, replacing tangent bundle by a principal $G$-bundle and replacing the procedure of \textit{restriction} of a \textit{global} algebraic evolution, by the procedure of \textit{extension} of a \textit{local} algebraic evolution in order to incorporate geometric structure as an additional component of an effective dynamics. 

If one \textit{assumes} that the principal $G$-bundle structure participates in an effective form of a temporal evolution, then this evolution shall be represented not by the standard liouvillean on $\H_{\phi(t)}$ alone, but by the standard liouvillean perturbed by the `gauge connection' operator $\ii\aaa_{\phi(t)}(\mathbf{A})$, which represents the change of vectors in the fibres $\H_{\phi(t)}$ caused by the fibrewise action of the group $G$ and the choice of an $\mathbf{A}$-section of a principal $G$-bundle $E$. (When the trajectory along the curve $c:[0,t]\ra\M(\N)$ with $c(0)=\phi(0)$ and $c(t)=\phi(t)$ is investigated as a source of memory effects, then $\mathbf{A}$ should be replaced by $\int_{\phi(0)}^{\phi(t)}\mathbf{A}$.)

For this purpose, we will use DJP perturbation approach, setting $\omega\in\M(\N)\subseteq\N^+_{\star0}$, $\H:=\H_\omega$, $\C:=\pi_\omega(\N)$, $Q:=\ii\aaa_\omega(\mathbf{A})$. We can define $\varsigma$ in two different ways. If $\N$ is equipped with a W$^*$-dynamical system structure $(\N,\RR,\alpha)$, then one can define $\varsigma$ globally in each fiber of the GNS bundle by means of
\begin{equation}
        \varsigma_t(\pi_\omega(x)):=\pi_\omega(\alpha_t(x))\;\;\forall x\in\N\;\;\forall t\in\RR\;\;\forall\omega\in\M(\N).
\label{boring.covariance.cond}
\end{equation}
Alternatively, if $\N$ is equipped with a quantum Poisson system $(\M(\N,\B),h)$ such that $\DF_\omega h\in\N^\sa$ $\forall\omega\in\M(\N,\B)$ and $\M(\N,\B)=\M(\N)$, then one can define $\varsigma$ pointwisely in each fiber by means of
\begin{equation}
	\varsigma_t(\pi_\omega(x)):=\ee^{\ii t\pi_\omega(\DF_\omega h)}\pi_\omega(x)\ee^{-\ii t\pi_\omega(\DF_\omega h)}\;\;\forall x\in\N\;\;\forall t\in\RR\;\;\forall\omega\in\M(\N).
\label{gen.Poisson.compatibility.cond}
\end{equation}
We will call these assuptions a \df{generalised Poisson compatibility condition} (PC$_2$). If $\B=\N^\sa$, $\M(\N,\N^\sa)$ is a submanifold of $\N^\sa_\star$, and the pair $(h,\alpha)$ satisfies \eqref{poisson.flow.compatibility}, then the above two definitions of $\varsigma$ agree. The difference between \eqref{gen.Poisson.compatibility.cond} and \eqref{boring.covariance.cond} (the latter corresponding to the covariance equation \eqref{covariance.equation.group.G}) indicates our approach to quantum dynamics, as being defined locally by the differential geometric properties of state space, instead of a global automorphism of an underlying algebraic structure. Alternatively, if $\M(\N)\subseteq\N^+_{\star0}$ is equipped with a quantum information manifold structure and a global vector field $\XXX_h\in\T\M(\N)$ such that $\XXX_h(\phi)\in\N^\sa$ $\forall\phi\in\M(\N)$, then $\varsigma$ can be defined by
\begin{equation}
	\varsigma_t(\pi_\omega(x)):=\ee^{\ii t\pi_\omega(\XXX_h(\omega))}\pi_\omega(x)\ee^{-\ii t\pi_\omega(\XXX_h(\omega))}\;\;\forall x\in\N\;\forall t\in\RR\;\forall\omega\in\M(\N).
\end{equation}

Thus, under some relatively weak conditions (affiliation of $\ii\aaa_\omega(\mathbf{A})$ with $\pi_\omega(\N)\subseteq\BBB(\H_\omega)$ and essential self-adjointness of sums $K_\omega^\varsigma+\ii\aaa_\omega(\mathbf{A})$ and $K_\omega^\varsigma+\ii\aaa_\omega(\mathbf{A})-\ii J_\omega\aaa_\omega(\mathbf{A})J_\omega$ on the intersection of domains of their components), the local gauge structure can be incorporated in the redefinition of the standard liouvillean. If 
\begin{equation}
        K_\omega^{\varsigma,\ii\aaa(\mathbf{A})}=K_\omega^\varsigma+\ii\aaa_\omega\left(\mathbf{A}\right)-\ii J_\omega\aaa_\omega\left(\mathbf{A}\right)J_\omega
\label{lagrangean.eq}
\end{equation}
satisfies the above conditions, then we will call it a \df{local gauge liouvillean} at $\omega$. 

Let $(\M(\N),\mathbf{A},\aaa)$ be a local gauge model with $\M(\N)\subseteq\N^+_{\star0}$ and let $\varsigma$ be defined as above, either by a quantum Poisson system $(\M(\N,\B),h)$ with $\DF_\omega h\in\N^\sa$ $\forall\omega\in\M(\N,\B)$ and $\M(\N,\B)=\M(\N)$, or by a W$^*$-dynamical system $(\N,\RR,\alpha)$ with $\alpha_\star^t(\M(\N))\subseteq\M(\N)$ $\forall t\in\RR$. Let $\rho_\star$ denote $w^h$ or $\alpha_\star$, respectively.
 If for every $\omega\in\M(\N)$ there exists a family of local gauge liouvilleans
\begin{equation}
        K_\omega^\varsigma+\ii\aaa_\omega\left(
				\mathbf{A}\right)-\ii J_\omega\aaa_\omega\left(
				\mathbf{A}\right)J_\omega,
\label{local.gauge.liouvillean}
\end{equation}
parametrised by $t\in\RR$, then the quadruple $(\M(\N),\varsigma,\mathbf{A},\aaa)$ will be called a \df{local gauge liouvillean model}. 
From the above construction we see that the W$^*$-dynamical system $(\pi_\omega(\N),\RR,\varsigma^{\ii\aaa(\mathbf{A})})$ may not correspond to any W$^*$-dynamical system on $\N$. The description of temporal evolution in terms of $\varsigma^{\ii\aaa(\mathbf{A})}$ is `local' in the sense that it is provided inside of each fibre of the GNS bundle independently. 

\ifvargaugecompile 

\else 
\footnote{If $G_1$ and $G_2$ are groups, and $G_1$ acts on $G_2$ by means of $\cdot:G_1\times G_2\ni(g_1,g_2)\mapsto g_1\cdot g_2\in\G_2$, then a \df{semi-direct product} $G_1\ltimes G_2$ is defined as a group with elements given by pairs $(g_1,g_2)\in G_1\times G_2$, and composition given by $(g_1,g_2)(g_3,g_4)=(g_1g_3,g_2(g_1\cdot g_4))$.}
\fi 

\ifvargaugecompile 
\subsection{Local source liouvilleans\label{local.source.liouv.section}}
\else 
\subsubsection{Local source liouvilleans\label{local.source.liouv.section}}
\fi 

In principle, apart from the `internal' dynamics (implemented by the evolution $\varsigma^t$) and the kinematic local gauge structure, the effective dynamics can also depend on some controlled `external' constraints. We will assume that these constraints can be specified in terms of external `sources', which can generally be represented by the variations $\delta(\phi(x))$ of expectation values. These variations can be decomposed into two parts: the variations $(\delta\phi)(x)$ of states, and the variations $\phi(\delta x)$ of operators. The constraints on changes $\delta\phi$ can be handled by restricting the form of the model $\M(\N)$ (for a geometric approach, see \cite{Mitchell:1967,Jaynes:1993,Favretti:2007,Kostecki:2013:PhD}). On the other hand, note that $\delta x$ can be in principle arbitrary, so it can also depend on $\phi$, and it may not arise as an infinitesimal change generated by a global automorphism of $\N$ (thus, it cannot be described by the setting of derivations of C$^*$-algebras). We will implement the perturbations $\delta x$ of elements $x$ of (a local GNS representation of) a W$^*$-algebra $\N$ by means of state dependent perturbations of liouvilleans. In this sense, the constraints on changes of operators will be handled by local (state dependent) additional terms modifying liouvillean evolution.

For this purpose, apart from local gauge structure on $\M(\N)\subseteq\N_\star^+$, we introduce also \df{local source term}, defined as a fibrewise family of operators
\begin{equation}
        (\lambda,H):\M(\N)\ni\omega\mapsto\lambda(\omega)H(\omega)\in(\Lin(\H_\omega))^\sa,
\end{equation}
with $\lambda(\omega)\in\RR$ called \df{local source strength} and $H(\omega)\in(\Lin(\H_\omega))^\sa$ called \df{local source operator}. If $\lambda(\omega)$ is independent of $\omega$, then it will be called \df{global source strength}. The $(2n+1)$-tuple
\begin{equation}
        (\M(\N),(\lambda_1,H_1),\ldots,(\lambda_n,H_n))
\label{local.source.model}
\end{equation}
will be called \df{local source model} if{}f $(\lambda_i,H_i)$ is a local source term for each $i\in\{1,\ldots,n\}$. If \eqref{local.source.model} is a local source model, $\M(\N)\subseteq\N^+_{\star0}$, $\varsigma$ is defined as in Section \ref{local.gauge}, and for $\omega\in\M(\N)$ the DJP perturbation of $K_\omega^\varsigma$ by $\lambda_1(\omega)H_1(\omega)+\ldots+\lambda_n(\omega)H_n(\omega)$ exists, then a unique self-adjoint extension of an essentially self-adjoint operator
\begin{equation}
        K_\omega^{\varsigma,\lambda_1H_1,\ldots,\lambda_nH_n}=K_\omega^\varsigma+\lambda_1(\omega)H_1(\omega)+\ldots+\lambda_n(\omega)H_n(\omega)-J_\omega(\lambda_1(\omega)H_1(\omega)+\ldots+\lambda_n(\omega)H_n(\omega))J_\omega
\end{equation}
will be called \df{local source liouvillean} at $\omega$. If a local source liouvillean exists for each $\omega\in\M(\N)$, then the $2(n+1)$-tuple $(\M(\N),\varsigma,\lambda_1,H_1,\ldots,\lambda_n,H_n)$ will be called \df{local source liouvillean model}.

Let us note that a local source $\lambda_i(\omega)H_i(\omega)$ at $\omega$ should be understood not as the $i$-th type ``interaction source'' \textit{localised} at $\omega$, but as a strength-and-action of the $i$-th type ``interaction source'' \textit{perceived} at location $\omega$.

Let $(\M(\N),\mathbf{A},\aaa)$ be local gauge model. Let $(\M(\N),\lambda_1,H_1,\ldots,\lambda_n,H_n)$ be local source model. Let $\varsigma$ be such as defined in Section \ref{local.gauge}. If, for a given $t\in\RR$ and $\omega\in\M(\N)$, there exists a DJP perturbation of a standard liouvillean $K_\omega^\varsigma$, given by the unique self-adjoint extension of an essentially self-adjoint operator
\begin{equation}
        \Lcal(\omega,t):=K_\omega^\varsigma+\ii\aaa_\omega\left(
				\mathbf{A}\right)-\ii J_\omega\aaa_\omega\left(
				\mathbf{A}\right)J_\omega+\sum_{i=1}^n\left(\lambda_i(\omega)H_i(\omega)-J_\omega\lambda_i(\omega)H_i(\omega)J_\omega\right),
\end{equation}
then $\rpktarget{EXT.LIOUV}\Lcal(\omega,t)$ will be called a \df{local liouvillean operator} at $(\omega,t)$. 
If $\Lcal(\omega,t)$ exists for all $t\in\RR$ and all $\omega\in\M(\N)$, then the $(4+2n)$-tuple
\begin{equation}
        (\M(\N),\varsigma,\mathbf{A},\aaa,\lambda_1,H_1,\ldots,\lambda_n,H_n)
\end{equation}
will be called an \df{local liouvillean model}. In the special case, all of operators $H_1,\ldots,H_n$ can be determined by the elements $h_1,\ldots,h_n$ of a W$^*$-algebra $\N$, with
\begin{equation}
        H_i(h_i):\M(\N)\ni\omega\mapsto H_i(\omega):=\pi_\omega(h_i)\in\BH^\sa.
\label{linear.sources.in.GNS.bundle}
\end{equation}
This allows, in particular, for a fibrewise representation of a `global gauge' action $G_0\ra\Aut(\N)$ of some Lie group $G_0$ (not necessarily related to $G$), whenever $\{h_i\}\subseteq\N$ are the generators of the representation of $G_0$ in $\Aut(\N)$. This observation can be generalised to $l$ subsets of $\{h_1,\ldots,h_n\}$ playing the role of generators of $l$ representations of $l$ Lie groups $G_l\ra\Aut(\N)$.

\ifvargaugecompile 
\subsection{Case study: The BLP perspective on nonlinear quantum fields}
\else 
\subsubsection{Case study: The BLP perspective on nonlinear quantum fields}
\fi 

Restriction of considerations from topological spaces $\M(\N)$ to BLP manifolds $\M(\N,\B)$ allows us to equip the operator algebraic approach with an additional differential geometric content, 
using Fr\'{e}chet derivatives of smooth functions on $\B_\star$ in the role of differential forms. In this Section we will investigate the possibility of interpretation of these forms as nonlinear quantum fields (understood in quite formal sense).

If $f\in\CIF(\B_\star;\RR)$ and $\phi\in\M(\N,\B)$, then $\DF_\phi f=\ddd_\phi f\equiv\ddd f(\phi)\in\T^\ct_\phi\M(\N,\B)$. Thus, if $\M(\N,\B)\subseteq\N^+_{\star0}$, $f\in\CIF(\M(\N,\B);\RR)$ satisfies $\DF_\omega f\in\N^\sa$ $\forall\omega\in\M(\N,\B)$, and $\varsigma$ is defined as in the previous two Sections, then one can consider local source liouvilleans determined by the perturbation
\begin{equation}
         K_\omega^\varsigma+\lambda(\omega)\left(\pi_\omega(\DF_\omega f)-J_\omega\pi_\omega(\DF_\omega f)J_\omega\right),
\end{equation}
for some family $\lambda(\omega)\in\RR$ $\forall\omega\in\M(\N,\B)$. As opposed to a function used for definition of $\varsigma$ and $K^\varsigma_\omega$, $f$ is not considered as a generator of a Poisson flow, and it is allowed to be arbitrary rescaled by $\lambda(\omega)$ at each point. 

Every $x\in\B$ can be represented again as a smooth function on $\B_\star$ by means of $\B_\star\ni\omega\mapsto\omega(x)\in\RR$, allowing to consider elements of $\B$ arising from multiple Fr\'{e}chet differentiation,
\begin{equation}
        \DF_{\omega_n}(\phi_{n-1}(\DF_{\omega_{n-1}}(\cdots(\phi_1(\DF_{\omega_1}f)))))\in\B,
\label{multiple.point.frechet}
\end{equation}
for $\omega_1,\ldots,\omega_n,\phi_1,\ldots,\phi_{n-1}\in\B$. These derivatives can be added and multiplied as elements of $\B$, and any of the resulting elements of $\B$ can be subjected to a representation in the GNS bundle as a local source operator. However, despite multiple application of Fr\'{e}chet differentiation, objects of type \eqref{multiple.point.frechet} are (just) elements of $\T^\ct_{\omega_n}\B_\star$. This leads us to ask whether it is possible to introduce higher order tensors on $\B_\star$, which could be used as source terms acting on the GNS bundle. The natural candidates for this purpose are \df{$(n,m)$-tensor fields} over $\B_\star$, defined pointwisely as
\begin{equation}
        \XXX_{k_1}(\phi)\boxtimes\cdots\boxtimes\XXX_{k_n}(\phi)\boxtimes\ddd f_1(\phi)\boxtimes\cdots\boxtimes\ddd f_m(\phi),
\end{equation}
which are the elements of
\begin{equation}
\left(\bigboxtimes^n\T_\phi\B_\star\right)\boxtimes\left(\bigboxtimes^m\T^\ct_\phi\B\right)       
\iso\left(\bigboxtimes^n\B_\star\right)\boxtimes\left(\bigboxtimes^m\B\right),
\end{equation}
where $\boxtimes$ denotes the tensor product considered in an algebraic sense (that is, without taking topological completion) and the dependence on $\phi$ is assumed to be smooth. The contraction at $\phi$ of an $(n,m-1)$-tensor field with an $(n,m)$-tensor field by means of the componentwise application of the duality $\duality{\cdot,\cdot}_{\B_\star\times\B}$ gives a one-form at $\phi$, which belongs to $\B$:
\begin{align}
        &\duality{\ddd l_1(\phi)\boxtimes\cdots\boxtimes\ddd l_n(\phi)\boxtimes\XXX_{h_1}(\phi)\boxtimes\cdots\boxtimes(\cdot)_i\boxtimes\cdots\XXX_{h_m}(\phi),\XXX_{k_1}(\phi)\boxtimes\cdots\boxtimes\XXX_{k_n}(\phi)\boxtimes\ddd f_1(\phi)\boxtimes\cdots\boxtimes\ddd f_m(\phi)}\nonumber\\
        &=\duality{\XXX_{k_1}(\phi),\ddd l_1(\phi)}\cdots
        \duality{\XXX_{k_n}(\phi),\ddd l_n(\phi)}
        \duality{\XXX_{h_1}(\phi),\ddd f_1(\phi)}
        \duality{\XXX_{h_{i-1}}(\phi),\ddd f_{i-1}(\phi)}
        \duality{\XXX_{h_{i+1}}(\phi),\ddd f_{i+1}(\phi)}\nonumber\\
        &\cdots\duality{\XXX_{h_m}(\phi),\ddd f_m(\phi)}\cdot\ddd f_i(\phi)=:\lambda(\phi)\ddd f_i(\phi).
\end{align}
When subjected to representation as a source term, $\lambda(\phi)$ is a natural candidate for a local source strength of a local source operator $\pi_\phi(\ddd f_i(\phi))$. We will (tentatively) call the local source operators of this type \df{quantum fields}. One can also introduce the antisymmetric wedge product on vectors and covectors (one forms) and define the corresponding contraction to 1-form and its source term representation in an analogous way. In particular, for $\B\iso\N^\sa$, a vector field $\XXX_k\in\T\M(\N,\B)$ can be represented in terms of a GNS fibre bundle $\H\M(\N,\B)$ as a family
\begin{equation}
        \M(\N,\B)\ni\phi\mapsto\omega_{\XXX_k}(\phi)\in\BBB(\H_\phi)_\star^\sa\iso\schatten_1(\H_\phi)^\sa
\end{equation}
determined by
\begin{equation}
\duality{\XXX_k(\phi),\ddd f(\phi)}_{\T_\phi\N_\star^\sa\times\T^\ct_\phi\N_\star^\sa}=:\duality{\omega_{\XXX_k}(\phi),\pi_\phi(\ddd f(\phi))}_{\BBB(\H_\phi)_\star^\sa\times\BBB(\H_\phi)^\sa}\;\;\forall\phi\in\M(\N)\;\forall f\in\CIF(\N_\star^\sa;\RR).
\end{equation}
The GNS representation of $(0,n)$-tensor fields with $n>1$ is also possible, however it can be provided in different ways. For example, given a $(0,2)$-tensor field $(\ddd f\boxtimes\ddd k)(\phi)$, it is possible to represent it as:
\begin{align}
        \pi_\phi(\ddd f(\phi))\otimes\pi_\phi(\ddd k(\phi))&\in\BBB(\H_\phi)\otimes\BBB(\H_\phi),\\
        \pi_\phi(\ddd f(\phi))\otimes\pi_\phi(\ddd k(\phi))&\in\BBB(\H_\phi\otimes\H_\phi),\\
        \pi_{\phi\otimes\phi}((\ddd f\boxtimes\ddd k)(\phi))&\in\BBB(\H_{\phi\otimes\phi}),\label{two.form.represent}
\end{align}
where $(\H_{\phi\otimes\phi},\pi_{\phi\otimes\phi},\Omega_{\phi\otimes\phi})$ is the GNS representation of $\N\otimes\N$ in the state $\phi\otimes\phi\in(\N\otimes\N)_\star^+$. It seems that the representation \eqref{two.form.represent} preserves most precisely the geometric content of $(0,2)$-tensor field, so we feel tempted to consider it as a preferred construction. However, this leads us to construction of a whole family of fibre bundles of Hilbert spaces over the manifold $\M(\N,\B)$. It is unclear at this stage whether this phenomenon should be considered as a virtue or as a failure. For $n$-ary tensor product $\phi\otimes\cdots\otimes\phi$ the corresponding fibre bundle of $\H_{\phi\otimes\cdots\otimes\phi}$ spaces, with $\phi$ varying over $\M(\N,\B)$, will be called a \df{$\bigotimes^n$-GNS bundle}, and denoted $\rpktarget{GNS.n}(\bigotimes^n\H)\M(\N,\B)$. For any $(0,n)$-tensor field $\ddd f_1\boxtimes\cdots\boxtimes\ddd f_n$ on $\M(\N,\B)$ with $n\leq\dim\M(\N,\B)$ there exists a unique representation
\begin{equation}
        \pi_{\phi\otimes\cdots\otimes\phi}\left((\ddd f_1\boxtimes\cdots\boxtimes\ddd f_n)(\phi)\right)\in\BBB(\H_{\phi\otimes\cdots\otimes\phi}).
\end{equation}
The same is true for any \df{$n$-form field}, defined as a section of the fibre bundle of an antisymmetric wedge product $\bigwedge^n\T^\ct\B_\star$, because $\bigwedge^n\T_\phi^\ct\B_\star\subset\bigboxtimes\T^\ct_\phi\B_\star$. As a result, each smooth section of $\bigwedge^n\T^\ct\M(\N,\B)$ can be represented as a family of bounded operators,
\begin{equation}
        \pi_{\phi\otimes\cdots\otimes\phi}:\bigwedge^n\T^\ct_\phi\B_\star\ni x\mapsto\pi_{\phi\otimes\cdots\otimes\phi}(x)\in\BBB(\H_{\phi\otimes\cdots\otimes\phi}),
\end{equation}
acting fibrewise over the $\bigotimes^n$-GNS bundle $(\bigotimes^n\H)\M(\N,\B)$. We will call such family a \df{quantum $n$-form field} over $\M(\N,\B)$.

The constant function on $\N_\star^\sa$, $\hat{\lambda}:\N_\star^\sa\ni\omega\mapsto\hat{\lambda}(\omega):=\lambda\in\RR$, is a geometric representation of an algebraic element of a center of $\N$, $\lambda\II\in\zentr_\N\subseteq\N$, in terms of an element of a smooth algebra, $\hat{\lambda}\in\CIF(\N_\star^\sa;\RR)$. Such function on $\N_\star^\sa$ will be called a \df{global charge}. From this it follows that, provided $\B\iso\N^\sa$, each globally constrained source strength is a global charge.

The set of quantum field one-forms in $\T^\ct\B_\star$, considered under its restriction to some quantum model $\M(\N,\B)\subseteq\N_\star^+$, can be equipped with the additional structure of a Lie algebra $\mathfrak{h}$, determined by the structure constants $\epsilon_\mathfrak{h}^{abc}$ of its adjoint representation by means of
\begin{equation}
        [(\ddd_\phi f)^a,(\ddd_\phi k)^b]_\mathfrak{h}=\sum_c\epsilon_\mathfrak{h}^{abc}(\phi)(\ddd_\phi h)^c\;\;\;\forall\phi\in\M(\N,\B).
\label{structure.Lie.forms}
\end{equation}
Here $(\ddd_\phi f)^a$ denotes a Lie algebra representation map $\mathfrak{h}\ra(\T^\ct\B_\star)$ at $\phi\in\M(\N,\B)$. Representation of these forms on the fibres of the GNS bundle gives
\begin{equation}
        [\pi_\phi((\ddd_\phi f)^a),\pi_\phi((\ddd_\phi k)^b)]_\mathfrak{h}=\sum_c\epsilon_\mathfrak{h}^{abc}(\phi)\pi_\phi((\ddd_\phi h)^c)\;\;\;\forall\phi\in\M(\N,\B).
\label{structure.Lie.operators}
\end{equation}
Note that the Lie algebra structure given by \eqref{structure.Lie.operators} is a priori independent of any possible principal $G$-bundle structure of $\H\M(\N,\B)$ or a principal $G$-bundle $E\ra\M(\N,\B)$ represented in terms on $\H\M(\N,\B)$ by means of a local gauge liouvillean. In order to keep the same relationship between source terms $\lambda_i(\phi)\pi_\phi((\ddd_\phi f_i)^a)$ in each fibre $\H_\phi$, one has to set $\lambda_i$ to be given by a global charge. We will call the source terms $\pi_\phi((\ddd_\phi f_i)^a)$ \df{local gauge quantum fields}, while the corresponding local source models with global source strengths $\lambda_i$ will be called \df{local gauge quantum field models}. If an extended liouvillean model $(\M(\N,\B),\mathbf{A},\aaa,\lambda_1,H_1,\ldots,\lambda_n,H_n)$ is equipped with an affine connection $\nabla$ such that $(\M(\N,\B),\mathbf{A},\aaa,\nabla)$ is a gauge geodesic propagation model, while $(\M(\N,\B),\lambda_1,H_1,\ldots,\lambda_n,H_n)$ is a quantum field model with $H_i(\phi)=\pi_\phi((\ddd_\phi f_i)^{a_i})$ $\forall i\in\{1,\ldots,n\}$, then the $(4+2n)$-tuple
\begin{equation}
        (\M(\N),\mathbf{A},\aaa,\nabla,\lambda_1,(\ddd f_1)^{a_1},\ldots,\lambda_n,(\ddd f_n)^{a_n})
\end{equation}
can be called a `quantum field model with gauge geodesic propagation'. If $G$ and $\aaa$ are chosen to agree with the Wigner classification theorem, then such model describes a family of quantum fields together with a quantum particle geodesic propagation. However, while the availability of these constructions is a quite remarkable fact, it is yet unclear how they could be translated to the usual objects of quantum field theory.


\ifvargaugecompile 
\subsection{Local liouvillean instruments and correlation functions}
\else 
\subsubsection{Local liouvillean instruments and correlation functions}
\fi 

The temporal evolution of a local liouvillean model is completely described by the fibrewise evolution
\begin{equation}
        \xi(\omega,t)=\ee^{-\ii t\Lcal(\omega,t)}\Omega_\omega,
\label{lagrangean.movement}
\end{equation}
which is a generalisation of \eqref{liovillean.movement} taking local gauge and local source structures on $\M(\N)\subseteq\N^+_{\star0}$ into account. The corresponding propagator (transition amplitude) between initial state $\omega$ and the final state $\phi$ reads
\begin{equation}
	\s{\Omega_\phi,V_{\phi,\omega}\xi(\omega,t)}_{\H_\phi}=\s{\Omega_\phi,V_{\phi,\omega}\ee^{-\ii t \Lcal(\omega,t)}\Omega_\omega}_{\H_\phi}.
\label{liouville.propagator}
\end{equation}
If $\Lcal$ is determined only by a given Poisson system $(\M(\N,\B),h)$ with $\DF_\omega h\in\N^\sa$ $\forall\omega\in\M(\N,\B)$, then the propagator \eqref{liouville.propagator} reads
\begin{equation}
	\s{\Omega_\phi,V_{\phi,\omega}\ee^{-\ii t\pi_\omega(\DF_\omega(h))}}_{\H_\phi}
	=\s{\Omega_\phi,\ee^{\ii t\log(J_\phi J_{\phi,\omega})}\ee^{-\ii t\pi_\omega(\DF_\omega(h))}\Omega_\omega}_{\H_\phi}.
\label{liouv.prop}
\end{equation}

We will now show that this evolution can be expressed as an instrument acting on $\N_\star^+$ and parametrised by $t\in\RR$. Let us choose some $\psi\in\N^+_{\star0}$. Then the elements of each fibre of the GNS bundle over $\N^+_{\star0}$ can be uniquely mapped to $\H_\psi$ by means of the standard unitary transition operator $V_{\psi,\omega}=J_\psi J_{\psi,\omega}=J_{\psi,\omega}J_\omega$, which preserves the positive cones
\ifvargaugecompile 
 (see Section \ref{standard.liouvilleans.section}).
\else 
 (see Section \ref{relative.modular.theory.section}).
\fi 
 Hence, at each value of $t\in\RR$ and at each $\psi\in\N^+_{\star0}$, the set
\begin{equation}
        \bigcup_{\omega\in\M(\N)}\{V_{\psi,\omega}\xi(\omega,t)\}\subseteq\H_\psi^+
\end{equation}
represents completely the evolution in a fibre bundle $\H\M(\N)$ that is defined by means of a local liouvillean operator. Using the bijective norm continuous homomorphism $\stdembed^\natural_\psi:\H_\psi^+\ra\N_\star^+$ (defined as $\stdembed^\natural_\pi$ for $\pi=\pi_\psi$), we can represent the mapping
\begin{equation}
        \RR\ni t\mapsto\bigcup_{\omega\in\M(\N)}\left\{V_{\psi,\omega}\xi(\omega,t)\right\}\subseteq\H_\psi^+
\end{equation}
as a temporal evolution of subsets of $\N_\star^+$,
\begin{equation}
        t\mapsto\stdembed^\natural_\psi\left(\bigcup_{\omega\in\M(\N)}\left\{V_{\psi,\omega}\xi(\omega,t)\right\}\right)\subseteq\N_\star^+.
\end{equation}
\ifvargaugecompile 
The 
\else 
It follows that the 
\fi 
 family of mappings
\begin{equation}
        \RR\ni t\mapsto\left\{\III_{\Lcal,\psi}(t):\M(\N)\ni\omega\mapsto(\III_{\Lcal,\psi}(t))(\omega):=\stdembed^\natural_\psi(V_{\psi,\omega}\ee^{-\ii t\Lcal(\omega,t)}\Omega_\omega)\in\N_\star^+\right\}
\end{equation}
\ifvargaugecompile 
%
\else 
is an instrument, which
\fi 
 will be called a \df{local liouvillean instrument} (this name may be a bit deceiving, because of nonlocality inherent in the $V_{\phi,\omega}$ operation. The uniqueness of standard unitary transition for each pair of elements of $\N^+_{\star0}$ together with bijectivity of $\stdembed^\natural_\psi$ implies
\begin{equation}
        \stdembed_\psi^\natural V_{\psi,\omega}\xi=\stdembed_\varphi^\natural V_{\varphi,\omega}\xi\;\;\;\forall\xi\in\H_\omega\;\forall\varphi,\omega\in\N^+_{\star0},
\end{equation}
hence an extended liouvillean instrument does not depend on the choice of $\psi$. In what follows we will denote it by $\rpktarget{LIOUV.INSTR}\III_{\Lcal(t)}$. Due to bijectivity of $\stdembed^\natural_\psi$ there is an equivalence between the evolution generated on the GNS bundle by the family of extended liouvillean operators and the evolution generated on $\N_\star^+$ by the extended liouvillean instrument. Thus, one can consider the extended liouvillean operator structure over a given model (including local gauge and local source structures, as well as the isometry $\alpha_\star$ or a Poisson flow $w^h$) as \textit{auxiliary} tools allowing to define suitable extended liouvillean instrument, but otherwise devoid of any foundational meaning.

\ifvargaugecompile 
\else 
\fi 

The GNS bundle allows to construct the $n$-point correlation functions, whenever all quantum states under consideration are faithful. Let $\N$ be a W$^*$-algebra, let $\phi_0,\phi_1,\ldots,\phi_n\in\N^+_{\star0}$ and let $x_1,\ldots,x_n\in\N$. Then we can define the \df{time independent $n$-point correlation function} as
\begin{equation}
        \e{x_1(\phi_1)\cdot\ldots\cdot x_n(\phi_n)}_{\phi_0}:=\s{\Omega_{\phi_0},V_{\phi_0,\phi_1}\pi_{\phi_1}(x_1)\cdots V_{\phi_{n-1},\phi_n}\pi_{\phi_n}(x_n)V_{\phi_n,\phi_0}\Omega_{\phi_0}}_{\H_{\phi_0}}.
\label{time.independent.corr.fun}
\end{equation}
If $\M(\N)\subseteq\N^+_{\star0}$ is $m$-dimensional, $\phi_0,\phi_1,\ldots,\phi_n\in\M(\N)$ and $\theta:\M(\N)\ra\RR^m$ is a coordinate system on $\M(\N)$, then \eqref{time.independent.corr.fun} can be expressed in terms of $\theta$ as
\begin{align}
\e{x_1(\theta_1)\cdot\ldots\cdot x_n(\theta_n)}_{\phi(\theta_0)}=&\left\langle\Omega_{\phi_0},V_{\phi(\theta_0),\phi(\theta_1)}\pi_{\phi(\theta_1)}(x_1)\cdots\right.\nonumber\\&\left.\cdots V_{\phi(\theta_{n-1}),\phi(\theta_n)}\pi_{\phi(\theta_n)}(x_n)V_{\phi(\theta_n),\phi(\theta_0)}\Omega_{\phi(\theta_0)}\right\rangle_{\H_{\phi(\theta_0)}},
\end{align}
where $\theta_0,\theta_1,\ldots,\theta_n\in\RR^m$ and $\phi_i=\phi(\theta_i):=\theta^{-1}(\theta_i)$. Constructions provided in this
\ifvargaugecompile 
 paper 
\else 
 section
\fi 
 allow us to define also the \df{time dependent $n$-point correlation functions} as
\begin{align}
        \e{x_1(\phi_1,t_1)\cdot\ldots\cdot x_n(\phi_n,t_n)}_{\phi_0}:=&\left\langle\Omega_{\phi_0},V_{\phi_0,\phi_1}\ee^{+\ii t\Lcal(\phi_1,t_1)}\pi_{\phi_1}(x_1)\ee^{-\ii t\Lcal(\phi_1,t_1)}\cdots\right.\nonumber\\&\left.\cdots V_{\phi_{n-1},\phi_n}\ee^{+\ii t\Lcal(\phi_n,t_n)}\pi_{\phi_n}(x_n)\ee^{-\ii t\Lcal(\phi_n,t_n)}V_{\phi_n,\phi_0}\Omega_{\phi_0}\right\rangle_{\H_{\phi_0}},
\label{time.dependent.corr.fun}
\end{align}
where $t_1,\ldots,t_n\in\RR$ and $\Lcal(\phi,t)$ is an extended liouvillean. If a reformulation in terms of a coordinate system $\theta$ (defined above) is possible, then \eqref{time.dependent.corr.fun} can be expressed as
\begin{align}
        \e{x_1(\theta_1,t_1)\cdot\ldots\cdot x_n(\theta_n,t_n)}_{\phi(\theta_0)}
				=\left\langle\Omega_{\phi(\theta_0)},V_{\phi(\theta_0),\phi(\theta_1)}
				\ee^{+\ii t\Lcal(\phi(\theta_1),t_1)}\pi_{\phi(\theta_1)}(x_1)\ee^{-\ii t\Lcal(\phi(\theta_1),t_1)}\cdots\right.\nonumber\\\left. \cdots V_{\phi(\theta_{n-1}),\phi(\theta_n)}\ee^{+\ii t\Lcal(\phi(\theta_n),t_n)}\pi_{\phi(\theta_n)}(x_n)\ee^{-\ii t\Lcal(\phi(\theta_n),t_n)}V_{\phi(\theta_n),\phi(\theta_0)}\Omega_{\phi(\theta_0)}\right\rangle_{\H_{\phi(\theta_0)}}
\end{align}
Due to different values taken by the components of $\Lcal$ operator at different points $(\phi,t)$, the evolution determined by \eqref{lagrangean.movement} and \eqref{time.dependent.corr.fun} does not have to be unitary. It is so only when the dynamics and perturbations in all fibres of the Hilbert bundle are the same.

Equation \eqref{time.dependent.corr.fun} describes how the predictive time dependent content of a quantum model $\M(\N)$ can be determined using the representation of geometric structures on $\M(\N)$ in terms of the algebraic structures on the GNS bundle. However, let us note that this equation is only an example of the variety of possible definitions of the time dependent correlation functions that could be constructed with the help of local liouvilleans and the GNS bundle. Moreover, one could carry the above constructions also for MCP bundle, obtaining different quantitative results. The identification of the proper construction should be based on a more detailed analysis of backwards compatibility with other approaches. In next Section we will approach the derivation of the path integral analogue of the propagator \eqref{liouville.propagator}.
\ifvargaugecompile 
\else 
\subsubsection{Quantum lagrangeans}
\fi 
\section{Quantum histories\label{quantum.histories.section}}
In order to solve the problems of `measurement' and `time' in quantum theory Griffiths \cite{Griffiths:1984}, Omn\`{e}s \cite{Omnes:1988:I,Omnes:1988:II,Omnes:1988:III,Omnes:1989:IV,Omnes:1990,Omnes:1992,Omnes:1994}, and Gell-Mann and Hartle \cite{GellMann:Hartle:1990:I,GellMann:Hartle:1990:II,GellMann:Hartle:1990:III,Hartle:1991:I,Hartle:1991:II,GellMann:Hartle:1993,Hartle:1995} have developed  `consistent histories' approach to quantum theory. Isham and Linden \cite{Isham:1994,Isham:Linden:1994,Isham:Linden:1995,Isham:Linden:SS:1997,Isham:Linden:Schreckenberg:1994} have proposed a modification of this approach, called the (continuous-time) `history projection operator' approach, which was developed later by Savvidou and Anastopoulos \cite{Savvidou:1999,Savvidou:1999:PhD,Anastopoulos:Savvidou:2002,Anastopoulos:2001,Anastopoulos:Savvidou:2003}. 

In Section \ref{histories.propositions.section} we will recall the elementary mathematical structure of the Isham--Linden approach, in Section \ref{Savvidou.action.operator} we will discuss Savvidou's construction \cite{Savvidou:1999,Savvidou:1999:PhD} of an action operator within this setting, while in Sections \ref{geometric.phase.histories} and \ref{Hilbert.space.geometry.path.integrals} we will follow the Anastopoulous--Savvidou analysis of the relationship of this framework with the geometric structures on the spaces of pure quantum states, and the Daubechies--Klauder \cite{Daubechies:Klauder:1985,Klauder:1988,Watson:Klauder:2002} continous-time regularised coherent states phase space path integration. Sections \ref{histories.propositions.section}-\ref{Hilbert.space.geometry.path.integrals} do not contain new results. Their aim is to lead us to a refined geometric perspective on the relationship between the Daubechies--Klauder formula and the local liouvilleans. In Section \ref{algebraic.action.operator.section} we will apply the local liouvillean approach to the Falcone--Takesaki construction of noncommutative flow of weights in order to construct the algebraic analogue of Savvidou's histories action operator. The discussion of limitations of this construction in the face of the results of Section \ref{from.local.gauge.to.lagrangeans.section} will lead us to construction of W$^*$-geometric generalisation of the Daubechies--Klauder path integration in Section \ref{algebraic.quantum.histories}.
\subsection{Propositions and evolution\label{histories.propositions.section}}
The starting point of the history projection operator approach is consideration of a \textit{history} $\varpi$ of abstract `propositions' $(P_{t_1},{P}_{t_2},\ldots,{P}_{t_n})$ about a quantum theoretic model, assigned to an ordered finite sequence  $(t_1,t_2,\ldots,t_n)$, where $t_1<t_2<\ldots<t_n$, and $t$ is interpreted as `time' parameter. Following the ideas of Mittelstaedt \cite{Mittelstaedt:1977} and Stachow \cite{Stachow:1980,Stachow:1981}, Isham \cite{Isham:1994} has proposed to specify this entity by the projection operator
\begin{equation}
        P_\varpi:=P_{t_1}\otimes P_{t_2}\otimes\ldots\otimes P_{t_n},
\end{equation}
acting on the Hilbert space 
\begin{equation}
{\cal{V}}_n:=\bigotimes_{i=1}^n\H_{t_i}:=\H_{t_1}\otimes\H_{t_2}\otimes\ldots\otimes\H_{t_n},
\end{equation}
where ${P}_{t_i}$ is a projection operator in the $n$-th copy of the Hilbert space $\H_{t_i}:=\H$ of a given quantum model. The history $\varpi$ of nonunitary propositions as well as the description of the unitary dynamics are contained in the \textit{class operator} on ${\cal{V}}_n$, defined as \cite{GellMann:Hartle:1990:I}
\begin{equation}
    C_\varpi:=U(t_0,t_1){P}_{t_1}U(t_1,t_2){P}_{t_2}\cdots U(t_{n-1},t_n){P}_{t_n}U(t_n,t_0),
\end{equation}
where $U(t_i,t_{i+1})=\ee^{-\ii\int_{t_{i}}^{t_{i+1}}\dd tH}$ are unitary evolution operators between times $t_i$ and $t_{i+1}$, acting on the Hilbert space ${\H}_{t_{i+1}}$ and generated by a self-adjoint hamiltonian operator $H$. For a dynamics generated by the hamiltonian $H$ with an initial state described by the density operator $\rho$, \textit{the probability of a history $\varpi$} is defined as
\begin{equation}
        p(\varpi;\rho,H):=\tr_{{\cal{V}}_n}(C^*_\varpi\rho C_\varpi).
\end{equation}
Using this equation, for two given histories $\varpi$ and $\vartheta$, one defines the \textit{histories functional}\footnote{For historical reasons, this object is usually called `decoherence functional'. However, such name suggests that the quantum histories formalism necessary involves the `decoherence approach to quantum measurement' semantics, which is not true. For this reason we choose to change the name of this mathematical object to much more neutral with respect to possible semantics.} \cite{Griffiths:1984}
\begin{equation}
\hf_{\rho,H}:\BBB({\cal{V}}_n)\times\BBB({\cal{V}}_n)\ni(P_\varpi,P_\vartheta)\mapsto\tr_{{\cal{V}}_n}(C^*_\varpi\rho C_\vartheta)\in\CC,
\label{decoherencefunctional}
\end{equation}
which, by definition, depends on $\rho$ and $H$. It satisfies, for $P_\varpi\leq\II-P_{\varkappa}$, 
\begin{align}
                \hf(P_\varpi,P_\varpi)&\geq0,\\
								\hf(P_\varpi,P_\vartheta)&=\hf(P_\vartheta,P_\varpi)^*,\\
								\hf(0,P_\varpi)&=0,\\
								\hf(\II,\II)&=1,\\
								\hf(P_\varpi+P_{\varkappa},P_\vartheta)&=\hf(P_\varpi,P_\vartheta)+\hf(P_{\varkappa},P_\vartheta).
\label{decoherence.rep.prop}
\end{align}
Description of a quantum theoretic model provided in terms of the class operator and histories functional is intended to serve as a single replacement for dualistic description of temporal behaviour of model in terms of Schr\"{o}dinger's 
and von Neumann--L\"{u}ders' 
equations.

The \textit{Born--Jordan--Dirac--Heisenberg} (BJDH) \textit{algebra}  \cite{Born:Jordan:1925,Born:Heisenberg:Jordan:1926,Dirac:1925}, generated by the \textit{canonical commutation relations}
\begin{equation}
        [\qq,\qq]=0,\;\;\;
        [\pp,\pp]=0,\;\;\;
        [\qq,\pp]=\ii\II,
\label{BJDH.CCR}
\end{equation}
is extended in the quantum histories framework to an algebra generated by the relations:
\begin{equation}
        [\qq_{t_i},\qq_{t_j}]=0,\;\;\;
        [\pp_{t_i},\pp_{t_j}]=0,\;\;\;
        [\qq_{t_j},\pp_{t_k}]=\ii\delta_{jk}\II,
\label{heisenberg}
\end{equation}
where the operators $\qq_{t_i}$ and $\pp_{t_i}$ are considered as operators defined in the Schr\"{o}dinger picture for different moments $t_i$ of time, acting on the Hilbert space $\H_{t_i}$, while $\II$ is a unit element of that algebra. In order to formulate an extension of this formalism to the case of a continuous time $t\in\RR$, Isham and Linden \cite{Isham:Linden:1995} have changed the above relations to the form of the so-called \textit{history algebra}:
\begin{equation}
        \begin{array}{c}
                [\qq_f,\qq_g]=0,\;\;
								[\pp_f,\pp_g]=0,\;\;
								[\qq_f,\pp_g]=\ii\int_{-\infty}^{+\infty}\dd t f(t)g(t)\II,
        \end{array}
\label{histories.cr}
\end{equation}
where $\qq$ and $\pp$ are operator valued distributions, $f,g\in L_2(\RR,\dd t)$, $\pp_f:=\pp(f)$, $\qq_f:=\qq(f)$.\footnote{Correspondingly, in the histories approach to quantum field theory one considers the `field' operator-valued distributions $\qq$ and $\pp$ which act on a subspace of $L_2(\RR^3,\dd^3x)$, where the parameter $\vec{x}\in\RR^3$ is interpreted as representing a three-dimensional `space', and extends the canonical commutation relations at single point of time with the additional dependence on time dimension handled by the histories algebra \cite{Isham:Linden:1995,Isham:Linden:SS:1997,Savvidou:1999}. When the dependence on the functions in $L_2(\RR^3,\dd^3x)$ is made implicit, these relations read
\begin{equation}
        [\qq_{t_1}(\vec{x}_1),\qq_{t_2}(\vec{x}_2)]=0,\;\;[\pp_{t_1}(\vec{x}_1),\pp_{t_2}(\vec{x}_2)]=0,\;\;
        [\qq_{t_1}(\vec{x}_1),\pp_{t_2}(\vec{x}_2)]=\ii\delta(t_1-t_2)\delta^3(\vec{x}_1-\vec{x}_2)\II.
\end{equation}
This means that the three-dimensional histories commutation relations are actually three-plus-one-dimensional  canonical commutation relations.}

Isham and Linden \cite{Isham:Linden:1994,Isham:Linden:SS:1997} (see also \cite{Savvidou:1999}), have shown that these commutation relations may be represented on the Hilbert \textit{continuous history space} 
\begin{equation}
        {\cal{V}}:=\bigotimes_{t\in\RR}\H_t:=\bigotimes_{t\in\RR}({L_2}\left(\RR,\dd x)\right)_t.
\end{equation}
The `continuous tensor product' space $\bigotimes_{t\in\RR}\H_t$ is defined to be the symmetric Fock--Cook space \cite{Fok:1932,Cook:1953}
\begin{equation}
        \FFF[\H]=\FFF[L_2(\RR,\dd x)]=\bigoplus_{n=0}^\infty\sym_n\left(\bigotimes_{k=0}^{n}L_2(\RR,\dd x)\right),
\end{equation}
where $\bigotimes_0L_2(\RR,\dd x):=\CC$,
\begin{equation}
        \sym_n:\xi_1\otimes\ldots\otimes\xi_n\mapsto\frac{1}{\sqrt{n!}}\sum_{s\in\SS(n)}\xi_{s(1)}\otimes\ldots\otimes\xi_{s(n)}
\end{equation}
and $\SS(n)$ is the group of permutations of the set $\{1,\ldots,n\}$. The history of $(t_1,\ldots,t_n)$ is represented by the vector of $\FFF[\H]$ generated by the action of $n$ `creation operators' 
\begin{equation}
\begin{array}{c}        b_{t_i}^*(f):\FFF[\H]\ni\sym_n(\xi_1\otimes\ldots\otimes\xi_n)
        \mapsto\sqrt{n+1}\sym_{n+1}(f\otimes\xi_1\otimes\ldots\otimes\xi_n)\in\FFF[\H],
        \end{array}
\end{equation}
which, together with the `annihilation operators'
\begin{equation}
        \begin{array}{c}        b_{t_i}(f):\FFF[\H]\ni\sym_n(\xi_1\otimes\ldots\otimes\xi_n)
        \mapsto\frac{1}{\sqrt{n}}\sum^n_{k=1}f\cdot\sym_{n-1}(\xi_1\otimes\ldots\otimes\xi_{k-1}\otimes\xi_{k+1}
        \otimes\ldots\otimes\xi_n)\in\FFF[\H],
        \end{array}
\end{equation}
satisfy
\begin{equation}
        \begin{array}{c}
        {}[b_{t_i}(f),b_{t_j}(g)]=0=[b^*_{t_i}(f),b^*_{t_j}(g)],\;\;
        {}[b_{t_j}(f),b^*_{t_k}(g)]=\int_{-\infty}^{+\infty}\dd tf(t)g(t)\II
        \end{array}
\end{equation}
on the common domain $\D\subset\H$. Moreover, it is assumed that there exists a unit vector $\Omega\in\D:=\Span\{b^*_t(f_1)\cdots b^*_t(f_n)\Omega\mid\forall f_1,\ldots,f_n\in\H\}$ such that 
\begin{equation}
        b_t(f)\Omega=0\;\forall t\in\RR\;\forall f\in\H.
\end{equation}
These assumptions define the Isham--Linden representation of the history algebra (\ref{histories.cr}) to be the Fock--Cook representation. The spectral projectors of this representation of histories algebra are interpreted \cite{Isham:Linden:1995} as propositions about the temporal histories of a given quantum theoretic model.

For a given hamiltonian $H_t$, the self-adjoint histories hamiltonian operator in the Schr\"{o}dinger picture is defined as
\begin{equation}
        H_\kappa:=\int_{-\infty}^{+\infty}\dd t\kappa(t) H_t,
\label{Hkappa}
\end{equation}
where $\kappa(t)\in L_2(\RR,\dd t)$ is a function which `smears' $H_t$ in time. The histories algebra generates the commutation relations with this hamiltonian.

The Araki theorem \cite{Araki:1960} states the existence and uniqueness of hamiltonian operator in the Fock--Cook representation if this operator (if unsmeared) has a form 
\begin{equation}
H_t(\pp_t,\qq_t)=\frac{1}{2}\int\dd x \pp_t^2(x)+\tilde{H}_t(\qq_t).
\end{equation}
For example, for a given `harmonic oscillator' hamiltonian operator $H_t=\frac{\pp_t^2}{2m}+\frac{m\mathrm{w}^2}{2}\qq_t^2$ acting on the Fock--Cook space ${\cal{V}}$, the representation of the history algebra is constructed through the `annihilation operator' $b_t$, which takes the form
\begin{equation}
        b_t=\sqrt{\frac{m\mathrm{w}}{2}}\qq_t+\ii\sqrt{\frac{1}{2m\mathrm{w}}}\pp_t,
\end{equation}
with the commutation relations
\begin{equation}
        [b_{t_i},b_{t_j}]=0,\;\;[b_{t_i},b_{t_j}^*]=\delta(t_i-t_j)\II.
\end{equation}
By the Araki theorem, the Fock--Cook representation of the history algebra in ${\cal{V}}$ is uniquely selected by the requirement that the operator $H_\kappa$ (\ref{Hkappa}) exists in this representation. In nonsmeared version, $H_t=\mathrm{w}b_t^*b_t$. This leads to the following commutation relations \cite{Savvidou:1999}:
\begin{equation}
        [H_\kappa,\qq_f]=-\frac{\ii}{m}\pp_{\kappa f},\;\;\;[H_\kappa,\pp_f]=\ii\mathrm{w}^2\qq_{\kappa f},\;\;\;[H_\kappa,H_{\kappa'}]=0.
\end{equation}

Anastopoulos \cite{Anastopoulos:2001} has shown that the construction of the continuous history Hilbert space ${\cal{V}}$ and the representation of the history algebra in ${\cal{V}}$ can be provided also for nonquadratic hamiltonians using the coherent states representation \cite{Schroedinger:1926:Naturwiss,Glauber:1963}. However, such representation lacks any characterisation of its uniqueness.
\subsection{Savvidou's action operator\label{Savvidou.action.operator}}
The important property of the histories approach, discovered by Savvidou \cite{Savvidou:1999}, is the existence of the self-adjoint \textit{quantum action operator}, acting on ${\cal{V}}$ and defined as
\begin{equation}
        S_{\lambda,\kappa}:=\int_{-\infty}^{+\infty}\dd t(\lambda(t)\pp_t\dot{\qq}_t-\kappa(t)H_t),
\label{quantumactionoperator}
\end{equation}
by an analogy to a Hamilton--Jacobi action functional in classical mechanics theory
\begin{equation}
        S_{\mathrm{HJ}}=\int_{t_2}^{t_1}dt(\pp_\Gamma(t)\dot{\qq}_\Gamma(t)-H_\Gamma(t)),
\label{Hamilton.Jacobi.action}
\end{equation}
where $\pp_\Gamma,\qq_\Gamma,H_\Gamma\in C^\infty(\Gamma)$, and $\Gamma$ is a classical mechanics phase space. In both these equations, the dot symbol denotes the differentiation $\frac{\dd}{\dd s}$ with respect to to time parameter $s$ of the evolution generated by hamiltonian, that is (in the quantum theory)
\begin{equation}
        \qq_t:=\qq_{t}(s):=\ee^{\ii sH_t}\qq_t\ee^{-\ii sH_t},
\end{equation}
\begin{equation}
        \dot{\qq}_t:=\frac{\dd}{\dd s}\qq_{t}(s).
\end{equation}
Savvidou has shown that there also exists the \textit{Liouville operator}
\begin{equation}
        V:=\int_{-\infty}^{+\infty}\dd t\widetilde{V}_t:=\int_{-\infty}^{+\infty}\dd t(\pp_t\dot{\qq}_t)
\end{equation}
which is self-adjoint on ${\cal{V}}$. Hence, for $\lambda(t)\equiv 1$, one may express the action operator as
\begin{equation}
        S_\kappa=V-H_\kappa=\int_{-\infty}^{+\infty}dt\pp_t\dot{\qq}_t-\int_{-\infty}^{+\infty}\dd t\kappa(t)H_t
\label{action}
\end{equation}
with the following commutation relations:
\begin{equation}
        [S_\kappa,H_{\kappa'}]=\ii H_{\dot{\kappa}'},\;\;[S_\kappa,V]=-\ii H_{\dot{\kappa}},\;\;[V,H_\kappa]=-\ii H_\kappa.
\end{equation}
For $\kappa(t)\equiv 1$ the histories quantum theory reduces to ordinary quantum theory, which (for $H:=\int_{-\infty}^{+\infty}\dd t H_t$) is reflected in the commutators $[V,H]=0$ and $[V,S]=0$. 

The operator $V$ acts on $b_t$ in the following way \cite{Savvidou:1999}:
\begin{equation}
        \ee^{\ii rV}b_{f(t)}\ee^{-\ii rV}=b_{f(t+r)}.
\end{equation}
Moving to the Heisenberg picture, one can compare the action of $V$, $H_t$ and $S$:
\begin{align}
        \ee^{\ii rV}b_{t}(s)\ee^{-\ii rV}&=b_{t+r}(s),\label{V.hist.auto}\\
        \ee^{\ii rH_t}b_{t}(s)\ee^{-\ii rH_t}&=b_{t}(s+r),\label{H.hist.auto}\\
        \ee^{\ii r S}b_{t}(s)\ee^{-\ii r S}&=b_{t+r}(s+r),\label{S.hist.auto}
\end{align}
where the `smeared' operator $S_\kappa$ acts by an automorphism
\begin{equation}
        \ee^{\ii rS_\kappa}b_{f(t)}\ee^{-\ii rS_\kappa}=b_{\Sigma_r(f)},
\end{equation}
where $\Sigma_r$ is an unitary operator acting on $\zeta\in L_2(\RR,\dd t)$ by 
\begin{equation}
        (\Sigma_r\zeta)(t):=\ee^{-\ii\mathrm{w}\int_{t}^{t+r}\dd r'\kappa(t+r')}\zeta(t+r).
\end{equation}
For not smeared $b_t$ this can be formally written as
\begin{equation}
        \ee^{\ii rS_\kappa}b_t\ee^{-\ii rS_\kappa}=\ee^{-\ii\mathrm{w}\int_t^{t+r}\dd r'\kappa(t+r')+r\frac{\dd}{\dd t}}b_t,
\label{s.h.auto2}
\end{equation}
where the self-adjoint generator $S_\kappa$ on $\FFF[L_2(\RR,\dd t)]$ corresponds to an action of self-adjoint $\sigma_\kappa$ on $L_2(\RR,\dd t)$ given by
\begin{equation}
        \sigma_\kappa\zeta(t):=-\left(-\mathrm{w}\kappa(t)-\ii\frac{\dd}{\dd t}\right)\zeta(t).
\label{s.h.auto}
\end{equation}
The map $\RR\ni r\mapsto \ee^{\ii rS_\kappa}$ is a weakly continuous representation of a one-parameter family of unitary operators. In the same way the action of the automorphism generated by the Liouville operator $V$
\begin{equation}
        \ee^{\ii rV}b_{f(t)}\ee^{-\ii rV}=b_{f(t+r)}
\end{equation}
corresponds to the action of 
\begin{equation}
(v_r\zeta)(t):=\ee^{r\frac{\dd}{\dd t}}\zeta(t)=\zeta(t+r)
\label{l.h.auto}
\end{equation}
on $\zeta\in L_2(\RR,\dd t)$.

Hence, $V$ transforms $b_t$ from time $t$, related with the Hilbert space $\H_t$, to time $t+r$, related with the Hilbert space $\H_{t+r}$ (strictly speaking, $V$ transforms the support of the operator valued distribution). This is, by definition, purely kinematical operation, which does not depend on the hamiltonian $H_t$. On the other hand, $H_t$ generates the unitary evolution of the system in the single space $\H_t\subset{\cal{V}}$. The action operator (\ref{action}) joins together these two types of transformations. This is in some sense analogous to the Hamilton--Jacobi formulation of classical mechanics, in which the Hamilton--Jacobi action functional (\ref{Hamilton.Jacobi.action}) is the generator of a canonical transformation of the classical mechanical model from one time to another. 

Savvidou \cite{Savvidou:1999} suggests that these two operators ($V$ and $H_\kappa$) are related respectively with two different types of time evolution: the nonunitary `reduction' (L\"{u}ders' rule) related with subsequent propositions $P_t$, and the ordinary hamiltonian evolution (Schr\"{o}dinger's equation) given by operators $\ee^{\ii sH_t}$. However, the projection operators $P_t$, class operators $C_\varpi$, and histories functional $\hf_{\rho,H}$ were used neither in derivation of the kinematical evolution related with the Liouville operator $V$, nor in derivation of the quantum action operator $S_\kappa$. This is reflected in the apparent unitary character of the corresponding temporal evolutions generated by $V$ and $S_\kappa$. Hence, so far this suggestion has been ungrounded. In the next two subsections we will follow Anastopoulos and Savvidou on their way of reintroduction of nonunitary elements in quantum histories formalism.
\subsection{Geometric phase as a trace of a history\label{geometric.phase.histories}}
There exists a class of quantities in quantum theory, which do not correspond to any element of an algebra of operators, but are nevertheless very closely related to quantitative results of experimental procedures. One of the important representatives of this class is the \textit{geometric phase}. It reflects the geometric structure of the Hilbert space. Mathematically, it is defined as a holonomy of the Berry connection on the Hopf bundle \cite{Simon:1983}. The Hopf bundle is a $U(1)$ principal bundle of the Hilbert space $\H$ over the projective Hilbert space $\PH\subset\H$. In other words, it is a principal bundle of the subspaces $\H_\PP$ of vectors in $\H$ over the space of generating rays of $\H$. The \textit{Berry connection} $\nabla^\PP$ is defined as a $U(1)$ connection $1$-form on the Hopf bundle, induced naturally by an inner product on the Hilbert space $\H$, and given in the coordinate-free form by
\begin{equation}
        \nabla^\PP:=\s{\cdot,\ddd\cdot}_\H:\H\times\H\ni(\zeta,\xi)\mapsto\ii\s{\zeta,\ddd\xi}_\H\in\CC.
\end{equation}
The geometric phase (called also the Pancharatnam--Berry phase \cite{Pancharatnam:1956,Berry:1984}) is then defined as
\begin{equation}
        \ee^{\ii\theta[\gamma]}:=\ee^{-\int_{\zeta\in\pathgamma}\s{\zeta,\ddd\zeta}_\H},
\label{geom.phase.def}
\end{equation}
where $\pathgamma$ is a closed path in $\PH$. We will denote a path generated by family of vectors $\RR\ni t\mapsto\zeta(t)\in\H$ by $\zeta(\cdot)$. In case of open paths $\zeta(\cdot)$ in $\PH$ it was shown in \cite{Aharonov:Anandan:1987} and \cite{Samuel:Bhandari:1988} that the geometric phase (called also the Aharonov--Anandan phase) is given by
\begin{equation}
        \ee^{\ii\theta[\pathgamma]}=\ee^{-\int_{t_0}^{t_1}\dd t\s{\zeta(t),\frac{\dd}{\dd t}\zeta(t)}_\H}\s{\zeta_{t_0},\zeta_{t_1}}_\H.
\label{generalberryphase}
\end{equation}

Consider now an initial vector $\zeta(t=0)$, the final vector $\zeta(t=r)$, equal to the initial one up to phase, and the unitary time evolution $U(s)$ on $\H$, described by the solution of the Schr\"{o}dinger equation with the hamiltonian $H$,
\begin{equation}
        U(r):\zeta(t=0)\mapsto\zeta(t=r):=\ee^{-\ii rH}\zeta(t=0),
\end{equation}
which acts along a loop $\pathgamma(t)$, $t\in[0,r]$, on the space $\PH$. The phase on the Hopf bundle is then transformed into
\begin{equation}
        \begin{array}{l}
                \ee^{\int_0^r\dd t\s{\zeta(t),(-\frac{\dd}{\dd t}-\ii H)\zeta(t)}_\H}=
												\ee^{-\int_0^r \dd t\s{\zeta,\frac{\dd}{\dd t}\zeta}_\H}\ee^{-\ii\int_0^r\dd t\s{\zeta(t),H\zeta(t)}_\H}=
        \ee^{\ii\theta[\pathgamma]}\ee^{-\ii\int_0^r\dd t\s{\zeta(t),H\zeta(t)}_\H}.
  \end{array}    
\label{hopfbundlephase}
\end{equation}
The first term, given by the geometric phase, does not refer to dynamics generated by the hamiltonian $H$, and  reflects purely geometric structure of the kinematical Hilbert space $\H$.

Following Anastopoulos and Savvidou \cite{Anastopoulos:Savvidou:2002}, one can analyse the geometric phase from the perspective of quantum histories. Consider first a quantum model with a hamiltonian $H=0$. For a given history $\varpi$ of the projections $({P}_{t_0},{P}_{t_1},\ldots,{P}_{t_n})$, where ${P}_{t_i}$ is a projection onto one dimensional vector subspace of $\H$ spanned by $\zeta_{t_i}$, the trace of the class operator is
\begin{equation}
        \tr_{{\cal{V}}_n}(C_\varpi)=\s{\zeta_{t_0},\zeta_{t_n}}_\H\s{\zeta_{t_1},\zeta_{t_0}}_\H\s{\zeta_{t_2},\zeta_{t_1}}_\H\cdots\s{\zeta_{t_n},\zeta_{t_{n-1}}}_\H,
\label{trace.class.operator.omega}
\end{equation}
while the corresponding histories functional
\begin{equation}
        \begin{array}{c}        
				\hf_{|\zeta_{t_0}\rangle\langle\zeta_{t_0}|,H=0}(P_{\zeta(\cdot)},P_{\xi(\cdot)})=\s{\zeta_{t_n},\zeta_{t_{n-1}}}_\H\cdots\s{\zeta_{t_1},\zeta_{t_0}}_\H
        \cdot\s{\zeta_{t_0},\xi_{\tilde{t}_1}}_\H\cdots\s{\xi_{\tilde{t}_{m-1}},\xi_{\tilde{t}_m}}_\H
        \end{array}
\label{eq.Bargmann}
\end{equation}
is a $(n+m+1)$ Bargmann invariant \cite{Bargmann:1964,Simon:Mukunda:1993,Anastopoulos:Savvidou:2003}. Assuming that $\delta t:=\sup_i\{|t_i-t_{i-1}|\}\approx\O(\frac{1}{n})$, one can approximate $\{\zeta_{t_i}\}_{i=1}^n$ by the path $\zeta(t)$ on $\PH$, and for large $n$ this gives
\begin{align}
        \log\tr_{{\cal{V}}_n}(C_\varpi)&=\log\s{\zeta_{t_0},\zeta_{t_n}}_\H-\sum_{i=1}^{n}\log\s{\zeta(t_i),\zeta(t_{i-1})}_\H\\
&=\log\s{\zeta_{t_0},\zeta_{t_n}}_\H-\sum_{i=1}^{n}\s{\zeta(t_i),\zeta(t_i)-\zeta(t_{i-1})}_\H+\O(n^{-2}),
\end{align}
hence
\begin{equation}
        \lim_{\delta t\ra 0}\log\left(\tr_{{\cal{V}}_n}(C_\varpi)\right)
				= \log\s{\zeta_{t_0},\zeta_{t_n}}_\H-\int_{\zeta_{t_0}}^{\zeta_{t_n}}\s{\zeta(t),\dd\zeta(t)}_\H,
\end{equation}
where the last term is the Stieltjes integral. Comparing this result with equation \eqref{generalberryphase}, one can see that for any path which allows for the definition of the Stieltjes integral, the trace of a class operator is equal to a geometric phase \eqref{generalberryphase}:
\begin{equation}
        \tr_{{\cal{V}}}(C_\varpi) = \ee^{\ii\theta[\zeta(\cdot)]}.
\label{traceless.berry}
\end{equation}
Hence, for a given history $\varpi$, its corresponding geometric phase is defined by the trace of a class operator. Observing that $C_\varpi$ is used in the definition \eqref{decoherencefunctional} of the histories functional, one can rewrite the latter in terms of the geometric phase:
\begin{equation}
        \hf_{\rho_{t_0},H=0}(P_{\zeta(\cdot)},P_{\xi(\cdot)})=\s{\zeta(t_0),\rho_{t_0}\xi(t_0)}_\H\s{\zeta(t_n),\xi(t_n)}_\H\ee^{-\int\limits_{t_0}^{t_n}\dd t\s{\zeta(t),\frac{\dd\zeta(t)}{\dd t}}_\H-\int\limits_{t_0}^{t_n}\dd t\s{\xi(t),\frac{\dd}{\dd t}\xi(t)}_\H}.
\label{geometric.phase.in.decoherence.functional}
\end{equation}
For a quantum theoretic model with a nonzero hamiltonian $H$ the histories functional is equal to \cite{Anastopoulos:Savvidou:2002}:
\begin{equation}
\begin{array}{c}        
	\hf_{\rho_{t_0},H}(P_{\zeta(\cdot)},P_{\xi(\cdot)})=\s{\zeta(t_0),\rho_{t_0}\xi(t_0)}_\H\s{\zeta(t_n),\xi(t_n)}_\H\ee^{\ii\e{S^*[\zeta(\cdot)]}+\ii\e{S^*[\xi(\cdot)]}},
\end{array}
\label{dd1}
\end{equation}
where
\begin{equation}
        \begin{array}{c}
        \e{ S^*[\zeta(\cdot)]}:=\int_{t_0}^{t_n}\dd t\s{\zeta(t),\left(\ii\frac{\dd}{\dd t}-H\right)\zeta(t)}_\H
                = \ii\int_{t_0}^{t_n}\dd t\s{\zeta(t),\frac{\dd\zeta(t)}{\dd t}}_\H-\int_{t_0}^{t_n}\dd t\s{\zeta(t),H\zeta(t)}_\H.
        \end{array}
\label{act}
\end{equation}
This agrees with the earlier result of Isham and Linden \cite{Isham:Linden:1995}, who have constructed a special case of histories functional $\hf_{\rho,H}$. Using the continuous time projection operator on ${\cal{V}}$ corresponding to coherent states and using the technical assumption of $t_0\ra-\infty$ and $t_n\ra+\infty$, they have obtained
\begin{equation}
\begin{array}{c}
        \hf_{\rho,H}(P_{\zeta(\cdot)},P_{\xi(\cdot)})=\s{\zeta(t_0),\rho_{t_0}\xi(t_0)}_\H\ee^{\int_{t_0}^{t_n}\left(\s{\zeta(t),\frac{\dd\zeta(t)}{\dd t}}_\H-\s{\xi(t),\frac{\dd\xi(t)}{\dd t}}_\H\right)}\ee^{\ii\int_{t_0}^{t_n}(\s{\xi(t),H\xi(t)}_\H-\s{\zeta(t),H\zeta(t)}_\H)},
        \end{array}
\end{equation}
where $\int_{t_0}^{t_n}\s{\zeta(t),\dd\zeta(t)}$ is a Stieltjes integral. In order to compare the equation (\ref{dd1}) with the action equation (\ref{action}), consider the Schr\"{o}dinger representation of the histories algebra (\ref{heisenberg}), provided by the operators $\pp_t=-\ii\frac{\partial}{\partial x_t}$ and $\qq_t=x_t$, acting on the space $L_2(\RR,\dd x_t)$. Then $\widetilde{V}_t=\pp_t\dot{\qq}_t=-\ii\frac{\dd}{\dd s}$, and the equation (\ref{act}) can be written in the form
\begin{equation}
    \e{S^*[\zeta(\cdot)]}=-\int_{t_0}^{t_n}\dd t\s{\zeta(t),(\pp_t\dot{\qq}_t+H)\zeta(t)}_\H
\label{complexaction}
\end{equation}
\textit{if} it is assumed that $t=s$ and $\frac{\dd}{\dd t}=\frac{\dd}{\dd s}$. This equation shows that the nonhamiltonian part of the action $S$ is reflected in the geometric phase. 

According to Anastopoulos and Savvidou, the complete form of histories functional can be reconstructed by summing over all paths $\zeta(\cdot)$ and $\xi(\cdot)$ that are compatible with the given histories $\varpi$ and $\vartheta$ respectively:
\begin{equation}
        \hf_{\rho,H}(P_\varpi,P_\vartheta)=\sum_{\zeta(\cdot)\subset\varpi}\sum_{\xi(\cdot)\subset\vartheta}\hf_{\rho,H}\left(P_{\zeta(\cdot)},P_{\xi(\cdot)}\right).
\label{dd2}
\end{equation}
In face of the above results, they conclude, that \cytat{the knowledge of the geometric phase---for a set of histories and of the automorphism that implements the dynamics---is sufficient to fully reconstruct the decoherence [histories] functional---and hence all the probabilistic content of the [histories approach to quantum] theory} \cite{Anastopoulos:Savvidou:2002}. In other words, the histories approach provides the complete description of temporal behaviour of quantum theoretic models using two levels of description: the unitary action automorphism and the histories functional, which incorporates the nonunitary changes of geometry of the Hilbert space related with the sequences of projection operators taken into consideration.

However, this result is not completely clear. The functional $\e{S^*[\zeta(\cdot)]}$, despite the suggestive notation, is not the expectation value of the adjoint of the action operator $S_\kappa$ (\ref{quantumactionoperator}), unless $\int_{-\infty}^{+\infty}\dd t$ is interchangeable with $\s{\zeta(t),\cdot\;\zeta(t)}_\H$, and the smearing function $\kappa$ is introduced consistently at some stage of derivation of (\ref{complexaction}). Moreover, the assumption $t=s$ is not justified by any reasons other than \textit{ad hoc} decision. It is unsatisfactory that in order to derive the relationship of two different temporal evolutions with the geometric phase one has to set the values of corresponding time parameters to be identical. There also remains the question to what extent several different technical assumptions used in the construction of $S_\kappa$ and $\e{S^*[\zeta(\cdot)]}$ are essential for the final results and conclusions. Moreover, it is problematic to what extent the operator $V$ can be related with the `external time' \textit{without} introducing the family of projections $P_{t_i}$ and the object $\e{S^*[\zeta(\cdot)]}$. By definition, the operators $V$ and $S_\kappa$ refer only to continuous number of copies of the same Hilbert space, but without any reference to projections or `measurements'. 

Finally, these results are based on the arbitrary choice of the particular Fock--Cook (or coherent states) representation of histories version of the BJDH commutation relations. If the number of degrees of freedom of the algebra is finite, then the Stone--von Neumann \cite{Stone:1930,vonNeumann:1931} theorem guarantees that the Schr\"{o}dinger representation of the Weyl form of the BJDH algebra of canonical commutation relations is a unique, up to unitary equivalence, irreducible representation. However, in the infinite-dimensional case there exists uncountable many different unitarily inequivalent irreducible representations of this algebra \cite{Garding:Wightman:1954:ccr,Wightman:Schweber:1955}. Hence, the choice of a particular representation provides a nontrivial decision problem and should be justified by some argument, but there is no such argument at sight. In order to resolve these problems we have to move to a more general approach to quantum theory, the algebraic approach.
\subsection{Hilbert space geometry and coherent state path integrals\label{Hilbert.space.geometry.path.integrals}}
In this subsection we will discuss the basic aspects of the geometric approach to the formalism of the Hilbert space based quantum theory \cite{Strocchi:1966,Marsden:1968:generalized,Kibble:1979,Cirelli:Lanzavecchia:Mania:1983,Simon:1983,Cirelli:Lanzavecchia:1984,Heslot:1985,Page:1987,Aharonov:Anandan:1987,Anandan:Aharonov:1990,Cirelli:Mania:Pizzocchero:1990,Gibbons:1992,Cirelli:Mania:Pizzocchero:1994,Hughston:1995,Hughston:1996,Schilling:1996,Field:1996,Field:1997,Brody:Hughston:1998,Ashtekar:Schilling:1999,Cirelli:Gatti:Mania:1999,Brody:Hughston:2001,Chruscinski:Jamiolkowski:2004,Bengtsson:Zyczkowski:2006} and its relationship with the description of temporal behaviour of quantum theoretic models in the Hilbert space based quantum histories approach.

Every complex Hilbert space $\H$ can be considered as the real Hilbert space of double dimension, equipped with a complex structure operator $\jbold^\H:\H\ra\H$ such that $(\jbold^\H)^2=-\II$ and $\s{\xi,\jbold^\H\zeta}_\H=\ii\s{\xi,\zeta}_\H$ \cite{Stueckelberg:1960}. The decomposition of the inner product on $\H$ into real and imaginary parts,
\begin{align}
        \re\s{\xi,\zeta}_\H&=:\frac{1}{2}\gbold^\H(\xi,\zeta),\\
				\im\s{\xi,\zeta}_\H&=:\frac{1}{2}\wbold^\H(\xi,\zeta),
\end{align}
equips the real Hilbert space $\H$ with the structure of the nondegenerate positive definite real inner product $\gbold^\H$ and nondegenerate closed two-form $\wbold^\H$. They turn, respectively, to a riemannian and a symplectic structure on the manifold $\PP\H$, with $\H$ understood as a tangent space over $\PP\H$. The complex structure $\jbold^\H$ imposes the relationships 
\begin{align}
        \gbold^\H(\xi,\zeta)&=\wbold^\H(\xi,\jbold^\H\zeta),\\
        \nabla^{\gbold^\H}\jbold^\H&=0,
\end{align}
where $\nabla^{\gbold^\H}$ is a covariant derivative on $\H$ associated with $\gbold^\H$. These two equations imply that the triple $(\gbold^\H,\wbold^\H,\jbold^\H)$ equips $\H$ in the structure of the K\"{a}hler manifold. If $\H\iso\CC^{n+1}$ and some orthonormal basis $\{e_a\}$ in $\H$ is chosen, $a\in\{0,\ldots,n\}$, then the inner product on $\H$ can be denoted (using abstract index notation)
\begin{equation}
        \s{\xi,\zeta}_\H=\xi_a^*\zeta^a,
\end{equation}
while the infinitesimal equations for riemannian metric $\gbold^\H$ and symplectic form $\wbold^\H$ read
        \begin{align}
        \ddd\tilde{s}^2&=\gbold_{ab}\ddd{\zeta^a}^*\otimes\ddd\zeta^b,\\
        \tilde{w}^\H&=\wbold^\H_{ab}\ddd\zeta^a\wedge\ddd\zeta^b.
\end{align}
The projection of these structures to the projective space $\PH$, provided in finite-dimensional case by $z^a:=\zeta^a/\zeta^0$, $a\in\{1,\ldots,n\}$, induces a metric $\ddd s_{\PH}^2$ and a $U(1)$ connection one-form $A_{\PH}$,
\begin{align}
        \ddd s^2_{\PH}&:=\frac{1}{1+z_a^*z^a},\\
				A_{\PH}&:=\ii z_a^*dz^a.
\end{align}
The space $\PH$ has the structure of the compact K\"{a}hler manifold, the metric $\ddd s^2_{\PH}$ is the Fubini--Study metric, while the connection one-form $A_{\PH}$ is the Berry connection. In finite dimensional case, 
\begin{equation}
        \H\iso\CC^{n+1}\limp\PH\iso\CC\PP^n\iso S^{2n+1}/U(1),
\end{equation}
while the Fubini--Study metric on $\PH$ is given explicitly by
\begin{equation}
        \gbold^{\mathrm{FS}}_{ab}=\frac{\s{\zeta,\zeta}_\H\delta_{ab}-\zeta_{(a}\zeta^*_{b)}}{\ab{\s{\zeta,\zeta}_\H}^2},
\label{Fubini.Study.equation}
\end{equation}
where the round brackets denote the symmetrisation of indices. The space $\CC\PP^n$ has a symmetry group of dimension $n(n+2)$, which is generated by a family of $n(n+2)$ Killing vector fields.

This framework allows a geometric description and reconsideration of the structure of the Hilbert space based framework of quantum theory. In particular, the self-adjoint operators on $\H$, generating the unitary Schr\"{o}dinger evolutions on $\H$, correspond to such smooth functions on the K\"{a}hler manifolds $\PH$ that preserve the K\"{a}hler structure (that is, their hamiltonian vector fields are also the Killing vector fields). These hamiltonian functions on $\PH$ are given by the normalised expectations $\s{\zeta,H\zeta}_\H/\s{\zeta,\zeta}_\H$ of the corresponding self-adjoint operators $H$ on $\H$. Moreover, the geodesic distance $d_{\gbold^{\mathrm{FS}}}$ with respect to the Fubini--Study metric determines the transition probability between two vectors,
\begin{equation}
        p(\zeta|\xi)=\ab{\s{\zeta,\xi}_\H}^2=\cos^2(d_{\gbold^{\mathrm{FS}}}(\zeta,\xi)),
\end{equation}
thus
\begin{equation}
	d_{\gbold^{\mathrm{FS}}}(\zeta,\xi)=\arccos(\tr_\H(P_\zeta P_\xi)),
\label{Fubini.Study.distance}
\end{equation}
where $P_\xi$ and $P_\zeta$ are projection operators on the $1$-dimensional subspaces of $\H$ that are linearly spanned by $\xi$ and $\zeta$, respectively.

An interesting additional geometric structure can be introduced using the coherent vectors representation. Let there be given a group $G$ together with its irreducible unitary representation $G\ni g\mapsto U(g)\in\BH$. Then, for a given choice of a normalised reference vector $\zeta\in\H$ (specified, for example, as the vector invariant under the maximal compact subgroup of $G$), one can define the Hilbert space vectors $U(g)\zeta\in\H$, introduce the equivalence relation
\begin{equation}
        g_1\sim g_2\iff\exists \ee^{\ii\lambda}\in\CC\;\;\;U(g_1)\zeta=\ee^{\ii\lambda}U(g_2)\zeta,
\end{equation}
and define the homogeneus quotient space $\Gamma:=G/\sim$. The space $\Gamma$ is a parameter space that defines and labels the \textit{coherent vectors} of $\H$ by \cite{Schroedinger:1926:Naturwiss,Glauber:1963,Perelomov:1972,Perelomov:1986}
\begin{equation}      
        \iota_\Gamma:\Gamma\ni z\mapsto U(z)\zeta\in\PH.
\end{equation}
Using $\iota_\Gamma$, one can pullback the geometric objects from $\PH$ to $\Gamma$, equipping $\Gamma$ with the symplectic, riemannian and affine structure:
        \begin{align}
        \ddd s^2_\Gamma&:=\n{\ddd z}^2-\ab{\s{z,\ddd z}_\H}^2=\ab{\s{\ddd z,\ddd z}_\H}^2-\ab{\s{z,\ddd z}_\H}^2,\\
        A_\Gamma&:=\ii\s{z,\ddd z}_\H,\\
				w_\Gamma&:=\ddd A_\Gamma,
        \end{align}
where $\dd$ denotes the exterior derivative on $\Gamma$, and $w_\Gamma$ is a symplectic structure on $\Gamma$ if it is nondegenerate. If the space $\Gamma$ is interpreted as the `phase space', then $\iota_\Gamma$ is intepreted as a map from `phase space' to `space of rays'.

Anastopoulos and Savvidou \cite{Anastopoulos:Savvidou:2003} have used these results in order to uncover the relationship between the quantum histories and the metric structure on the projective Hilbert space. Using coherent vectors $z\in\H$, they derive
   \begin{align}
        \s{z,z+\delta z}&=1+\s{z,\partial_az^a}_\H\delta z^a+\frac{1}{2}\s{z,\partial_a\partial_bz}_\H\delta z^a\delta z^b+\O(\delta z^3)\\
				&=\exp\left(\ii A_a(z+\frac{1}{2}\delta z)\delta z^a-\frac{1}{2}\gbold^{\mathrm{FS}}_{ab}\delta z^a\delta z^b\right)+\O(\delta z^3).
        \end{align}
For $\delta z_k=z_{k+1}-z_k$, the equation (\ref{eq.Bargmann}), written in the form
\begin{equation}
        \hf_{H=0}(P_\varpi,P_\vartheta)=\prod_k\s{z_k,z_k+\delta z_k}_\H,
\end{equation}
leads to second-order approximation
\begin{equation}
        \hf_{H=0}(P_\varpi,P_\vartheta)=\exp\left(\ii\sum_k(z_k+\frac{1}{2}\delta z_k)\delta z_k-\frac{1}{2}\sum_k\delta s^2_k\right).
\end{equation}
If the paths $z(\cdot)$ are continuous and the variations $\delta z_k$ are bounded ($\ab{\delta z_k^a}<\epsilon$ and $\epsilon\ra 0$), then this equation converges to the expression (\ref{geom.phase.def}) on the geometric phase. However, if the paths $z(\cdot)$ cannot be considered as continuous (or differentiable) functions of $t$, then the approximation of the histories functional for the cut-off of the scale of $t$ given by $\frac{1}{\upsilon}$ leads to \cite{Anastopoulos:Savvidou:2003}
\begin{equation}
        \hf_{H=0}(P_\varpi,P_\vartheta)=\exp\left(\ii\int_\pathgamma A_\Gamma-\frac{1}{2\upsilon}\int_\pathgamma \dd t\gbold^{\mathrm{FS}}_{ab}(z(t))\dot{z}^a\dot{z}^b\right).
\label{histories.functional.with.metric}
\end{equation}
This result is very closely related to the Daubechies--Klauder approach \cite{Daubechies:Klauder:1985,Klauder:1988,Klauder:1995,Klauder:2011}, who introduced exact continuous-time regularised coherent vectors propagator for the phase space path integral, and proved that under mild assumptions on hamiltonian (square and quadric integrability, see e.g. \cite{Klauder:2003} for a brief statement of those) one has
\begin{align}
        \s{z(t=s),\ee^{-\ii Hs}z(t=0)}_\H&=\lim\limits_{\upsilon\ra+\infty}\int \DD z(\cdot)
				\ee^{\left(\ii\int_\pathgamma A_\Gamma\right)}
				\ee^{\left(-\ii\int_0^s\dd t\;h(z(t))\right)}
				\ee^{-\frac{1}{2\upsilon}\left(\int_0^s\dd t\;g_{ab}(z(t))\dot{z}^a\dot{z}^b\right)}\label{Klauder.path.integral}\\
				&=2\pipi\lim\limits_{\upsilon\ra+\infty}\ee^{\upsilon s/2}\int\tmu_{\mathrm{W}}^\upsilon(\pp_\Gamma,\qq_\Gamma)\ee^{\ii\int(\pp_\Gamma\dd\qq_\Gamma-H(\pp_\Gamma,\qq_\Gamma)\dd t)}\label{DK.Wiener.expression},
\end{align}
where $h(z(t))$ is a hamiltonian function\footnote{In the context of our paper, we consider it to be defined by $h(z(t)):=\s{z(t),Hz(t)}_\H$ for a given self-adjoint hamiltonian operator $H$ on $\H$.} on $\Gamma$, while $\tmu_{\mathrm{W}}^\upsilon(\pp_\Gamma,\qq_\Gamma)$ is a pinned Wiener measure on a phase space $\Gamma$. Moreover, this formulation is covariant under canonical transformations of phase space coordinates, what is not the case for most of other approaches to quantisation, including the Schr\"{o}dinger quantisation 
\begin{equation}
        \pp_\Gamma,\qq_\Gamma\in C^{\infty}(\Gamma)\rightsquigarrow-\ii\frac{\partial}{\partial x},x:L_2(\RR,\dd\lambda)\ra L_2(\RR,\dd\lambda),
\end{equation}
the Born--Jordan--Dirac--Heisenberg quantisation 
\begin{equation}
        \pp_\Gamma,\qq_\Gamma\in C^{\infty}(\Gamma)\rightsquigarrow \pp,\qq:\ell_2(\NN)\ra\ell_2(\NN),\;[\qq,\pp]=\ii\II,
\end{equation}
as well as the lattice formulation of phase space version of Feynman path integral (see a discussion in \cite{Klauder:2003}). All these methods of quantisation depend on the particular choice of the phase space coordinates, what makes these prescriptions incomplete, because descriptions which are considered to be canonically equivalent on the level of phase space become unitarily inequivalent on the level of the Hilbert space, and there is provided no procedure solving the problem of choice of unique description among inequivalent ones. The above result shows that the metric structure on the Hilbert space (and the corresponding metric structure on $\Gamma$) provides an important conceptual and mathematical element of the quantum theory. 

It is also interesting to note that for finite value of $\upsilon$ the propagator (\ref{Klauder.path.integral}) is not longer unitary \cite{Klauder:1995}. From the perspective of histories approach to quantum theory, this means that the metric structure on the Hilbert space allows (some sort of) quantification of the nonunitary (and noncontinuous) temporal behaviour. This observation should be furnished by an additional result of Klauder and Maraner \cite{Klauder:Maraner:1997}, who showed that the usual definition of dynamics on phase space by means of Hamilton's variational principle,
\begin{equation}
        \delta\int \dd t(\theta_a\dot{\xi}^a-h(\xi(t)))=0,
\label{hamilton.variational}
\end{equation}
where $\omega_{ab}=\partial_a\theta_b-\partial_b\theta_a$ is a symplectic form on the phase space, while $\xi$ are arbitrary phase space coordinates, is equivalent to the variational principle
\begin{equation}
        \delta\int\dd t(\theta_a\dot{\xi}^a+\frac{1}{2}\lambda\gbold^{\mathrm{FS}}_{ab}(\xi(t))\dot{\xi}^a\dot{\xi}^b)=0
\label{Klauder.Maraner.metric.variation}
\end{equation}
under constraint 
\begin{equation}
	\det\left(\gbold^{\mathrm{FS}}_{ab}(\xi)\right)=h^{-2n}(\xi)
\label{Klauder.Maraner.prior}
\end{equation}
and in the limit $\lambda\ra0$, where $\gbold^{\mathrm{FS}}_{ab}(\xi)$ is a riemannian metric on the phase space, $2n$ is the dimension of this space, while $\lambda\in\RR$ is an arbitrary scale factor. This result was derived in the context of phase space $\Gamma$, but nothing forbids us from applying it to $\PH$, with the hamiltonian function provided by normalised expectation of hamiltonian operator and with the riemannian metric provided by the Fubini--Study metric of $\PH$. The equation (\ref{hamilton.variational}) takes then the form of variation of the equation (\ref{act}),
\begin{equation}
        \delta\e{S^*[\zeta(\cdot)]}=\delta\int_{t_0}^{t_n}\dd t\s{\zeta(t),\left(\ii\frac{\dd}{\dd t}-H\right)\zeta(t)}_\H=0,
\end{equation}
which gives the Schr\"{o}dinger equation. This leads to a question whether some modification of the variational principle (\ref{Klauder.Maraner.metric.variation}) on $\PH$ could result in an interesting form of the temporal behaviour of quantum theoretic models? In particular, it seems that for not vanishing metric term, provided by the finite values of $\lambda$ corresponding to finite values of $\upsilon$ in (\ref{histories.functional.with.metric}) and (\ref{Klauder.path.integral}), the resulting temporal behaviour would be nonunitary.

These observations will play an important guiding role in generalisation of the elements of histories approach to an algebraic context.

\subsection{Case study: Algebraic action operator and the limits of unitarity\label{algebraic.action.operator.section}}

Given a W$^*$-algebra $\N$ and $\psi\in\W_0(\N)$, consider a crossed product $\N\rtimes_{\sigma^\psi}\RR$, defined as the von Neumann algebra acting on the Hilbert space $L_2(\RR,\dd t;\H)\iso\H\otimes L_2(\RR,\dd t)$ and generated by the operators $\pi_{\sigma^\psi}(x)$ and $u_\RR(t)$, which are defined by\rpktarget{PI.SIGMA} \rpktarget{URR}
\begin{align}   
(\pi_{\sigma^\psi}(x)\xi)(t)&:=\sigma_{-t}^\psi(x)\xi(t),
\label{pi.sigma}\\
        (u_\RR(t_2)\xi)(t_1)&:=\xi(t_1-t_2),
\label{lambda.tau}
\end{align}
for all $x\in\N$, $t,t_1,t_2\in\RR$, $\xi\in L_2(\RR,\dd t;\H)$, see e.g. \cite{vanDaele:1978}. These two operators satisfy the covariance equation
\begin{equation}
        u_\RR(t)\pi_{\sigma^\psi}(x)u_\RR^*(t)=\pi_{\sigma^\psi}(\sigma_t^\psi(x)).
\label{covariance.eqn}
\end{equation}
The equation \eqref{pi.sigma} can be written as
\begin{equation}
        (\pi_{\sigma^\psi}(x)\xi)(t)=\Delta_\psi^{\ii t}x\Delta_\psi^{-\ii t}\xi(t)=\ee^{-\ii K_\psi t}x\ee^{\ii K_\psi t}\xi(t),
\end{equation}
where $K_\psi$ is a modular hamiltonian of the modular operator $\Delta_\psi$. So, the covariance equation \eqref{covariance.eqn} translates between the family of unitaries that partially generate the crossed product algebra $\N\rtimes_{\sigma^\psi}\RR$ and the modular automorphism of the underlying von Neumann algebra $\N$:
\begin{equation}
        u_\RR(t)\pi_{\sigma^\psi}(x)u_\RR(t)^*=
        \pi_{\sigma^\psi}(\ee^{-\ii t K_\psi}x\ee^{\ii t K_\psi}).
\label{modular.histories.covariance}
\end{equation}

Using the uniqueness of the standard representation up to unitary equivalence, Falcone and Takesaki \cite{Falcone:Takesaki:2001} (see \cite{Kostecki:2013} for a pedagogical introduction) proved that the map $\N\mapsto\N\rtimes_{\sigma^\psi}\RR$ extends to a functor $\VNCore$ from the category $\VNIso$ of von Neumann algebras with $*$-isomorphisms to its own subcategory $\VNsfIso$ of semi-finite von Neumann algebras with $*$-isomorphisms. The functoriality $\CanVN:\WsIso\ra\VNIso$ of Kosaki's construction \cite{Kosaki:1980:PhD} of canonical representation $\pi_\C$ of any W$^*$-algebra $\C$ turns the assignment
\begin{equation}
        \C\mapsto\pi_\C(\C)=:\N\mapsto\core=\widetilde{\pi_\C(\C)}
\end{equation}
to a functor
\begin{equation}
        \WstarCore:\WsIso\ra\VNsfIso,
\end{equation}
where $\WstarCore:=\VNCore\,\circ\,\CanVN$, while $\WsIso$ consists of W$^*$-algebras and $*$-isomorphisms. For any W$^*$-algebra $\N$, the object $\WstarCore(\N)\in\Ob(\VNsfIso)$ will be called \df{canonical core} of $\N$ and denoted $\core$. By equipping the canonical core von Neumann algebra $\core$ of the countably finite W$^*$-algebra $\N$ with the choice of some $\omega\in\N^+_{\star0}$, we obtain a unitary isomorphism $\core\iso\N\rtimes_{\sigma^\omega}\RR$. 

The operator $u_\RR(t)$, when considered as an operator on $L_2(\RR,\dd t)$, takes the form
\begin{equation}
        u_\RR(r)=\ee^{-\ii r\tilde{V}},\;\;\;
				\tilde{V}:=-\ii\frac{\dd}{\dd t}.
\label{explicit.form}
\end{equation}
So, the analogue of a quantum `histories liouvillean' automorphism \eqref{l.h.auto} is naturally present in the structure of a unitary representation of the canonical core algebra. From the covariance equation \eqref{covariance.eqn} it follows that this automorphism of $\N\rtimes_{\sigma^\omega}\RR$ uniquely corresponds to the modular automorphism of $\N$. Hence, the pair $(\N,\omega)$ uniquely determine a W$^*$-dynamical system $(\N,\RR,\sigma^\omega)$. But there might be also given another description of a temporal behaviour related with the same algebra $\N$, provided by some group of $*$-automorphisms $\alpha:\RR\ra\Aut(\N)$. If $\alpha$ is continuous in the weak-$\star$ topology, then one has to consider the coexistence of two W$^*$-dynamical systems: $(\N,\RR,\sigma^\omega)$ and  $(\N,\RR,\alpha)$. While $\sigma^\omega$ is completely determined by the properties of $\N$ and $\omega$, the $*$-automorphism $\alpha$ is arbitrary. Using familiar terminology, one can say that $\sigma^\omega$ is a `kinematic' automorphism, while $\alpha$ is a `causal' automorphism. These characteristics of $\sigma^\omega$ and $\alpha$, together with an equation \eqref{explicit.form} lead us to propose to consider $\sigma^\omega$ as an algebraic replacement of the `histories liouvillean' automorphism \eqref{V.hist.auto} and to consider $\alpha$ as an algebraic replacement of the `histories hamiltonian' automorphism \eqref{H.hist.auto}. We will join these two separated automorphisms into one automorphism, representing the `complete temporal behaviour' of the quantum theoretic model $(\N,\alpha,\omega)$, and forming an algebraic replacement for the `histories action' automorphism \eqref{S.hist.auto}. 

We do not require the model $(\N,\alpha,\omega)$ to be a `quantum dynamical system' (in the sense of \cite{Pillet:2006}) with respect to $\alpha:\RR\ra\Aut(\N)$, because we do not need to (and do not want to) assume the invariance of $\omega$ with respect to $\alpha$. In fact, instead of declaring invariance of $\omega$ with respect to $\alpha$, we will use $\alpha$ in order to construct a new algebraic state $\phi$. Consideration of the derivations of these $*$-automorphisms together with the corresponding hamiltonians is also not useful here, because of the lack of a unique characterisation of unbounded generators of $*$-automorphisms in terms of corresponding self-adjoint hamiltonians (see \cite{Kostecki:2013} for more discussion and further references on this). 

The standard liouvillean of $\sigma^\omega$ is given by its modular hamiltonian $K_\omega=-\log\Delta_\omega$. We will denote by $L_\alpha$ the standard liouvillean of $\alpha$ in the GNS representation $(\H_\omega,\pi_\omega,\Omega_\omega)$. If
\begin{equation}
        L_{\alpha,\omega}:=L_\alpha+K_\omega+J_\omega K_\omega J_\omega
\label{updated.action.liouvillean}
\end{equation}
is essentially self-adjoint on $\dom(K_\omega)\cap\dom(L_\alpha)\cap\dom(J_\omega K_\omega J_\omega)$, and if $K_\omega+L_\alpha$ is essentially self-adjoint on $\dom(K_\omega)\cap\dom(L_\alpha)$, then
\begin{equation}
        u_{\alpha,\omega}(x)(t):=\ee^{\ii t(K_\omega+L_\alpha)}x\ee^{-\ii t(K_\omega+L_\alpha)} \;\;\forall x\in\pi_\omega(\N),
\label{updated.action.automorphism}
\end{equation}
is a weak-$\star$ continuous $*$-automorphism of $\pi_\omega(\N)$, and $L_{\alpha,\omega}$ is its standard liouvillean. Hence, $(\pi_\omega(\N),\RR,u_{\alpha,\omega})$ is a W$^*$-dynamical system with a corresponding crossed product $\pi_\omega(\N)\rtimes_{u_{\alpha,\omega}}\RR$. Moreover, if $L_\alpha$ is bounded, then the DFFA convergent perturbation expansions hold:
 \begin{align}
                u_{\alpha,\omega}(x)&=
                        \sum_{n=0}^\infty \ii^n
                        \int_{0\leq t_n\leq\cdots\leq t_1\leq t}
                        \dd t_1\cdots \dd t_n
                {}[\alpha_{t_n}(K_\omega),
                        {}[\ldots,
                                {}[\alpha_{t_1}(K_\omega),
                                \alpha_t(x){}]
                        \ldots{}]
                {}],\\
                E_{\alpha,K_\omega}(t):=
                        \ee^{\ii t(K_\omega+L_\alpha)}\ee^{-\ii tL_\alpha}&
												=\sum_{n=0}^\infty \ii^n
                        \int_{0\leq t_n\leq\cdots\leq t_1\leq t}
                        \dd t_1\cdots \dd t_n
        \alpha_{t_n}(K_\omega)\cdots\alpha_{t_1}(K_\omega).
\end{align}

Hence, under some relatively weak conditions, the modular automorphism $\sigma^\omega$ and the additional $*$-automorphism $\alpha$ form together a unique automorphism $u_{\alpha,\omega}$ with its own self-adjoint liouvillean $L_{\alpha,\omega}$. If one thinks of $\alpha$ as an algebraic analogue of an automorphism generated by the `interaction hamiltonian', then the automorphism $u_{\alpha,\omega}$ can be considered as a `correction' of $u_\alpha$ by means of an associated $U(1)$ connection $1$-form $[K_\omega,\,\cdot\,]$ on the Hilbert bundle of Hilbert spaces $\H_\omega$ over the image of real line $\RR$ in $\M(\N)$. Given $L_\alpha$ and $K_\omega$, we can define also the operator
\begin{equation}
	L_{\omega,\alpha}:=K_\omega+L_\alpha+J_\omega L_\alpha J_\omega,
\label{dual.std.action.liouvillean}
\end{equation}
with the conditions on essential self-adjointness analogous to the case of $L_{\alpha,\omega}$. This operator is not interpretable as a standard liouvillean of $\alpha$ perturbed by a $U(1)$ connection form. However, as we will see below, it also encodes some very interesting information.

By an analogy with the Hilbert space based histories approach to quantum theory we will call $L_{\alpha,\omega}$ an (\textit{algebraic}) \df{action operator} and will call $u_{\alpha,\omega}$ an (\textit{algebraic}) \df{action automorphism}. We will call $L_{\omega,\alpha}$ a \df{dual action operator}.

Recall that in the Hilbert space based histories approach the action operator was a generator of a complete unitary temporal behaviour of a given quantum theoretic model, including not only the `internal' temporal unitary changes related to the fixed Hilbert space, but also the `external' temporal unitary changes between two different Hilbert spaces (in fact, that formalism was limited to the continuous one-parameter family of identical copies of the same Hilbert space). In the algebraic approach, the change of a Hilbert space corresponds to the change of an algebraic state, and it implies the corresponding change of a representation of an underlying W$^*$-algebra. In order to strengthen the relationship between the Hilbert space based histories and our algebraic approach, we will show how the action operator is related to the change between two representations or between two different algebraic states.

In order to do this, we need to use one more result of Derezi\'{n}ski, Jak\v{s}i\'{c} and Pillet \cite{DJP:2003}. They show that if $\omega$ is a faithful Kubo--Martin--Schwinger state with respect to a $*$-automorphism $u$ with a parameter $\beta$, then under the assumptions used previously for derivation of $*$-automorphism $u_t^Q$ and its standard liouvillean $L_Q$, and assuming additionally that $\n{e^{-\beta Q/2}\Omega_\omega}_{\H_\omega}<\infty$, 
\begin{align}
        \Omega_Q&:=\ee^{-\beta(L_u+Q)/2}\Omega_\omega,\\
				\omega_Q(\cdot)&:=\s{\Omega_Q,\cdot\Omega_Q}/\n{\Omega_Q}^2_{\H_\omega}
\end{align}
satisfy
\begin{enumerate}
\item[0)] $\Omega_\omega\in\dom(\ee^{-\beta(L_u+Q)/2})$,
\item[1)] $\Omega_Q\in\stdcone_\omega$ is cyclic and separating for $\pi_\omega(\N)$,
\item[2)] $\omega_Q$ is KMS with respect to $u^Q$ and $\beta$,
\item[3)] $\log\Delta_{\Omega_Q}=-\beta L_Q$ and 
\begin{equation}
        \log\Delta_{\Omega_Q,\Omega_\omega}=-\beta L_u-\beta Q.
\label{log.updated.rel.mod.op}
\end{equation}
\end{enumerate}
By the Takesaki theorem, the faithful state $\omega$ on a W$^*$-algebra is always KMS with respect to $\sigma^\omega$ with $\beta=1$. Hence, under the assumptions allowing for the construction of the dual algebraic action operator $L_{\omega,\alpha}$, and assuming also that $\n{e^{-(K_\omega+L_\alpha)/2}\Omega_\omega}<\infty$, it holds that
\begin{equation}
        \phi(\cdot):=\frac{\s{\ee^{-(K_\omega+L_\alpha)/2}\Omega_\omega,(\,\cdot\,)\ee^{-(K_\omega+L_\alpha)/2}\Omega_\omega}_{\H_\omega}}{\n{\ee^{-(K_\omega+L_\alpha)/2}\Omega_\omega}^2_{\H_\omega}}
\label{updated.action.omega}
\end{equation}
is KMS with respect to $u_{\omega,\alpha}$ with $\beta=1$. Hence, $u_{\omega,\alpha}$ is a modular automorphism of $\pi_\omega(\N)$ with respect to $\phi$. This is a very interesting result, because it means that while $(u_{\alpha,\omega},L_{\alpha,\omega})$ play the role of an action automorphism and an action operator with respect to the pair $(\N,\omega)$, $(u_{\omega,\alpha},L_{\omega,\alpha})$ play the role of a modular automorphism and a modular hamiltonian with respect to the pair $(\N,\phi)$. Hence, under the assumptions
\begin{enumerate}
\item $\omega$ is a faithful normal algebraic state on a W$^*$-algebra $\N$,
\item $L_\alpha$ is a standard liouvillean of $*$-automorphism $\alpha$ of $\N$ affiliated with $\pi_\omega(\N)$,
\item $K_\omega+L_\alpha$ is essentially self-adjoint on $\dom(K_\omega)\cap\dom(L_\alpha)$,
\item $K_\omega+L_\alpha-J_\omega L_\alpha J_\omega$ is essentially self-adjoint on $\dom(K_\omega)\cap\dom(L_\alpha)\cap\dom(J_\omega L_\alpha J_\omega)$,
\item $\n{\ee^{-(K_\omega+L_\alpha)/2}\Omega_\omega}_{\H_\omega}<\infty$,
\end{enumerate}
the $*$-automorphism $\alpha$ can be always assimilated as a part of the modular automorphism $\sigma^{\phi}$ that is uniquely specified by an `updated' algebraic state $\phi$ \eqref{updated.action.omega}. In other words, the $*$-automorphism forming a `causal' part of an algebraic quantum action automorphism can be considered as a constitutive element of a `kinematic' temporal behaviour, just with respect to another algebraic state.

This way Savvidou's construction of the Liouville and action operators acting on the symmetric Fock--Cook Hilbert space $\FFF[\H]$ and generating two corresponding types of unitary temporal evolution becomes replaced by the algebraic construction of liouvillean and action operators generating the $*$-automorphisms of corresponding representations of canonical core W$^*$-algebra $\core$. The main structures of each approach, the Fock--Cook Hilbert space $\FFF[\H]$ and the Falcone--Takesaki W$^*$-algebra $\core$, are constructed in a functorial way from the corresponding underlying ingredients of the given quantum theoretic model: the Hilbert space $\H$ and the W$^*$-algebra $\N$, respectively (for a functorial description of Fock space construction see e.g. \cite{Blute:Panangaden:Seely:1993}). Both approaches show that every quantum theoretic model is generically equipped with two different types of \textit{unitary} temporal evolution: the `kinematic' automorphism and the `causal' automorphism. Moreover, while the particular quantitative form of the latter evolution can be arbitrarily postulated, the quantitative form of the former is determined by the particular (quantitative) representation of an abstract algebra that is used in the given model. Both approaches enable to incorporate these two different unitary temporal evolutions into a single unifying unitary `action' evolution. Both approaches enable also to describe the generators of the `kinematic' and `action' evolutions in terms of operators acting on the `temporal' space $L_2(\RR,\dd t)$. In the case of the Hilbert space based approach, all these automorphisms are generated by the corresponding hamiltonian operators, while in the case of algebraic approach the quantitative representations of all these automorphisms are generated by the corresponding standard liouvillean operators.

However, apart from these similarities, there are also important differences between those two approaches. In particular, the representation of the histories algebra on the Fock--Cook space is unique, up to unitary equivalence, only for hamiltonians which have a form specified by the Araki theorem \cite{Araki:1960}. For a general hamiltonian there is no possibility to guarantee the uniqueness (up to unitary equivalence) of the Fock--Cook representation of the histories algebra of the Fock--Cook Hilbert space $\FFF[\H]$. In contrast to this, the representation of a core algebra $\core$ in terms of a crossed product algebra $\N\rtimes_{\sigma^\omega}\RR$ acting on $\H_\omega\otimes L_2(\RR,\dd t)$ is uniquely determined, up to unitary equivalence, by any particular choice of a state $\omega\in\N^+_{\star0}$, \textit{which is considered as part of the definition of the model}. Moreover, while in both approaches the initial `dynamic' unitary automorphism can be postulated as an arbitrary additional component of the model, only in the algebraic approach can the resulting (dual) `action' automorphism be considered as purely `kinematic' (modular) automorphism, related to the change of the algebraic state. The change of unitary description of the temporal behaviour of the quantum theoretic model $(\N,\omega)$ equipped with an additional `unitary' $*$-automorphism $\alpha$ is completely determined by the quantum theoretic model $(\N,\omega)$ and the map $\omega\mapsto\phi$, \textit{which can be considered as a part of the definition of the model}. There is no corresponding result of such type in the Hilbert space based approach to quantum histories. We consider these two results as an important suggestion in favour of the change of perspective on the role of unitary temporal behaviour of quantum theoretic models. Stating it briefly, instead of \textit{postulating the hamiltonian} as an independent component of quantum theoretic model and later perturbing it (what seems to be the only method within the frames of the Hilbert space based approach to mathematical foundations of quantum theory), an algebraic approach allows to \textit{derive the liouvillean} that characterises the unitary temporal behaviour, given the information about change of state. The change between two identical quantitative Hilbert spaces equipped with the same quantitative representation of the operator algebra becomes replaced by the change between two different (but faithful) algebraic states which correspond to two different (but unitarily equivalent) quantitative representations. 

In the similar way as in Savvidou's Hilbert space based formulation: 1) when the generators of `kinematic' and `causal' automorphisms are joined into the new `action' generator $K_\omega+L_\alpha$, the reference to two \textit{different} temporal parametrisations of $\alpha$ and $\sigma^\omega$ disappears (the choice of rescaling of the time parameter between $\alpha_t$ and $\sigma^\omega_s$ was implicitly set above to be $t=s$, however any scalar relationship $t=\lambda s$, $\lambda\in\RR$, will work, and any of such choices corresponds to the choice of a specific section of a $U(1)$ bundle for the $-[\log\Delta_\omega,\,\cdot\,]$ connection); 2) the resulting description of temporal behaviour is an unitary automorphism which does not possess any explicit relationship with the von Neumann--L\"{u}ders nonunitary type of temporal behaviour. In consequence, the above construction is insufficient to deal with the problem of algebraic reformulation of the Anastopoulos--Savvidou histories description of the geometric phase. It seems that the idea of construction of localised unitary evolution without taking into account more specific information about the changes of local geometry of quantum state spaces is just not enough.

In particular, the restriction of description of temporal behaviour of quantum models to $*$-au\-to\-mor\-ph\-isms implies the preservation of spectrum: if $\alpha$ is a $*$-automorphism of a C$^*$-algebra $\C$, then
\begin{equation}
        \alpha((z\II-x)^{-1})=(z\II-\alpha(x))^{-1}\;\forall x\in\C\;\forall z\in\CC.
\label{automorphisms.preserve.spectrum}
\end{equation}
The \textit{decision} that description of temporal behaviour of quantum theoretic models should be provided in terms of the $*$-automorphisms removes \textit{a priori} the possibility to describe the changes of the eigenvalues in time. This restriction is imposed by the `spectral principle' which is a part of an idealistic ontological interpretation of a quantum theoretic formalism. However, it is too strong for many practical purposes.\footnote{In consequence, the range of applicability of the `unitary' framework is usually extended by the use of additional mathematical tools and techniques, like parameter fitting or renormalisation, which are explicitly nonunitary, but are not considered as part of the content of the quantum theoretic model.} We do not see any reasons for accepting this situation other than wish of securing the validity of some very particular interpretation.\footnote{According to this interpretation, the eigenvalues of operators can be specified with infinite precision (at least in principle) by the quantitative results of experimental procedures, hence they have ontological meaning, and correspond to the `possessed properties' of `ontological quantum systems', as opposed to `postulated properties' of `quantum theoretic models'. Unfortunately, this idealistic ontological interpretation does not apply to any actual experimental situation without \textit{additional} techniques of processing of the quantitative results of experimental procedures which render the fundamental assumption of this interpretation false (or at least meaningless).} In order to develop the framework which bypasses the double standards of dealing with description of experimental information and temporal behaviour, one has to consider the nonunitary description of temporal behaviour as a valid constituent of the structure of quantum theoretic models. The nonunitary changes of quantitative representation can be determined in an algebraic approach by nonunitary changes of algebraic states. Hence, in order to provide such nonunitary description, one has to introduce some method of `updating' the algebraic state that corresponds to a specified information.

Note that in the current Section we can replace the use of a standard liouvillean of a \textit{global} W$^*$-dynamical system $(\N,\RR,\alpha)$ by a \textit{local} quantum Poisson system $(\M(\N,\B),h)$, using the perturbations of a local liouvillean $\pi_\omega(\DF_\omega h)$ by $K_\omega$ (and, dually, $K_\omega$ by $\pi_\omega(\DF_\omega h)$). This localises the linearity of a flow to a tangent space, allowing for a nonlinear generating function for the `causal' part of the dynamics. As a result, the above discussion can be applied to \df{local action operator} and its dual. Yet, the localisation does not change the qualitative conclusions drawn from the above discussion, so we have chosen to keep the presentation in maybe a bit more familiar global language. Because $\B$ is a Banach Lie algebra and $\M(\N,\B)$ is constructed as a Banach Lie--Poisson submanifold, locally the generators of causal dynamics will be always linear, so will be the flow determined by the Lie--Poisson bracket $\{\cdot,\cdot\}$. 

Hence, in order to get nonlinear contributions to the effective dynamics, some other geometric structure, beyond $\pi_\omega(\DF_\omega h)$ and $K_\omega$, has to be used. In particular, from the discussion in Section \ref{Orlicz.Br\`{e}gman.section} it follows that, given a bundle of the GNS Hilbert spaces over a trajectory of faithful normal states, a natural parallel transport operator is given by the standard unitary equivalence $V_{\phi,\psi}$. The corresponding connection $\nabla^{1/2}$ is a Levi-Civita connection of the Wigner--Yanase riemannian metric, and the local geodesic `free fall' along $\nabla^{1/2}$ corresponds to a norm projection in the (standard representation) Hilbert space, associated to a local continuous-time projective measurement. In this sense, the connection $\nabla^{1/2}$ \textit{locally} implements this what was an original intention of the nonhamiltonian part of the histories functional, as exposed by the equations \eqref{trace.class.operator.omega}, \eqref{eq.Bargmann}, \eqref{traceless.berry}, and \eqref{geometric.phase.in.decoherence.functional}. In a discussion of Savvidou's action operator in Section \ref{geometric.phase.histories} we have noticed that it does not restore this aspect of histories functional. Because the above algebraic action operator provides an exact algebraic generalisation of Savvidou's formulation, it shares the same feature. One can think of Savvidou's `Liouville' operator $V$ and modular hamiltonian $K_\omega$ as generators of `intrinsic' kinematic automorphisms of, respectively, a single Hilbert space $\H$ or a single W$^*$-algebra $\N$. These should be taken into account when one provides a \textit{spatial representation} of the `intrinsic' causal automorphism of $\H$ or $\N$, respectively, in terms of a bundle of copies of $\H$ or $\N$ over a real line $\RR$. However, neither $\ee^{\ii s V}$ or $\sigma^\omega_s$ can be understood as representing the \textit{changes} between \textit{quantitatively distinct} Hilbert spaces, corresponding to different measurements. In case of the Hilbert space based histories, this change requires to use the Berry connection, while in the algebraic framework (as implemented systematically in Sections \ref{local.geodesic.propagation} and \ref{from.local.gauge.to.lagrangeans.section}) this requires to use the connection $\nabla^{1/2}$, corresponding to the parallel transport operator $V_{\phi,\psi}$.\footnote{As opposed to $K_\omega$, $V_{\phi,\psi}$ cannot be used in the perturbation of the standard liouvillean, because it is a mapping between two different standard representations, not an operator acting on a single Hilbert space.}

\subsection{W$^*$-geometric quantum histories\label{algebraic.quantum.histories}}

In \cite{Kostecki:2014,Hellmann:Kaminski:Kostecki:2016} we showed that the nonunitary change of quantum states due to L\"{u}ders' rule (and other rules, see also \cite{MunkNielsen:2015}) is a special case of the constrained minimisation of the quantum relative entropy functional $D_0$. Moreover, the local smooth geometry of the quantum models can be derived (under mild conditions) as the subsequent terms of the Taylor expansion of any smooth information distance $D$. This leads to the idea \cite{Kostecki:2007:qht,Kostecki:2008:aqh,Kostecki:2010:AIP} to use quantum relative entropy as a general tool of generating nonunitary evolution of quantum states that takes into account the geometric structure of the quantum model.

Taking into account the above discussion, we consider the connection $\nabla^{D_\Psi}$ derived from a Br\`{e}gman distance $D_\Psi$ to be the appropriate replacement for the Berry connection used in the Anastopoulos--Savvidou analysis, as well as for our own use of $K_\omega$ above. However, we are unfortunately lacking the mathematical structure that would allow us to practically use other connection then $\nabla^{1/2}$, thus below we will consider only this possibility. On the other hand, the affine Killing hamiltonian vector field used in Sections \ref{geometric.phase.histories} and \ref{Hilbert.space.geometry.path.integrals} can be replaced by an arbitrary hamiltonian function $h$ on $\M(\N)$, provided the latter is equipped with a BLP manifold structure. Those two substitutions allow us to state the W$^*$-geometric versions of the formulas \eqref{hopfbundlephase} and \eqref{act}. On the differential geometric level (and ignoring for a moment a functional analytic incompatibility between BLP, GNS, and quantum information geometric manifold structures), the effective dynamics is described by the $1$-form
\begin{equation}
	{\cal F}=\ddd h(\phi)-\ddd_{\nabla^{1/2}}(\phi),
\label{causal.inferential.form}
\end{equation}
where $\ddd_{\nabla^{1/2}}$ is a connection form of the Levi-Civita connection $\nabla^{1/2}$. This formula states that local causal dynamics and local inferential dynamics participate to the same extent in the effective local dynamics. Hence, neither inference nor causality is considered as more fundamental. The form ${\cal F}$ can be considered as a localisation of the causal inference instrument \eqref{global.hamiltonian.entropic.map} that does not impose the ordering on composition of causal and inferential dynamics. 

 In order to generalise the additional regularising riemannian term in \eqref{histories.functional.with.metric} and \eqref{Klauder.path.integral}, let us consider the expansions
\begin{equation}
	D(\phi+\varepsilon v,\phi)=\frac{\varepsilon^2}{2}\gbold^D_{ab}(\phi)v^av^b+\mathcal{O}(\varepsilon^3)
\end{equation}
and \cite{Rodriguez:2003}
\begin{equation}
	D_\gamma(\phi+\varepsilon v,\phi)=\frac{\varepsilon^2}{2}\gbold^{D_\gamma}_{ab}(\phi)v^av^b+\frac{\varepsilon^3}{6}(\Gamma^{\nabla^0}_{abc}+\Gamma^{\nabla^\gamma}_{abc}+\Gamma^{\nabla^1}_{abc})v^av^bv^c+\mathcal{O}(\varepsilon^4),
\end{equation}
where $\Gamma^\nabla_{abc}$ are the Christoffel symbols of the corresponding connections. Setting $\varepsilon ^2=\frac{1}{\upsilon}$ suggests us to use the quantity
\begin{equation}
	P^{\pathgamma,\epsilon}_{k,1,1}:=\ee^{-k\int_\pathgamma \dd t D\left(\phi(t)+\varepsilon\frac{\dd\phi(t)}{\dd t},\phi(t)\right)}\sqrt{\det(\gbold^D)}
\label{local.entropic.prior}
\end{equation}
as a generalised regulariser, where $k\in\RR^+$ is a constant. We interpret this object as a \textit{local} quantum entropic prior: an expression for a local prior measure representing user's ignorance about the choice of propagation between neighbouring states along a specific trajectory $\pathgamma:[0,s]\ra\M(\N)$. See Section \ref{entropic.priors.section} for a discussion of entropic priors in the commutative case. More specifically, \eqref{local.entropic.prior} is a localised quantum version of the $P_{k,1,1}$ prior. For $D_{1/2}(\sigma,\rho)=\frac{1}{2}\n{\sqrt{\sigma}-\sqrt{\rho}}^2_\H$, this corresponds to integrating against a local gaussian measure. The global Jeffreys prior $\sqrt{\det(\gbold^{D_{1/2}})}$ appears already in the Klauder--Maraner formula \eqref{Klauder.Maraner.metric.variation}, as a constant \eqref{Klauder.Maraner.prior}, which sets a relationship between local measure of uncertainty of inference and local generator of causal dynamics.

Thus, we propose to generalise the formula \eqref{Klauder.path.integral} to
\begin{align}
	\lim\limits_{\varepsilon\ra+0}
	\int\mathcal{D}\phi(\cdot)
	&\ee^{\ii\int_\pathgamma\dd t\s{\Omega_{\phi(t)},\ddd_{\nabla^{1/2}}(\phi)\Omega_{\phi(t)}}_{\H_{\phi(t)}}}
	\ee^{-\ii\int_\pathgamma\dd t \s{\Omega_{\phi(t)},\pi_{\phi(t)}(\DF_{\phi(t)}h)\Omega_{\phi(t)}}_{\H_{\phi(t)}}}\nonumber\\
	&\cdot\ee^{-k\int_\pathgamma \dd t D\left(\phi(t)+\varepsilon\frac{\dd\phi(t)}{\dd t},\phi(t)\right)}\sqrt{\det(\gbold^D)}.
\label{RPK.propagator}
\end{align}
As discussed in Section \ref{local.quantum.dynamics.intro.section}, there are some legitimate reasons to believe that at least at the level of the second order approximation of an entropic prior, the above formula can receive an exact foundation by means of stochastic integration process. Yet, without a proof of this conjecture, the formula \eqref{RPK.propagator} has now a status of a heuristic proposal. However, most of the applications of path integrals in theoretical physics have precisely the same status (the exactness of the Daubechies--Klauder formula \eqref{Klauder.path.integral} is more an exception than a rule).

The differences between the formulas \eqref{RPK.propagator} and  \eqref{liouv.prop} correspond to the standard differences between algebraic and path integral formulations. Both formulations admit introducing additional local gauge and source terms, so they can be used to study various applied models. Taking a closer look at the Daubechies--Klauder formula \eqref{Klauder.path.integral}, one may note that the left side of this equation is formulated without taking into consideration the possible changes of the GNS representation along the states, because the coherent vector states are considered to be defined in a single Hilbert space. If such changes would be considered (as we do it here), then an operator $\ee^{-\ii Hs}$ in \eqref{Klauder.path.integral} should be multiplied from left by a corresponding standard unitary transition operator. This leads us to the conjecture:
\begin{equation}
	\eqref{liouv.prop} = \eqref{RPK.propagator},
\label{SS}
\end{equation}
\textit{if} the left hand side of this equation is evaluated in terms of the MCP, instead of the GNS, Hilbert space. While this conjecture is quite heuristic, it seems to be a legitimate candidate for a W$^*$-geometric analogue of the Daubechies--Klauder propagator formula \eqref{Klauder.path.integral}. A development of a suitable stochastic calculus allowing for an exact mathematical treatment of \eqref{RPK.propagator}, as well as the proof that the proposed construction of MCP Hilbert space is well defined, are the necessary conditions to approach the problem of proving this conjecture. Yet, in Section \ref{new.local.quantum.foundations}, based on the discussions in Sections \ref{local.infogeometry.histories.intro.section}, \ref
{Orlicz.Br\`{e}gman.section}, and \ref{section.dynamics.renormalisation}, we will propose another, more geometric approach to the equivalence intended behind the formula \eqref{SS}, without requiring equality on the level of Hilbert bundles.
\subsection{Appendix: Entropic priors on statistical models\label{entropic.priors.section}}
To simplify the notation, whenever we will use the coordinate-dependent formulas in this Section, we will assume that the statistical model $\M:=\M(\X,\mho(\X),\tmu)\subseteq L_1(\X,\mho(\X),\tmu)^+$ is equipped with a global coordinate system $\theta:\Theta\ra\M$, where $\Theta\subseteq\RR^n$ is open.

The entropic prior $P_{k,\alpha,\beta}$ is defined in order to provide the general \cytat{statistical representation of the [notion of the] vacuum of information in a given hypothesis space} \cite{Rodriguez:1999}. Every probability measure encodes some knowledge, hence the notion of the `vacuum of information' has also to refer to some given knowledge which defines it. The `vacuum of information' is relative to the given information manifold, and as such it is defined to depend on the invariant volume measure on the information manifold, the Jeffreys prior \cite{Jeffreys:1946}
\begin{equation}
	\J(\theta)=\sqrt{\ab{\det\gbold(\theta)}}\dd\theta_1\land\ldots\land\theta_n,
\end{equation}
which distributes prior probability over all hypothesis space, as well as on the initial reference density $p_0(x,\theta)=p_0(x|\theta)P(\theta)$ on $\X\times\Theta$ which sums up all additional reference knowledge (e.g., the quantitative results of previous experimental procedures) which will be encoded into the structure of the vacuum of information.

In particular, when the reference knowledge consists only of model-independent information encoded in the density $m(x)$, then the reference density factorises to $p_0(x,\theta)=m(x)P(\theta)$. The entropic prior build with respect to such factorisation encodes the `vacuum of information' regarding the dependence between the parameters $\Theta$ of the model $\M$ and the initial knowledge $m(x)$ about the data ${\X}$. 

In the nonparametric formulation based on $D_\gamma$ entropies for $\gamma\in[0,1]$, the \df{entropic prior} is defined as such $P_{k,\alpha,\beta}$ which minimises the functional \cite{Rodriguez:2003}
\begin{equation}
        \inf_P(k\int P(p)D_\alpha(p,p_0)+D_\beta(P,\J)),
\label{xep}
\end{equation}
where $(\alpha,\beta)\in[0,1]\times[0,1]$, the distance $D_\alpha$ is calculated over $\X\times\Theta$, the distance $D_\beta$ is calculated over $\Theta$, and the scalar $k\geq0$ parametrises the preference of $P$ over $\J$ with respect to the reference density $p_0$. In the parametric formulation, the functional minimised in \eqref{xep} reads 
\begin{equation}
        k D_\alpha(p(x|\theta)P(\theta),p_0(x|\theta)P(\theta))+D_\beta(P(\theta),\sqrt{\ab{\det\gbold(\theta)}}\dd\theta).
\end{equation}
By definition, the entropic priors are minimisers of the estimation by expected loss (decision) functional $D_\alpha$ under the constraint that the $D_\beta$-distance of entropic prior $P$ from volume measure $\J$ does not exceed some constant value. In other words, they express the degree of confidence in the reference distribution, relatively to degree of confidence in volume measure invariance of $P$.

The general solution of the above minimisation problem takes the form 
\begin{equation}
	P_{k,\alpha,\beta}=\tilde{P}_{k,\alpha,\beta}\J,
\end{equation}
where $\tilde{P}_{k,\alpha,\beta}(\theta)$ is a scalar density which, up to normalisation, is equal to \cite{Rodriguez:2003,Snoussi:2005,Snoussi:2007}
\begin{equation}
\tilde{P}_{k,\alpha,\beta}(\theta):=
\left\{
\begin{array}{ll}
        1+k(1-\beta)D_\alpha(p_\theta,p_0)^{-\frac{2}{1+\beta}}
				&:\;\beta\neq1,\\
				\exp(-k D_\alpha(p_\theta,p_0))
				&:\;\beta=1,
\end{array}
\right.
\label{entropic.priors}
\end{equation}
under the condition that
\begin{equation}
        k_{min}:=\inf\{k\geq 0\mid\textstyle\int P_{k,\alpha,\beta}<\infty\}
\end{equation}
exists and $k\geq k_{min}$. If $k_{min}=0$, then ${P}_{0,\alpha,\beta}$ is Jeffreys prior. On the other hand, ${P}_{\infty,\alpha,\beta}$ is the Dirac delta concentrated on $p_0$. Moreover, $\tilde{{P}}_{k,\alpha\in\{0,1\},\beta\in[0,1[}$ is a multivariate Student $t$ distribution, $\tilde{{P}}_{k,\alpha,0}$ is a generalised multivariate Cauchy distribution, $\tilde{{P}}_{k,\alpha,1}$ is a minimum $D_\alpha$ prior density, and 
\begin{equation}
        \tilde{{P}}_{k,1,1}(p)=\ee^{-kD_1(p,p_0)}
\end{equation}
is a maximum relative entropy (= minimum $D_1$ distance) prior density. If $p_0$ is taken to be the Bernoulli--Laplace uniform prior, then $\tilde{{P}}_{1,1,1}(p)=\ee^{-\entropy_{GS}(p)}$ is a maximum Gibbs--Shannon entropy density (Jaynes prior \cite{Jaynes:1957}). A maximum Gibbs--Shannon entropy density can be also recovered as $\tilde{{P}}_{\infty,1,1}(p)$ if $p_0$ is an element of the exponential family. When no reference distribution (no background information) is specified, then ${P}_{k,\alpha,\beta}$ reduces trivially to Jeffreys' prior. If the reference measure $p_0$ does not belong to a manifold $\Q$ on which the minimisation procedure generating the entropic prior is evaluated, then $k\left(1-\beta\right)$ factor for $\beta\neq1$ is replaced by a more general scalar quantity $\tilde{k}$, dependent on the projection of $p_0$ on $\Q$ (for details, see \cite{Snoussi:2005}).

If $p_0$ is a maximum entropy distribution obtained under some given constraints, then the entropic prior quantifies to what extent the densities other than $p_0$ (within a given model) are less probable or less reliable. The reliability of $p(x|\theta)$ other than $p_0(x|\theta)$ decreases exponentially with the deviation of $p(x|\theta)$ from $p_0(x|\theta)$, and the sensitivity for this exponential decrease is controlled by the constant $k$. In general, the larger $k$ is, the stronger is the impact of reference distribution (assumed background information) on the inference provided with respect to the `vacuum of information' ${P}(\theta)$. If the reference hypothesis is built up from knowledge independent of the model (encoded in $m(x)$), then the larger $k$ is, the more preference is given to this independent knowledge. On the other hand, the smaller $k$ is, the more inference based on ${P}(\theta)$ will depend on distributions other than $p_0(x|\theta)$, so it becomes easier for the eventual `noise' in constraints to be taken by inference to be a `signal'. So, while $k\ra 0$ smoothens the prior, $k\ra\infty$ sharpens it. Jeffreys' and Jaynes' (maximum Gibbs--Shannon entropy) priors are just two extreme points of this scale.

The entropic priors $P_{k,1,1}$ are the only entropic priors on probabilistic manifold which are based on the measure of distance which is coordinate invariant, local, consistent for independent subsystems and additive. This is a characterisation of $D_1$ as a unique distance functional on the space of normalised probability densitites used for the purpose of probability updating, as provided in \cite{Shore:Johnson:1980,Shore:Johnson:1981,Johnson:Shore:1983}. Hence, they are the unique priors which encode the notion of coordinate invariant, local, additive, and independent subsystem consistent `vacuum of information'. For more discussion on the topic of entropic priors, see \cite{Skilling:1989,Caticha:2001:a,Caticha:Preuss:2003,Caticha:Preuss:2004,Rodriguez:1991,Rodriguez:2003,Snoussi:MohhamadDjafari:2003,Snoussi:2005}.
\section{Information theoretic renormalisation\label{information.theoretic.renormalisation}}
In this section we will analyse another application of quantum information geometry for the purpoes of inference over quantum models. We start from discussion of the Jaynes--Mitchell source theory \cite{Mitchell:1967,Jaynes:1985:scattering,Jaynes:1993,Grandy:1987,Grandy:2008}, which describes the general continuous changes of information states of an exponential model driven by the sources of information. Next, we discuss Favretti's \cite{Favretti:2007} information geometric generalisation of this theory to the setting of dually flat information manifolds. It allows for a strictly geometric implementation of the idea of renormalisation of dynamics by reduction of dimensionality of the model by fixing the control parameter, which is provided on the space of information states (as opposed to the space of functions or operators). Our original contribution amounts to an observation that the Jaynes--Mitchell--Favretti approach is canonically related to the use of Br\`{e}gman distances, so it can be used to \textit{locally} approximate information dynamics on an arbitrary manifold of quantum states. We discuss how this setting allows to use the departure of local geometry from the dually flat smooth geometry (generated by quantum Br\`{e}gman distances) as the geometric description of multiparameter nonlinear quantum control and renormalisation problems. We also introduce another type of geometric renormalisation of inferential dynamics of quantum states, which describes situations where none of  specific control (covariate) parameter is fixed, but the quantum model is subjected to the action of completely positive maps. This procedure is based on the use of $D_\fff$ distances as well as associated \textit{contraction coefficients}, introduced by Ruskai et al \cite{CIRRSZ:1993,Choi:Ruskai:Seneta:1993,Ruskai:1994,Lesniewski:Ruskai:1999}.
\subsection{Jaynes--Mitchell source theory\label{section.JM.source.theory}}
Consider first an arbitrary statistical model $\M(\boole)$ over finite boolean algebra\footnote{For the reasons of mathematical fanciness, we occasionally consider the sets $\M(\boole)\subseteq L_1(\boole)^+$ of finite positive measures over localisable boolean algebras $\boole$, but this is completely equivalent to consideration of localisable measure spaces $\M(\X,\mho(\X),\tmu)\subseteq L_1(\X,\mho(\X),\tmu)^+$.} $\boole$ (with $m\in\NN$ denoting the number of elements of $\boole$), a set $\{f_k\}_{k=1}^{n}$ of functions $f_k:\boole\ra\RR$ with $n\in\NN$, and a change in the expectation $\e{f_k}_p$, caused by the independent changes in both $f_k(x_i)=:f_k^i$ and $p(x_i)=:p_i$,
\begin{equation}
        \delta\e{f_k}=\sum_{i=1}^mp_i\delta f_k^i+\sum_{i=1}^mf_k^i\delta p_i.
\label{change.of.expectation}
\end{equation}
If $f_k$ depends on some additional parameters $r=(r_1,\ldots,r_l)$, such that
\begin{equation}
        \delta f_k(x_i,r)=\sum_{j=1}^{l}\frac{\partial f_k(x_i,r)}{\partial r_j}\delta r_j,
\label{add.param}
\end{equation}
then the first term of \eqref{change.of.expectation} reads
\begin{equation}
        \sum_{i=1}^mp_i\delta f_k^i=\e{\delta f_k}_p=\e{\sum_{j=1}^l\frac{\partial f_k}{\partial r_j}\delta r_j}_p=:\delta W_k.
\label{work.def}
\end{equation}
We denote the second term of \eqref{change.of.expectation} by $\delta Q_k$, so
\begin{equation}
        \delta Q_k:=\sum_{i=1}^mf_k^i\delta p_i=\delta\e{f_k}_p-\e{\delta f_k}_p,
\label{Q.def}
\end{equation}
which gives
\begin{equation}
        \delta\e{f_k}_p=\delta W_k+\delta Q_k.
\label{gen.1st.law.1}
\end{equation}

Consider now an \df{exponential family} defined as an $n$-dimensional parametric probabilistic manifold \cite{Darmois:1935,Koopman:1936,Pitman:1936}\rpktarget{MEXP}
\begin{equation}
        \M_{\mathrm{exp}}(\X,\mho(\X),\tmu):=
        \{
                p(\xx,\theta):=
                        \exp(
                                -\log Z(\theta)-\textstyle\sum_{i=1}^n\theta^if_i(\xx)
                        )\mid
        \theta:=(\theta^1,\ldots,\theta^n)\in\Theta\subseteq\RR^n
        \},
\label{commutative.exponential.model}
\end{equation}
where $f_i:\X\ra\RR$ are assumed to be arbitrary functions, linearly independent of each other and of the constant function $1$ (this guarantees that $\theta\mapsto p(\theta)$ is one-to-one and that the matrix $\gbold_{ij}$ is invertible \cite{Wehrl:1978}),
\begin{equation}
        \log Z(\theta):=\log\int_\X \tmu(\xx)\exp\left(-\sum_{i=1}^n\theta^if_i(\xx)\right)
\label{partition.function.eq}
\end{equation}
is a factor arising from normalisation condition $\int_\X\tmu(\xx)p(\xx,\theta)=1$, called a \df{Massieu functional} \cite{Massieu:1869,Massieu:1876}, while $\Theta\subseteq\RR^n$ is supposed to be such open set that the integral in \eqref{partition.function.eq} converges. The study of geometric properties of this family provided an original stimulus for development of information geometry \cite{Chencov:1966,Chencov:1972,Efron:1978,Amari:1982}. In particular, Chencov found \cite{Chencov:1966,Chencov:1968} that the finite dimensional exponential families are geodesic surfaces of $\nabla^0$-connections and admit the generalised pythagorean equation \eqref{generalised.pythagore.Bregman} for the Kullback--Leibler distance. 

If $\dim\X=:m<\infty$, then $\int_\X\tmu(\xx)k(\xx)=\sum_{j=1}^mk(\xx_j)$ for any $k:\X\ra\RR$. In such case $\M_{\mathrm{exp}}(\X,\mho(\X),\tmu)$ can be characterised in terms of the Gibbs--Jaynes \cite{Gibbs:1902,Jaynes:1957} procedure of maximisation of the \df{Gibbs--Shannon entropy} \cite{Gibbs:1902,Shannon:1948,Shannon:Weaver:1949}
\begin{equation}
        \entropy_\GS(p):=-\sum_{j=1}^mp(\xx_j)\log p(\xx_j)
        \rpktarget{SGS}
\end{equation}
subject to constraints $F(p)$ given by 
\begin{equation}
\left\{
\begin{array}{l}
\sum_{j=1}^mp(\xx_j)1=1,\\
\sum_{j=1}^mp(\xx_j)f_i(\xx_j)=\eta_i,
\end{array}
\right.
\label{exp.family.constraints}
\end{equation}
with $\eta:=(\eta_i)\in\Xi\subseteq\RR^n$. The maximum value attained by $\entropy_\GS$ for a given $(\eta_i)$ (or, equivalently, for a given $(\theta^i)$), reads
\begin{equation}
        \entropy_\GS(p(\theta))=\log Z(\theta)+\sum_{i=1}^n\theta^i\eta_i.
\label{max.ent.equation}
\end{equation}

If $p$ belongs to an exponential family with $\lambda_k:=\theta^k$, $k\in\{1,\ldots,n\}$, then the corresponding change in entropy reads
\begin{align}
\delta\entropy_\GS&=\delta\log Z(\lambda)+\delta\left(\sum_{k=1}^n\lambda_k\e{f_k}_p\right)\notag\\
&=-\frac{1}{Z}\left(\sum_{k=1}^n\delta\lambda_k f_k^i+\sum_{k=1}^n\lambda_k\delta f_k^i\right)\ee^{-\sum_{k=1}^n\lambda_k f_k^i}+\sum_{k=1}^n\delta\lambda_k\e{f_k}_p+\sum_{k=1}^n\lambda_k\delta\e{f_k}_p\notag\\
&=\sum_{k=1}^n\lambda_k(\delta\e{f_k}_p-\e{\delta f_k}_p)=\sum_{k=1}^n\lambda_k\delta Q_k.
\label{entropy.and.work}
\end{align}
Due to \eqref{add.param}, $\sum_{k=1}^n\lambda_k\delta Q_k(\e{f_{\widetilde{k}}}_p,r)$ is an exact differential of $\entropy_\GS(\e{f_{\widetilde{k}}}_p,r)$, even if $\delta Q_k(\e{f_{\widetilde{k}}}_p,r)$ is not an exact differential of any function. Thus, \eqref{entropy.and.work} is equivalent to
\begin{equation}
        \delta\entropy_\GS=\sum_{k=1}^n\lambda_k\delta\e{f_k}_p-\sum_{k=1}^n\sum_{j=1}^l\lambda_k\e{\frac{\partial f_k}{\partial r_j}}_p\delta r_j
\label{gen.1st.law.2}
\end{equation}
and
\begin{equation}
        \left.\frac{\partial\entropy_\GS(\e{f_k}_p,r)}{\partial\e{f_k}_p}\right|_{r=\mathrm{const}}=\lambda_k.
\label{gen.1st.law.3}
\end{equation}
These results are completely analogous to the first law of equilibrium thermodynamics. The change of `information work' $W_k$ is dependent only on the changes of quantity $f_k$. The change of `information heat' $Q_k$ and of absolute entropy $\entropy_\GS$ depends only on the change of information state provided by probability $p$. While having the same mathematical form, the above results are completely independent of thermodynamics and hold for any exponential family. The first law of equilibrium thermodynamics is just a \textit{special case} of the above result.

Let us now consider a three dimensional exponential family
 \begin{equation}
        \M_{\mathrm{exp}}(\X,\mho(\X),\tmu;\Theta):=\left\{p(\lambda_A,\lambda_B,\lambda_C)=\frac{1}{Z}\ee^{-\lambda_Af_A(x)-\lambda_Bf_B(x)-\lambda_Cf_C(x)}\mid(\lambda_A,\lambda_B,\lambda_C)\in\Theta\right\},
\end{equation}
where $f_A,f_B,f_C\in L_\infty(\X,\mho(\X),\tmu)$, and $\Theta\subseteq\RR^3$ is some fixed open set. Let the change of information be described by $\e{f_A}_p\ra\e{f_A}_p+\delta\e{f_A}_p$ with the additional conditions that the possible changes of $\e{f_B}_p$ are left unconstrained ($\delta\lambda_B=0$ but we allow $\delta\e{f_B}_p\neq0$), and it is known that $\e{f_C}_p$ does not change ($\delta\e{f_C}_p=0$ but we allow $\delta\lambda_C\neq0$). The quantity $\e{f_A}_p$ is called a `driving variable'. Thus, we consider a \textit{source-and-response problem} with an \textit{additional control variable}:
\begin{align*}
\delta\e{f_A}&=0,\;\;\delta\lambda_A\neq0\;\;\mbox{`driving variable' (source parameter)}\\
\delta\e{f_B}&\neq0,\;\;\delta\lambda_B=0\;\;\mbox{`information heat bath' (response parameter)}\\
\delta\e{f_C}&=0,\;\;\delta\lambda_C\neq0\;\;\mbox{`control variable' (additional source)}
\end{align*}
Following Mitchell and Jaynes, will now provide an answer to a question: how the presence of the second source affects the relationship between first source and the response parameter?

Given \textit{some} finite dimensional statistical model $\M(\boole)$ parametrised by a coordinate system $\lambda:\M(\boole)\ra U\subseteq\RR^n$ with $n:=\dim(\M(\boole))$, then the general form of the variation of an expectation functional $\e{f}_p$ for some element $p(\lambda_0)\in\M(\boole)$ reads
\begin{equation}
        \delta\e{f}_p
        :=\e{f}_{p(\lambda)}-\e{f}_{p(\lambda_0)}
        =\sum_{i=1}^\infty
        \frac{1}{i!}\sum_{(j_1,\ldots,j_i)}
        \left.
        \frac{\partial^i\e{f}_{p(\lambda)}}%
        {\partial\lambda_{j_1}\cdots\partial\lambda_{j_i}}
        \right|_{\lambda=\lambda_0}
        \delta\lambda_{j_1}\cdots\delta\lambda_{j_i}.
        \label{infinite.perturbation.covariance.expansion}
\end{equation}
The first order term of \eqref{infinite.perturbation.covariance.expansion} (corresponding to the linear character of  variation) reads
\begin{equation}
        \delta\e{f}_p=
        \sum_{j=1}^n\frac{\partial\e{f}_p}{\partial\lambda_j}
        \delta \lambda_j.
        \label{first.order.covariance.expansion}
\end{equation}
In the case of exponential model $\M_{\mathrm{exp}}(\boole;\Theta)$, from the equations \eqref{first.order.covariance.expansion} and \begin{equation}
	K_{ij}=\frac{\partial^2\log Z(\lambda)}{\partial\theta^i\partial\theta^j}=-\frac{\partial\e{f_i}_p}{\partial\lambda^j}=-\frac{\partial\e{f_j}_p}{\partial\lambda^i},
\end{equation}
it follows that the relationship between `fluxes of information' $\delta\e{f_k}_p$ and `forces of information' $(-\delta\lambda_k)$ can be determined in the first (linear) order by the covariance matrix\footnote{In this terminology $\lambda_k$ play the role of the `potentials of information', but this should not be confused with the `scalar potentials' $\Psi$ and $\Psi^\lfdual$ on hessian manifolds, such as $-\log Z(p)$ and $\entropy_{\mathrm{GS}}(p)$ (which play the role of information discrimination functionals).}
\begin{equation}
        \left(\begin{array}{c}\delta\e{f_A}_p\\\delta\e{f_B}_p\\\delta\e{f_C}_p\end{array}\right)=-\left(\begin{array}{ccc}
K_{AA}&K_{AB}&K_{AC}\\
K_{BA}&K_{BB}&K_{BC}\\
K_{CA}&K_{CB}&K_{CC}    \end{array}\right)\left(\begin{array}{c}\delta\lambda_A\\\delta\lambda_B\\\delta\lambda_C\end{array}\right).
\label{fff}
\end{equation}
Hence, 
\begin{equation}
        \delta\lambda_C=-\frac{K_{CA}}{K_{CC}}\delta\lambda_A,
\end{equation}
which means that $\delta\lambda_C$ and $\delta\lambda_A$ are not independent of each other. This is also reflected in the second equation following from \eqref{fff}, namely 
\begin{equation}
        \delta\e{f_A}_p=-\delta\lambda_A\left(K_{AA}-\frac{K^2_{AC}}{K_{CC}}\right),
\end{equation}
which is equivalent to
\begin{equation}
        \frac{\delta\e{f_A}_p}{\delta\lambda_A}=\frac{\partial\e{f_A}_p}{\partial\lambda_A}-\frac{K^2_{AC}}{K_{CC}}.
\end{equation}
If we \textit{decide} to consider only the variables $\e{f_A}_p$ and $\e{f_B}_p$ (removing $\e{f_C}_p$ from the \textit{definition} of the problem), then the covariance matrix of the problem takes the form
\begin{equation}
\left(\begin{array}{c}\delta\e{f_A}_p\\\delta\e{f_B}_p\end{array}\right)=-\left(\begin{array}{cc}
K_{AA}&K_{AB}\\
K_{BA}&K_{BB}
\end{array}\right)\left(\begin{array}{c}\delta\lambda_A\\\delta\lambda_B\end{array}\right).
\label{kv1}
\end{equation}
From the assumption that there are no additional parameters $r$ of control associated with the element $f_A$ of the abstract algebra (that is, $\e{\delta f_A}_p=0$), it follows that $\delta\e{f_A}_p=\delta Q_A$. In such case the above equation turns into
\begin{equation}
        \delta\e{f_B}_p=\frac{K_{BA}}{K_{AA}}\delta Q_A.
\end{equation}
Hence, the changes of $\e{f_B}_p$ are driven by the `source of information' $\delta Q_A$. We will call the corresponding evolution of probability distribution $p(\lambda_A,\lambda_B,\lambda_C)\in\M_{\mathrm{exp}}(\boole;\Theta)$ an `information driving'. The number of different variables is not limited to three, but three variables are sufficient to describe all possible \textit{types} of constraints. Mitchell \cite{Mitchell:1967} has shown that the readjustment of expectation values of functions $\{f_k\}$ under driving caused by sources of information can be described the following \textit{equivalent} principles:
\begin{itemize}
\item[i)] expectations uncorrelated with driven variables remain unchanged,
\item[ii)] Lagrange multipliers of unconstrained variables remain unchanged,
\item[iii)] $\entropy_\GS$ is re-maximised under new values of constraints.
\end{itemize}

Now we move to the problem of renormalisation of sources, which amounts to removing the variable $C$ from the definition of the model, while keeping it as a constraint in the allowed transformations of variables (information flows). Consider again the covariance matrix \eqref{fff}, with the constraint $\delta\e{f_C}_p=0$. A direct calculation shows that the relationships between `fluxes' and `forces' of information related with $A$ and $B$ can be completely described by the covariance matrix
\begin{equation}
\left(\begin{array}{c}\delta\e{f_A}_p\\\delta\e{f_B}_p\end{array}\right)=-\left(\begin{array}{cc}
\tilde{K}_{AA}&\tilde{K}_{AB}\\
\tilde{K}_{BA}&\tilde{K}_{BB}
\end{array}\right)\left(\begin{array}{c}\delta\lambda_A\\\delta\lambda_B\end{array}\right),
\label{kv3}
\end{equation}
where 
\begin{equation}
        \left\{
                \begin{array}{l}
                        \tilde{K}_{AA}:=K_{AA}-K_{AC}K_{CC}^{-1}K_{CA}\\
                        \tilde{K}_{AB}:=K_{AB}-K_{AC}K_{CC}^{-1}K_{CB}\\
                        \tilde{K}_{BA}:=K_{BA}-K_{BC}K_{CC}^{-1}K_{CA}\\
                        \tilde{K}_{BB}:=K_{BB}-K_{BC}K_{CC}^{-1}K_{CB}.
                \end{array}
        \right.
\end{equation}
The covariance matrix \eqref{kv3} can be thought of as a `renormalised' version of the covariance matrix \eqref{kv1}, where the dependence on an additional correlated information related to variable $C$ is taken into account. Assuming again that $\delta\lambda_B=0$ and $\e{\delta f_A}_p=0$, the predicted change of $\e{f_B}_p$ due to the action of the source $\delta Q_A$ takes the form
\begin{equation}
        \delta\e{f_B}_p=\frac{\tilde{K}_{BA}}{\tilde{K}_{AA}}\delta Q_A=\left(\frac{K_{BA}}{K_{AA}}-\frac{K_{BC}}{K_{CC}}\frac{K_{CA}}{K_{AA}}\right)\delta \tilde{Q}_A,
\label{renormalized.equation}
\end{equation}
where
\begin{equation}
        \delta \tilde{Q}_A:=\frac{\delta Q_A}{(1-R_{AC}^2)}
\label{reqf}
\end{equation}
is the `renormalised information source strength', while 
\begin{equation}
        R_{AC}:=\frac{K_{AC}}{(K_{AA}K_{CC})^{\frac{1}{2}}}
\end{equation}
is the correlation coefficient. In other words, the additional constraint ($\delta\e{f_C}_p=0$) imposed on the information related to an additional variable that is correlated with the driving variable ($\e{f_A}_p$) is observed in `renormalisation' of the action of the driving source ($\delta Q_A$) on the dimensionally `reduced' system of variables (without $\e{f_C}_p$):
\begin{equation}
        \delta\e{f_B}_p=\frac{\tilde{K}_{BA}}{\tilde{K}_{AA}}\frac{\delta Q_A}{1-R_{AC}^2}.
\label{reduced.system}
\end{equation}
Now, if $R_{AC}$ has a spectral radius smaller than $1$, one can expand the renormalisation factor in \eqref{reqf} and \eqref{reduced.system},
\begin{equation}
        (1-R_{AC}^2)^{-1}=\sum_{n=0}^\infty(R_{AC}^2)^n=1+R^2_{AC}+R^4_{AC}+\ldots\;.
\label{Jaynes.expansion}
\end{equation}
Defining the `propagators' $\mathcal{G}_{ij}:=-K_{ij}K_{jj}^{-1}$, one can expand \eqref{renormalized.equation} in the form
\begin{equation}
        \begin{array}{c}
\delta\e{f_B}_p=(\mathcal{G}_{BA}-\mathcal{G}_{BC}\mathcal{G}_{CA}+\mathcal{G}_{BA}\mathcal{G}_{AC}\mathcal{G}_{CA}-\mathcal{G}_{BC}\mathcal{G}_{CA}\mathcal{G}_{AC}\mathcal{G}_{CA}+\ldots)\delta Q_A.
        \end{array}
\label{expanded.change}
\end{equation}

Thus, the dimensional reduction of the information model which removes from the scope the correlated constrained variable changes the description of information flow, which can be recasted in terms of perturbative series of propagators between the `sources' of driving variables and `information fluxes' of driven variables (`sinks') that are mediated by the ``virtual'' (removed) variable. Comparison of \eqref{first.order.covariance.expansion} with \eqref{infinite.perturbation.covariance.expansion} leads us to note, following Jaynes, that the above effects appear at the \textit{first} level of perturbative expansion in powers of information source strength. In consequence, the corresponding classification of approximated results is provided by the degree of fine tuning of the available information. This brings a clear meaning to the perturbative expansion and renormalisation as the process of classification of approximated description of \textit{the quantitative effects of change of information} with respect to the degree of quantitative refinement of this information (which is given by information source strength). This approximation does not refer to any additional `theoretical' or `physical' dimensional constant parameters and keeps the values and meaning of the constants defining experimental response scales, etc., to be fixed by definition and not entering the scene. Thus, there is also no need for `renormalisation' of these constants, avoiding the conceptual problems which are always caused by such procedure.
\subsection{Favretti's dually flat geometrisation\label{Favretti.section}}
Now we turn to reformulation and generalisation of the Jaynes--Mitchell source theory provided by Favretti \cite{Favretti:2007}. Suppose that $\M(\boole)$ is a probability manifold with $\dim\M(\boole)=n\in\NN$, equipped with the pair of coordinate systems $(\theta,\eta):\M(\boole)\ra\Theta\times\Xi\subseteq\RR^n\times\RR^n$. Let the information about trajectory $p(t)\in\M(\boole)$ be specified as the constraints expressed in terms of both coordinate systems: 
\begin{equation}
\left\{
        \begin{array}{l}
                F_1(\theta(p),t)=0,\\
                F_2(\eta(p),t)=0.
        \end{array}
\right.
\label{constraints.JF}
\end{equation}
Favretti shows that under additional assumption that $\M(\boole)$ is equipped also with a riemannian metric $\gbold$ and a pair of affine connections $(\nabla^\theta,\nabla^\eta)$ such that $(\M(\boole),\gbold,\nabla^\theta,\nabla^\eta)$ is a dually flat manifold with a dually flat coordinate system given by $(\theta,\eta)$ (see Section \ref{distances.NS.geom.section}), the implicit function theorem allows one to describe geometrically the evolution $p(t)$ quantitatively, in terms of one of these coordinate systems.

Let the scalar potential functions determined by the above dually flat geometry be denoted by $\Psi(\theta):=\Psi\circ\theta$ and $\Psi^\lfdual(\eta):=\Psi^\lfdual\circ\eta$, where $\Psi^\lfdual$ is a Fenchel dual of $\Psi$ with respect to \eqref{selfduality.Rn}. Consider the diagram
\begin{equation}
        \xymatrix{&\ar[dl]_{\eta}\M(\boole)\ar[dr]^{\theta}&\\
\Xi\ar@<0.5ex>[rr]^{\LFtrafo^{-1}_\Psi}\ar@<0.5ex>[d]^{\pi_A^\Xi}&&\Theta\ar@<0.5ex>[ll]^{\LFtrafo_\Psi}\ar@<0.5ex>[d]^{\pi_B^\Theta}\\
\Xi_A\ar@<0.5ex>[u]^{(\pi^\Xi_A)^{-1}}&&\Theta_B\ar@<0.5ex>[u]^{(\pi^\Theta_B)^{-1}},}
\label{favretti.diag}
\end{equation}
where $\LFtrafo^{-1}_\Psi:\Xi\ra\Theta$ and $\LFtrafo_\Psi:\Theta\ra\Xi$ are the Legendre transforms given by smooth diffeomorphisms, which are expressed in coordinate-dependent way as
\begin{align}
        \theta^i=(\LFtrafo^{-1}_\Psi(\eta))^i&=\frac{\partial}{\partial\eta_i}\Psi^\lfdual(\eta)=:\partial^i\Psi^\lfdual(\eta),\label{LFT.MJF.Phi}\\
        \eta_i=(\LFtrafo_\Psi(\theta))_i&=\frac{\partial}{\partial\theta^i}\Psi(\theta)=:\partial_i\Psi(\theta),
\label{LFT.MJF.Psi}
\end{align}
while 
\begin{align}
        \pi_A^\Xi:\Xi\ni\eta&\mapsto\eta_A\in A\subseteq\RR^k,\label{first.project.fv}\\
        \pi_B^\Theta:\Theta\ni\theta&\mapsto\theta^B\in B\subseteq\RR^{n-k},
\end{align}
are projections with 
\begin{align}
\eta&=(\eta_A,\eta_B)\in\RR^k\times\RR^{n-k},\\
\theta&=(\theta^A,\theta^B)\in\RR^k\times\RR^{n-k}.\label{last.project.fv}
\end{align}
The maps $\pi_A^\Xi$ and $\pi_B^\Theta$, when equipped with particular values at their codomain (denoted here, respectively, by $\bar{\eta}_A\in\Xi_A\subseteq\RR^k$ and $\bar{\theta}^B\in\Theta_B\subseteq\RR^{n-k}$), provide an example of the constraints \eqref{constraints.JF}:
\begin{equation}
\left\{\begin{array}{l}
\eta_A(p(t))=\pi^\Xi_A(\eta(p(t)))=\bar{\eta}_A,\\
\theta^B(p(t))=\pi^\Theta_B(\theta(p(t)))=\bar{\theta}^B.
\end{array}\right.
\end{equation}
The subspaces $\Xi_A$ and $\Theta_B$ denote, respectively, the range of the values $\bar{\eta}_A$ and $\bar{\theta}^B$ of the constraints $\pi_A^\Xi$ and $\pi_B^\Theta$. The fibres corresponding to these projections are given by
\begin{align}
        \M_\Xi(\bar{\eta}_A)&:=(\pi_A^\Xi)^{-1}(\bar{\eta}_A)=\{\eta\in\Xi\mid\eta_A=\bar{\eta}_A\}\subseteq\Xi,\\
\M_\Theta(\bar{\theta}^B)&:=(\pi^\theta_B)^{-1}(\bar{\theta}^B)=\{\theta\in\Theta\mid\theta^B=\bar{\theta}^B\}\subseteq\Theta,
\end{align}
and they induce the corresponding leaves of a pair of foliations of $\M(\boole)$ by
\begin{align}
        \M(\bar{\eta}_A)&:=\{p\in\M(\boole)\mid(\pi_A^\Xi\circ \eta)(p)=\bar{\eta}_A\},\\
        \M(\bar{\theta}^B)&:=\{p\in\M(\boole)\mid(\pi_B^\Theta\circ \theta)(p)=\bar{\theta}^B\},
\end{align}
with
\begin{equation}
        \bigcup_{\bar{\theta}^B\in\Theta_B}\M(\bar{\theta}^B)=\M(\boole)=\bigcup_{\bar{\eta}_A\in\Xi_A}\M(\bar{\eta}_A).
\end{equation}
Using the orthogonality \eqref{dual.metric.flatness} of the coordinate systems $\theta^j$ and $\eta_i$, Favretti shows that the tangent space at the point $p\in\M(\bar{\theta}^B)\cap\M(\bar{\eta}_A)$ has the following orthogonal decomposition
\begin{align}
        \T_p\M(\boole)&=\T_p\M(\bar{\theta}^B)\oplus\T_p\M(\bar{\eta}_A)=\Span\{\partial_1,\ldots,\partial_k\}\oplus \Span\{\partial^{k+1},\ldots,\partial^n\}.
\end{align}
For any $a\in\Xi_A$ and $b\in\Theta_B$, the leaves $\M(a)$ and $\M(b)$ are, respectively, $\nabla^\eta$- and $\nabla^\theta$- autoparallel submanifolds of $\M(\boole)$, hence they are called \df{mutually dual foliations}. Favretti observes that this allows to consider the evolution $t\mapsto p(t)$ geometrically, as a horizontal lift with respect to an integrable Ehresmann connection.

Let us now assume that $\M(\boole)$ is an $(n+m)$-dimensional dually flat probability manifold equipped with the projections \eqref{first.project.fv}-\eqref{last.project.fv}, as well as with an additional projection generated by
\begin{equation}
        \pi^{\Xi}_C:\Xi\ni\eta\mapsto\eta_C\in \Xi_C\subseteq\RR^m,
\end{equation}
where $\eta=(\eta_A,\eta_B,\eta_C)\in\RR^k\times\RR^{n-k}\times\RR^m$. One can introduce the foliation $\M(\bar{\eta}_C)$, corresponding to the constraint $\eta_A-\bar{\eta}_A=0$, in the same way as before. In such case the pairs of mutually dual foliations are given by $\M(\bar{\eta}_A,\bar{\eta}_C)$, $\M(\bar{\theta}^B)$ and $\M(\bar{\eta}_C)$, $\M(\bar{\theta}^A,\bar{\theta}^B)$. However, $\M(\bar{\eta}_C)$, $\M(\bar{\theta}^B)$ are not mutually dual.

This setting allows for the geometric generalisation of the source renormalisation procedure in the following form. Let temporal evolution $t\mapsto p(t)$ satisfy the constraints  $\theta^B(t)=\bar{\theta}^B$ and $\eta_C(t)=\bar{\eta}_C$, that is,
\begin{equation}
\left\{\begin{array}{l}
        \theta^B(p(t))=(\pi_B^\Theta\circ \theta)(p(t))-\bar{\theta}^B=0,\\
        \eta_C(p(t))=(\pi_C^\Xi\circ \eta)(p(t))-\bar{\eta}_C=0,
\end{array}\right.
\label{constr.F}
\end{equation}
then
\begin{equation}
        p(t)\in\M(\bar{\theta}^B)\cap\M(\bar{\eta}_C).
\end{equation}
These conditions rephrase the Jaynes--Mitchell conditions $\delta\lambda_B=0$ and $\delta\e{f_C}=0$ in the information geometric terms. Now one can find what is the form of evolution determined by these constraints, if changes of information are specified by the temporally driven `response' parameters $\eta_A=\eta_A(t)$ (which corresponds to the `driving variable' $\e{f_A}$). The constraints \eqref{constr.F} on the evolution can be restated using \eqref{LFT.MJF.Phi} in the form
\begin{equation}
        \partial^B\Psi^\lfdual(\eta_A(t),\eta_B(t),\bar{\eta}_C)-\bar{\theta}^B=0.
\label{linear.evolution.favretti}
\end{equation}
Favretti has shown that the implicit function theorem applied to \eqref{linear.evolution.favretti} implies the existence of a smooth map 
\begin{equation}
        h:\Xi_A\times\Xi_C\ni(\eta_A,\eta_C)\mapsto\eta_B\in\RR^{n-k}
\end{equation}
such that
\begin{align}
        \eta_B(t)&=h(\eta_A(t),\bar{\eta}_C),\\
        \dot{\eta}_B(t)&=\partial^A h(\eta_A(t),\bar{\eta}_C)\dot{\eta}_A(t),\label{eta.B.dot}\\
        h(\eta_A(t),\bar{\eta}_C)&=
                \left.
                -\left(
                        (\partial^B\partial^B\Psi^\lfdual)^{-1}
                        \partial^A\partial^B\Psi^\lfdual
                \right)
                \right|
                        _{\eta_B=h(\eta_A(t),\bar{\eta}_C)}.
\end{align}
By the assumption of dual flatness, this gives also
\begin{equation}
        \dot{\eta}_B(t)=
                (\tilde{\Psi}_{,BA}(\tilde{\Psi}_{,AA})^{-1})
                        |_{\theta=\bar{\theta}}\dot{\eta}_A(t)
        =:\tilde{\mathcal{G}}_{BA}(\theta)\,\dot{\eta}_A(t),
\label{inferred.favr.evol}
\end{equation}
where
\begin{align}
                (\Psi^\lfdual)^{,ij}&:=
                        \partial^i\partial^j\Psi^\lfdual\equiv
                        \frac{
                                \partial^2\Psi^\lfdual(\eta)
                        }{
                                \partial\eta_i\partial\eta_j
                        }=
                        \gbold^{ij}(\eta),\label{g.phi.hessian}\\
                \Psi_{,ij}&:=
                \partial_i\partial_j\Psi\equiv
                \frac{
                        \partial^2\Psi(\theta)
                }{
                        \partial\theta^i\partial\theta^j
                }=
                \gbold_{ij}(\theta),\label{g.psi.hessian}
\end{align}
and
\begin{align}
        \bar{\theta}&:=
                \LFtrafo^{-1}_\Psi(
                        \eta_A,
                        h(
                                \eta_A(t),
                                \bar{\eta}_C
                        ),
                        \bar{\eta}_C
                ),
        \\
        \tilde{\Psi}_{,BA}&:=
                \Psi_{,BA}-
                \Psi_{,BC}
                (\Psi_{,CC})^{-1}
                \Psi_{,CA},
        \\
        \tilde{\Psi}_{,AA}&:=
                \Psi_{,AA}
                (\II-R^2_{AC}),\label{Psi.prim.AA}
        \\
        R_{AC}^2&:=
                (\Psi_{,AA})^{-1}
                \Psi_{,AC}
                (\Psi_{,CC})^{-1}
                \Psi_{,CA}.
\end{align}
The equation \eqref{inferred.favr.evol} can be written more explicitly as
\begin{equation}
        \dd\eta_B(p(t))=\gbold_{BA}(p(t))\frac{1}{\II-R^2_{AC}(p(t))}\left(\gbold_{AA}(p(t))\right)^{-1}\dd\eta_A(p(t)).
\label{JMF.renormalisation.of.metric}
\end{equation}
From \eqref{Psi.prim.AA} it follows that
\begin{equation}
        (\tilde{\Psi}_{,AA})^{-1}=(\II-R_{AC}^2)^{-1}(\Psi_{,AA})^{-1},
\end{equation}
hence, if $R_{AC}$ has a spectral radius smaller than $1$, one can use \eqref{Jaynes.expansion}, which leads to the perturbative expansion in terms of corrections that come from interaction with the additional source,
        \begin{align}
        \tilde{\mathcal{G}}_{BA}&=\tilde{\Psi}_{,BA}(\II-R_{AC}^2)^{-1}(\Psi_{,AA})^{-1}=\mathcal{G}_{BA}-\mathcal{G}_{BC}\mathcal{G}_{CA}+\mathcal{G}_{BA}\mathcal{G}_{AC}\mathcal{G}_{CA}+\ldots,
\end{align}
where
\begin{equation}
        \mathcal{G}_{ij}:=\Psi_{,ij}(\Psi_{,jj})^{-1}
\end{equation}
For $\dd\eta_B=\dot{\eta}_B(t)\dd t$, $\dd\eta_A=\dot{\eta}_A(t)\dd t$, the above expression takes the form
\begin{equation}
        \dd\eta_B=(\mathcal{G}_{BA}-\mathcal{G}_{BC}\mathcal{G}_{CA}+\mathcal{G}_{BA}\mathcal{G}_{AC}\mathcal{G}_{CA}-\ldots)\dd\eta_A,
\label{diff.renorm.eq}
\end{equation}
which is a generalisation of \eqref{expanded.change}. Hence, the additional constraint $\dd\eta_C=0$ acts as a source of information, which imposes nontrivial corrections in the relationship between the evolution of $\dd\eta_A$ and $\dd\eta_B$, that are perturbatively described by equation \eqref{diff.renorm.eq}. Note that the implicit function theorem does not provide an explicit form of the function $h$. Thus, one may need to integrate the equation \eqref{eta.B.dot}. The equation \eqref{diff.renorm.eq} provides a perturbative approximation of \eqref{eta.B.dot}, which can be subjected to integration. This might be called a `perturbative renormalisation' or `inferential scattering' of $\mathcal{G}_{BA}$. 
\subsection{Br\`{e}gman distance and nonlinear quantum control\label{section.dynamics.renormalisation}}
An important feature of the Jaynes--Mitchell theory is that it allows to consider not only the `source' (`input', `configuration') and `response' (`output', `registration') variables, but also the `control' (`covariate') variables, defined as fixed parameters of the model. The constraints imposed by these fixed variables can be factored out from the relationship between causes and effects, but at the price of `renormalisation' of the source terms. It amounts to reduction of the dimensionality of the model (removing the dimensions described by control parameters) and subsequent rescaling of the remaining source terms by the `renormalisation factors'. Thus, one can eliminate control parameter from the \textit{model construction} at the price of renormalisation of the source terms that determine the \textit{information dynamics}. This procedure is \textit{nonperturbative} and \textit{geometric}, but under certain conditions it can be expanded in the perturbative series of corrections. Quite remarkably, the renormalisation factor that appears at the \textit{first order} of expansion in powers of source strength can be perturbatively expanded in an infinite series of corrections, which contain \textit{all orders} of interaction effects with the `virtual' source terms that can be associated with the factored-out `control' variables.\footnote{This phenomenon was first observed in the Heims--Jaynes analysis of the gyromagnetic effect \cite{Heims:Jaynes:1962}, and appeared later also in Jaynes' analysis of the Rayleigh acoustic scattering \cite{Jaynes:1985:scattering,Jaynes:1993} and (independently) in  Schwinger's source theory \cite{Schwinger:1969,Schwinger:1970}.}

Besides generalisation from exponential families to dually flat manifolds, the information geometric framework introduces important conceptual change: the `source', `response', and `control' variables are no longer associated with particular functions on the sample space, but rather with the particular coordinate variables on the information model. As discussed in Section \ref{qig.foundations.intro.section}, all these variables form specific examples of \textit{observables} in our approach to quantum foundations. As a consequence, renormalisation of sources can be considered \textit{exactly} as a transformation of information models that amounts to `coarse graining' and subsequent rescaling. The coarse graining provides a a reduction of dimension of the model that is preserving the operational definitions of the coordinate variables on a submodel, but at the price of redefinition of their functional relationship by means of change of the local geometry of the model from dually flat to curved one.  

In principle, the change of an information state $\phi\in\M(\boole)$ or $\phi\in\M(\N)$, associated with an integrable real function $f$ over $\boole$ or an operator $f\in\N^\sa$, respectively, can be specified in three different ways: by means of $\delta(\phi(f))$, by means of $\phi(\delta f)$, or by means of a \df{source term} $(\delta\phi)(f)=:\delta Q_f$. The main insight of the source theory is that the changes specified by source terms have the direct operational meaning whenever the model $\M(\boole)$ is equipped with a pair of dually flat coordinate systems. In such case, the changes $(\delta\phi)(\cdot)$ can be reexpressed in terms of corresponding changes of source-and-response parameters. However, the change of information driven by the change of one of source-and-response variables leads to change of other variables that are correlated with it. 

This can be interpreted as a cause-and-effect relationship, but under the condition that `causes' and `effects' are understood as inputs and outputs of correlation relationships, respectively. This is \textit{different} from the meaning assigned to these terms in other sections of this work. In general, there are possible at least two clearly distinct perspectives on what the `causes' and `effects' are. From the purely operational perspective, any reproducible relationship between configuration and response parameters of description of experimental situation deserves to be called a causal relationship, and any predictively verifiable inferential procedure relating them is considered as a satisfying method of the theoretical modelling of causality (see e.g. \cite{Hinkelmann:Kempthorne:2005}). On the other hand, from the ontologically flavoured perspective, the `causes', `effects', and their relationships are theoretical notions, which may indirectly correspond to epistemic parameters and predictively verifiable relationships between them, derived from some inductive procedure (see e.g. \cite{Pearl:2000}). In this Section we chose the former terminology (speaking of `epistemic causality', because the term ``inferential causality'' would probably cause, nomen omen, more confusion), while in the rest of this work we consider causality and inference as a priori independent theoretical constructs, but without attribution of any ontological claims. From the inferential perspective, `causes' are just the same as `configurations'. 

If one of the parameter spaces $\{\Theta,\Xi\}$ can be considered as a space of `causes' (`configurations'), the other becoming a space of `effects' (`registrations'). The role of the Legendre transform $\LFtrafo_\Psi:\Theta\ra\Xi$ is to associate effects with causes (and vice versa).\footnote{Note that this notion of `causality' belongs strictly to a theoretical layer of scientific inquiry. Without specification of some particular epistemic semantics it is not related in any specific way with the experimental `effects' and `causes'.}  This allows to use $\LFtrafo_\Psi$ (and ``epistemic'' `cause-and-effect' interpretation) in order to analyse changes of effects following (correlatively, inferentially) from the changes of causes, as well as changes of causes following from the changes of effects. These two issues are known, respectively, as \textit{forward} and \textit{backward induction problems}. Let us also note that the dually flat geometry always satisfies the relationships \eqref{LFT.MJF.Phi} and \eqref{LFT.MJF.Psi} \textit{as well as} \eqref{g.phi.hessian} and \eqref{g.psi.hessian}. Hence, it also satisfies
\begin{equation}
        \gbold_{ij}^\Psi(\theta)=\frac{\partial\eta_j}{\partial\theta^i}.
\label{local.causality.metric}
\end{equation}
The equations \eqref{LFT.MJF.Psi} and \eqref{local.causality.metric} assert that
\begin{itemize}
\item a Legendre transform $\LFtrafo_\Psi$ governs the relationship between causes and effects,
\item a riemannian metric $\gbold^\Psi$ governs the relationships between \textit{changes} of causes and \textit{changes} of effects.
\end{itemize}
From the perspective discussed in the Section \ref{local.infogeometry.histories.intro.section}, $\LFtrafo_\Psi$ defines the `system of epistemic causality' of an individual user (the perspective of a fixed measurement frame), relating the `configurations' and `registrations' in such way that the discrimination function on the space of configurations defines the discrimination function on the space of registration. On the other hand, $\gbold^\Psi$ defines the `local system of epistemic causality' \textit{on} the information manifold, allowing to translate between different local users. Hence, the source theoretic renormalisation can be interpreted as a perturbation of the local system of \textit{epistemic} causality due to presence of the nonzero sources of information.

Now let us observe that, as discussed in Section \ref{distances.NS.geom.section}, every dually flat manifold determines an associated Br\`{e}gman distance $D_\Psi$. While our notation in Section \ref{Favretti.section} indicates this fact, it was left unnoticed by previous authors. Also, we note, following the discussion in Section \ref{Orlicz.Br\`{e}gman.section}, that the framework of dually flat manifolds and associated Br\`{e}gman distances is applicable \textit{locally} to \textit{any} quantum manifolds, as long as one defines the manifold structure using a specific Br\`{e}gman function. The extension from commutative to quantum dually flat geometries is straightforward. The notation applied by us in Section \ref{Favretti.section} keeps the correct order of multiplications, so the quantities used in this Section can be interpreted as operators as well.

The equation \eqref{local.causality.metric} is a geometric equivalent of the linear case \eqref{first.order.covariance.expansion} of the expansion \eqref{infinite.perturbation.covariance.expansion}. In order to obtain higher-order terms of \eqref{infinite.perturbation.covariance.expansion}, one needs to consider information models that are not dually flat. Hence, one can in principle \textit{begin} with an \textit{arbitrary} quantum information model $\M(\N)$, equipped with a Br\`{e}gman distance $D_\Psi$ which defines the local `ideal' dually flat manifold structure $(\M(\N),\gbold^{D_\Psi},\nabla^{D_\Psi},(\nabla^{D_\Psi})^\nsdual)$, and consider the emergence of the nonzero curvature of the effective riemannian geometry $(\M(\N),\tilde{\gbold})$ as a result of presence of the additional source (control) terms that are `renormalised out' by the transition $\gbold\mapsto\tilde{\gbold}$. It is quite interesting that the departure from the dually flat geometry and constant curvature of a model implies the presence of additional information sources operating at different points. From the perspective of analogy with general relativity, we can say that \textit{sources curve the geometry of a quantum information manifold}. Combining this with our discussion of the role of quantum riemannian metric in the Daubechies--Klauder formula (see Sections \eqref{local.infogeometry.histories.intro.section} and \eqref{algebraic.quantum.histories}), we can conclude that the JMF renormalisation leads to the redefinition of the \textit{local} prior measure used for the path integration. Interpreting the local prior measure as an information theoretic analogue of mass, we can say that this process encodes dependence on additional sources by the change of the geometry of a model, which is in turn reflected in the renormalisation of an information theoretic local mass, and a corresponding point-dependence of the zero-point energy.


We will use the term \df{br\`{e}gmanian renormalisation} to refer to a local renormalisation of $(\M,D_\Psi)$ using JMF source theory. More generally, let us observe (following Lauritzen \cite{Lauritzen:1987:statistical:manifolds}), that the Norden--Sen geometry captures the description of information geometry only up to third order of Taylor series expansion of information distance, which in principle allows to develop higher--order differential tensor theories of information geometry, more general than the Norden--Sen geometry. Thus, it is plausible that the higher order source renormalisation terms may also possess a complete geometric representation, but requiring to use higher order tensor geometries arising from the Taylor expansion of the Br\`{e}gman distance as the referential object subjected to renormalisation.

The `source term' defined as above has different meaning than the `source term' introduced in Section \ref{local.source.liouv.section}. Yet they are complementary. The former corresponds to a perturbation $\delta\theta$ of a coordinate system $\theta:\phi\mapsto\theta(\phi)=\phi(x)$ under constraint $\phi(\delta x)=0$. This description rests on the assumption that all relevant local information which has to be taken under consideration is completely specified by means of the variations $\delta(\phi(x))$ and $(\delta\phi)(x)$. The latter corresponds to perturbation $\delta x$ of an element $x$ of (a local GNS representation of) a $W^*$-algebra $\N$ by means of state dependent perturbation of liouvillean. Our approach allows $\delta x$ to be arbitrary, so it can also depend on $\phi$, and may not arise as an infinitesimal change generated by a global automorphism of $\N$. These two different uses of a single notion are compatible and complementary in the sense provided by the equations \eqref{work.def}-\eqref{gen.1st.law.1}: while the `sources' of Sections \ref{section.JM.source.theory} and \ref{Favretti.section} generalise the notion of `heat sources', the `sources' of Section \ref{local.source.liouv.section} generalise the notion of `work sources'. In our work we view `work sources' (respectively, `heat sources') as the geometric perturbation of the geometry of causal evolution (respectively, inferential evolution).

\subsection{Contraction coefficients}

The Br\`{e}gmannian renormalisation answers the question about the behaviour of the constraints of inference (and resulting information dynamics) under \textit{dimensional} reduction of information model due to the presence of constant control parameters. However, given any information model and constraints of inference, there appears also another renormalisation-type question: what is the behaviour of these objects under coarse grainings? Because the constraints may involve information geometric quantities (for example, the `two point correlation function' $K_{xy}(\rho)=\int_0^1\dd\lambda\tr_\H(\rho^\lambda x\rho^{1-\lambda}y)$ is an evaluation of the quantum Bogolyubov--Kubo--Mori riemannian metric $\gbold^{D_1}_\rho$ on the pair of tangent vectors $x,y\in\T_\rho\M(\N)$), this is related to the question about behaviour of information geometric quantities under completely positive maps. Restriction to quantum $D_\fff$-geometries, where $\fff$ is an operator convex function defining the $D_\fff$ distance, secures the Markov monotononicity of $\gbold^{D_\fff}$ and $\nabla^{D_\fff}$, but this does not extend naturally to every geometric quantity on $\M(\N)$ that can be built using these objects and their derivatives. 

In the commutative case Chencov \cite{Chencov:1969,Chencov:1972} has defined the Markov monotone connections as such affine connections $\nabla^{D_\fff}$ that for any Markov map $T$ the image of a $\nabla^{D_\fff}$-geodesic line on $\M(\boole)$ belongs to a $\nabla^{D_\fff}$-geodesic line on $T_\star(\M(\boole))$ as its interval or its point, while an affine parameter of this line remains, up to rescaling, an affine parameter of the $\nabla^{D_\fff}$-geodesic line in the image \cite{Chencov:1972}. Thus, the behaviour of any trajectory along a given $\nabla^{D_\fff}$-geodesic under the action of Markov maps is characterised by their invariance properties under \textit{coarse graining} by preduals of Markov maps and \textit{rescaling} by an affine parameter. This leads to a question whether it is possible to find a suitable analogue of an affine parameter for  arbitrary quantum information model $\M(\N)$ which would allow for some sort of control over the mutual behaviour of $\M(\N)$, its information geometry, and information dynamics under coarse grainings. More specifically, we need to find some scalar \df{contraction coefficient} $\rpktarget{CONTR}\contrcoeff$, which globally characterises the geometry of $\M(\N)$ \textit{and} is Markov monotone, $\contrcoeff(\M(\N))\geq\contrcoeff(T_\star(\M(\N))$, and then use it in order to rescale the constraints of inference on $\M(\N)$. 

Some examples of contraction coefficients $\contrcoeff(T_\star)$ were provided in the case when $\dim\M(\N)<\infty$, with semi-finite $\N$ by Lesniewski and Ruskai \cite{Lesniewski:Ruskai:1999}, following earlier works \cite{CIRRSZ:1993,Choi:Ruskai:Seneta:1993,Ruskai:1994}:
\begin{align}
        \contrcoeff_{D_\fff}(T_\star)&:=
                \sup_{
                                        \omega,\phi\in\M(\N),
                                        }
                \left\{
                        \frac{
                                        D_\fff(
                                                        T_\star(\omega),
                                                        T_\star(\phi)
                                                )
                                }{
                                        D_\fff(
                                                        \omega,
                                                        \phi
                                                        )
                                }
                \mid
                \omega\neq\phi
                \right\}                                
                                ,\\
        \contrcoeff_{\gbold^{D_\fff}}(T_\star)&:=
                \sup_{
                                        \phi\in\M(\N)
                                        }
                \left\{
                \sup_{
                                        u\in\T_\phi\M(\N)
                                        }
                \left\{
                        \frac{
                                        \gbold^{D_\fff}_{
                                                        T_\star(\phi)
                                                }(
                                                        T_\star(u),T_\star(u)
                                                )
                                        }{
                                                \gbold^{D_\fff}_{\phi}(u,u)
                                        }
                \right\}                                        
                \right\},
                                        \\
        \contrcoeff_{d_{\gbold^{D_\fff}}}(T_\star)&:=
                \sup_{\omega,\phi\in\M(\N)}
                \left\{
                        \frac{(d_{\gbold^{D_\fff}}(T_\star(\omega),T_\star(\phi)))^2}{(d_{\gbold^{D_\fff}}(\omega,\phi))^2}
                        \mid
                        \omega\neq\phi
                \right\}
                        ,
\end{align}
where 
\begin{equation}
        d_{\gbold^{D_\fff}}(\omega,\phi):=\inf_{c\in C}\left\{\int_0^1\dd t\sqrt{\gbold^{D_\fff}_{c(t)}(\dot{c}(t),\dot{c}(t))}\right\},
\end{equation}
and $C$ is defined as a class of all smooth curves $c:[0,1]\ni t\mapsto c(t)\in\M(\N)$ such that $c(0)=\omega$ and $c(1)=\phi$. Apart from Markov monotonicity of the above coefficients, Lesniewski and Ruskai proved that these coefficients are convex in $T_\star$, and satisfy
\begin{equation}        
        1\geq\contrcoeff_{D_\fff}(T_\star)\geq\contrcoeff_{\gbold^{D_\fff}}(T_\star)\geq\contrcoeff_{d_{\gbold^{D_\fff}}}(T_\star).
\end{equation} 

Now, let the inferential quantum dynamics be given by $D_\fff$  entropic projection on $\M(\N)$, with the constraints $\Q\subseteq\M(\N)$ specified in terms of lower semi-continuous convex function $F:\M(\N)\ra\,]-\infty,+\infty]$,
\begin{equation}
        \M(\N)\ni\omega\mapsto\arginf_{\phi\in\M(\N)}\left\{D_\fff(\omega,\phi)+F(\phi)\right\}\in\M(\N).
\label{renormalise.this.dynamics}
\end{equation}
With the nontrivial examples of contraction coefficients at hand, we can propose to control the behaviour of constraints $\Q$ under coarse grainings $T_\star$ by means of \df{markovian renormalisation semi-group} transformation
\begin{equation}
        F(\phi)\mapsto\frac{1}{\contrcoeff(T_\star)}F(T_\star(\phi)),
\label{renorm.eq}
\end{equation}
which amounts to subsequent coarse graining and rescaling of constraints. The choice of a particular form \eqref{renorm.eq} of transformation of constraints can be justified either by appealing to arguments and insights based on ordinary renormalisation semi-group theory or by recalling another result of Lesniewski and Ruskai:
\begin{equation}
        \contrcoeff_{D_\fff}(T_\star)\neq\contrcoeff_{\gbold^{D_\fff}}(T_\star)\iff\exists \omega\neq\phi\;\;\mbox{such that}\;\; \frac{1}{\contrcoeff_{D_\fff}(T_\star)}D_\fff(T_\star(\omega),T_\star(\phi))=D(\omega,\phi).
\label{renorm.Dfff}
\end{equation}
In such case, the invariant $\contrcoeff_{D_\fff}(T_\star)$ contains a complete information about the behaviour of distance $D_\fff(\omega,\phi)$ under rescaling by coarse grainings $T_\star$. In view of \eqref{renorm.Dfff}, the aim of rescaling \eqref{renorm.eq} is to obtain \textit{the same form of information dynamics} (for a given initial state) independently of the coarse graining. In consequence, we will say that the quantum dynamics \eqref{renormalise.this.dynamics} is in a \df{fixed point} of markovian renormalisation semi-group transformation with respect to a contraction coefficient $\contrcoeff$ if{}f, given an initial state $\omega\in\M(\N)$, $\omega\neq\phi$, the equations
\begin{equation}
\left\{
\begin{array}{l}
\contrcoeff(T_\star)F(\phi)=F(T_\star(\phi)),\\
\contrcoeff(T_\star)D(\omega,\phi)=D_\fff(T_\star(\omega),T_\star(\phi))
\end{array}\right.
\label{renorm.sg.fixed.point}
\end{equation}
hold for any $T_\star$ on $\M(\N)$. If \eqref{renorm.sg.fixed.point} is not satisfied, then the action of \eqref{renorm.eq} generates the `flow' of forms of dynamics along the `trajectory' of semi-group of markovian morphisms.

As these examples show, quantum information geometry provides quantitative tools allowing to develop various renormalisation procedures for quantum inference, which possess explicit conceptual and quantitative meaning. In particular, the br\`{e}gmannian and markovian renormalisation procedures reflect, respectively, two different problems: coarse graining and rescaling of the \textit{solution} of dynamical (inferential) problem and coarse graining and rescaling of the \textit{definition} of dynamical (inferential) problem.\footnote{Consider $y=f(x)$, where $f$ is an arbitrary function. It is clear that the procedure used to control the quality of approximation of the initial data $x$ does not need to correspond to the procedure used to control the quality of approximation on the space of solutions of this equation.} We refer to \cite{Beny:Osborne:2012,Beny:Osborne:2013,Beny:Osborne:2014} for another (and more developed) approach to quantum information geometric renormalisation based on the use of markovian morphisms $T_\star$ (see also \cite{DeBrota:2015} for a pedagogical overview).


\pagebreak
\section*{Acknowledgments}
%
%
\begin{flushright}
{\begin{spacing}{0.7}
\foreignlanguage{russian}{\scriptsize \textit{Пока жива,\\
Я могу стараться на лету не упасть,\\
Не разучиться мечтать... любить...}\\
О.В. Яковлева}
\end{spacing}}
\end{flushright}
{\small This research was supported in part by Perimeter Institute for Theoretical Physics. Research at Perimeter Institute is supported by the Government of Canada through Industry Canada and by the Province of Ontario through the Ministry of Research and Innovation. This research was also partially financed by the National Science Center of the Republic of Poland (Narodowe Centrum Nauki) through the grant number N N202 343640. Early versions of this work were developed as a part of research supported by the grants: \textit{Mistrz} 2007 grant of Foundation for Polish Science for Jerzy Lewandowski and his students, Quantum Geometry and Quantum Gravity grants 1955 and 2706 of European Science Foundation, and 182/N QGG/2008/0 grant of Ministerstwo Szkolnictwa Wyższego i Nauki. I acknowledge substantial reliance on the knowledge resources provided for free by Library Genesis (\href{http://gen.lib.rus.ec}{http://gen.lib.rus.ec}) and Sci-Hub (\href{http://sci-hub.cc}{http://sci-hub.cc}). I believe that providing unrestricted free access to the results of scientific research is a matter of an elementary respect to other human beings. This paper grew out of the heuristic considerations in the preprints \cite{Kostecki:2007:qht,Kostecki:2008:aqh,Kostecki:2010:aqh}, motivated in turn by the ideas of Stanis{\l}aw Lem \cite{Lem:1974,Lem:1981} (information can be exchanged into energy and create a black hole), Jean Baudrillard \cite{Baudrillard:1980,Baudrillard:1981} (symbolic systems of information gravitate), and Antoni K\k{e}pi\'{n}ski \cite{Kepinski:1972:Rytm,Kepinski:1972:Schizofrenia} (time of entropic processing and time of energy flow are \textit{a priori} independent). I am indebted to Stanis{\l}aw Woronowicz for introducing me to the theory of noncommutative flow of weights, which allowed me to make some of these ideas mathematically sound. I want to warmly thank: C\'{e}dric B\'{e}ny, Pavel B\'{o}na, Dylan Butson, John DeBrota, Jan Derezi\'{n}ski, Paolo Gibilisco, Frank Hellmann, Carlos Guedes, Anna Jen\v{c}ov\'{a}, Wojtek Kami\'{n}ski, Jerzy Kijowski, Jacek Kope\'{c}, Władysław A. Majewski, W{\l}odek Natorf, Dmitri\u{\i} Pavlov, Claude-Alain Pillet, Carlos C. Rodr\'{\i}guez, David Sherman, Ray Streater, and Karol \.Zyczkowski for discussions, Marco Favretti for sharing his unpublished preprint \cite{Favretti:2007}, Jerzy Lewandowski for his trust in, and support of, my nonorthodox research, Chris Isham for his interest in, and comments on, the early version of this paper, Fotini Markopoulou, Olaf Dreyer, Jamie Vicary, Cecilia Flori, and Tomasz Kołodziejski for their generous hospitality in London, Nottingham, Oxford, and Berlin during the periods of intense work on this project, Basia Baranowska, Janka Spicha-Konarzewska, and Aneta P\k{e}ka{\l}a for sharing their life, love and support, and my flatmates Teresa Szczepi\'{n}ska, Rafa{\l} Gr\"{o}ger, and Agnieszka Gronowicz, as well as medical teams of \textit{Szpital Kliniczny Dzieci{\k{a}}tka Jezus} and \textit{Centralny Szpital Kliniczny MSW}, for rescuing me from the carbon monoxide brain death.}
\section*{References}
\addcontentsline{toc}{section}{References}
{%
\scriptsize

All Cyrillic titles and names were transliterated from the original papers and books. For the Latin transliteration of the Cyrillic script (in references and surnames) we use the following modification of the system GOST 7.79-2000B: {\foreignlanguage{russian}{ц}} = c, \foreignlanguage{russian}{{ч}} = ch, \foreignlanguage{russian}{х} = kh, \foreignlanguage{russian}{ж} = zh, \foreignlanguage{russian}{ш} = sh, \foreignlanguage{russian}{щ} = shh, {\foreignlanguage{russian}{ю}} = yu, {\foreignlanguage{russian}{я}} = ya, {\cyrrm{\"{e}}} = \"{e}, {\foreignlanguage{russian}{ъ}} = `, {\foreignlanguage{russian}{ь}} = ', {\foreignlanguage{russian}{э}} = \`{e}, {\foreignlanguage{russian}{й}} = \u{\i}, with an exception that names beginning with {\foreignlanguage{russian}{Х}} are transliterated to H. For Russian texts: {\cyrrm{y}} = y, {\cyrrm{i}} = i; for Ukrainian: {\cyrrm{i}} = y, i = i, \"{\i} = \"{\i}. Note: All links provided in references link to the free access files. Files and digital copies that are subject to any sort of restricted access were not linked. See \href{http://michaelnielsen.org/polymath1/index.php?title=Journal_publishing_reform}{michaelnielsen.org/polymath1/index.php?title=Journal\_publishing\_reform} for the reasons why.

\begingroup
\raggedright
\bibliographystyle{rpkbib}
\renewcommand\refname{\vskip -1cm}

\endgroup           
}%
{\vskip 1cm}
\noindent {\scriptsize \texttt{This text is distributed under CC BY-SA 4.0 license.}}

\end{document}